\documentclass[11pt,a4paper]{article}
\usepackage{epsfig}
\usepackage{latexsym}

\newcommand{\bi}{\begin{itemize}}
\newcommand{\ei}{\end{itemize}}
\newcommand{\be}{\begin{equation}}
\newcommand{\ee}{\end{equation}}

\newcommand{\bea}{\begin{eqnarray}}
\newcommand{\eea}{\end{eqnarray}}
\newcommand{\beastar}{\begin{eqnarray*}}
\newcommand{\eeastar}{\end{eqnarray*}}

\newcommand{\lav}{\left\langle}
\newcommand{\rav}{\right\rangle}

\newcommand{\eq}[1]{~(\ref{#1})}
\newcommand{\eqq}[2]{~(\ref{#1},\ref{#2})}

\newcommand{\order}{{{\mathcal O}}}
\newcommand{\ie}{{\it i.e.}}
\newcommand{\eg}{{\it e.g.}}

\newcommand{\nv}{{\bf n}}
\newcommand{\rv}{{\bf r}}
\newcommand{\kv}{{\bf k}}
\newcommand{\T}{^{\rm T}}
\newcommand{\peq}{p\eql}
\newcommand{\eql}{_{\rm eq}}
\newcommand{\irr}{^{\rm irr}}
\newcommand{\mca}{^{\rm MCA}}
\newcommand{\et}{\eta}
\newcommand{\s}{\sigma}

\newcommand{\dE}{\Delta E}
\newcommand{\eps}{\epsilon}
\newcommand{\tw}{t_{\rm w}}
\newcommand{\cre}{b^\dagger}
\newcommand{\ann}{b}

\newcommand{\uv}{{\bf u}}
\newcommand{\omegav}{\mbox{\boldmath$\omega$}}
\newcommand{\hov}{\mbox{\boldmath$\hat\omega$}}
\newcommand{\etal}{{\em et al}}
\newcommand{\teff}{T_{\rm eff}}
\newcommand{\deriv}[1]{{\partial\over\partial{#1}}}
\newcommand{\ww}{w}
\newcommand{\zz}{f_i}
\newcommand{\eff}{_{\rm eff}}
\newcommand{\dyn}{_{\rm dyn}}

\begin{document}

\title{Glassy dynamics of kinetically constrained models}

\author{F Ritort$^\dag$ and P Sollich$^\ddag$\\
$^\dag$Department of Physics, Faculty of Physics, 
University of Barcelona,\\
Diagonal 647, 08028 Barcelona, Spain; {\tt ritort@ffn.ub.es}\\
$^\ddag$Department of Mathematics, King's College London,\\
Strand, London WC2R 2LS; {\tt peter.sollich@kcl.ac.uk}}

\date{17 October 2002}

\maketitle

\begin{abstract}
\noindent
We review the use of kinetically constrained models (KCMs)
for the study of dynamics in glassy systems. The characteristic feature
of KCMs is that they have trivial, often non-interacting,
equilibrium behaviour but interesting slow dynamics due to
restrictions on the allowed transitions between configurations. The
basic question which KCMs ask is therefore how much glassy physics can
be understood without an underlying ``equilibrium glass transition''.
After a brief review of glassy phenomenology, we describe the main
model classes, which include spin-facilitated (Ising) models,
constrained lattice gases, models inspired by cellular structures such
as soap froths, models obtained via mappings from interacting systems
without constraints, and finally related models such as urn,
oscillator, tiling and needle models. We then describe the broad range
of techniques that have been applied to KCMs, including exact
solutions, adiabatic approximations, projection and mode-coupling
techniques, diagrammatic approaches and mappings to quantum systems or
effective models. Finally, we give a survey of the known results for
the dynamics of KCMs both in and out of equilibrium, including topics
such as relaxation time divergences and dynamical transitions,
nonlinear relaxation, aging and effective temperatures,
cooperativity and dynamical heterogeneities, and finally
non-equilibrium stationary states generated by external driving.  We
conclude with a discussion of open questions and possibilities for
future work.
\end{abstract}

\tableofcontents

\section{Introduction}
\label{intro}

After many decades of research our theoretical understanding of the
glass transition remains substantially incomplete.  Ideally, a
comprehensive theory should explain all thermodynamic and kinetic
properties of glasses, both at the macroscopic and the mesoscopic
level. It should also be consistent with the wealth of experimental
data which 
has been accumulated in the past century, and to which ongoing work
is continuing to add.

Theoretical approaches to the glass transition range between two
extremes. At one end of the spectrum are microscopic theories, which
start from first principles (\eg\ Newton's equations for classical
particles). To arrive at predictions that can be compared to
experiment, rather drastic mathematical approximations are then
required, whose physical meaning can be difficult to assess. One of
the most successful theories of this kind is the mode-coupling theory
(MCT, see references at the end of this introduction), which predicts
a dynamical arrest in sufficiently supercooled liquids that arises
from the nonlinear interaction of density fluctuations.  On the other
extreme are phenomenological theories which incorporate a set of
basic ingredients chosen on the grounds of physical intuition as most
relevant for glass transition dynamics. Predictions are normally
easier to derive from such theories, and conceptual ideas can be
tested relatively directly. This flexibility is also a disadvantage,
however: phenomenological theories can be difficult to disprove if
they can always be extended or modified to account for new data. Among
the best-known theories in this group are the free volume theories
developed by Flory and Cohen, the entropic theories due to Adam,
Gibbs and Di Marzio and the energy landscape approach introduced by
Stillinger and Weber.

The models we discuss in this review have a character intermediate
between these two extremes. Similarly to the phenomenological
approaches, they use effective variables which are normally of
mesoscopic character, \eg\ averages of particle density over
suitably small coarse-graining
volumes, and are chosen on an intuitive basis as most
directly responsible for glassy dynamics. On the other hand, as in the
microscopic theories, a Hamiltonian (or energy function) and
appropriate dynamical evolution equations are explicitly defined, and
one attempts to predict the behaviour of the model on this basis,
without further approximation if possible.

The above category of models is still rather rich. The basic variables
can be discrete or continuous, for example, and the energy function
may contain pairwise potentials or higher-order interactions. The
dynamics are normally constrained only to obey detailed balance
w.r.t.\ the specified energy function, and this leaves considerable
freedom when defining a model. The energy function may even include
quenched disorder, and it has been shown that \eg\ appropriate
spin-glass models can reproduce much of the phenomenology of
structural glasses such as window glass. As expected, the more
complicated the energy function, the more complicated also the static
(equilibrium) behaviour of the resulting models; spin-glass models,
for example, exhibit nontrivial ergodicity breaking transitions at low
temperature.

The philosophy of the {\em kinetically constrained models} (KCMs)
which we discuss in this review is to simplify the modelling approach
further by considering models with essentially trivially equilibrium
behaviour; the simplest models of this type in fact have energy
functions without any interactions between the mesoscopic variables
considered. In other words, KCMs ask the question: how much glassy
physics can we understand without relying on nontrivial equilibrium
behaviour? Instead, KCMs attempt to model glassy dynamics by
introducing ``kinetic constraints'' on the allowed transitions between
different configurations of the system, while preserving detailed
balance. (As we will see in detail below, the easiest method of
implementing this is to forbid transitions between certain pairs of
configurations.) Since it is now widely recognized that the glass
transition is a dynamical phenomenon, such a focus on dynamics
certainly makes sense. Of course, the simplicity of the energy
function of KCMs means that one would not expect them to reproduce the
behaviour of supercooled liquids and glasses under all conditions;
instead, they should capture those aspects of their behaviour which
are predominantly caused by dynamical slowing-down.  One obvious
aspect ignored by KCMs is crystallization: real glass-forming liquids
can crystallize if cooled sufficiently slowly through the melting
point. However, it is widely believed that the existence of a
crystalline phase is not crucial for the behaviour of glasses and
supercooled liquids; this view is supported by the fact that
spin-glass models, where the analogue of an ordered crystalline phase
is suppressed through quenched disorder in the energy function,
nevertheless display many features characteristic of glasses. By
disregarding crystallization effects, the KCM approach therefore
avoids unnecessary complications in glass modelling and focuses on
the key dynamical mechanisms for glassy behaviour.

It is worth addressing already at this point another possible
objection to the KCM approach. By construction, since all the
``interesting'' features of KCMs arise from the dynamical rules, a
relatively minor change in these rules can alter the resulting
behaviour quite dramatically; we will see examples of this below in
the difference between models with directed and undirected
constraints. This lack of ``robustness'' may appear undesirable, and
contrasts with models with more complicated energy functions where the
location and character of {\em equilibrium} phase transitions is
normally unaffected by the precise dynamics chosen. However, as
explained above, KCMs should be regarded as effective mesoscopic
models which encode in their dynamics the complex interactions of an
underlying microscopic model (see Sec.~\ref{model:effective} for
simplified instances of this kind of mapping). In this view, a change
in the dynamical rules corresponds to a nontrivial modification of the
underlying microscopic model, \eg\ by adding new interaction terms to
the energy function, and it makes sense that this should have a
significant effect on the resulting behaviour.

Initially introduced in the early Eighties by Fredrickson and
Andersen, KCMs have recently seen a resurgence in interest. Due to
their simplicity, many questions can be answered in detail, either
analytically or by numerical simulation, and so KCMs form a useful
testbed for our understanding of the key ingredients of glassy
dynamics. We feel it is time now to gather the existing results, to
analyse what we have learnt from recent work on KCMs, and to assess
the successes and drawbacks of the KCM approach. The topics that we
discuss will be inspired both by experimental issues surrounding the
glass transition, and by theoretical questions that have wider
relevance to the field of non-equilibrium statistical mechanics.

The scope of this review is as follows. The core KCMs are the
spin-facilitated Ising models pioneered by Fredrickson and Andersen, and the
kinetically constrained lattice gases introduced by Kob and Andersen,
and J\"ackle and coworkers. We have attempted to be comprehensive in
our coverage of the literature on these and closely related models, up
to a cutoff date around the end of 2001. Nevertheless, omissions will
undoubtedly have occurred, and we apologize in advance to any
colleagues whose work we may have overlooked. There is also a range of
models which do not strictly speaking belong into the KCM category but
which we felt were sufficiently closely related to merit
inclusion. For these models we have only tried to give a
representative cross-section of publications. Finally, to the vast
literature in the general area of glassy dynamics we can only
give a few pointers here. A summary of early experiments and theories
of glasses can be found in
\eg~\cite{NYAS76,NYAS81,Donth81,Brawer83,ParRaoRao83,NYAS86,Scherer86}.
The state of the
art in theory and experiment as of 1995 is reviewed in a series of
papers~\cite{Angell95,Greer95,Hodge95,FriRic95,Stillinger95};
Refs.~\cite{EdiAngNag96,Trieste,Debenedetti96} give more recent accounts.
Moving on to more specific topics, there are a number of reviews of
MCT, \eg~\cite{BenGoetSjo84,Goetze91,GoetSjoe92,GoetSjoe95}, while
Ref.~\cite{Jaeckle86} contains a good overview of the more
phenomenological glass theories. For some of the earliest work on
aging~\cite{Struik78} is a good resource; a very recent review of
fluctuation-dissipation theorem violations in aging systems can be
found in~\cite{CriRit02b}. Ref.~\cite{BouCugKurMez98} provides an
in-depth discussion of modern theories of disordered systems and spin
glasses and their relation to ``older'' glass theories, and
Ref.~\cite{Cugliandolo02} gives a recent and wide-ranging overview of
theoretical approaches to glassy dynamics. The topic of
dynamical heterogeneities in glasses is reviewed
in~\cite{Ediger00,Sillescu99}; and Ref.~\cite{DebSti01} surveys the
energy landscape approach to glassy dynamics. Finally, on KCMs in
particular, the reviews~\cite{Jaeckle86,Fredrickson88,Palmer89}
provide excellent guides to work on these models done up to the end of
the 1980s. The proceedings of a recent workshop on
KCMs~\cite{Barcelona} complement this with surveys of current work,
and we will refer below to a number of articles from this volume as
useful sources of further detail.

We wrote this review with two groups of readers in mind: ``quick''
readers, who may be new to the field of KCMs and want to get an
overview of the most important models, results and open questions; and
``experts'' who already work on aspects of KCMs but are interested in
a comprehensive survey of other research in the area. Accordingly,
there are two different routes through this review. Quick readers
could read Sec.~\ref{basics}, where we give some background on glass
phenomenology and important topics in glassy dynamics;
Sec.~\ref{allmodels}, where we define the various KCMs and related
models and summarize the most important results; and
Sec.~\ref{conclusion}, which contains our conclusions and an outlook
towards open questions for future work. Expert readers, on the other
hand, may only need to refer to Sec.~\ref{basics} to acquaint
themselves with our notation, and to browse Sec.~\ref{allmodels} for
the definitions of the models we discuss. For them, the more detailed
sections that follow should be of most interest: in
Sec.~\ref{techniques} we review the broad range of numerical and
analytical techniques that have been used to study KCMs, while
Sec.~\ref{res} provides a comprehensive survey of the results
obtained.

\section{Basics of glassy dynamics}
\label{basics}

In this section we outline some basic issues in glassy dynamics to set
the scene for the questions that have been studied using
KCMs. Sec.~\ref{basics1} contains a sketch of important experimental
phenomena, including the all-important pronounced slow-down in the
dynamics as temperature is lowered. In Sec.~\ref{basics2} we review
how dynamics in the stationary regime---where a liquid is already
supercooled past its melting point, but still in metastable
equilibrium---can be characterized using correlation and response
functions. Sec.~\ref{intro:fdt} generalizes this to the glass regime,
where equilibrium is no longer reached on accessible timescales;
correlation and response then become two-time quantities because of
aging effects, and can be useful for defining so-called effective
temperatures. In Sec.~\ref{intro:landscape} we review the energy
landscape approach to understanding glassy dynamics, whose usefulness
for KCMs has recently been investigated in some
detail. Sec.~\ref{intro:cooperativity} introduces the issues of
dynamical lengthscales and heterogeneities, and in
Sec.~\ref{other_systems} we briefly mention some other systems
exhibiting glassy dynamics.

\subsection{Some experimental phenomena}
\label{basics1}

\begin{figure}
\begin{center}
\epsfig{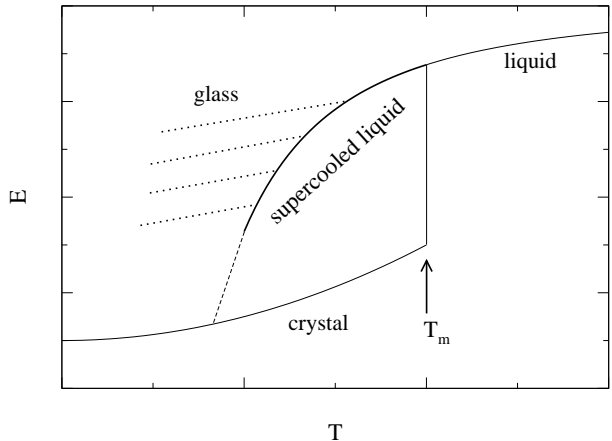}
\end{center}
\caption{Schematic plot of energy $E$ vs temperature $T$, showing the
liquid line continued into the supercooled regime; no singularities
appear at the melting temperature $T_m$. Also shown are, for four
different cooling rates, the deviations which occur as the system
falls out of equilibrium at a cooling-rate dependent glass transition
temperature $T_g$. A naive extrapolation of the supercooled liquid
line, shown by the dashed line, could suggest a thermodynamic glass
transition at a lower (Kauzmann) temperature.
\label{intro_energy}
}
\end{figure}

The standard experimental procedure for generating a glass is to take
a liquid well above its melting temperature and cool it down quickly
enough to avoid crystallization. On cooling through the melting
temperature $T_m$, the liquid is initially in a metastable equilibrium
state---the true equilibrium state being the crystal---and therefore
referred to as supercooled. On timescales much shorter than those
required for crystallization processes to occur, the properties of the
supercooled liquid at a given temperature are stationary, \ie\
independent of time. They are also smoothly related to those of the
genuine equilibrium liquid above $T_m$, so that a plot of \eg\ the
energy of a supercooled liquid against temperature would show no
unusual behaviour as $T_m$ is crossed (see Fig.~\ref{intro_energy}).

As cooling proceeds, the dynamics in the supercooled liquid slows
down, often very rapidly. At some temperature $T_g$, the longest
relaxation timescales of the supercooled liquid therefore begin to
exceed the experimental timescale set by the inverse $1/r$ of the
cooling rate $r$. The system then falls out of its (metastable)
equilibrium and becomes a glass proper, whose properties evolve slowly
with time even at constant temperature; the plot of \eg\ energy vs $T$
begins to deviate markedly from the supercooled line at $T_g$ (see
Fig.~\ref{intro_energy}). As defined, it is clear that the glass
transition temperature $T_g$ depends on the cooling rate, being the
temperature where the longest relaxation times $\tau$ are of order
$1/r$. In line with the expectation that the dynamics slow down as
temperature is lowered, $T_g$ is observed to decrease when the cooling
rate $r$ is reduced. The actual dependence $T_g(r)$ is generally
logarithmic, corresponding to an exponential temperature variation of
relaxation timescales; see below.

One of the most striking experimental manifestations of the dynamical
slow-down in supercooled liquids is the temperature dependence of the
viscosity $\eta$. One can write $\eta=G\tau$, where $G$ is the shear
modulus and $\tau$ is the relaxation time (more precisely, the
integrated relaxation time for shear stress relaxation; see
Sec.~\ref{basics2}). Since $G$ is only weakly temperature dependent,
$\eta$ therefore gives a direct measure of a typical relaxation
timescale $\tau$ in supercooled liquids. The point where $\eta$ reaches
the value $10^{13}$ Poise (= $10^{12}$ Pa s) is often used to define the
glass transition temperature $T_g$ operationally; given typical values
of $G$, this corresponds to relaxation timescales $\tau$ of the order of
hundreds of seconds or more.
\begin{figure}
\begin{center}
\epsfig{file=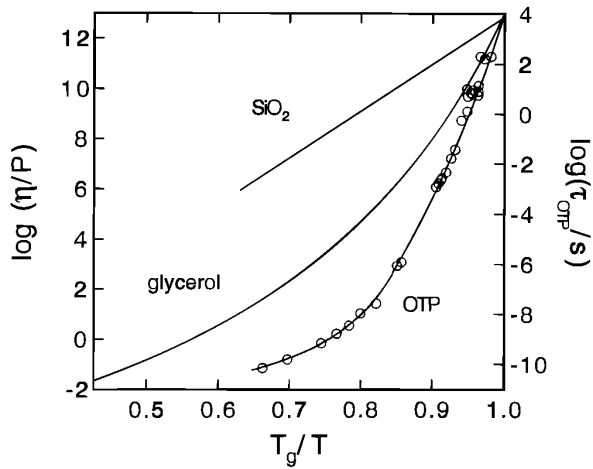, width=8cm}
\end{center}
\caption{Angell plot of log-viscosity (essentially log-relaxation
time, see right axis) against $T_g/T$. In this representation, strong
glass-formers such as SiO$_2$ with their Arrhenius dependence of
timescales on temperature give straight lines, while the
super-Arrhenius divergence of timescales in fragile glasses (\eg\
glycerol) leads to curved plots.
\label{intro_visco}
}
\end{figure}
In so-called {\em strong} liquids, of which silica (SiO$_2$) is an
example, $\tau$ as determined from viscosity measurements increases
according to an {\em Arrhenius} law
\be
\tau \sim \exp\left(\frac{B}{T}\right)
\label{eq4S22}
\ee
which corresponds to thermal activation over a---possibly
effective---barrier $B$; here and throughout we set $k_{\rm
B}=1$. In an ``Angell plot'' of log-viscosities
against $T_g/T$, as shown in Fig.~\ref{intro_visco}, this gives a
straight line. 
A more pronounced timescale increase is referred to as 
super-Arrhenius or superactivated---we use both terms
interchangeably---and occurs in the so-called {\em fragile}
supercooled liquids. They can
show a dramatic growth in $\tau$, of up to 15 orders of magnitude, over
a temperature interval as narrow as 10\% of $T_m$. This increase is
commonly fitted by the Vogel-Tamman-Fulcher (VTF)
law~\cite{Vogel21,Fulcher25,TamHes26}
\be
\tau \sim \exp\left(\frac{A}{T-T_0} \right)
\label{eq3S22}
\label{VTF}
\ee
This suggests a divergence of $\tau$ at some nonzero temperature
$T_0$, though it has been argued that this is difficult to justify
from microscopic models~\cite{Anderson79}. An exponential inverse
temperature square (EITS) law
\be
\tau\sim\exp\left(\frac{A}{T^2}\right)
\label{EITS}
\ee
which exhibits no such divergence can provide an equally good fit to
data for many systems~\cite{Baessler87,RicBaes90}. In its most general
form, one can write the relaxation time increase of fragile
supercooled liquids as an Arrhenius law with an effective barrier
$B(T)$ that increases as $T$ decreases,
\be
\tau \sim \exp\left(\frac{B(T)}{T} \right)
\label{eq5S22}
\ee
Over the experimental time window, with longest accessible times of
the order of hours or days, $B(T)$ then increases by at most a factor
of around five while $\tau$ itself increases by many orders of
magnitude. This limited range of $B(T)$ makes it clear why it is
almost impossible to distinguish, on the basis of experimental data,
between the VTF and EITS laws or indeed other possible superactivated
fitting forms for $\tau(T)$. Theories have been proposed to link the
drastic slowing-down in fragile supercooled liquids to (near-)
singularities in their thermodynamic properties; an early and still
hotly debated example is the proposal by Adam and
Gibbs~\cite{AdaGib65} that the effective activation barrier scales
as the inverse of the entropy of the configurational degrees of
freedom. We will not dwell on this point here, but return to the issue
of how configurational entropies can be defined in
Sec.~\ref{intro:landscape}.

Within the supercooled regime discussed above one can define a further
characteristic temperature $T_c$ at which relaxation processes begin
to take place in two temporally separated stages, with relaxation
functions developing shoulders that eventually grow into plateaux (see
Sec.~\ref{basics2}). The longest relaxation timescales $\tau$ in this
regime, of order $10^{-6}$ s, are already large compared to their
values in the liquid but still small relative to the timescales at
$T_g$. In this temperature region the growth of $\tau(T)$ can often be
fitted with an apparent power-law divergence at nonzero temperature,
as suggested by
MCT~\cite{BenGoetSjo84,Goetze91,GoetSjoe92,GoetSjoe95}.

At the transition from the supercooled liquid to the glass, one
observes experimentally a drop in the specific heat over a narrow
temperature interval; the location of this drop defines the so-called
calorimetric glass transition temperature. Intuitively, the change in
specific heat corresponds to the effective freezing of those slow
degrees of freedom which fall out of equilibrium at the glass
transition. In a plot of energy versus temperature, it corresponds to
a change in slope from a larger value in the supercooled regime to a
rather smaller value for the glass. (Superficially, the specific heat
jump resembles the behaviour at second-order phase transitions with
vanishing specific heat exponent $\alpha$, but in this latter case the
specific heat actually {\em increases} as $T$ is lowered.) Notice that our
terminology above is appropriate for systems at constant volume; we
use this since all models discussed below are of this
type. Experiments are normally carried out at constant
pressure. Instead of energy, the relevant thermodynamic potential
whose temperature derivative gives the specific heat is then the
enthalpy. For simplicity, we will continue to refer to the
constant-volume situation below.

\begin{figure}
\begin{center}
\epsfig{file=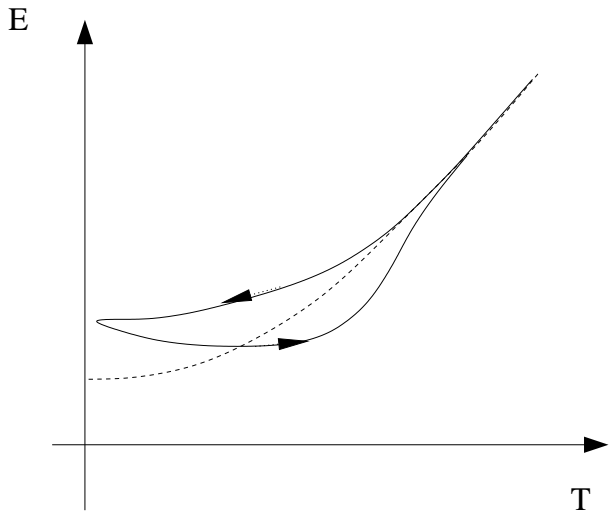, width=8cm}
\end{center}
\caption{Schematic hysteresis plot for a heating-cooling cycle. On
cooling, $E$ remains above the supercooled line (dashed) as the system
falls out of equilibrium; on reheating, $E$ remains low and crosses
underneath the supercooled line before rejoining it in a steep rise.
\label{intro_hyste}
}
\end{figure}
Further experimental illustration of the non-equilibrium nature of the
glass state is provided by interesting hysteresis effects in
heating-cooling cycles. As explained above, on cooling the energy will
initially follow the supercooled line but then depart from it at
$T_g$, with a concomitant drop in the specific heat. On further
cooling, the energy remains above the supercooled line. If the system
is then heated back up through $T_g$, however, the energy increases
initially very slowly and actually crosses below the supercooled line,
rejoining it by a steep increase at a temperature slightly above the
original $T_g$ (see Fig.~\ref{intro_hyste}). In the specific heat this
increase shows up as a pronounced peak. The crossing of the energy
below the supercooled line is a characteristic non-equilibrium effect
which reveals that the glass retains a strong memory of its
temperature history.

To rationalize the complexities of non-equilibrium behaviour, it is
tempting to look for a description of glasses in terms of a few
effective thermodynamic parameters. For example, the dynamics in a
glass at fixed low temperature can be so slow that quantities such as
the energy are effectively constant. One could then experimentally
define a ``fictive temperature''~\cite{Tool46} as that for which the
(extrapolated) energy of the supercooled liquid has the value measured
in the glass. However, the same procedure applied to a different
experimental quantity, such as density, will not necessarily give the
same fictive temperature (see \eg~\cite{Scherer86}), so that the
physical meaning of such assignments remains unclear. More recently,
it has been argued~\cite{CugKurPel97} that two-time correlation and
response functions may be more appropriate for defining effective
temperatures; this proposal is discussed in more detail in
Sec.~\ref{intro:fdt}.

\subsection{Stationary dynamics: Correlation and response}
\label{basics2}

We next describe some of the correlation and response or relaxation
functions that can be used to probe the stationary (\ie\ equilibrium,
though metastable) behaviour of supercooled liquids. Many of these
correspond to experimentally measurable quantities, and are therefore
key quantities which one would like to predict from theoretical
models.

Let us denote by $\phi(t)$ any observable quantity which can evolve in
time after applying a given perturbation $h(t)$. For instance, $\phi$
could be the polarization of a supercooled liquid and the
corresponding perturbation $h$ the electric field, or $\phi$ could be
the volume and the perturbation $h$ a change in pressure. Suppose the
system is in equilibrium at $t=0$, \ie\ $\phi(0)=\phi\eql$, from
which time a perturbation $h$ is applied and held constant. For small
$h$, the deviation of $\phi(t)$ from its equilibrium value then defines
the linear response function to a step perturbation,
\be
\chi(t) = \frac{\phi(t)-\phi\eql}{h}
\label{chi_eq}
\ee
The long-time limit $\chi\eql=\chi(t\to\infty)$ of this then also
gives the equilibrium susceptibility. Thinking of the response as the
relaxation from an original perturbed state to a new equilibrium
state, one can also define the relaxation function
\be
\psi(t)=1-\frac{\chi(t)}{\chi\eql}
\label{eq1S22}
\ee
which is normalized to one at $t=0$ and decays to zero for
$t\to\infty$. Analogues of $\chi$ and $\psi$ also exist for large
perturbations which drive the system far from equilibrium. An extreme
example would be a sudden lowering (``quench'') of temperature from
the supercooled into the glass regime, with the corresponding
nonlinear relaxation function describing the out-of-equilibrium
relaxation of the energy.

Equally relevant for experiments are correlation functions of
fluctuating quantities; density fluctuations, for example, can be
measured by scattering techniques. The equilibrium
autocorrelation function of observable $\phi$ is defined as
\be
C(t) = \lav \phi(t)\phi(0) \rav - \phi\eql^2
\label{C_eq}
\ee
and obeys $C(t)=C(-t)$ from time-translation invariance (TTI). It is
related to the linear response function $\chi(t)$ by the
fluctuation-dissipation theorem (FDT)~\cite{Reichl80}, which states
that for $t>0$
\be
\deriv{t}\chi(t) = R(t) = -\frac{1}{T}\deriv{t}C(t)
\label{eq_fdt}
\ee
Here $R(t)$ is the impulse response, \ie\ the response of $\phi(t)$ to
a perturbation $h\delta(t)$. In integrated form the FDT reads
$C(t)=T[\chi\eql-\chi(t)]$. Eq.\eq{eq1S22} then shows that
$C(t)=C(0)\psi(t)$ with $C(0)=T\chi\eql$, so that the relaxation
function also gives the time evolution of the correlations: in
equilibrium, fluctuations decay with the same time-dependence whether
occurring spontaneously or induced by an applied perturbation.

The FDT can also be expressed in the frequency-domain, where it
relates the linear response to oscillatory perturbations to the power
spectrum of equilibrium fluctuations. The time- and
frequency-dependent quantities can of course be expressed in terms of
each other; experimentally, the latter are often more easily
accessible, while theoretical work tends to focus on the former. From
linearity, the response to a small oscillatory perturbation $h(t)=\Re
[h\exp(i\omega t)]$ is $\phi(t)=\int_{-\infty}^t dt'\ R(t-t')h(t')$ with
$R(t)=\partial\chi(t)/\partial t$ the impulse response as before. After an
integration by parts one then has
$\phi(t)=\Re[\hat\chi(\omega)h\exp(i\omega t)]$
with
\be
\hat\chi(\omega)=i\omega \int_{0}^{\infty}dt\ \chi(t)e^{-i\omega t}
= \chi\eql - i\omega\chi\eql \int_{0}^{\infty}dt\ \psi(t)e^{-i\omega t}
\label{eq6S22}
\ee
(Formally, an infinitesimal negative imaginary part should be added here to
$\omega$ to make all integrals convergent; physically this corresponds
to a very slow switching on of the oscillatory perturbation.) The
complex susceptibility $\hat\chi(\omega)$ can be written as
$\hat\chi(\omega) = \hat\chi'(\omega) - i\hat\chi''(\omega)$ where
$\hat\chi'$ is the in-phase or reversible part of the response and
$\hat\chi''$ is the out-of-phase or dissipative part. The related
fluctuation quantity is the power spectrum, which gives the amplitude
of fluctuations of frequency $\omega$ and can be expressed as the
temporal Fourier transform of the correlation function
\be
S(\omega)=\int_{-\infty}^{\infty}dt\ C(t)e^{-i\omega t}
\label{eq9S22}
\ee
The FDT\eq{eq_fdt} together with $C(t)=C(-t)$ then relates the power
spectrum of fluctuations to the dissipative part of the reponse,
according to
\be
S(\omega)=2T\hat\chi''(\omega)/\omega
\label{freq_FDT}
\ee

At high temperatures, relaxation functions are often simple
(``Debye'') exponentials, $\psi(t)=\exp(-t/\tau)$, giving a
dissipative response
$\hat\chi''(\omega)=\chi\eql\omega\tau/(1+\omega^2\tau^2)$ with a
single maximum at the peak frequency $\omega=1/\tau$, and a power
spectrum $S(\omega)\sim1/(1+\omega^2\tau^2)$ of Lorentzian shape.  It
was observed already in 1854 by Kohlrausch~\cite{Kohlrausch54}, and
later by Williams and Watts, that relaxation functions decay
non-exponentially in supercooled liquids at low temperatures, and can
often be fitted by a stretched exponential or Kohlrausch-William-Watts
(KWW) function
\be
\psi(t)=\exp(-at^b)
\label{eq2S22}
\ee
with a stretching parameter $b<1$. This can be thought of as a
superposition of exponential relaxations with a broad spectrum of
relaxation timescales; in the frequency domain, the corresponding
dissipative response $\hat\chi''(\omega)$ therefore shows a broad
maximum. The value of the stretching exponent $b$ typically decreases with
temperature, reaching values around 0.5 at $T_g$. The value of $a$
decreases rapidly with $T$, corresponding to a large increase in the
typical relaxation time.  It should be noted that fits to experimental
data, which cover a limited range of timescales where $\psi(t)$ is
often not yet small compared to unity, cannot exclude a crossover to
simple exponential behaviour for much longer times. Nevertheless, the
ubiquity of stretched exponential relaxation in supercooled liquids
suggests that this is a generic feature of glassy dynamics which
theory needs to be able to predict.  An interesting issue is whether
the observed stretching arises from an average over a heterogeneous
spatial structure, with different local regions having very different
relaxation times, or whether the relaxation dynamics is intrinsically
non-exponential but homogeneous; we return to this point in
Sec.~\ref{intro:cooperativity}.

We have already hinted that one can obtain a relaxation time $\tau$
from the relaxation function $\psi(t)$. A number of different
definitions have been used; broadly one would hope that they give
qualitatively similar values, though we will see counterexamples below.
Common procedures for defining $\tau$ are:
\begin{itemize}

\item
The {\em instantaneous relaxation time}, defined as the time at which
the relaxation function has decayed to $1/e$ of its initial value,
$\psi(t=\tau)=1/e$. This time is simple to measure and therefore
favoured by experimentalists.

\item
The {\em integrated relaxation time}, defined as
$\tau=\int_0^{\infty}dt\, \psi(t)$. This is mostly used in theoretical
analysis; its use in experiment would require a fit for the long-time
behaviour to carry out the time-integration.

\item
The {\em fitting time}, which is defined as the timescale parameter
appearing in an appropriate fit of the relaxation function. For a KWW
fit\eq{eq2S22}, for example, one can write $\psi(t)=\exp[-(t/\tau)^b]$
with $\tau=a^{-\frac{1}{b}}$.

\end{itemize} 
Notice that all definitions coincide for an exponential relaxation
function, $\psi(t)=\exp(-t/\tau)$. 

Many relaxation functions in supercooled liquids actually display
behaviour more complicated than described above, requiring the
definition of several relaxation times. For example, the relaxation of
density fluctuations in sufficiently supercooled liquids (as defined
further below) proceeds in two stages. The initial decay of $\psi(t)$
is to a nonzero plateau value. Physically, this {\em
$\beta$-relaxation} process is thought to correspond to the localized
motion of particles in the structural ``cages'' formed by their
neighbours; the corresponding relaxation time $\tau_\beta$ normally
increases in an Arrhenius fashion as $T$ is decreased. On a much
longer timescale $\tau_\alpha$, the relaxation function then decays
from the plateau to zero, and only the long-time part of this {\em
$\alpha$-relaxation} is well described by a stretched exponential. The
$\alpha$-relaxation dominates the integral $\int_0^\infty dt\,
\psi(t)$ of the relaxation time, so that the integrated relaxation
time $\tau$ is of the same order as $\tau_\alpha$. It is this
timescale that increases strongly as temperature is lowered, with the
temperature dependence discussed in Sec.~\ref{basics1}. (MCT in fact
predicts that $\tau_\alpha$ genuinely diverges at some nonzero
temperature~\cite{BenGoetSjo84,Goetze91,GoetSjoe92,GoetSjoe95}.) In
the frequency domain, the presence of two relaxation processes with
widely separated timescales means that the dissipative response
$\hat\chi''(\omega)$ has two maxima around the inverses of the
$\alpha$- and $\beta$-relaxation times.

We finish this section by mentioning two important examples of
correlation functions. In a system consisting of a number of particles
with position vectors $\rv_a$, the Fourier component with wavevector
$\kv$ of the local density is $\phi_{\kv}=\sum_a
\exp(i\kv\cdot\rv_a)$, up to a constant prefactor which we ignore. As
long as the system remains ergodic, particles are equally likely to be
anywhere inside the system volume at equilibrium, so that
$\phi_{\kv}^{\eql}=0$ for nonzero $\kv$. The correlation function of
$\phi_{\kv}$ is therefore, using the obvious generalization of\eq{C_eq} to
complex observables,
\be
C(\kv,t) = \lav \phi_{\kv}(t)\phi_{-\kv}^*(0)\rav = 
\sum_{ab} \lav e^{i\kv\cdot[\rv_a(t)-\rv_b(0)]} \rav
\label{c}
\ee
This ``coherent'' correlation function (also known as dynamic structure
factor or intermediate scattering function) can be measured using dynamic
light scattering experiments, for example. On large lengthscales, \ie\ for
small $\kv$, and for long times density fluctuations should relax
diffusively and so one expects $C(\kv,t) \sim \exp(-D\kv^2 t)$. This
relation can be used to deduce from knowledge of $C(\kv,t)$ for small $\kv$
and large $t$ the value of the {\em collective diffusion} constant $D$
controlling the relaxation of long-wavelength density fluctuations. A
self-correlation analogue of $C(\kv,t)$ can also be defined,
as the sum of correlation functions for the single-particle observables
$\exp(i\kv\cdot \rv_a)$,
\be
C_s(\kv,t) = \sum_{a} \lav e^{i\kv\cdot[\rv_a(t)-\rv_a(0)]} \rav
\label{cs}
\ee
For small $\kv$ and long $t$ this correlation function, referred to as
the intermediate self-scattering function, should again
behave as $C_s(\kv,t)\sim \exp(-D_s\kv^2t)$. Since $C_s(\kv,t)$ only
measures correlations of each particle with itself, however, the
diffusion constant $D_s$ entering here is the one for {\em
self-diffusion}, and determines the long-time mean-square displacement
of individual particles according to $\lav [\rv_a(t)-\rv_a(0)]^2 \rav
= 6D_s t$. Notice that $C_s(\kv,t)$ is the Fourier transform of the 
so-called self-part of the van Hove correlation function,
\be
G_s(\rv,t)=\sum_{a}\lav\delta(\rv_a(t)-\rv_a(0)-\rv)\rav
\label{gs}
\ee
The latter is conventionally normalized by dividing by the total number of
particles, so that $G_s(\rv,t=0)=\delta(\rv)$; the second moment $\int
d\rv\, \rv^2 G_s(\rv,t)$ then gives the mean-square particle displacement as a
function of time $t$.

\subsection{Out-of-equilibrium dynamics: two-time quantities 
and effective temperatures}
\label{intro:fdt}

When supercooled liquids are cooled to sufficiently low temperatures,
their longest relaxation times will become comparable and eventually
exceed experimental timescales. The system is then referred to as a
glass. It no longer reaches (metastable) equilibrium on accessible
timescales and instead {\em ages}: its properties depend on the
waiting time $\tw$ elapsed since the glass was prepared, \eg\ by a
temperature quench. We review in this section how correlation and
response functions are generalized to two-time quantities in the aging
regime. We also discuss how out-of equilibrium correlation and
response can be used for defining effective temperatures. This
suggestion first appeared in the context of mean-field spin-glass
models but has since found much wider application; see
\eg~\cite{BouCugKurMez98} for a review.

The two-time autocorrelation function of an observable $\phi$ is
defined, in a natural generalization of\eq{C_eq}, as 
\be
C(t,\tw)=\lav \phi(t)\phi(\tw)\rav -\lav \phi(t) \rav \lav \phi(\tw) \rav
\label{corr}
\ee
Similarly, one can define a two-time impulse response function
\[
R(t,\tw)=\left.\frac{\delta \lav \phi(t)\rav}{\delta h(\tw)}\right|_{h=0}
\]
which gives the linear response of $\phi(t)$ to a small impulse
$h(t)=h\delta(t-\tw)$ in the conjugate perturbation at time $\tw$. The
step response is then given by
\be \chi(t,\tw)=\int_{\tw}^t\!  dt'\, R(t,t')
\label{switch_on}
\ee
and tells us how $\phi$ responds to a small constant field switched on
at time $\tw$.

\begin{figure}
\begin{center}
\epsfig{file=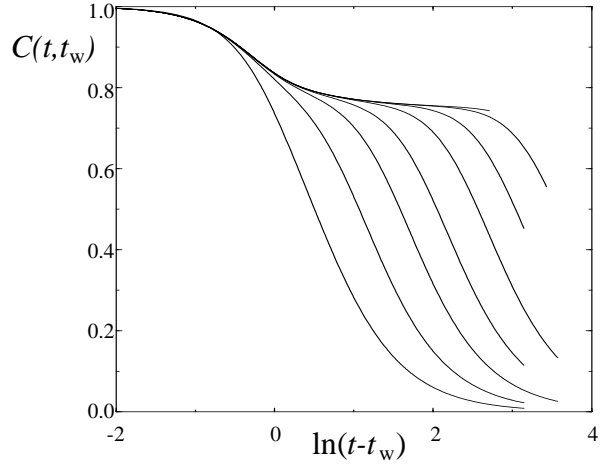, width=8cm}
\end{center}
\caption{Typical shape of a two-time correlation function, plotted as
a function of $t-\tw$ (in log-scale) with system age $\tw$
increasing from left to right. Notice the two separate relaxation
processes: the short-time part of the relaxation is independent of
$\tw$ and obeys time-translation invariance while the long-time decay
from the plateau takes place on a timescale growing with $\tw$.
\label{aging_LP}
}
\end{figure}
Now, in {\em equilibrium}, $C(t,\tw)=C(t-\tw)$ by time-translation
invariance (TTI); the same will be true of $R$ and $\chi$ and
FDT\eq{eq_fdt} holds. A parametric ``FDT plot'' of $\chi$ vs.\ $C$ is thus
a straight line of slope $-1/T$. In an aging system such as a glass, on the
other hand, correlation and response functions will be nontrivial functions
of both their arguments. A generic scenario for the behaviour of the
correlation function is depicted in Fig.~\ref{aging_LP}: the initial
($\beta$-)part of the relaxation takes place on a timescale which---for
large enough $\tw$---is independent of the age $\tw$. In this regime
$C(t,\tw)$ is a function of $t-\tw$ only and thus obeys TTI. The long-time
($\alpha$-)relaxation, on the other hand, takes part on ``aging
timescales'' growing with $\tw$; the most straightforward case where
$\tau_\alpha\sim \tw$ is often referred to as simple aging.

The out-of-equilibrium, two-time correlation and response functions are not
expected to obey FDT; to quantify this one can define an
FDT violation factor $X(t,\tw)$ through~\cite{CugKur93,CugKur94}
\be
\label{eqn:non_eq_fdt}
-\deriv{\tw} \chi(t,\tw)= R(t,\tw) = \frac{{X(t,\tw)}}{T}\deriv{\tw}C(t,\tw)
\ee
One may wonder why derivatives w.r.t.\ $\tw$ are used here rather than
$t$; in equilibrium the two choices are equivalent since all functions
depend only on $t-\tw$. However, derivatives w.r.t.\ $t$ would make
rather less sense in the out-of-equilibrium regime, since only the
$\tw$-derivative of $\chi(t,\tw)$ is directly related to the impulse
response $R(t,\tw)$; physically, this corresponds to 
causality of the response. Adopting therefore the
definition\eq{eqn:non_eq_fdt}, one sees that values of $X$ different
from unity mark a violation of FDT. In glasses, these can persist even
in the limit of long times, indicating strongly non-equilibrium
behaviour even though one-time observables of the system---such as
energy and entropy---may be evolving only extremely slowly.

\begin{figure}
\begin{center}
\epsfig{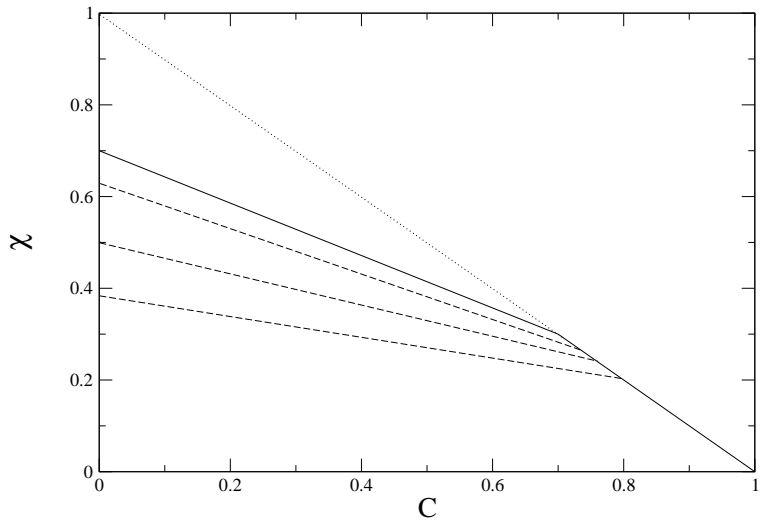}
\end{center}
\caption{Schematic FDT plots of step response $\chi$ versus
correlation $C$ in glassy systems. The dotted line shows the
equilibrium slope of $-1/T$. The FDT plot first follows this line as
the correlation function decays from its initial value, but then
crosses over to a non-equilibrium part, as shown by the dashed lines
for a series of increasing waiting times $\tw$ (bottom to top). The
(negative inverse) slope of this part of the plot can be used to
define an effective temperature $T\eff$, which in this example is $>T$
and decreases as the system ages. In some exactly solvable mean-field
models a nontrivial limiting plot is approached for long times (solid line).
\label{fdt}
}
\end{figure}
Remarkably, the FDT violation factor for several {\em mean field}
models~\cite{CugKur93,CugKur94} assumes a special form at long times:
Taking $\tw\to\infty$ and $t\to\infty$ at constant $C=C(t,\tw)$,
$X(t,\tw)\to X(C)$ becomes a (nontrivial) function of the single
argument $C$. If the equal-time correlator $C(t,t)$ also approaches a
constant $C_0$ for $t\to\infty$, it follows that
\be
\chi(t,\tw)=\frac{1}{T}\int_{C(t,\tw)}^{C_0}
\!dC\, X(C)
\label{mf_limit}
\ee
Graphically, this limiting non-equilibrium FDT relation is obtained by
plotting $\chi$ vs $C$ for increasingly large times; from the slope
$-X(C)/T$ of the limit plot, an {\em effective
temperature}~\cite{CugKurPel97} can be defined as $\teff(C)=T/X(C)$.
Typical FDT plots are shown in Fig.~\ref{fdt}.

In the most general aging scenario, a system displays dynamics on
several characteristic timescales, one of which may remain finite as
$\tw\to\infty$, while the others diverge with $\tw$; the case with one
finite timescale and one growing with $\tw$ is illustrated in
Fig.~\ref{aging_LP}. If these different timescales become infinitely
separated as $\tw\to\infty$, they form a set of distinct `time
sectors'; in mean field, $\teff(C)$ can then be shown to be {\em
constant} within each such sector~\cite{CugKur94}. In the short time
sector ($t-\tw=O(1)$), where $C(t,\tw)$ decays from $C_0$ to some
plateau value, one generically has quasi-equilibrium with $\teff=T$,
giving an initial straight line with slope $-1/T$ in the FDT plot. The
further decay of $C$ (on aging time scales $t-\tw$ that grow with
$\tw$) gives rise to one of three characteristic shapes: (i) In models
which statically show one step replica symmetry breaking (RSB), \eg\
the spherical $p$-spin model~\cite{CugKur93}, there is only one aging
time sector and the FDT plot exhibits a second straight line, with
$\teff>T$ (see Fig.~\ref{fdt}). (ii) In models of coarsening and
domain growth, \eg\ the
$O(n)$ model at large $n$, this second straight line is flat, and
hence $\teff=\infty$~\cite{CugDea95}. (iii) In models with an infinite
hierarchy of time sectors (and infinite step RSB in the statics, \eg\
the SK model) the FDT plot is instead a continuous
curve~\cite{CugKur94}.

$\teff$ has been interpreted as a timescale-dependent non-equilibrium
temperature, and within mean field has been shown to display many of
the properties associated with a thermodynamic
temperature~\cite{CugKurPel97}. For example (within a given time
sector), it is the reading which would be shown by a thermometer tuned
to respond on that time scale. Furthermore---and of crucial importance
to its interpretation as a temperature---it is independent of the
observable $\phi$ used to construct the FDT plot~\cite{CugKurPel97}.
While this picture is theoretically well established only in mean
field models, nontrivial FDT plots have recently also been found in
many non-mean field systems including KCMs. A number of open questions
remain, however, over whether these FDT relations can be used to
defined meaningful effective temperatures (see Sec.~\ref{res:fdt} for
details). A unique $\teff$ may not result for arbitrary observables
$\phi$, for example, and one may have to restrict attention to a
suitable class of ``neutral'' observables. Also, in some cases the
slope of the FDT plot is not constant in a given time sector but
changes when $t-\tw$ is changed by a factor of order one, while a
meaningful $\teff$ should be insensitive to such changes.

We finish this section with a brief discussion of the most appropriate
representation of FDT plots in non-mean field systems, which can be
somewhat subtle~\cite{SolFieMay02}. For mean field systems the
existence of a limiting relation\eq{mf_limit} between response $\chi$
and correlation $C$ ensures that parametric plots of $\chi$ versus $C$
converge, for long times, to a limiting FDT plot whose negative slope
directly gives $X(C)/T$. Eq.\eq{mf_limit} implies that the plots can
be produced either with $t$ as the curve parameter, holding the
earlier time $\tw$ fixed, or vice versa. The first version is more
convenient and therefore normally
preferred~\cite{CugKur93,CugKur94}. In general, however, the
definition\eq{eqn:non_eq_fdt} ensures a slope of $-X(t,\tw)/T$ for a
parametric $\chi$-$C$ plot {\em only} if $\tw$ is used as the
parameter, with $t$ being fixed. If the equal-time correlator $C(t,t)$
varies with $t$, then ``raw'' FDT plots at increasing $t$ may also grow
or shrink in scale, indefinitely if $C(t,t)\to0$ or $\to\infty$. It is
therefore helpful to ``attach'' the plots to a specific point, either
by showing $\chi(t,\tw)$ vs $\Delta C(t,\tw) =
C(t,t)-C(t,\tw)$~\cite{BuhGar02b} to get a plot through the origin, or
by plotting the normalized values $\tilde{\chi}(t,\tw)=\chi(t,\tw)/C(t,t)$
and $\tilde{C}(t,\tw)=C(t,\tw)/C(t,t)$ to get curves passing through
$(\tilde{C}=1,\tilde{\chi}=0)$~\cite{SolFieMay02}. If a limiting
plot exists for $t\to\infty$, this then means that $X$ becomes a function
of only $\Delta C$ or $\tilde{C}$ in the limit. Either $t$ or $\tw$
can be used as the curve parameter in such a situation, but the
reference value of the correlator must still be $C(t,t)$ rather than
$C(\tw,\tw)$ to maintain the link between $X(t,\tw)$ and the slope of
the FDT plot.

\subsection{Energy landscape paradigms}
\label{intro:landscape}

An interesting take on glassy behaviour is provided by viewing the
dynamics ``topographically'', as an evolution in a very rugged
$3N$-dimensional (if there are $N$ particles) potential energy
landscape~\cite{Goldstein69}. This point of view was taken up in the
early Eighties by Stillinger and Weber (SW)~\cite{StiWeb82} and has
since been further
developed~\cite{Stillinger95,Stillinger95b,DebSti01}; for a selection
of references on successful applications of the framework to
Lennard-Jones glasses see also Ref.~\cite{KobSciTar00}. The basic idea of
SW was to split configuration space into the basins (or valleys, or
inherent structures (IS)) of the energy landscape.  Each basin can be
defined as the set of configurations that map onto the same
configuration in a steepest descent (zero temperature) dynamics on the
energy; because this mapping is deterministic, it splits configuration
space into non-overlapping basins. Each one can be labelled by a
representative configuration, taken as the one of minimum energy
$e_{\rm IS}$ (per particle, say) within the basin. The number density
of basins as a function of $e_{\rm IS}$ will be exponential in system
size, ${\cal N}(e_{\rm IS})=\exp[Ns_c(e_{\rm IS})]$, and by doing the
sum over configurations ${\cal C}$ basin by basin the partition
function can be written as ($\beta=1/T$)
%
\be
Z
=\sum_{\rm IS}\sum_{{\cal C}\in {\rm IS}}e^{-\beta E({\cal C})}
=\int de_{\rm IS}\, e^{N[s_c(e_{\rm IS})-\beta e_{\rm IS}-\beta
\Delta f(\beta,e_{\rm IS})]}
\label{landscape1}
\ee
Here the term
\be
\Delta f(\beta,e_{\rm IS})
=-\frac{T}{N}\ln \sum_{{\cal C}\in {\rm IS}}e^{-\beta [E({\cal
C})-Ne_{\rm IS}]}
\label{landscape2}
\ee
effectively measures the width of a given basin, being a within-basin
free energy relative to the bottom $e_{\rm IS}$ of the basin. We have
assumed that $\Delta f(\beta,e_{\rm IS})$ is the same for all basins
with the same $e_{\rm IS}$; otherwise a more general definition would
be needed in place of\eq{landscape2}. 
We have also written discrete sums over configurations ${\cal C}$,
rather than integrals as would be appropriate for classical particle
systems, in anticipation of the discrete configuration spaces of most
KCMs. The above description naturally introduces the concept of
configurational entropy or complexity of inherent structures,
$s_c(e_{\rm IS})$, and this has been argued to be more relevant to
glassy dynamics than the standard thermodynamic entropy over all
configurations originally contemplated by Adam, Gibbs and Di
Marzio~\cite{GibDim58,AdaGib65}. The reason is that $s_c$ as defined
above excludes all ``trivial'' contributions to the entropy arising
from local excitations within a given basin. In a supercooled liquid
these would correspond to small vibrations of the particles around
their average positions. Assuming that these vibrations are similar in
the supercooled liquid and the crystal, one can alternatively view
$s_c$ as the difference between the entropies of a supercooled
liquid and of a crystal at the same temperature.

Looking ahead, we note that in models for glassy dynamics where
configuration space is discrete (which includes most KCMs) the
dynamics remains stochastic even at zero temperature: there can be
many equivalent directions in configuration space that lead to the
same energy decrease. The boundaries between basins in configuration
space determined by the $T=0$ dynamics then become ``soft'', and would
need to be specified in terms of the probability of a given
configuration being assigned to a specific basin. This complicates the
calculation of the within-basin free energies $\Delta f$. However, the
``bottom'' of each basin remains unambigous and corresponds to a
configuration which will not evolve at $T=0$, so that the
configurational entropy $s_c(e_{\rm IS})$ can be defined and
calculated as before.

A promising recent refinement of the SW approach is to define the
configurational entropy by counting basins with the same {\em free
energy} $f=e_{\rm IS} + \Delta f$ rather than the same $e_{\rm
IS}$~\cite{MarCriRitRoc01}. This makes sense because the equilibrium
weight of each basin is $\exp(-\beta f)$ rather than $\exp(-\beta
e_{\rm IS})$; the additional factor $\exp(-\beta\Delta f)$ correctly
accounts for the different weight of narrow and wide basins. (In
mean-field spin glasses, equal weight is similarly assigned to basins
of equal free energy~\cite{FraVir00}. In this case the division of
configuration space is more clear cut, however, since the different
basins are separated by energy barriers that diverge in the
thermodynamic limit and thus correspond to genuine thermodynamic
states. See~\cite{BirMon00,CriRitRocSel00} for further discussion.)
Since the effective width $\Delta f$ of a basin depends on
temperature, so does the configurational entropy $s_c(f,\beta)$
defined in this way.

Developing the SW approach in a different direction, one may wonder
about the rationale for splitting configuration space according to
basins defined by steepest-descent dynamics. For example, if two
adjacent basins are separated by a low energy barrier then at nonzero
temperature it will make more sense to regard them as a single basin
which the system will explore on short timescales. Biroli and
Kurchan~\cite{BirKur01} proposed that one should therefore replace the
notion of basins with metastable states, \ie\ collections of
configurations within which the system equilibrates on a given
timescale $t_*$ (and at a given temperature $T$). This leads to a {\em
timescale-dependent} definition of the configurational entropy, which
is physically very plausible: \eg\ on infinite timescales $t_*$ the
system must equilibrate over the whole of configuration space and so
the configurational entropy must vanish. To get a meaningful result
for the configurational entropy, the timescales for equilibration
{\em inside} metastable states and for transitions {\em between} such
states must 
be well separated, with $t_*$ chosen to lie between them. (As
explained above, in mean-field systems the metastable states are
normally genuine thermodynamic states with transition times between
them that diverge in the thermodynamic limit; a nontrivial
configurational entropy is thus obtained even for $t_*\to\infty$. For
further discussion of these and related issues
see~\cite{CriRit01,CriRit02a,CriRit02b}.)

Closely related to inherent structures are ideas that have arisen out
of attempts to describe the dynamics of granular media under external
tapping or vibration (see Sec.~\ref{other_systems}) by an effective
equilibrium statistical mechanics. Edwards (see \eg~\cite{EdwOak89})
proposed that an appropriate statistical ensemble would be a flat
(microcanonical) distribution over all {\em blocked} (or ``jammed'')
configurations of
a granular system with given volume or energy etc. The logarithm of
the number of such configurations then defines an ``Edwards entropy'',
from which analogues of \eg\ temperature and pressure can be
derived~\cite{EdwOak89}. The connection to IS follows from the fact
that after a tap on its container, a granular material relaxes to some
blocked configuration where no particle can move further; since
thermal energies are irrelevant in granular materials (see
Sec.~\ref{other_systems}), this corresponds to the steepest-descent or
$T=0$ dynamics used to define IS. Biroli and Kurchan~\cite{BirKur01}
proposed that the notion of an Edwards measure could be extended to
generic glassy systems, where \eg\ nonzero temperature will play a
role, by generalizing it to a flat distribution over metastable states
of a given lifetime $t_*$. One intriguing, and largely open, question
is under what circumstances the effective temperatures derived from
Edwards measures (or analogously from configurational entropies) match
those used to rationalize out-of-equilibrium fluctuation-dissipation
violations (see Sec.~\ref{intro:fdt}).

\subsection{Dynamical lengthscales, cooperativity and heterogeneities}
\label{intro:cooperativity}

An obvious question to ask about glassy dynamics is whether the
dramatic slow-down of the dynamics is correlated with a corresponding
increase in an appropriately defined lengthscale. Critical
slowing-down around second-order phase transitions, for example, is
correlated with the divergence of a static correlation length. In
supercooled liquids, the consensus is that there is no growing {\em
static} lengthscale, since
\eg\ the static structure---as measured by the amplitude of density
fluctuations---changes only negligibly while relaxation timescales
grow by orders of magnitude. (KCMs take this insight to extremes, by
assuming that static correlations are entirely absent.) Any growing
lengthscale in glassy dynamics must therefore be of {\em dynamic}
origin, and as such rather more difficult to define unambiguously.

One route to the definition of a dynamical lengthscale is via the idea
of cooperative motion, which goes back to at least Adam and
Gibbs~\cite{AdaGib65}. In a system of densely packed (spherical, say)
particles, for example, motion of one particle over a distance
comparable to its diameter should require many of its neighbours to move in
concert in order to create a space big enough for the particle to move
into. There is support for this theoretically appealing idea. In
simulations of particles interacting via Lennard-Jones
potentials~\cite{DonDouKobPliPooGlo98,DonGloPooKobPli99}, for example,
the most mobile particles were found to ``follow each other around''
along string-like clusters. Limitations on computer time mean that
such simulations only probe the temperature regime where relaxation
timescales are still relatively short compared to those at $T_g$.
Experiments, however, allow longer timescales to be accessed. For
example, recent work~\cite{WeeCroLevSchWei00} on colloidal glasses (dense
suspensions of spherical colloid particles) found that the most mobile
particles---defined as having moved furthest on an appropriately
chosen timescale---form extended clusters, with neighbouring fast
particles moving predominantly in parallel directions, \ie\
cooperatively. The cluster size distribution was observed to be broad,
so that a precise definition of a cooperativity lengthscale would have
been difficult, but typical clusters sizes were found to be on the
order of tens of particles. (There was also some evidence that the
structure of the largest clusters was fractal, with fractal dimension
$\approx 2$.)

The above results show that the idea of cooperativity is closely
linked to the appearance of dynamical heterogeneities, \ie\ the
existence of local regions in a material with very different
relaxation timescales.  The existence of such heterogeneities is also
suggested by the non-exponential character of relaxation functions in
supercooled liquids and glasses, though the alternative of
intrinsically non-exponential but homogeneous dynamics is equally
possible (see Sec.~\ref{basics2}). Standard experimental quantities
such as the intermediate scattering function\eq{c} measure spatial averages
and so do not directly reveal heterogeneities.  However, more refined
experimental techniques such as multidimensional nuclear magnetic
resonance~\cite{HeuWilZimSpi95,TraWilHeuFenSchSpi98},
photobleaching~\cite{InoCicEdi95,CicEdi95} and dielectric
measurements~\cite{SchBohLoiCha96} do give access to local quantities
and provide support for the existence of dynamical heterogeneities;
for a recent review see~\cite{Ediger00}. The size of the
heterogeneities, \ie\ of local regions with a well-defined relaxation
timescale, provides an alternative definition of a dynamical
lengthscale. How this is related to the cooperativity length is not
obvious, however; Ediger~\cite{Ediger00} argues that the latter must
be smaller than the size of the heterogeneous regions, on the grounds
that cooperativity makes sense only among particles relaxing on
comparable timescales. In addition to a lengthscale, heterogeneities
also define a timescale: in an ergodic system, every local region must
eventually sample the whole ensemble of local relaxation times.  Thus,
heterogeneities must have a certain finite lifetime, over which the
local relaxation time remains approximately constant before switching
to a new value. Of particular interest is the ratio $Q$~\cite{Heuer97}
of this lifetime to the typical relaxation timescales within a local
region. In order for the local relaxation time to be well-defined, one
expects $Q\geq 1$. Some experiments (see \eg~\cite{Ediger00} for
review) do indeed give $Q$ of order unity, suggesting that the time in
which slow local structures lose memory of their relaxation time is of
the order of the relaxation time itself. More recently, values of $Q$
orders of magnitude larger have also been found, however. Experimental
results on the rotation of probe molecules in supercooled polymer
melts~\cite{DesVan01}, for example, show heterogeneities persisting
for times much longer than typical relaxation times. The switching of
local relaxation times was interpreted as due to rare, large-scale,
cooperative rearrangements of heterogeneities; interestingly, this
suggests that the associated cooperativity length is actually {\em
larger} than the size of heterogeneities, contrary to Ediger's
argument~\cite{Ediger00}.

It is clear even from the brief sketch above that the existence of
heterogeneities and dynamical lengthscales induced by \eg\
cooperativity remains an intriguing open problem in glassy
dynamics. We will see in Sec.~\ref{res:hetero} that KCMs can provide
considerable insight in this area, allowing different definitions of
dynamical lengthscales to be compared and cooperativity effects to be
investigated in detail.

\subsection{Glassy dynamics in other systems}
\label{other_systems}

So far in this overview of glassy dynamics we have focused on glasses
which are produced by the conventional route of cooling appropriate
``glass-forming'' liquids; essentially all liquids fall into this
category though the poorer glass-formers may require very high cooling
rates~\cite{Jaeckle86}. Glassy dynamics is a much more widespread
phenomenon, however; we have already mentioned polymer melts, which
become glassy at sufficiently low temperatures, and suspensions of
colloid particles, where glassy effects are induced by compression to
sufficiently large densities. The glass transition has indeed been
viewed as a special case of a more general ``jamming transition'' (see
\eg~\cite{LiuNag98}) which occurs in variety of systems including \eg\
dense granular materials such as sand. We highlight the latter case
here because KCMs have recently also been used as models of such
granular materials. As reviewed in \eg~\cite{JaeNagBeh96,LiuNag01},
these materials display a number of ``glassy'' features. An
interesting difference to conventional glasses is that thermal
excitation effects are negligible since $k_{\rm B}T$ at room
temperature is negligible compared to the energy required to lift a
grain of sand by its own diameter.  Effectively, one therefore has
$T=0$ and the dynamics is driven by external excitations such as
vibrations or vertical tapping of the container. Increasing or
decreasing the tapping intensity then corresponds to changing
temperature, and hysteresis effects appear in the density of the
material when the tapping intensity is modified cyclically. The
temporal increase of density at constant tapping intensity has also
received much attention, and is experimentally observed to have a very
slow, logarithmic dependence on time that is referred to as
logarithmic compaction.

\section{Overview of models}
\label{allmodels}

In this section we collect all KCMs and related models that are
covered in this review. The ``core'' KCMs are the spin-facilitated
models (Sec.~\ref{model:SFM}), which have inspired a number of
variations (Sec.~\ref{model:SFM_variations}), and the constrained
lattice gases discussed in Sec.~\ref{model:lattice_gases}. Closely
related are some models defined on hierarchical structures
(Sec.~\ref{related:hierarchical}); inspired by cellular structures
such as froths (Sec.~\ref{topological}); or obtained via mappings from
models with unconstrained dynamics (Sec.~\ref{model:effective}). All
models covered in these subsections have stochastic, Markovian
dynamics obeying detailed balance with respect to a trivial energy
function, and as their key ingredient explicit constraints forbidding
some local transitions between configurations. In the final
Sec.~\ref{model:other}, we gather other models which are not strictly
speaking KCMs according to this classification, but merit inclusion
because they share a number of features with KCMs.

\subsection{Spin-facilitated Ising models}
\label{model:SFM}

Spin-facilitated Ising models (SFM) were introduced in the early
Eighties in the seminal work of Fredrickson and
Andersen~\cite{FreAnd84,FreAnd85}. They can be formulated in terms of
$N=L^d$ two-state variables $n_i=0,1$ on a $d$-dimensional lattice,
normally chosen as cubic with side length $L$. Physically, an up-spin
$n_i=1$ represents a mobile, low-density region of a supercooled
liquid or glass, while $n_i=0$ models a less mobile region of
higher density. A generic energy function with nearest-neighbour
(n.n.) interactions is then the Ising Hamiltonian,
\be
E=-J\sum_{(i,j)}(2n_i-1)(2n_j-1) + \sum_{i}n_i
\label{eq1S311b}
\ee
The coefficient of the linear (``magnetic field'') term has been set
to unity and fixes the temperature scale, and its sign is chosen in
line with the intuition that at low temperatures most regions should
be of high density, $n_i=0$. The sum in the interaction term runs over
all distinct n.n.\ pairs. For $J>0$ this
term favours neighbouring regions to be in the same state, but we will
see shortly that this effect is unimportant, with most work on the
model focusing on the case $J=0$. The model with the energy
function\eq{eq1S311b} has no equilibrium phase transition due to the
presence of the nonzero field term, and at low temperatures the
concentration $c=\lav n_i\rav$ of up-spins or mobile regions tends to
zero. 

The key idea of Fredrickson and Andersen was that rearrangements in
any given region of the material should be possible only if there are
enough mobile low-density regions in the neighbourhood that can {\em
facilitate} the rearrangement. In the language of spins, a
rearrangement from low to high density or vice versa corresponds to a
spin-flip, and the facilitation constraint is formalized by requiring
that a spin can flip only if at least $f\geq 1$ of its n.n.s are in the
mobile state $n_i=1$. In line with much---though unfortunately not
all---notation in the literature we will call the resulting model the
$f,d$-SFM: the spin-facilitated (Ising) model on a $d$-dimensional cubic
lattice, with $f$ facilitating up-spins required for spin-flips.
Mathematically, its dynamical evolution is governed by a master
equation for the time-dependent probability $p(\nv,t)$ of being in a
given configuration $\nv=(n_1\ldots n_N)$,
\be
\deriv{t} p(\nv,t) = \sum_{\nv'} \ww(\nv'\to\nv)p(\nv',t) - \sum_{\nv'}
\ww(\nv\to\nv') p(\nv,t)
\label{master}
\ee
Here $\ww(\nv\to\nv')$ is the {\em rate} for a transition from $\nv$
to $\nv'$ ($\neq \nv$), defined such that in a small time interval
$dt$ the probability for this transition is $\ww(\nv\to\nv')\,dt$. The
only allowed transitions in the $f,d$-SFM are spin-flips. {\em Without
the kinetic constraint}, the rates for these would be given by
\be
\ww(n_i\to 1-n_i)=\ww_0(\Delta E)
\label{eq3S311}
\label{unconstr_rates}
\ee
Here $\dE$ is the change of the energy\eq{eq1S311b} in the transition
from $n_i$ to $1-n_i$, and $\ww_0(\dE)$ is a transition rate that
obeys detailed balance w.r.t.\ $E$. The Metropolis rule $w_0(\Delta
E)=\min(1,\exp(-\beta \Delta E))$ and Glauber dynamics $w_0(\Delta
E)=1/[1+\exp(\beta \Delta E)]$ are the most common choices; we set
$\beta=1/T$ throughout. We also adopt the convention that rates for
any transitions that are not explicitly listed are zero. The full
set of transition rates defined by\eq{eq3S311} is therefore
\be
\ww(\nv\to\nv') = \sum_i \delta_{\nv',F_i\nv} \ww(n_i\to 1-n_i)
\label{all_rates}
\ee
with $F_i$ the operator that flips spin $i$, $F_i\nv = (n_1\ldots
1-n_i\ldots n_N)$. Finally, in\eq{master} we have used a
continuous-time formulation which is convenient for theoretical
work. A discrete-time version would be as follows. Advance time in
discrete steps $1/N$. At each step, randomly select one of the $N$
spins, $n_i$ say, for a possible spin-flip. Accept this proposed
``move'' with probability proportional to $\ww(n_i\to 1-n_i)$,
otherwise reject it. In the thermodynamic limit $N\to\infty$, this
discrete-time algorithm leads to the same results as its
continuous-time counterpart, \ie\ it gives the same evolution of
$p(\nv,t)$ up to possibly a trivial rescaling of time (see
Sec.~\ref{numer:mc}).

Having set up the general framework for the dynamics, we now need to
incorporate the kinetic constraints. Define $\zz$ to be the number of
up-spin neighbours of spin $n_i$; Fredrickson and Anderson then
proposed to implement the kinetic constraint by modifying the
transition rates from\eq{eq3S311} to
\be
\ww(n_i\to 1-n_i)=\zz(\zz-1)\cdots (\zz-f+1)\ww_0(\Delta E)
\label{sfm_rates}
\ee
The new factor forces the rate to be zero whenever $\zz<f$. For $f=1$,
for example, this factor is simply $\zz$, which is zero for $\zz=0$
but nonzero for $\zz\geq 1$; for $f=2$ the kinetic constraint factor
$\zz(\zz-1)$ vanishes for $\zz=0$ or $\zz=1$ but is nonzero for
$\zz\geq 2$. Importantly, the fact that some rates are zero due to the
kinetic constraint does not break detailed balance, since a transition
and its reverse transition are always forbidden together. It is also
clear that the main effect of the kinetic constraint factor
$\zz(\zz-1)\cdots (\zz-f+1)$ is to set some rates to zero and thus
rule out the corresponding transitions. Its precise value for {\em
allowed} transitions should not affect the results qualitatively, and
one could equally define it so that it always equals unity for allowed
transitions~\cite{NakTak86b,NakTak86,GraPicGra93,GraPicGra97}. An
advantage for theoretical treatment of the form\eq{sfm_rates} is that
the constraint factor can be written relatively simply in terms of the
neighbouring spin variables,
\be
\ww(n_i\to 1-n_i) = \sum_{j_1\neq \ldots \neq j_f} n_{j_1} \cdots n_{j_f}
\ww_0(\dE)
\label{sfm_rates_formal}
\ee
where the site indices $j_1,\ldots, j_f$ are summed over the n.n.\
sites of spin $i$.

The origin of glassy dynamics in the $f,d$-SFM is easy to understand
intuitively. From the energy function\eq{eq1S311b} we see that at low
temperatures the equilibrium concentration $c\eql=\lav n_i\rav$ of
up-spins, \ie\ mobile regions, becomes small; for $T\to 0$, $c\eql\to
0$ since the field-term in the energy function forces all spins to
point down. Only a very small number of spins will then have $f$ or
more up-spin neighbours, while all other spins will be effectively
frozen until enough of their neighbours flip up. The kinetic
constraint thus creates a dynamical bottleneck, which becomes more
pronounced as the number $f$ of facilitating spins is increased.

We should stress that the variables $n_i=0,1$ in SFMs do not
correspond to particles, but merely to high and low values of an
appropriately coarse-grained density. This will be different in the
lattice gas models discussed in Sec.~\ref{model:lattice_gases}, where
$n_i=0$ and $n_i=1$ correspond to a particle and a hole, respectively,
and $\sum_i n_i$ represents the total particle number, a conserved
quantity. Notice also that in the lattice gases the glassy ``jammed''
regime of high density corresponds to $c=\lav n_i\rav$ close to one,
whereas for SFMs $c$ represents the concentration of mobile regions
and glassy features occur when $c$ becomes small. Finally, it is worth
pointing out that SFMs have often been formulated in terms of spin
variables taking the values $-1$ and $+1$ rather than $0$ and $1$. We
find the latter more convenient, especially since in SFMs the up- and
down-states do not represent equivalent physical states related by
symmetry.

It is clear from the above discussion that glassy dynamics in SFMs
will occur whenever the concentration of up-spins is small. As
anticipated above, the interaction term in the energy
function\eq{eq1S311b} is not necessary for this effect to occur, and
therefore most studies of the $f,d$-SFM have focused on the case of
the non-interacting energy function
\be
E=\sum_{i}n_i
\label{eq4S311}
\label{SFM_trivial_E}
\ee
Compared to\eq{eq1S311b} this produces completely trivial
thermodynamics, corresponding to free spins in a field. The
equilibrium concentration of up-spins is therefore
\be
c\eql=1/(1+e^{\beta})
\label{ceql_trivial}
\ee
and, as expected, becomes very small in the low-temperature limit of
large $\beta$. The energy change $\dE$ entering the unconstrained
transition rates $\ww_0(\dE)$ then also simplifies to $\dE=1-2n_i$,
and Glauber transition rates take the simple form
\be
\ww_0(\dE)= (1-c\eql)n_i+c\eql (1-n_i)
\label{Glauber}
\ee
Unless $J\neq 0$ is specified explicitly, we will always mean the
noninteracting case $J=0$ when referring to the $f,d$-SFM in the
following. From\eq{ceql_trivial}, either temperature or the up-spin
concentration $c\eql$ can then be used to specify the equilibrium state of
the system.

More recently, versions of SFMs with {\em directed constraints}
(sometimes also called asymmetric constraints) have
been introduced, mainly by J\"ackle and coworkers, and have proved to
be very useful in adding to our understanding of the original
SFMs. The new feature of models with directed constraints is that only
n.n.\ spins in specific lattice directions can act as
facilitators. Two such models have been considered in some detail. The
simplest is the {\em East model}, first proposed in~\cite{JaecEis91}
and later rediscovered~\cite{MunGabInaPie98}. The model is defined in
dimension $d=1$, with a spin allowed to flip only if the nearest
neighbour on the left is up. (The name ``East model'' derives from the
fact that in the original formulation of the model~\cite{JaecEis91}
the opposite convention was chosen for the direction of the
constraint, with facilitating neighbours assumed to be on the right,
\ie\ to the East.) The transition rates for spin-flips in the East
model are $\ww(n_i\to 1-n_i)=n_{i-1} \ww_0(\dE)$, which for the
trivial energy function\eq{eq4S311} and Glauber dynamics\eq{Glauber}
becomes
\be
\ww(n_i\to 1-n_i)=n_{i-1} [(1-c\eql)n_i+c\eql (1-n_i)]
\label{eq5S311}
\ee
The model is the directed version of the $1,1$-SFM; in the latter, an
up-spin neighbour either to the left or right can facilitate a
spin-flip, while in the East model a spin can never flip if its left
neighbour is down, whatever the state of its right neighbour. This
seemingly innocent modification 
actually has profound effects on the dynamics; see the
summary of results in Sec.~\ref{sfm:some_results}. On a square lattice
one can similarly define the directed analogue of the $2,2$-SFM,
called the North-East model, by requiring that a spin can flip only if
both its neighbours to the North and East are
up~\cite{ReiMauJaec92}. A weaker directionality constraint had earlier
been proposed by Reiter~\cite{Reiter91}, who considered a model where
a spin can flip if at least two neighbours in orthogonal
directions---\eg\ North and West, or South and West---are up.

For the East model and the $1,1$-SFM, a model which interpolates
between the two extreme cases of fully directed and undirected
constraints has also been considered very
recently~\cite{CriRitRocSel00,BuhGar01,BuhGar02}. The transition rates
can be chosen as, for example
\be
\ww(n_i\to 1-n_i)=(n_{i-1}+an_{i+1}) [(1-c\eql)n_i+c\eql (1-n_i)]
\label{asymmetric_SFM}
\ee
For $a=0$ and $1$ this gives the East model and the $1,1$-SFM,
respectively; for intermediate values of the parameter $a$ one has an
``asymmetric $1,1$-SFM'' where spins with an up-spin neighbour on the
right are able to flip but only with a rate reduced by the factor $a$.

Finally, models with directed constraints have also been defined on
more abstract structures, \eg\ Cayley
trees~\cite{ReiMauJaec92}. Starting from a root node, at each node the tree
branches into $a-1$ nodes on the next level down, so that each node is
connected to $a$ others, one above and $a-1$
below. Fig.~\ref{fig_FELIX_1} below shows an example with three levels
and $a=4$. The directed
$(a,f)$-Cayley tree model is then defined by the constraint that spins
can only flip if $f$ of the $a-1$ spins below them are up. An
undirected version of this model could also be contemplated, by
allowing spins to flip whenever {\em any} $f$ of their $a$ neighbours,
whether above or below, are up. To make sure that the root node also
has $a$ neighbours, it is then sensible to consider a Bethe lattice,
\ie\ a set of $a$ Cayley trees linked together at a common root
node. It has been argued~\cite{ReiMauJaec92}, however, that this
undirected variant has features very similar to the directed Cayley
tree model---with \eg\ blocking transitions, where a finite fraction
of spins are permanently frozen, occurring at the same up-spin
concentration---so we do not consider it further in the following.

\subsubsection{Interlude: Reducibility and ergodicity}
\label{intro:irred}

So far, we have naively assumed that the equilibrium behaviour of
KCMs---such as SFMs and their directed analogues---is described by the
usual Boltzmann distribution. In the presence of kinetic constraints,
this is not completely trivial, and one has to consider the
possibility that the dynamics might be {\em reducible}. We pause in
our overview of KCMs to discuss this issue, contrasting it with the
closely related though distinct question of ergodicity.

Recall that the Boltzmann distribution $p\eql(\nv)\sim \exp[-\beta
E(\nv)]$ is guaranteed to describe the unique equilibrium state, \ie\
the long-time limit of $p(\nv,t)$ for a finite system, under two
conditions: that the dynamics obeys detailed balance w.r.t.\ the
energy function $E$, and that the dynamics is
irreducible. Irreducibility means that the system can pass from any
configuration to any other by some finite number of ``allowed''
transitions, \ie\ transitions with nonzero rates. Pictorially, there
must be a path in configuration space from any one configuration to
any other. Notice that the definition of irreducibility refers to a
finite system, and only addresses the existence of paths and not the
(possibly very long) time it would take the system to traverse a given
path.

In systems without kinetic constraints and at nonzero temperature,
irreducibility is normally trivial. E.g.\ in Ising models with the
unconstrained spin-flip rates\eq{unconstr_rates}, any spin-flip is an
allowed transition, and one can get from any configuration to any
other with at most $N$ spin-flips. In KCMs, on the other hand, the
presence of the kinetic constraints can cause the dynamics to be {\em
reducible}, with configuration space splitting into mutually
inaccessible {\em partitions}. A partition can be constructed by
starting from some configuration $\nv$, adding all configurations that
are accessible from $\nv$ via allowed transitions, and iterating until
no new configurations are found. All configurations in the partition
are then mutually accessible: detailed balance ensures that if there
are paths from $\nv$ to $\nv_1$ and $\nv_2$, then the reverse path
from $\nv_1$ back to $\nv$ also exists and can be followed from there
to $\nv_2$. On the other hand, no paths exist that connect
configurations in different partitions. Thus, if the system is started
off in a configuration in a given partition, it will equilibrate to
the Boltzmann distribution in {\em that partition} only, while the
probability of being in states in other partition remains zero for all
time. 

A simple example of reducibility, with only two partitions, is
provided by the $1,d$-SFM. Clearly, the configuration with all spins
down, $n_i=0$, allows no transitions at all to other configurations
since no facilitating up-spins exist; it forms a partition on its
own. On the other hand, starting in any other
configuration, one can flip up the n.n.s of all up-spins and continue
this process until all spins are up. All these configurations are
therefore connected to the all-up configuration and form a single
partition which---since it contains the all-up state---is normally
referred t oas the high-temperature partition in the context of
SFM. (This is somewhat of a misnomer, since for the energy
function\eq{SFM_trivial_E} the equilibrium state with $c\eql=\lav
n_i\rav =1$ corresponds formally to $\beta=-\infty$, rather than
$\beta=0$.) The reducibility in this case is thus of a rather trivial
nature: the dominant high-temperature partition contains all
configurations except for a fraction $2^{-N}$ which vanishes for
$N\to\infty$. One can thus proceed to calculate equilibrium properties
as if the Boltzmann distribution extended over all configurations,
and we can call the system {\em effectively irreducible}.

\begin{figure}
\begin{center}
\epsfig{file=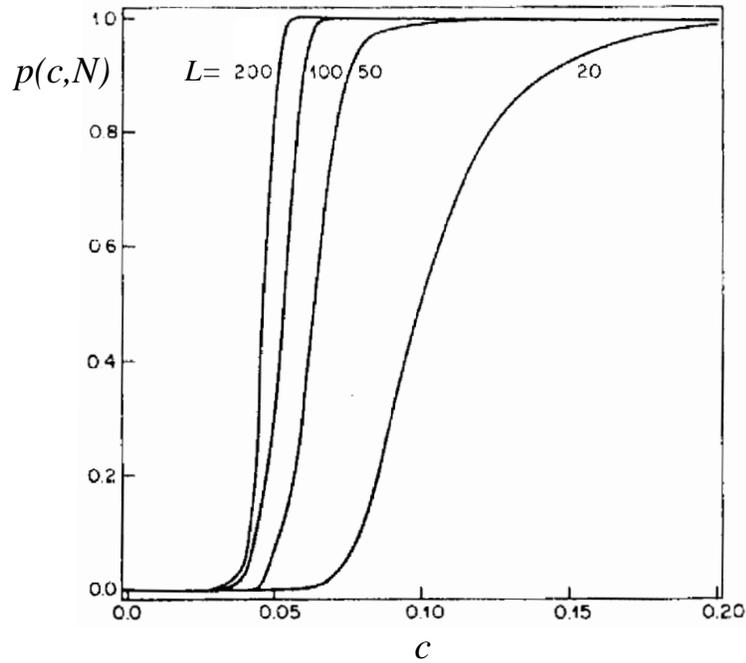, width=10cm}
\end{center}
%
\caption{Probability $p(c,N)$ for states with up-spin concentration $c$
to belong to the high-temperature partition in the $2,2$-SFM, for
different values of $L=N^{1/2}$ as indicated. Notice that the value
$c_*(N)$ above which $p(c,N)$ is close to one and the system thus
effectively irreducible decreases to zero only very slowly with $N$
(in fact logarithmically; see Sec.~\protect\ref{res:reduc}).
From~\protect\cite{FreBra86}.
\label{fig_irrec_sfm}
}
\end{figure}
To formulate the requirement for effective irreducibility more
generally, consider again SFMs with the trivial energy
function\eq{SFM_trivial_E}. The partition function for the
high-temperature partition can then be written down as~\cite{FreBra86}
\be
Z = \sum_{E=0}^N \frac{N!}{E!(N-E)!}  p(E/N,N) e^{-\beta E}
\ee
with $p(c,N)$ the fraction of configurations with up-spin
concentration $c=E/N$ that are in the high-temperature partition. The
naive partition function calculated over all states has the same form
but with $p(E/N,N)$ replaced by $1$. The two procedures for calculating $Z$
will give the same answers in the thermodynamic limit if $p(c,N)\to 1$
for $N\to\infty$ at the naive equilibrium up-spin concentration
$c=1/(1+e^{\beta})$ (see\eq{ceql_trivial}). For effective
irreducibility we would like this to hold at any nonzero temperature,
and will therefore define a system to be effectively irreducible if
$p(c,N)\to 1$ for $N\to\infty$ at any fixed $c>0$. Two comments are in
order here. First, effective irreducibility does not say anything
about the total {\em number} of configuration space partitions, which
in fact generically grows exponentially with system size; it merely
requires that the fraction of total configuration space {\em volume}
taken up by partitions other than the dominant (high-temperature)
partition must shrink to zero in the thermodynamic limit. Second, the
function $p(c,N)$ can exhibit strong finite-size effects. As explained
in more detail in Sec.~\ref{meth:reduc}, if one defines a threshold
up-spin concentration $c_*(N)$ above which a finite system is
effectively irreducible because $p(c,N)\approx 1$, then this will
often converge to zero only very slowly, \eg\ logarithmically in
$N$. One then has to be careful not to assume naive equilibrium
results to hold for arbitrarily low $c$ and finite $N$; the results
for $p(c,N)$ for the $2,2$-SFM shown in Fig.~\ref{fig_irrec_sfm}
illustrate this.

We stress once more that (effective) irreducibility, and the existence
of the corresponding (effectively) unique Boltzmann equilibrium
distribution, are static notions that tell us nothing about the
timescales involved. Since they relate only to the existence of paths
in configuration space, but not to the time it would take the system
to traverse these paths, time is effectively always taken to infinity
for finite $N$, \ie\ before the thermodynamic limit is invoked. This
contrasts with {\em ergodicity}: we will call a system ergodic if any
two configurations---with the exception of possibly a vanishingly
small fraction of configuration space---remain mutually accessible on
timescales that remain {\em finite} in the limit $N\to\infty$. Of
course, reducibility implies non-ergodicity, but the reverse is not
true. Another way of putting this is to say that irreducibility is
concerned with the existence of configuration space paths, whereas
ergodicity focuses on whether these paths retain sufficient
statistical weight in the thermodynamic limit~\cite{Jaeckle02}. A
simple example is the Ising ferromagnet in zero field and with
unconstrained Glauber dynamics. As explained above, the dynamics is
then irreducible for any $T>0$, but ergodicity is broken below the
critical temperature $T_c$, with states of positive and negative
magnetization mutually inaccessible on finite timescales. The
ergodicity breaking occurs here (as it does in general, though not for
KCMs; see below) at an equilibrium phase transition; at $T_c$, a
singular change in the equilibrium properties of the system occurs,
and the two ergodic components into which configuration space splits
have genuine meaning as different physical phases of the system. This
should be contrasted with the concept of reducibility, for which
temperature is irrelevant---the value of $T$
never changes an allowed transition
into a forbidden one, as long as $T>0$---and where the different
mutually inaccessible 
partitions of configuration space have no interpretation as
thermodynamic phases.

As reviewed in Sec.~\ref{res:reduc} below, most of the KCMs we will
consider in this review are effectively irreducible; the exceptions
are the Cayley-tree and the North-East models, which become strongly
reducible below a critical value $c_*$ of the up-spin concentration.
For the effectively irreducible models, equilibrium properties can be
calculated in the naive way, and the trivial energy functions used
ensure that there are no equilibrium phase transitions. An intriguing
question then poses itself: can these models nevertheless show {\em
dynamical transitions} where ergodicity is broken even though there is
no underlying thermodynamic transition? Such a transition could be
caused by a divergence of a relaxation time at nonzero temperature,
for example; see~\cite{Palmer82} for a detailed discussion. This
effect occurs in some mean-field spin glasses (see \eg\ the
review~\cite{BouCugKurMez98}) and is also predicted by approximations
for supercooled liquids such as
MCT~\cite{BenGoetSjo84,Goetze91,GoetSjoe92,GoetSjoe95}. For most KCMs
the evidence points towards the absence of a true dynamical
transition; we defer a detailed discussion of this point to
Sec.~\ref{res:dyntrans} below.

We end this section with a suggestion advanced in~\cite{KobAnd93} that
reducibility in KCMs may not be as important as it seems: one could
consider a weaker form of kinetic constraints where the notionally
forbidden transitions take place with a very small rate $1/\tau_0$. As
long as $\tau_0$ is finite, the connectivity of configuration space is
the same as that for an unconstrained model and so the dynamics is
trivially irreducible. On the other hand, the dynamical evolution of
the system should be independent of $\tau_0$ for times $t\ll\tau_0$,
so that the weakening of the constraint is irrelevant for the
behaviour on {\em finite} timescales.

\subsubsection{Some results for spin-facilitated models}
\label{sfm:some_results}

Having clarified the issue of reducibility, we now return to our
discussion of spin-facilitated models (SFMs). In this section we give
an illustrative overview of some of the key ideas and results for
SFMs, primarily for ``quick'' readers who do not wish to delve too
deeply into the details; ``expert'' readers can find the latter in
Sec.~\ref{res}.

\begin{figure}
\begin{center}
\epsfig{file=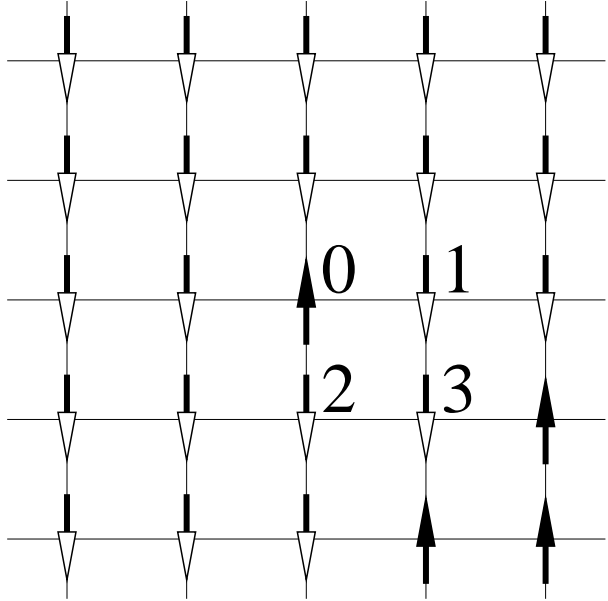, width=8cm}
\end{center}
\caption{An example configuration of a small region of an $f,2$-SFM.
The spin in the centre, labelled by 0, is in the up-state ($n=1$)
and can flip to $n=0$ in different ways depending on the
value of $f$. If $f=1$, all four n.n.s of spin 0 are mobile and spin 0
itself can flip down after any one of these has flipped up.
If $f=2$, on the other hand, spin 0 can only be flipped down in 
a more cooperative process, with the sequence of spin-flips $3\to 2
\to 1 \to 0$ or $3\to 1\to 2\to 0$.
\label{sfm_fig1}
}
\end{figure}

To start with, it is important to note that SFMs can be classified
into two broad families. In models with undirected constraints and
$f=1$ (one-spin facilitated models), relaxation occurs primarily by
the {\em diffusion of defects}, which in this case are isolated
up-spins, and there are close links to other defect-diffusion models,
\eg~\cite{Glarum60,Bordewijk75,ShlMon84,BenShl87}.  All other
models---\ie\ the $f,d$-SFMs with $f>1$ and the models with directed
constraints---require {\em cooperative} processes for relaxation to
occur. This distinction is important because the dynamical effects of
the kinetic constraints are very different in the two model families:
we will see that the models with diffusing defects show strong glass
behaviour, \ie\ an Arrhenius temperature dependence of relaxation
times, while the cooperative models exhibit much more pronounced
relaxation time increases resembling those in fragile glasses. To
understand the origin of this difference, consider an $f,2$-SFM in
equilibrium at low up-spin concentration $c=1/(1+e^\beta)\approx
e^{-\beta}$; we write $c$ instead of $c\eql$ here for brevity. We can
then think of the up-spins as defects in the ground state
configuration with all spins down. Because $c$ is small, a typical
defect is surrounded by down-spins as illustrated in
Fig.~\ref{sfm_fig1}. Let us focus on the relaxation of the central
defect, which proceeds by different mechanisms depending on whether
$f=1$ or $f\geq 2$.

\begin{figure}
\begin{center}
\epsfig{file=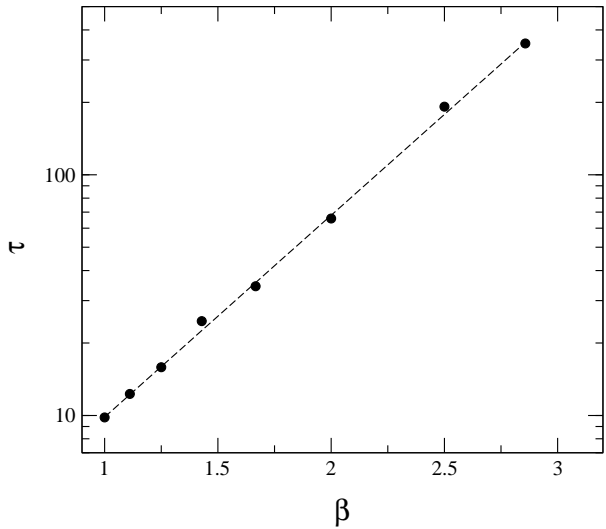, width=8cm}
\end{center}
\caption{Integrated relaxation time $\tau$ as function of $\beta=1/T$
in the $1,1$-SFM. The straight line fit is given by
$\tau=1.43\exp(1.93\beta)$, close to the theoretically expected
behaviour $\tau\sim \exp(2\beta)$ for low
$T$ (see Sec.~\ref{res:statdyn}). From~\protect\cite{CriRitRocSel00}.
\label{sfm_fig4} 
}
\end{figure}

If $f=1$, the central defect can facilitate an up-flip of any of its
neighbouring down-spins. From\eqq{sfm_rates}{Glauber}, the rate for this is
$\ww_0(\dE)=c$, with a corresponding Arrhenius timescale $1/c\approx
e^\beta$. Once a neighbouring spin points up, two different transitions can
happen, both with rate $\ww_0(\dE)=1-c\approx 1$: either the new up-spin
flips back down, or the original up-spin flips down. In the latter case,
the defect has effectively moved to the neighbouring site, and the
effective rate for this process is $c/2$ for small $c$. (The down-flips do
not contribute to this because they only take time of order one, while the
factor $1/2$ arises because the original defect will only move if it flips
down before the newly created up-spin does.) By a repetition of this
process, the defect can then move {\em diffusively} through the whole
lattice, with effective diffusion constant $D\eff=c/2$ if the lattice
constant is fixed to one; the same argument applies to $1,d$-SFMs in any
dimension $d$. The longest relaxation time in these models can be estimated
as the timescale on which diffusing defects encounter each other. With
typical distances between defects of order $l\sim c^{-1/d}$ this gives
\be
\tau \sim l^2/D\eff \sim c^{-1-2/d} \approx \exp[(1+2/d)\beta]
\label{defect_diffusion_tau}
\ee
demonstrating the Arrhenius temperature-dependence of relaxation times
anticipated above. Depending on the precise definition of the
relaxation time, Arrhenius behaviour with different effective
activation energies may result; see Sec.~\ref{res:statdyn} for
details. The {\em integrated} relaxation time, for example, is
estimated to scale as $\sim \exp(2\beta)$ in $d=1$, diverging less
slowly than the longest relaxation time\eq{defect_diffusion_tau};
Fig.~\ref{sfm_fig4} shows results for the former quantity in the
$1,1$-SFM.
Notice that the diffusive character of the dynamics in the SFMs with
diffusing defects is also visible in the out-of-equilibrium dynamics.
After a quench from equilibrium at high temperature, for example, the
average distance between up-spins increases with the characteristic
power law $l(t)\sim t^{1/2}$; see Fig.~\ref{sfm_fig3} below.

\begin{figure}
\begin{center}
\epsfig{file=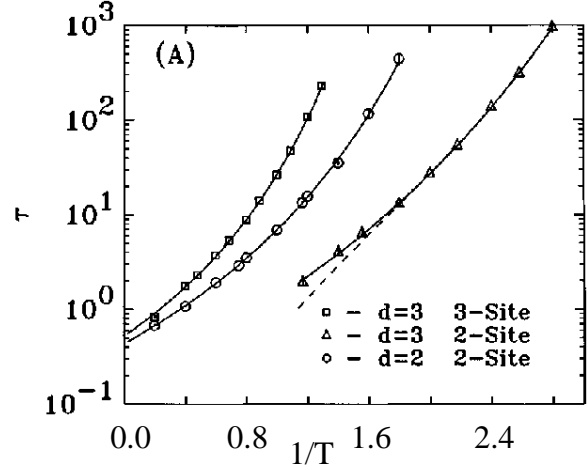, width=8cm}
\end{center}
\caption{Relaxation times in cooperative $f,d$-SFMs as a function of
$1/T$. Three cases are shown from left to right: $(f=d=3)$, $(f=d=2)$,
$(f=2,d=3)$. From~\protect\cite{GraPicGra93,GraPicGra97}.
\label{sfm_fig5}
}
\end{figure}

Now contrast the above analysis for $f=1$ with the cooperative case
$f>1$. Due to the stronger kinetic constraint, the central defect in
Fig.~\ref{sfm_fig1} cannot now on its own facilitate up-flips of its
neighbouring down-spins. It therefore remains itself unable to move
until a region of up-spins further away manages to flip up spins
in its neighbourhood and to propagate this up-spin ``wave'' until it
reaches the 
central up-spin. The example in Fig.~\ref{sfm_fig1} shows that this can
be highly cooperative process, requiring a significant number of
spin-flips to take place in the right order. While no simple scaling
argument for the relaxation time $\tau$ exists in this case, it is
clear that $\tau$ cannot scale as a fixed inverse power of $c$, since
the number of up-flips involved in the cooperative process grows as
$c$ decreases and the distance between defects
increases. Correspondingly, as a function of temperature one has a
superactivated timescale increase. Exemplary results
from~\cite{GraPicGra93,GraPicGra97} are shown in Fig.~\ref{sfm_fig5};
the curvature in the plot of log-relaxation time versus $1/T$ clearly
demonstrates the non-Arrhenius behaviour (and should be contrasted
with Fig.~\ref{sfm_fig4}). Beyond the general recognition that
the cooperative SFMs behave like fragile glasses, very little is known
about the precise form of the timescale increase at low temperature;
some studies have suggested that it might in fact be
doubly-exponential, $\tau \sim \exp[A\exp(1/T)]$ (see
Sec.~\ref{res:hetero}).
\begin{figure}
\begin{center}
\epsfig{file=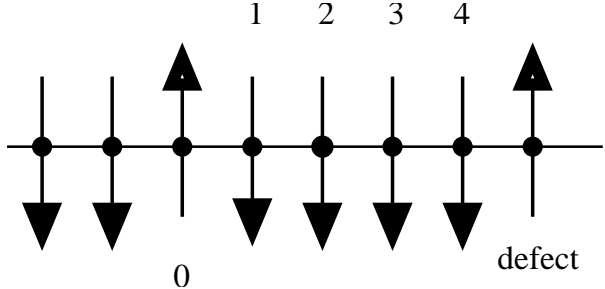, width=8cm}
\end{center}
\caption{Typical configuration of the East model at low temperatures,
where up-spins are separated by large domains of down-spins. The
up-spin labelled 0 can progressively activate (flip up) spins 1, 2, 3
and 4 until the defect on the right can be relaxed, \ie\ flipped
down. At low $T$, the relaxation proceeds via the route with the
smallest activation energy, \ie\ the smallest number of spins that are
simultaneously up. This is achieved by creating ``anchor'' spins (see
text for details): in this example, spin 1 can be flipped down once
spin 2 has been flipped up, and the relaxation can proceed from there
with spin 2 as the anchor.
\label{sfm_fig2}
}
\end{figure}
One example of a cooperative model where relaxation times {\em can} be
deduced by relatively simple arguments is the East model. A typical
equilibrium configuration at low $c$ is shown in
Fig.~\ref{sfm_fig2}. The defect on the left can progressively flip up
its neighbours to the right, and thus eventually relax the defect on
the right. One may suspect that this process requires all intermediate
spins to be flipped up, suggesting a relaxation rate $\tau \sim
(1/c)^{l-1} \approx \exp[(l-1)\beta]$ for defects a distance $l$
apart; the factor $l-1$ in the exponent just gives the energy barrier
arising from the additional up-spins that need to be created. In fact,
one can show that the relaxation process can be made more efficient if
some spins are flipped back down once ``anchoring'' spins between the
two defects have been flipped up. This process proceeds in a
hierarchical fashion, with anchors created successively at distances
1, 2, 4, \ldots to the right of the original defect, and requires a
maximum number of $k\approx \ln l/\ln 2$ up-spins at any one time; see
Sec.~\ref{meth:indint} for details. With typical distances between
defects of order $l\sim 1/c \approx \exp(\beta)$, this gives a
relaxation time $\tau \sim \exp(k\beta) \sim \exp(\beta^2/\ln 2) =
\exp[1/(T^2\ln 2)]$. This is an EITS law\eq{EITS} and gives the very
strong increase of $\tau$ at low temperatures that is typical of
fragile glasses.

Due to the cooperative nature of the dynamics, $f,d$-SFMs with $f\geq
2$ also show rather complex relaxation functions in their {\bf equilibrium
dynamics}. Stretched exponential behaviour has been found in spin
autocorrelation functions, for example; in the East model there is
evidence that the stretching may become extreme at low temperatures,
with the stretching exponent tending to 0 for $T\to 0$ (see
Sec.~\ref{res:statdyn}). The out-of-equilibrium dynamics is also
rather more intricate than in the models with diffusing defects
($f=1$); the East model again provides a simple example, with the
up-spin concentration after a quench decaying as an anomalous power
law $\sim t^{-T\ln 2}$ with a temperature-dependent exponent (see
Sec.~\ref{res:relax}).

Work on {\bf out-of-equilibrium correlation and response functions} of SFMs
and their variants is rather more recent, and we do not yet have a
coherent picture of fluctuation-dissipation theorem (FDT) violations
in these models and the corresponding effective temperatures; see
Sec.~\ref{res:noneq} for details. One complication is that response
functions in these models can be non-monotonic. In the $1,1$-SFM, for
example, only spins that are next to an up-spin are mobile and can
contribute to the response to an applied field; after a quench the
number of such spins decreases in time with the total up-spin
concentration. The response for any {\em given} spin increases with
time after the field has been switched on, but the decrease in the
number of spins that can respond makes the overall response
non-monotonic. In an appropriate representation, FDT plots for some
observables can nevertheless be well-behaved; a recent study for the
$1,1$-SFM found, surprisingly, that even trivial equilibrium FDT plots
can result~\cite{BuhGar02b} (though subtleties remain; see
Sec.~\ref{res:aging}). Whether FDT relations can be used to define
physically meaningful effective temperatures in these models remains
largely an open questions; static definitions of an effective
temperature (\eg\ via configurational entropies, see
Sec.~\ref{res:landsc}) do not appear to be useful.

SFMs have also yielded insights into the cooperative nature of glassy
dynamics, and the existence of {\bf dynamical heterogeneities}, with most
work having been done on the 2,2-SFM (see
Sec.~\ref{res:hetero}). Simulations have
confirmed~\cite{ButHar91,ButHar91b} the intuitive scenario described
above, showing regions of inactive sites which remain frozen until
``mobility is propagated'' to them via a coooperative sequence of spin
flips from active sites, \ie\ mobile spins, elsewhere in the
lattice. See Fig.~\ref{fig:ButHar} below. A number of definitions for
dynamical lengthscales have also
been investigated, one of them being the typical distance between the
(only vaguely defined) active sites referred to above. Relaxation
timescales were found to increase as a power law with this lengthscale
to good approximation, with a large exponent, giving very long
timescales even for modest dynamical lengths. Future work on SFMs
should help to identify more precise definitions of dynamical
lengthscales and shed more light on the role of dynamical
heterogeneities in glassy dynamics.

We mention finally that a number of recent studies have considered
SFMs as abstract models for {\bf granular dynamics}, studying the behaviour
under a sequence of ``taps'' (modelled by evolution at $T>0$, for
example, followed by relaxation at $T=0$). This approach has yielded
insights into logarithmic compaction (see Sec.~\ref{res:driven}). In
some circumstances the resulting non-equilibrium stationary states can
also be described in terms of effective equilibrium using Edwards
measures, but much remains to be done to rationalize when and why this
approach works.

\subsection{Variations on spin-facilitated models}
\label{model:SFM_variations}

SFMs have inspired a number of variations, which we review in the
present section.

A variation of the SFMs with added ``mean-field'' facilitation was
introduced in~\cite{PigKimFri99}: spins can flip if either at least
$f$ of their neighbours are up, {\em or} if the overall concentration
of up-spins in the system is greater than some threshold $c_{\rm
th}$. If the model without the added facilitation is reducible, such
as in the case of the 2,1-SFM, then the extended model has a sharp
dynamical transition when the concentration of up-spins reaches
$c_{\rm th}$; for lower concentrations (\ie\ lower $T$) the chain
splits into segments consisting of frozen and mobile spins,
respectively. A more detailed analysis of the dynamics has not been
performed, however; the model also goes somewhat against the
philosophy of KCMs by introducing a global restriction instead of a
local one. (Global constraints arise in some other models related to
KCMs, but are then motivated by global conservation laws; see
Sec.~\ref{model:other}.)

Variations on SFMs involving quenched disorder have also been
considered.  Schulz and Donth~\cite{SchDon94}, for example, considered
a $2,2$-SFM with locally varying quenched couplings $J_{ij}$ and
fields $h_i$. This gives corresponding locally varying timescales for
spin flips, thus broadening out the spectrum of the faster
$\beta$-processes which involve only a few spin flips. For the slow
$\alpha$-processes, on the other hand, which rely on cooperative flips
of a large number of spins, the local timescale variations tend to
average out and so a single dominant $\alpha$-timescale is retained.
Willart \etal\ considered an SFM in $d=2$ with $f=1.5$, defined
by assigning $f=1$ to a randomly chosen set of half the lattice sites,
and $f=2$ to the remaining sites~\cite{WilTetDes99}; again, these
assignments are quenched, \ie\ fixed during the course of a
simulation.

Schulz, Schulz and Trimper considered a model with two species coupled
together, spins and ``ion concentrations''~\cite{SchSchTri98}. The
motivation comes from the so called mixed mobile ion effect, a strong
nonlinear dependence of the conductivity in strong covalently bonded
glasses (such as SiO$_2$) on the composition ratio of two different
species of alkali ions included. In this model there are two two-state
variables at each lattice site, $n_i=0,1$ for immobile and mobile
regions as before and $r_i$ for the two types of cations. Introducing
a kinetic constraint for the $r_i$ similar to that for the $n_i$,
which forbids diffusion of ions in a locally homogeneous environment,
the authors indeed found the expected strong variation of the
diffusivity with the composition ratio of cations.

Schulz and Reineker~\cite{SchRei93} considered a variation on
$2,2$-SFMs that allows, beyond $n_i=0,1$, a third state at each
lattice site to model local ``vacancies''. These vacancies are
introduced to allow fast local relaxation processes, and so are
postulated to lift the kinetic constraint on all their neighbours;
vacancies are also allowed to diffuse through the lattice at some
constant rate by changing place with neighbouring up-spins. At low
temperatures the unconstrained relaxation near vacancies, with its
almost temperature-independent timescale, is faster
than the highly cooperative dynamics that does not rely on vacancies. 
This fast
process---whose existence can also be deduced from simple mean-field
approximations~\cite{PigSchTri99}---produces a plateau in correlation
functions and can be likened to the $\beta$-relaxation observed in
structural glasses~\cite{SchRei93}. A disadvantage is that also the
long-time (``$\alpha$'') behaviour loses its cooperative aspects and
becomes dominated by the diffusion of vacancies through the lattice,
with relaxation times exhibiting simple Arrhenius
behaviour~\cite{SchRei93}.

For ferromagnetic spin systems, many different kinds of kinetic
constraints have been considered, mainly for the Ising chain in zero
field with energy function $E=-J\sum_i \s_i\s_{i+1}$ in terms of
conventional spins $\s_i=\pm 1$. In fact, already Kawasaki dynamics,
where the only allowed transitions are the exchange of neighbouring
spins with opposite orientation and the up-spin concentration is
therefore conserved, can be thought of as a kind of kinetic
constraint. It does give rise to some glassy features, \eg\ freezing
into non-equilibrium domain structures when the system is cooled
sufficiently rapidly~\cite{CorKasSti91}. Skinner~\cite{Skinner83}, in
the context of an abstract model for polymer dynamics, considered
Glauber dynamics with the constraint that spins can flip only if they
have exactly one up and one down neighbour. This is equivalent to
evolution at constant energy, leading to a random walk of a fixed
number of domain walls that can neither cross nor annihilate. Because
of the fixed energy restriction, the model cannot be used to study
out-of-equilibrium relaxation, but Skinner~\cite{Skinner83} predicted
using an approximate calculation that the spin-spin autocorrelation
function in equilibrium at low temperatures should have a stretched
exponential decay $\sim \exp(-t^b)$ with exponent $b=1/2$. The value
of the exponent, which was later obtained rigourously as the true
asymptotic behaviour~\cite{Spohn89}, is related to the diffusive
motion of the domain walls and also appears in similar
defect-diffusion models~\cite{Glarum60,Bordewijk75}. (Intuitively, the
exponent $b=1/2$ arises since a given spin relaxes within time $t$ if
there are initially domain walls present within the diffusion distance $d\sim
t^{1/2}$, and the probability for this to be the case decays
exponentially with $d$.) The model can be extended by relaxing the
constraint; spins with two identical neighbours can then flip but at a
reduced rate. Numerical results again show a stretched exponential
decay, but with stretching exponent $b>1/2$~\cite{BudSki85}.

In Skinner's~\cite{Skinner83} model, spins are constrained to be
immobile if their two neighbours are either both up or both
down. Recently, Majumdar \etal~\cite{MajDeaGra01} considered a
weaker constraint where only spins with two {\em up-spin} neighbours
are prevented from flipping. While the energy function is still that of an
Ising chain in zero field, the kinetic constraint breaks the symmetry
between up- and down-spins and this has interesting consequences for
the coarsening behaviour at low temperatures which are described in
Sec.~\ref{res:relax}.

An approach opposite to that of Skinner has also been taken, by
considering an Ising chain where spin flips which do {\em not} change
the energy are forbidden; only spins with two equal neighbours can
flip. In the $T=0$ limit the only allowed transitions are then those
which lower the energy, \ie\ flips of up-spins sandwiched between two
down-spins or vice
versa~\cite{LefDea01,DeaLef01,PraBre01,DesGodLuc02}. This ``falling''
dynamics has also been considered on more complicated structures such
as ferromagnets on random graphs where each spin is linked to a fixed
number of randomly chosen neighbours~\cite{DeaLef01}. Even at $T>0$,
the constraint that the energy must change in a move is strong enough
to make the dynamics reducible; a domain of an even number of
up-spins surrounded by all down-spins, for example, can never be
eliminated. The main interest in these models therefore arises when
the falling dynamics is coupled with periodic excitations (\eg\
``tapping'' by random spin flips) that restore an element of
ergodicity; see Sec.~\ref{res:driven}.

\subsection{Lattice gas models}
\label{model:lattice_gases}

The spin-facilitated models discussed so far do not conserve the
number of up-spins, which model mobile low-density regions in the
material. But in structural glasses the overall particle number and
hence density is conserved. To model this situation more directly,
lattice gases with kinetic constraints have been defined. Here
particles occupy the sites of a finite-dimensional lattice and can
move to nearest neighbour sites according to some dynamical rules;
each site can be occupied by at most one particle.  In some sense
these are the simplest KCMs because all allowed configurations
have the same energy and the same Boltzmann weight so the energy
landscape is trivial. Kob and Andersen (KA)~\cite{KobAnd93} introduced
the simplest of these models, originally for particles on a cubic
lattice. Particles move to empty nearest neighbour sites with unit
rates, subject to the condition that the particle has fewer than $m$
occupied neighbour sites both {\em before and after} the move. (The
parameter $m$ as defined here is larger by one than that
of~\cite{KobAnd93}; our choice has the advantage that the same $m$
appears in the bootstrap percolation problem closely related to the
irreducibility of the KA model.) The restriction on the number of
neighbours after the move is necessary to ensure detailed balance. The
choice of $m$ determines the strength of the kinetic constraint. For
$m=6$, the model is unconstrained while for $m=3$ it would clearly be
strongly reducible: any set of eight particles occupying the sites of
a $2\times2\times 2$ cube could never move, all particles having at
least three neighbours whether or not sites around the cube are
occupied. KA chose the smallest value, $m=4$, which does not produce
such obvious reducibility effects, and this defines the standard KA
model. The intuition behind the kind of kinetic constraint imposed is
that if particles are ``caged in'' by having too many neighbours, they
will not be able to move. KA originally proposed the model to test the
MCT for supercooled
liquids~\cite{BenGoetSjo84,Goetze91,GoetSjoe92,GoetSjoe95}, which is
based mainly on this caging effect, but found surprisingly poor
agreement. The model has nevertheless been intensively studied, with
several variants proposed recently as reviewed below.

It is worth pointing out that if we let $n_i=1$ for sites $i$ that are
occupied by a particle and $n_i=0$ for those that are not, then the KA
model with $m=4$ is actually very similar to a 3,3-SFM with Kawasaki
dynamics; the exchange of neighbouring up- and down-spins is
equivalent to moving a particle. Notice however that the facilitating
states are now the ``holes'' $n_i=0$, corresponding to down-spins
rather than up-spins. (A second distinction is that in order to
preserve detailed balance, motion is allowed if the up-spin of the
pair to be exchanged has at least $6-m+1=3$ down-spin neighbours and
if the down-spin has at least 2 down-spin neighbours.) Due to the
facilitation by holes, the glassy regime in the KA model occurs at
high particle density $c=\lav n_i\rav \approx 1$, while in the SFMs it
corresponds to $c\approx 0$.  Finally, if a configuration in the KA
model is described using the $n_i$, it is clear that the particles are
treated as indistinguishable; this is appropriate for studying the
behaviour of density fluctuations, for example. However, for
observables related to self-diffusion, concerning \eg\ the average
displacement of a given particle over some time interval, particles
need to be distinguished one from the other and a configuration is
then specified by giving the position vector (or site number) for each
particle.

In the KA model as described above the total number of particles is
conserved, and since we have a lattice model so is therefore the
particle density. This makes it difficult to study aging effects where
the density evolves with time. One might wish to study, for example,
the behaviour following an instantaneous quench (or better crunch) to
a higher density of particles. An extension of the KA model for this
purpose was introduced in~\cite{KurPelSel97}. (See
also~\cite{PadRit97} for a related model with a global kinetic
constraint.) Here particle exchange with a reservoir is allowed in one
designated plane of the lattice. This can be thought of as the
``surface'', although periodic boundaries are maintained so that one
effectively has a slab of material between two parallel surfaces in
contact with the reservoir. The rates for eliminating and introducing
a particle in this layer are $e^{-\mu}$ and 1, respectively,
corresponding to a reservoir at chemical potential $\mu$ (with the
inverse temperature fixed to $\beta=1$). The dynamics obeys detailed
balance with respect to the energy function $E=-\mu\sum_i n_i$, and
the equilibrium particle density is
\be
c\eql=1/(1+e^{-\mu})
\label{grandcanonical_KA}
\ee
A crunch can be obtained by increasing $\mu$; to have meaningful results
for a bulk system one then needs to check, however, that the density does
not exhibit strong inhomogeneities. This grandcanonical version of the
KA-model can alternatively be thought of as canonical~\cite{Sellitto02},
with $1/\mu$ and $1/c$ respectively corresponding roughly to temperature
$T$ and energy $E$ in a system that is brought into the glassy region by
lowering temperature. A recent overview of relevant results for the
grandcanonical KA model can be found in~\cite{Sellitto02}. A significant
attraction of allowing particle exchange with a reservoir is that many more
configurations become mutually accessible, if necessary by first removing
particles one by one and then reinserting them; this significantly weakens
reducibility effects (see Sec.~\ref{res:reduc}).

\begin{figure}
\begin{center}
\epsfig{file=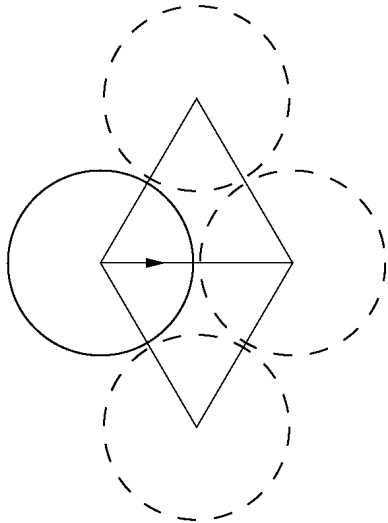, width=6cm}
\end{center}
\caption{Hop path in the triangular (two-vacancy-assisted) lattice
gas. The particle indicated by the full-line circle can hop to the
neighbouring site along the path indicated by the arrow only if the
three sites surrounded by the dashed circles are
free. From~\protect\cite{Jaeckle02}.
\label{FIG_LG2}
}
\end{figure}
A model similar to the KA lattice gas, on a two-dimensional triangular
lattice, was 
considered by~\cite{JaecKro94}; reviews of the main results for this
model, as well as the hard-square lattice gas (see
Sec.~\ref{model:other}) can be found in~\cite{Jaeckle97,Jaeckle02}.  The
constraint here is that the two sites which are nearest neighbours of
both the departure and the arrival site---in other words, the sites
adjoining the ``hop path''---are empty; see Fig.~\ref{FIG_LG2}. This
two-vacancy assisted hopping model, called triangular lattice gas below
for short, was found to display typical glassy features. However, if the
constraint is relaxed from two vacancies to one, these largely
disappear~\cite{JaecKro94}: pairs of vacancies can then diffuse freely
even in an otherwise fully occupied lattice, since each vacancy in a
pair can rotate around the other one. A generalization of two-vacancy
assisted hopping to a three-dimensional face-centred cubic lattice has
also been suggested~\cite{JaecKro94}; hopping is again between nearest
neighbour sites and vacancies are required on all four sites adjoining
the hop path. Since the hop path and two of the vacancies are within a
triangular lattice plane, in crystallographic $(111)$ orientation, the
constraint includes the one for the two-dimensional model and so would
be expected to lead to even more pronounced glassy effects. Finally, for
the two-dimensional two-vacancy assisted model an extension to particles
with orientational degrees of freedom has been
proposed~\cite{DonJac96}. The rules for translational motion are as
before. The kinetic restriction on rotational motion can best be
visualized if one thinks of the particles as ``lemons'', \ie\ hard discs with
small noses on opposite sides; a rotation of one such lemon by $\pi/3$
is allowed only if the two neighbouring sites located along the
direction between the old and the new orientation are empty (otherwise
the noses would get stuck). Orientations are randomly distributed in
equilibrium, but the kinetic constraint couples orientational
fluctuations to the translational motion.

A model similar to one-vacancy assisted hopping, but on a
$d$-dimensional hypercubic lattice, was studied
in~\cite{ArtSchTri98}. There, the hopping rate was taken to be
proportional to the number of vacancies on the n.n.\ sites
surrouding the hop path; hops are therefore allowed only if at
least one such vacancy is present. Unfortunately the analysis of this
model given in~\cite{ArtSchTri98} was flawed since it involved an
approximation which violates conservation of particle number.

Finally, the KA model has recently been generalized to include the
effects of gravity~\cite{SelAre00}. Here the energy of a configuration
is (setting particle mass and gravitational acceleration to unity)
\be
E = \sum_i h_i n_i
\ee
with $h_i$ the height of site $i$. The kinetic constraints are of the
same kind as in the KA model---in~\cite{SelAre00}, for example, a
b.c.c.\ (body-centred cubic)
lattice with $m=5$ was used---but the nonzero transition rates now
take energy changes into account when particles move up or down. Such
models are particularly useful for studying glassy effects in granular
materials, where gravity is important in driving phenomena such as
compaction. As explained in Sec.~\ref{other_systems}, the temperature
$T$ used in this context is not the thermodynamic one but should
rather be regarded as representing some external excitation of the
material, \eg\ by vibration or vertical tapping.

\subsubsection{Some results for lattice gas models}
\label{ka:some_results}

This section is again intended for quick readers and summarizes
important results for the constrained lattice gas models. Details can
be found in Sec.~\ref{res}.

\begin{figure}
\begin{center}
\epsfig{file=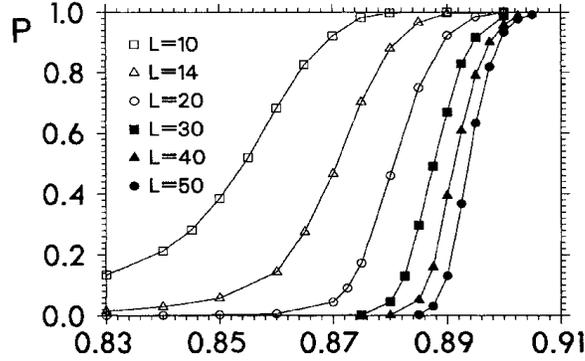, width=8cm}
\end{center}
\caption{Plot of the probability that a randomly chosen configuration
of the KA model has a backbone of frozen particles, against particle
density $c$ for a range of linear lattice sizes $L$ as
shown. From~\protect\cite{KobAnd93}.
\label{lg_fig1}
}
\end{figure}
The KA model has been studied mainly on cubic lattices in $d=3$, with
the constraint parameter set to $m=4$ (so that particles with four or
more neighbours are unable to move). As in the case of
spin-facilitated models (see Sec.~\ref{intro:irred}), one has to be
careful with {\bf reducibility effects} in finite-sized systems. The
analogue of a configuration belonging to the high-temperature
partition is, in a lattice gas, that all particles should be able to
move throughout the whole lattice eventually, \ie\ that no particles are
permanently blocked. KA~\cite{KobAnd93} suggested that one could
define a
``backbone'' of frozen particles by iteratively removing particles
from the system until all remaining particles are frozen; this in
principle only gives a lower bound on the number of frozen particles
in the system but in fact turns out to be an accurate approximation (see
Sec.~\ref{res:reduc}). An example of a backbone would be a $2\times2$
tube of particles that stretches across a finite system and, due to
periodic boundary conditions, connects back onto itself; there could also be
several such tubes with ``bridges'' between them etc.
Fig.~\ref{lg_fig1} shows the probability $p=p(c,L)$ for a random
configuration of density $c$ on a lattice of linear size $L$ to
contain a backbone. One sees that linear sizes of $L\approx20$ are
sufficent to have negligible reducibility effects up to $c=0.86$, but
that even for slightly larger densities (\eg\ $c=0.89$) much larger
systems are needed. This is
consistent with theoretical estimates. These are based on the close
link between the iterative process that defines the backbone and
so-called bootstrap percolation (see Sec.~\ref{meth:reduc}) and
predict that the threshold density $c_*(L)$ for which $p(c,L)=1/2$
converges to one only as $1-c_*(L)\sim 1/\ln(\ln L)$ or even more slowly;
conversely, the system size $L$ required for effective irreducibility
diverges as a {\em double} exponential of $1/(1-c)$ for $c\to 1$.

\begin{figure}
\begin{center}
\epsfig{file=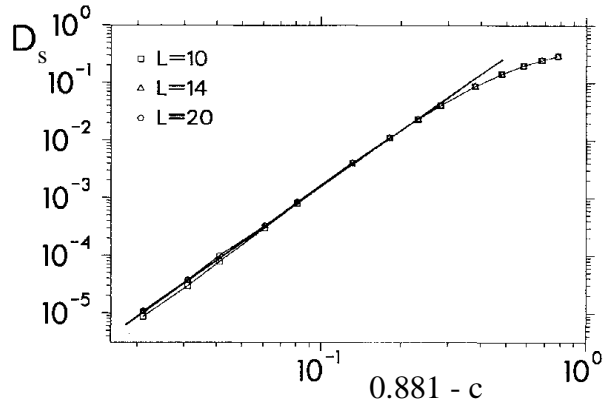, width=8cm}
\end{center}
\caption{Self-diffusion constant $D_s$ in the KA model as a function
of $c\dyn-c$, for system sizes $L$ as shown. From~\protect\cite{KobAnd93}.
\label{lg_fig2}
}
\end{figure}
To investigate the slowing down of the (equilibrium) dynamics with
increasing density, KA calculated the self-diffusion constant $D_s$
from the measured mean-squared displacement of particles as a function
of time; see after\eq{cs}. The results are shown in Fig.~\ref{lg_fig2}
and suggest that $D_s$ vanishes at $c\dyn \simeq 0.881$ according to a
power law $D_s \sim (c\dyn-c)^\phi$ with $\phi\simeq 3.1$. Data for
$D_s$ in the triangular lattice gas could also be fitted with the same
functional form~\cite{JaecKro94}, though there an alternative form $D_s \sim
\exp[-A/(1-c)]$ which predicts no dynamical transition at any $c<1$ also
provided a good fit; see Sec.~\ref{res:dyntrans}.

\begin{figure}
\begin{center}
\epsfig{file=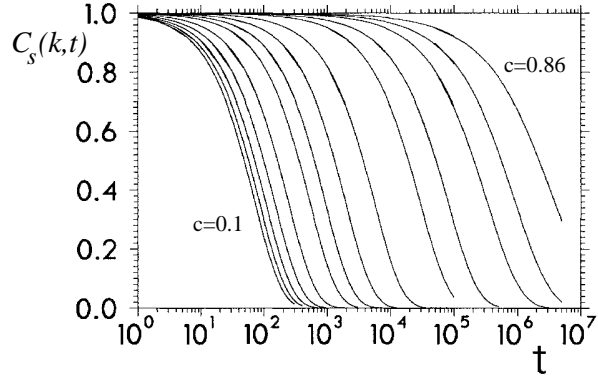, width=8cm}
\end{center}
\caption{Self-intermediate scattering function in the KA model plotted
against time, for the smallest possible ``wavevector'' $k=2\pi/L$ and
a range of densities $c$. Notice the
absence of a two-step relaxation with intermediate plateau even at
high density. From~\protect\cite{KobAnd93}.
\label{lg_fig3}
}
\end{figure}
{\bf Equilibrium relaxation functions} have also been studied for
constrained lattice gases. KA~\cite{KobAnd93} considered an analogue
of the intermediate self-scattering function $C_s$\eq{cs}, modified
suitably to take into account the lattice symmetry. The modification
consists in replacing $\exp(i\kv\cdot\Delta\rv)$ in\eq{cs} (where
$\Delta\rv \equiv \rv_a(t)-\rv_a(0)$ for short) by $(1/3)[\exp(ik\Delta
x)+\exp(ik\Delta y)+\exp(ik\Delta z)]$; in the corresponding van Hove
correlation function\eq{gs} this means that particle displacements are
measured in terms of the number of lattice planes traversed in the
$x$-, $y$- or $z$-direction. Typical results for the shortest ``wavevector''
$k=2\pi/L$ are shown in Fig.~\ref{lg_fig3} for a range of densities; KA
found that for the higher densities the long-time decay is well
described by a stretched exponential. At shorter times, the absence of
an intermediate plateau, and thus of a clear separation into
$\beta$- and $\alpha$-relaxation, is notable; KA argued that the $\beta$
relaxation was either absent or very weak (giving a plateau too close to the
initial value to be visible) because in the KA model particles either
diffuse or are completely stuck, rather than initially ``rattling'' in cages
formed by their neighbours. (Intriguingly, the {\em triangular}
lattice gas {\em does} exhibit two-step relaxations~\cite{KroJaec94};
see Sec.~\ref{res:statdyn}.) The density dependence of the relaxation
time $\tau$ extracted from the self-intermediate scattering function
depends on the wavevector $k$. KA~\cite{KobAnd93} found that for the
largest $k=\pi$, corresponding to distances of the order of the
lattice spacing, $\tau$ diverged as a power law
$\tau \sim (c_{\rm dyn}-c)^{-\phi'}$, with a value of $c\dyn\simeq
0.88$ compatible with that estimated from the self-diffusion constant
but with a different exponent $\phi'\simeq 5$.

\begin{figure}
\begin{center}
\epsfig{file=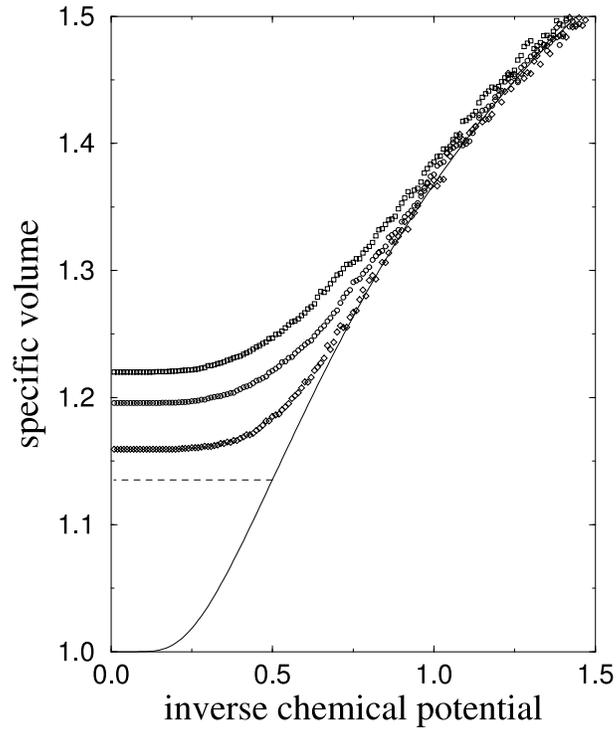, width=8cm}
\end{center}
\caption{Density as a function of chemical potential in the
grandcanonical KA model. (Plotted is the inverse density, \ie\ the
specific volume, versus the inverse chemical potential, to emphasize the
analogy with plots of $E$ vs $T$.) The analogue of a cooling run is
shown, where the chemical potential of the particle reservoir is
increased gradually over time; the curves from top to bottom
correspond to decreasing ``cooling'' rates. Notice that the system falls out of
equibrium when the inverse density approaches $1/c\dyn\approx1/0.881\approx
1.135$ (dashed line). From~\protect\cite{KurPelSel97}.
\label{lg_fig4}
}
\end{figure}
More recently, {\bf out-of-equilibrium effects} in constrained lattice gases
have also been investigated (see Sec.~\ref{res:noneq} for details),
using the versions of the KA model where either particle exchange with
a reservoir is allowed or the particle density can change under the
influence of gravity. Typical glassy features are observed. For
example, the analogue of cooling rate effects have been investigated
in the grandcanonical model by gradually increasing the reservoir
chemical potential~\cite{KurPelSel97}. Typical results are shown in
Fig.~\ref{lg_fig4}; as the system becomes more compressed, the density
appears to get stuck around $c\dyn$ even though the chemical potential
is increased further, demonstrating that the system falls out of
equilibrium. Cyclical compression and decompression then also lead to
hysteresis effects. Sudden ``crunches'', \ie\ increases in chemical
potential, have been studied as well and result in slow power-law
relaxation of the density. If during this relaxation mean-square
particle displacements and the conjugate two-time response function
are measured, one finds the remarkable result that the FDT plot has a
simple ``mean field'' form, consisting of two straight line segments;
and the slope of the out-of-equilibrium part can be understood on the
basis of an appropriate flat Edwards measure over frozen
configurations (see Sec.~\ref{res:landsc}). These exciting results are
only a beginning, however, and much remains to be done to
understand the origin of such apparent mean-field behaviour.

\subsection{Constrained models on hierarchichal structures}
\label{related:hierarchical}

Almost simultaneously with the first proposal of SFMs by Fredrickson
and Andersen, Palmer \etal introduced the idea of a whole {\em
hierarchy} of kinetic constraints, in a paper~\cite{PalSteAbrAnd84}
that has been instrumental in establishing the conceptual basis of
the field of KCMs.

\begin{figure}
\begin{center}
\epsfig{file=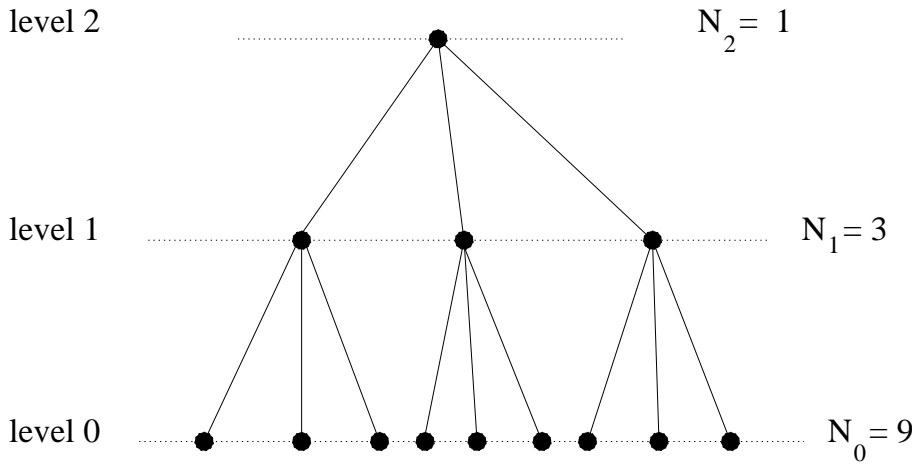, width=12cm}
\end{center}
\caption{A hierarchical tree with three levels $n=0,1,2$ containing
$N_0=9$, $N_1=3$ and $N_2=1$ spins respectively. The spins on level 0
have a short relaxation time $\tau_0$. Spins on level 1 relax more
slowly because each is constrained by the configuration of $\mu_0=3$
specific spins in the level below, as indicated by the solid
lines. The top spin on level 2 is similarly constrained by the
$\mu_1=3$ spins in level 1. The connectivity shown here is that of a
Cayley tree; Palmer \etal~\protect\cite{PalSteAbrAnd84} also allowed
for more general cases where a given spin can act as facilitator for
more than one spin in the next level above.
\label{fig_FELIX_1}
}
\end{figure}

In the model of~\cite{PalSteAbrAnd84} the microscopic degrees of
freedom are represented by spins that
live on a hierarchical tree (Fig.~\ref{fig_FELIX_1}) containing $N_l$
spins at levels numbered by $l=0, 1, \ldots$; $N_l$ decreases as one
moves up in the hierarchy with increasing $l$. Although initially
devoid of any microscopic interpretation the spins can be thought of
as representing cooperative regions in a glass of lengthscales
increasing with $l$, with the bottom level 0 corresponding to the
dynamics of single spins or particles. The relaxation time of large
regions should depend on that of smaller ones, and this is modelled by
assuming that a spin in level $l+1$ can relax only if a given set of
$\mu_l$ facilitating spins in level $l$ are in one particular
configuration out of the $2^{\mu_l}$ possible ones. If the spins are
assumed to be up or down with equal probability, the typical
relaxation time for the spins in level $l+1$ is
\be
\tau_{l+1}=2^{\mu_l}\tau_l
\label{eq1S315}
\ee
which gives
\be
\tau_l=\tau_0 \exp\left[(\ln 2) \sum_{k=0}^{l-1}\mu_k\right]
\label{eq2S315}
\ee
If $N$ is the total number of spins and $w_l=N_l/N$ is the fraction at
level $l$, then the equilibrium correlation function can be estimated
by averaging the autocorrelation over all spins, \ie\ over all levels, 
giving
\be
C(t)=\sum_{l=0}^{\infty}w_l\exp(-t/\tau_l)
\label{eq3S315}
\ee
for a hierarchy with infinitely many levels. (A formally similar
solution was also found by Ogielski and Stein in a model of particle
hopping on hierarchical structures~\cite{OgiSte85}.) Palmer
\etal~\cite{PalSteAbrAnd85} argued that since realistic systems are
not expected to have the assumed sharp partitioning into discrete
levels, one could equally or better regard $l$ as a continuous
variable and replace the sum in\eq{eq3S315} by an integral; see
also~\cite{Zwanzig85}. They investigated the asymptotic behaviour of
the correlation function for different choices of $w_l$ and
$\mu_l$~\cite{PalSteAbrAnd84}. In particular, assuming
$\mu_l=\mu_0/l^p$ and $w_{l}=w_0/\lambda^l$ (with $\lambda>1$) they
found stretched exponential behaviour for the correlation function.
The maximum relaxation time, obtained from\eq{eq2S315} for
$l\to\infty$, is $\tau_{\rm max}=\tau_0\exp[(\mu_0\ln 2)/(p-1)]$ which
is reminiscent of the VTF law\eq{VTF} if $p$ depends on temperature
and vanishes linearly with $T$ in the vicinity of $T_0$. If one
instead assumes that $\mu_l\propto w_l$ at all levels, then the
correlation function shows a slow logarithmic decay in an intermediate
time regime, independently of the precise $l$-dependence of
$w_l$~\cite{BrePra01}. The effects of heating and cooling cycles have
also recently been investigated in an appropriate
generalization~\cite{PraBre01b} of the hierarchical model.

The models discussed above have been very influential conceptually, in
emphasizing that glassy dynamics could be caused by kinetic constraints
linking a hierarchy of degrees of freedom. However, the large number of
parameters $\mu_l$ and $w_l$, which are difficult to assign on physical
grounds, is a drawback if one wants to make quantitative statements about
the behaviour of these models. We will therefore omit them from further
discussion. One interesting special case that we will cover, however, is
the $(a,a-1)$ Cayley tree model discussed in Sec.~\ref{model:SFM}. This can
be regarded as a concrete realization of the hierarchical scenario
discussed above: in the Cayley tree there is a finite number of levels
$l=0\ldots L$, with $N_l=(a-1)^{L-l}$, $\mu_l=a-1$, and the sets of spins
in each level that facilitate the relaxation of different spins in the
level above are chosen so that they do not overlap each other. Notice however
that the approach by Palmer \etal\ effectively fixes the up-spin concentration
to $c\eql=1/2$ (corresponding to infinite temperature), while this is
normally regarded as an important tunable parameter in the Cayley tree
models.

\subsection{Models inspired by cellular structures}
\label{topological}

\begin{figure}
\begin{center}
\epsfig{file=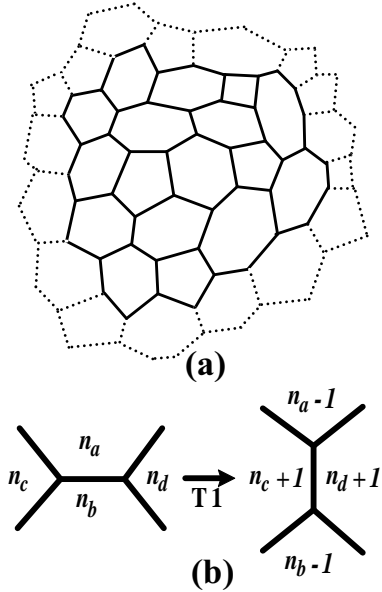, height=8cm}
\end{center}
%
\caption{(a) A topological froth. (b) A T1 or neighbour-switching
move, in which two cells gain one edge and two others lose one
edge. From~\protect\cite{AstShe99}.
\label{fig_FELIX_2}
}
\end{figure}

Kinetically constrained models have also been inspired by the study of
soap froths and other cellular patterns. These models incorporate
topological constraints as kinetic restrictions in the transition
rules. (They also have some similarities with tiling models, see
Sec.~\ref{related:tiling} but, especially in the lattice versions
discussed below, are more amenable to analytical investigation.) The
simplest cellular pattern in the plane is a hexagonal tiling, comprising
only six-sided cells that have six neighbours each. However, most
cellular structures in nature, such as froths and biological tissue,
are disordered. Aste and Sherrington~\cite{AstShe99} proposed a model
which only keeps track of the topology of the cellular structure, \ie\
of which cells are neighbours to each other; see
Fig.~\ref{fig_FELIX_2}(a). If cell $i$ has $n_i$ sides, then the average
value of $n_i$ is six from the Euler theorem; in the perfect hexagonal
arrangement, one even has $n_i=6$ for each individual cell. The
deviation from this arrangement can be characterized by an energy
function
\be
E = \sum_i (n_i-6)^2
\ee
which contains no interactions. The kinetic constraint arises from the
fact that the only allowed transitions are so-called T1 moves, where
four cells exchange neighbours; see Fig.~\ref{fig_FELIX_2}(b). Two cells
thus gain an edge, while the other two lose an edge, and a proposed
move is accepted with the usual Glauber probability
$1/[1+\exp(\beta\dE)]$. For high temperatures, the equilibrium
structure of the cellular pattern is disordered, with many cells with
$n_i\neq 6$. As $T\to 0$, on the other hand, only a small number of
pentagonal ($n_i=5$) and heptagonal ($n_i=7$) cells, effectively
defects in a hexagonal structure, are present. There are then very few
moves which do not increase the energy, and the dynamics becomes
dominated by activated processes. The only freely diffusing defect
structures are in fact 5-7 pairs of cells~\cite{DavShe00}, which can
annihilate when they meet or be absorbed by isolated pentagons or
heptagons.

The topological froth model discussed above has the complication that
its equilibrium behaviour is nontrivial to work out, requiring a sum
over all possible topological arrangement of cells. The situation is
clearer in a lattice analogue which has genuinely trivial equilibrium
behaviour. In this lattice model, three-state spins
$\s_i\in\{-1,0,1\}$ occupy the cells of a hexagonal
lattice~\cite{DavSheGarBuh01,SheDavBuhGar02}. The $\s_i$ correspond to
the local deviations $n_i-6$ from the optimal hexagonal structure in
the off-lattice model. The energy is therefore defined as
\be
E=\sum_i \s_i^2
\label{hexagonal}
\ee
The initial configuration is chosen so that $\sum_i \s_i=0$ (in analogy
with $\lav n_i\rav = 6$). The kinetic constraint is modelled on that
of the off-lattice model: the only allowed transitions are those in
which the spins in two neighbouring cells are increased by one and the
spins in their two common neighbours decreased by one, or vice
versa. Moves that would produce spins outside the range $-1,0,1$ are
of course forbidden. At low temperatures in this model, ``dimers''
composed of a 
$+1$-spin and a $-1$-spin can diffuse (in a zig-zag motion) across a
background of largely $0$-spins. These are the analogues of the 5-7
pairs of the topological model. They come in six different possible
orientations, and can annihilate with an ``anti-dimer'' of the
opposite orientation when the meet, or be absorbed by isolated
defects, \ie\ $\pm 1$-spins. Dimer diffusion dominates the fast
dynamics of the model. On longer, activated timescales, isolated
defects themselves can diffuse by creating freely diffusing dimers at
an energy cost of $\dE=2$. Overall, the model produces glassy
phenomena similar to the original off-lattice version.  A variant with
$E=-\sum_i \s_i^2$ has also been considered; this is still
non-interacting but has a highly degenerate ground state, leading to
subtle modifications of the low-temperature behaviour detailed
in~\cite{DavSheGarBuh01}. Finally, the model can be further simplified
by using a square rather than hexagonal lattice, without qualitatively
changing the behaviour~\cite{SheDavBuhGar02}.

It is clear from this overview that the above models inspired by
cellular structures can all be understood by mappings to defects which
can diffuse and ``react'' with each other; this reaction-diffusion
behaviour gives rise to characteristic power-laws in the relaxation
functions. The timescales involved are activated, so that these KCMs
are appropriate for modelling strong glasses. As explained for the
model\eq{hexagonal}, two separate timescales can be involved and give
two-step relaxations, as well as aging effects when the longest
timescale exceeds the experimental or simulation time window.

\subsection{Models with effective kinetic constraints}
\label{model:effective}

Recently it has been realized that there exists a class of models
which are conventionally formulated in terms of nontrivial
energy functions and unconstrained dynamics, but which can be mapped to
noninteracting defects with constrained dynamics;
see~\cite{Garrahan02} for a review. The simplest example is a Glauber
Ising chain, with energy function
\be
E = - J \sum_i \s_i \s_{i+1}
\ee
in terms of conventional spins $\s_i=\pm1$. One can introduce defect
variables $n_i = (1-\s_i\s_{i+1})/2$, with $n_i=0$ and $n_i=1$
corresponding to the absence and presence of a domain wall,
respectively. Importantly, the mapping from the $\s_i$ to the $n_i$
is one-to-one---subject to appropriate boundary conditions, \eg\ an
open chain with the left spin fixed---so that either set of variables
can be used to specify a configuration. The mapping to defect
variables has two consequences. On the one hand, the energy becomes
$E= 2J \sum_i n_i$ up to a constant, so that the defects are
noninteracting. On the other hand, while the dynamics is simple in
terms of spin flips, it is effectively constrained in terms of the
$n_i$ since only simultaneous changes of pairs of neighbouring $n_i$'s
are allowed.

More interesting than the trivial Ising chain example are higher
dimensional models. In $d=2$, there are two cases where a one-to-one
mapping to defects is possible, subject again to appropriate boundary
conditions~\cite{Garrahan02}. The ``triangle model'' has spins on a
triangular lattice with triplet interactions in the {\em downward}
pointing triangles only~\cite{NewMoo99,GarNew00}
\be
E=-J \sum_{ijk\in\bigtriangledown}
\s_i \s_j \s_k
\ee
The defect variables $n_i = (1-\s_i\s_j\s_k)/2$ live on the centres of
the downward triangles, which themselves form a dual triangular
lattice that is isomorphic to the original one; the energy is again
$E=2J\sum_i n_i$ up to a constant. Spin flip dynamics in the original
model implies that the only allowed transitions between defect
configurations are the inversions of three $n_i$ at the corners of any
{\em upward} pointing elementary triangle of the dual lattice. At low
temperatures, where most defects ($n_i=1$) are isolated, this
constraint slows the dynamics: flipping any triangle of defect
variables then leads to a state with an additional defect, and
requires an activation over the energy barrier $2J$. In fact, one can
show that there is a whole hierarchy of energy barriers, and an
associated hierarchy of slow timescales, arising from the relaxation
of defects arranged in the corners of equilateral triangles with side
length a power of two.  As discussed in detail in
Sec.~\ref{res:relax}, the situation is in fact very similar to that in
the East model, and the longest relaxation timescale exhibits an EITS
divergence characteristic of fragile glasses.

A second model in $d=2$ has plaquette interactions on a square
lattice~\cite{AlvFraRit96,Lipowski97b}
\be
E = - J \sum_{ijkl\in \Box} \s_i \s_j \s_k \s_l
\ee
with defect variables $n_i = (1- \s_i \s_j \s_k \s_l)/2$ sitting on
the dual square lattice, and elementary moves being the simultaneous
flipping of four defects around a plaquette of the dual lattice. The
seemingly innocent change of lattice structure from the triangle model
has a profound effect on the dynamics: two neighbouring defects along
one of the lattice directions can now diffuse freely along the
orthogonal direction, and the diffusion of these defect-pairs gives
the model strong glass characteristics, with timescales growing only
in an Arrhenius fashion as temperature is lowered. 

In three dimensions, models similar to those above could be
constructed on \eg\ an f.c.c.\ (face-centred cubic) lattice with
four-spin interactions on
downward-pointing tetrahedra, or a cubic lattice with eight-spin
interactions between spins around the elementary
cubes~\cite{Garrahan02}. We note in passing that closely related to
the $d=2$ plaquette model are the so-called gonihedric spin models,
which normally include additional two-spin interactions and exhibit
some glassy features as well as interesting metastability
effects~\cite{LipJoh00,LipJoh00b,LipJohEsp00,DimEspJanPra02}.

In terms of the defect variables, the static equilibrium behaviour of
the above models is of course trivial. In the following, we always
assume that the coupling $J$ in the original model is chosen so that
$E=\sum_i n_i$ in terms of the defect variables, giving again $c=\lav
n_i\rav = 1/(1+e^{\beta})$. The equilibrium properties of the
underlying spin system can be worked out from that of the $n_i$. In
particular, one finds that $\lav \s_i\rav = 0$ and that spin
correlation functions are nonvanishing only if they can be expressed
as a product of a finite number of defects (or, more precisely, of the
variables $2n_i-1\in \{-1,+1\}$). In the triangle model, for example, the
simplest non-vanishing correlation function is that of the three
spins at the corners of an elementary downward
triangle~\cite{GarNew00}.

It is clear that, in the description in terms of defect variables, the
triangle and plaquette models are quite similar to the lattice
versions of the topological froth described in the previous section:
only certain groups of the elementary variables are allowed to flip
together. An advantage is that these kinetic constraints are not
imposed ad hoc, but result naturally from the dynamics of the
underlying spin system. For uniformity of terminology, we will
normally refer to the 
defect variables $n_i$ as spins when no confusion with the variables
$\sigma_i$ of the underlying unconstrained spin system is possible.

\subsection{Related models without explicit kinetic constraints}
\label{model:other}

This section gives an overview of some models which do not strictly
have kinetic constraints but which in many cases share some features
of KCMs. In some cases the thermodynamics of these models may be not
trivial, in other cases it is very simple although the system may
present a critical point, for instance at zero temperature. A
simplifying feature of the models discussed in this section is that
the dynamics is normally trivially irreducible.

\subsubsection{Ordinary Ising models}
\label{related:ising}

Ordinary Ising models with ferromagnetic nearest-neighbour
interactions and {Glau\-ber} dynamics are the ``baseline'' models for
SFMs. In spite of the absence of kinetic constraints, they display
some features associated with glassy dynamics, especially when
quenched to low temperatures near or below their critical points; see
\eg~\cite{GodLuc02} for a recent overview. The simplest example is the
one-dimensional Glauber chain, for which many exact results were
already obtained by Glauber himself~\cite{Glauber63}; in fact a fully
exact diagonalization of the master equation can be obtained via a
mapping to free fermions~\cite{Felderhof71}. The critical point is at
$T_c=0$, where the model coarsens by diffusion and annihilation of
domains walls, with the typical domain size growing as $l(t) \sim
t^{1/2}$. Two-time correlation and response functions obey simple
scaling with $\tw/t$, or equivalently with $l(\tw)/l(t)$ (see
\eg~\cite{PraBreSan97,GodLuc00,LipZan00}) as expected from general
arguments for coarsening models~\cite{Bray94}. A nontrivial FDT plot is
obtained in the limit of long times~\cite{GodLuc00,LipZan00} but is
nontrivially dependent on the observable
considered~\cite{SolFieMay02,MaySol02}. In equilibrium at low but
nonzero temperatures, relaxation functions also show stretched
exponential behaviour at intermediate times~\cite{BrePra96};
hysteresis effects are found when the system is heated and cooled
cyclically (see Sec.~\ref{res:heatcool}). In higher dimensions,
finally, coarsening at $T=T_c>0$ and below need to be distinguished
and give different scaling relations for two-time
quantities~\cite{GodLuc00b,GodLuc02}.

\subsubsection{Urn models}
\label{model:urn}

This category of models has recently received considerable attention.
Urn models do not contain local kinetic constraints; instead a
conservation law acts as a global constraint leading to cooperative
behaviour. Their equilibrium properties are very simple and usually
independent of the dimensionality. Like KCMs---and in contrast to
other models with standard second-order phase transitions, such as the
Ising models discussed above---they do not have a large equilibrium
correlation length at low temperatures. Instead, they show a
condensation transition, at either zero or nonzero $T$.

Urn models generally are comprised of a number $M$ of urns or boxes and $N$
particles distributed among these boxes. A configuration is specified
through the occupancies $n_r$ in the boxes $r=1\ldots M$. Each set of
such occupation numbers has assigned to it a degeneracy factor which
encodes whether the particles are regarded as distinguishable or
indistinguishable. The original urn model, introduced by
Ehrenfest~\cite{Ehrenfest90} at the beginning of the 20th century to
prove that thermal equilibrium in the microcanonical ensemble
corresponds to the maximum entropy state, has $M=2$ urns and a large
number $N$ of particles.  The Backgammon model~\cite{Ritort95},
which stimulated renewed interest in urn models, instead considers the
limit $N,M\to\infty$ with $N/M=\rho$ held constant; cooperative glassy
behaviour then appears if the energy function is defined
appropriately. Many other aspects of the model and a number of
variations have since been
studied~\cite{FraRit95,GodBouMez95,FraRit96,GodLuc96,FraRit97,GodLuc97,
Lipowski97,MurKeh97,PraBreSan97c,DroGodCam98,GodLuc99,AroBhaPra99,
GodLuc01,GodLuc02b,LeuRit02}.

In the most general formulation, the energy function of urn models is
written as
\be
E=\sum_{r=1}^MF(n_r)
\label{eq1S322} 
\ee
where $F(x)$ is an arbitrary function subject only to the condition
that it must yield a well defined thermodynamics.  To show the type of
global constraint present in this type of model it is useful to work
out the partition function in the canonical ensemble,
\be
{Z}_{\rm c}(N,M)=\sum_{\{n_r\}}D(\lbrace
n_r\rbrace)\exp\left[-\beta\sum_{r=1}^MF(n_r)\right]
\delta\left(\sum_{r=1}^Mn_r-N\right)
\label{eq2S322} 
\ee
where $D(\lbrace n_r\rbrace)$ is the degeneracy factor
$D=\prod_{r=1}^Md(n_r)$, with $d(n_r)=1/n_r!$ for distinguishable
particles and $d(n_r)=1$ for indistinguishable ones. The interesting
aspect of\eq{eq2S322} is the fact that, although the
energy function\eq{eq1S322} is noninteracting, particle conservation as
expressed through the (discrete) delta function makes the
thermodynamics nontrivial and allows phase transitions to occur. To
actually work out the partition function, one switches to the
grandcanonical ensemble to eliminate the global constraint, which yields
\be
{Z}_{\rm gc}=\sum_{N=0}^{\infty} z^N{Z}_{\rm c}(N,M)=
e^{MG(\beta,z)}
\label{eq3S322} 
\ee
with $z=\exp(\beta\mu)$ the fugacity and $G(\beta,z)$ given by
\be
G(\beta,z)=\ln\sum_{n=0}^{\infty} z^nd(n)\exp[-\beta F(n)]
\label{eq4S322}
\ee
The equation of state in this type of models relates the three
variables $T$, $z$, and density $\rho=N/M$ by
\be
\rho=\frac{\lav N\rav}{M}=\frac{\partial G(\beta,z)}{\partial \ln z}
\label{eq5S322}
\ee
The equilibrium properties discussed up to now are the same in the
canonical and grandcanonical ensembles. Non-equilibrium properties
differ substantially, on the other hand, and we are interested only in
the canonical case where the global constraint induces cooperativity.

The allowed transitions in the dynamics of urn models are moves of
individual particles from a ``departure'' to an ``arrival'' box; move
proposals are accepted according to the conventional Metropolis rule
which depends on the energy change $\dE$ in the move. (We discuss the
dynamics here in the framework of a discrete-time Monte Carlo
simulation.) To fully define the dynamics one still needs to specify
how departure and arrival boxes are chosen. For the case of
distinguishable particles, a departure box is picked with probability
proportional to its occupation number $n_r$; this is equivalent to
choosing a {\em particle} to move at random. For indistinguishable
particles, each box has the same probability of being chosen as the
departure box. Godr{\`{e}}che and Luck~\cite{GodLuc01,GodLuc02b}
called these two types of dynamics ``Ehrenfest class'' and ``Monkey
class'', respectively. The arrival box is always picked at random from
all boxes connected to the departure box; which boxes are connected
defines the geometry of the model. Simplest is the mean-field
geometry, where all boxes are connected to each other. It can yield
exactly solvable models and has been the focus of most recent
work. More complicated is the short-range case, where boxes are
located on a finite-dimensional lattice and particles can be moved
between neighbouring boxes only.

In summary, urn models are defined by specifying (a) the energy
function $F$, (b) whether particles are distinguishable or not,
corresponding to Ehrenfest class and Monkey class dynamics
respectively, and (c) the geometry of connections between boxes. The
resulting behaviour is very rich, and even a change in only one of the
features (a)-(c) can change the dynamics completely.  Two specific urn
models which lead to interesting glassy dynamics are as follows.

\begin{enumerate}

\item The Backgammon model has mean-field geometry, particles are
distinguishable and $F(n)=-\delta_{n,0}$ so that the local energy is
either $-1$ or $0$ depending on whether the box is empty or not. This
model shows a $T=0$ condensation transition---where all particles
gather in one box---and typical relaxation timescales increasing in
Arrhenius fashion as $T$ is lowered. The interesting feature of this
model is that the system can evolve without having to surmount energy
barriers: relaxation can always proceed by moves (those which move
particles into boxes that are already occupied) which do not increase
the energy. Instead there are entropy barriers, created by a
bottleneck in the number of such escape ``directions'' from a given
configuration. This bottleneck appears because at low $T$ only a few
boxes are occupied; moves where a particle lands in an occupied box
are then very rare.

\item Zeta-urn models also have mean-field geometry, but particles are
indistinguishable (Monkey class dynamics) and $F(n)=\ln(n+1)$. This
model shows a $T>0$ condensation transition with a $T-\rho$ phase
diagram where a critical line separates regions of different dynamical
behaviour. Much effort has been gone into the analytical description of
the dynamics along this critical line.

\end{enumerate}

Urn models are schematic approaches which allow a number of general
questions about non-equilibrium dynamics to be investigated
explicitly. In particular, the Backgammon model and its variants are cases
where slow dynamics is determined by the presence of entropy barriers
that slow down the dynamics (see above): the system has to attempt
many moves before finding a downhill direction in energy. The main
difference to models dominated by energy barriers is that the latter
arrest completely at $T=0$ after short-time relaxations are complete,
whereas models with entropy barriers relax even at $T=0$; this
relaxation dominates also the dynamics at $T>0$ until a crossover at
long times where activation effects come into play. Interestingly, the
configurational bottleneck created by the entropy barriers induces a
typical relaxation time~\cite{Ritort95} of activated (Arrhenius)
character. Much work remains to be done on urn models with nontrivial
spatial structure, to understand for example whether they might
display cooperative effects reminiscent of those found in real glasses
(resulting in \eg\ fragile behaviour of relaxation timescales).

\subsubsection{Oscillator models}
\label{model:osc}

Another family of models which share some similarities with
kinetically constrained models are oscillator models. These are
mean-field models comprising an ensemble of uncoupled linear
oscillators with Monte Carlo dynamics. These models have neither local
nor global kinetic constraints. Nevertheless, they share some
similarities with KCMs in that there is no
interaction among oscillators---making the thermodynamics trivial,
with no phase transition even at zero-temperature---while the dynamics
is glassy due to the dynamical rules. In this respect they are simpler
than the urn models discussed above where, in some cases, a
condensation transition may take place due to the effective
interaction induced by particle conservation. The slow-down in the
dynamics at low temperatures and long times is caused by the low rate
at which proposed Monte Carlo moves are accepted; this low acceptance
rate could loosely be viewed as a kinetic ``constraint'' generated by the
dynamics itself.

Originally, oscillator models were introduced indirectly in the
analysis of the Monte Carlo dynamics of the spherical
Sherrington-Kirkpatrick model, which can be mapped to a set of
disordered harmonic
oscillators~\cite{BonPadParRit96a,BonPadParRit96b}. The ``oscillator
model'' proper is obtained by simplifying this to an ensemble of
identical harmonic oscillators~\cite{BonPadRit98}. It is defined by
the energy function
\be
E =\frac{K}{2}\sum_{i} x_i^2
\label{eq1S323}
\ee
where the $x_i$ are the real-valued displacement variables of the $N$
oscillators and $K>0$ is a Hooke constant. The equilibrium properties
are trivial due to the absence of interactions, but the Monte Carlo
dynamics couples the oscillators in a nontrivial way. Moves are
proposed according to
\be
x_i\to x'_i=x_i+\frac{r_i}{\sqrt{N}}
\label{eq2S323}
\ee
where the $r_i$ are Gaussian random variables with zero mean and
variance $\Delta^2$, and are accepted according to the usual
Metropolis rule. Each move is a parallel update of the whole set of
oscillators. Both the energy function\eq{eq1S323} and the dynamics as
defined by\eq{eq2S323} are invariant under rotations in the
$N$-dimensional space of the $x_i$. This symmetry makes the dynamics
exactly solvable, so that questions about \eg\ aging, effective
temperatures and FDT violations can be answered analytically.

At low temperatures the oscillator model displays slow dynamics, as
can be easily understood from the following argument. For small $T$
the equilibrium energy, which from equipartition is $E=NT/2$, is very
small. Correspondingly, equilibrium configurations are located in a
small sphere around the point
$x_i=0$, with radius of order $R=({NT}/{K})^{1/2}$. This sphere
shrinks to zero as $T\to 0$, so that the vast 
majority of new configurations proposed according to\eq{eq2S323} fall
outside, producing very small Metropolis acceptance probabilities
$\exp(-\beta \dE)$. In fact, at $T=0$ the system never reaches
equilibrium, and the radius of the configuration space sphere explored
by the system shrinks to zero logarithmically in time.

Some of the main known results for the oscillator model (see
\eg~\cite{Garriga02}) are as follows. The relaxation time shows
Arrhenius behaviour at low temperatures; as in KCMs,
this occurs even though there are no static interactions.
At low temperatures {\em aging} effects occur, with correlation
functions and responses showing simple aging scaling with
$(t-\tw)/\tw$ up to subdominant logarithmic corrections. The effective
temperature defined via the fluctuation-dissipation ratio
(Sec.~\ref{intro:fdt}) can be computed analytically. Surprisingly,
even in the out-of-equilibrium dynamics at $T=0$ it is linked to the
time-dependent value of the energy by the equipartition relation
$E(\tw)=NT\eff(\tw)/2$.

A number of variants of the oscillator model have been considered, all
sharing the feature that oscillators do not interact. For example,
Nieuwenhuizen and coworkers~\cite{Nieuwenhuizen98b,LeuNie01b,LeuNie01}
studied a model with (spherical) spin variables in addition to
oscillators. The new variables are used to mimic fast relaxation
processes not contained in the original formulation; this imposed
separation 
into slow and fast degrees of freedom mimics the $\alpha$- and
$\beta$-relaxation processes in supercooled liquids. We will not
detail results for this model below, but refer to~\cite{LeuNie02} for
a recent overview.

\subsubsection{Lattice gases without kinetic constraints}

We discussed in Sec.~\ref{model:lattice_gases} the KA model, a lattice
gas with a trivial energy function but glassy dynamics produced by local
kinetic constraints. The converse approach, where glassy behaviour
results from unconstrained dynamics but nontrivial interactions
between the particles has of course also been explored. A simple example
is the so-called hard-square lattice gas, where particles moving on a
square lattice are not allowed to occupy n.n.\ sites. This interaction
results naturally if the particles are visualized as hard squares
oriented at $45^{\rm o}$ degrees to the lattice axes and with
side length $\sqrt{2}a$, $a$ being the lattice constant. This model has
a nontrivial equilibrium phase transition at particle density $\approx
0.37$, above which particles are located preferably on one of two
sublattices. As the maximum density of $1/2$ is approached, the
dynamics becomes very slow and shows glassy features. 
Since we focus in this review on models with essentially trivial
thermodynamics, we will only touch on results for the hard-square
lattice gas occasionally and refer the interested reader
to~\cite{ErtFroJaec88} and the recent review~\cite{Jaeckle02} for
details.

The unconstrained baseline version of the KA model, a lattice gas
without interactions---except for the standard hard-core repulsion
that allows at most one particle per site---and without kinetic
constraints has been studied under the name of ``sliding block
model''. The name arises from the children's puzzle, where blocks can
be slid around only by moving them into a neighbouring hole. The
interesting limit is normally that of very low vacancy concentration;
the vacancies then just perform random walks. The movement of the
particles, however, is nontrivial, and the typical displacement of a
given particle shows a stretched exponential increase with time at
short times. We refer to~\cite{AjaPal90} for simulations and
references to earlier theoretical work on this type of model.

\subsubsection{Tiling models}
\label{related:tiling}

For completeness we now discuss tiling models. This is a slight
departure from
our overall philosophy since in these models the energy function chosen
leads to nontrivial equilibrium behaviour. Nevertheless, the following
attractive features make them worthy of a brief mention: (a) crystalline
phases can be included in addition to amorphous ones, (b) irreducibility is
trivial to establish and (c) the idea of a cooperative length scale is
included right from the beginning.  We focus on the $j$-tiling model
introduced by Stillinger and Weber~\cite{StiWeb86} and developed in detail
by Weber, Fredrickson and Stillinger~\cite{WebFreSti86}.

Tiling models are systems made up of non-overlapping square tiles, which can
fragment into smaller tiles or, conversely, be joined together into a
larger one. Consider a square lattice with $N=L^2$ sites and periodic
boundary conditions, covered without gaps by non-overlapping square tiles
of all possible side lengths $j=1\ldots L$. Let $n_j$ denote the number of
squares of size $j\times j$; these numbers satisfy the global constraint
\be
\sum_{j=1}^Lj^2n_j=N
\label{eq1S325}
\ee
The ideal amorphous packing of particles is represented by a single
tile of size $L$, while between smaller tiles it is assumed that there
is a strain energy cost proportional to the contact length, arising
from a mismatch in the particle packing in neighbouring
tiles. This gives the energy function
\be
E=2\lambda\sum_{j=1}^Ljn_j
\label{eq2S325}
\ee
A crystal phase can be added by designating tiles of a certain size
$j_0\times j_0$ as crystalline and adding a term $-\mu n_{j_0}$ to
$E$. The equilibrium behaviour of this model cannot be solved
exactly even at infinite temperature, but a perturbation expansion
around $\beta\lambda=-\infty$ gives a first order phase transition to the
configuration with a single macroscopic tile around $\beta\lambda_c=0.27\pm
0.1$.  This is confirmed by series expansions and mean-field Flory
approximations~\cite{StiWeb86} and transfer matrix calculations and
upper bound estimates~\cite{BhaHel87}.

The dynamics of tiling models is made interesting by kinetic
constraints on the possible fragmentation and aggregation processes.
One possibility (``minimal aggregation'') is to allow tiles of
side length $pq$ to divide into $p^2$ tiles of side length $q$ if and
only if $p$ is the smallest prime factor (larger than 1) of $pq$. The
corresponding rule applies to the reverse aggregation
process. Aggregation and fragmentation rates are given by the standard
Metropolis rule; in addition, however, a slowing down of the dynamics
for large tiles is implemented by including an additional factor of
$\alpha^{2pq(p-1)}$ in the rates, with $\alpha\le 1$. In an
alternative version of the dynamics (``boundary shift''), tiles of
side length $(p+1)$ fragment into a tile of side length $p$ and an
$L$-shaped band of $(2p+1)$ unit tiles~\cite{WebSti87}. Both types of
dynamics are trivially irreducible since all configurations can be
transformed into the one with all unit tiles. It is not clear,
however, which dynamical rules are most appropriate for modelling
glasses, and this may be one reason why tiling models have received
much less attention than SFMs. Some generic glassy features have
nevertheless been found~\cite{WebFreSti86,WebSti87}. Energy-energy
autocorrelation functions in equilibrium, for example, can be fitted
to stretched exponentials (but progressively cross over to power-law
decays at long times as the glassy regime is approached). Typical
relaxation times derived from these correlation functions show
superactivated temperature-dependences; cooling rate effects on the
energy have also been observed. Beyond this brief overview, we do not
consider tiling models further in this review; nonetheless, as
explained at the beginning of this section, they have some attractive
features which may make them worth revisiting in future work.

There are a number of extensions of the $j$-tiling models in the
literature. E.g.\ Bhattacharjee~\cite{Bhattacharjee89} (see
also~\cite{RaoBha92}) considered the equilibrium behaviour of a model with
an additional term of the form $\sum_j j^{\alpha}n_j$ in the energy
function. Random tiling
models~\cite{Widom93,Kalugin94,JarJoh96,DegNie96,Kalugin97} or Wang tiles
for quasicrystals have also been studied, though again with a focus on
equilibrium; an exception for the latter case is the analysis of the
dynamics without kinetic constraints in~\cite{LeuPar00}.

\subsubsection{Needle models}
\label{related:needle}

Models of thin needle-shaped particles interacting only via hard
(excluded-volume) interactions and subject to Newtonian or diffusive
Brownian dynamics may not appear related to KCMs at first sight. They
do have interesting glassy dynamics accompanied by trivial equilibrium
behaviour, however, and are therefore included here.

Frenkel and Maguire~\cite{FreMag81,FreMag83} investigated the Newtonian
dynamics of a gas of infinitely thin, hard rods of length $L$ at number
densities $\sim 1/L^3$. Since the average excluded volume for rods of zero
diameter is zero, all static properties of the system are those of an ideal
gas. The equilibrium dynamics are nontrivial, but diffusion constants and
autocorrelation functions vary smoothly with density even at large
normalized densities $L^3\rho$, so that the model does not present
pronounced glassy features. Edwards, Evans and
Vilgis~\cite{EdwEva82,EdwVil86} considered the same model at finite but
still small needle diameter $D\ll L$, and larger densities $\rho$ of order
$1/(DL^2)\gg 1/L^3$. They argued that in this regime the rotational
diffusion of needles is so strongly suppressed that they can effectively
only translate along their axis. Because of the nonzero $D$, other needles
will impede this one-dimensional diffusive motion, however; each needle can
only move if enough of its neighbours move out of the way. A
self-consistency argument then suggests that the diffusion constant
decreases to zero at some finite value of $DL^2\rho$, and that on
approaching this value relaxation times should diverge~\cite{EdwEva82}; a
more sophisticated version of the theory can also reproduce a VTF-like
divergence of the inverse diffusion constant. Unfortunately, in roughly the
same density regime an equilibrium phase transition occurs to a state of
nematic ordering~\cite{Onsager49}, where the needles align with each other
rather than being randomly oriented as assumed in the calculation. The
glassy phenomena predicted by Edwards \etal\ would therefore be observable
only after a sufficiently fast density increase which avoids this
transition.

To eliminate the possibility of equilibrium phase transitions, other models
postulate that the needles are fixed to a crystal lattice so that only
rotational motion is allowed. A number of variants have been considered,
including attaching the midpoints of the needles to an
f.c.c.~\cite{RenLowBar95} or b.c.c.~\cite{JCBCTFLCF02} lattice, or their
endpoints to a cubic or square lattice~\cite{ObuKobPerRub97}; in the last
case the motion of the needles was assumed to take place in the
(three-dimensional) half-space to one side of the lattice plane. In the
limit of vanishing needle diameter, assumed throughout, equilibrium
properties are again trivial. The dynamics can become glassy, however, for
ratios $L/a$ of needle length and lattice constant above order one: the
motion of each needle is then restricted by those around it, leading to
``orientational caging''.

\subsubsection{Models without detailed balance}
\label{related:nobalance}

Since our main concern is with models with trivial equilibrium
behaviour but interesting dynamics, we will not discuss models without
detailed balance, for which the nature of any stationary state can be
highly nontrivial. A recent review of some models of this type can be
found in~\cite{Stinchcombe01}. Here we only give a few examples that
are closely related to KCMs. In SFMs for example, detailed balance
will be broken if the kinetic constraint only operates on spin flips
in one direction, \eg\ for flipping an up-spin down, while the reverse
transition is unconstrained. Stationary states are then determined by the
competition between the constraints and the structure of the
energy function~\cite{Trimper99}. Halpern~\cite{Halpern99,Halpern00}
introduced a ``cluster-facilitated'' variant of SFMs, where the
kinetic constraint is that the cluster containing the spin to be
flipped and its nearest neighbours must contain at least $f$
up-spins. This means that up-spins require only $f-1$ facilitating
up-spin neighbours, while down-spins need $f$. Again, this asymmetry
destroys detailed balance; for $f=1$, for example, the only stationary
distribution is the one which assigns probability one to the configuration
with all spins down. However, at sufficiently high temperatures
long-lived metastable states with nonzero up-spins concentrations can
exist.

Schulz and Reineker also considered a model for the irreversible
growth of a crystalline phase into a glass~\cite{SchRei95}. With
$n_i=0,1$ to represent immobile and mobile regions as before, the
model of~\cite{SchRei95} effectively introduces a third spin state
$n_i=-1$ into the $2,2$-SFM, to model regions with local crystalline
ordering. The kinetic constraint remains as before (at least two
up-spin neighbours are required for a spin to flip,
$n_i=0\leftrightarrow n_i=1$) but an additional irreversible process
$n_i=1\to n_i=-1$ subject to the same constraint is postulated to
model crystallization. Crystal formation from immobile regions,
$n_i=0$, is not allowed. One can now consider an equilibrium
configuration of the $2,2$-SFM at some up-spin concentration $c\eql$,
on a lattice with periodic boundary conditions in the $x$-direction
(say). If this configuration is ``seeded'' with a crystalline surface
by setting all spins with vertical coordinate $y=0$ to $n_i=-1$, then
this crystalline phase will grow irreversibly into the ``glass'' phase
at $y>0$. At long times the average height of the interface will grow
linearly with time; the crystal phase behind the interface is not
homogeneous but contains inclusions of liquid- and solid-like
regions. The fluctuations of the interface height across the sample
also define a roughness, whose scaling behaviour can be used to define
a characteristic lengthscale of cooperative behaviour. The results
agree broadly with those found for the conventional $2,2$-SFM
using other definitions; see Sec.~\ref{res:hetero}.

Finally, many non-detailed balance variations of Glauber dynamics in the
one-dimensional ferromagnetic Ising chain have been studied. A recent
example is~\cite{GonDeHTag01}, where detailed balance is broken by imposing
transition rates that only depend on the left neighbour of the spin to the
flipped.

\section{Techniques}
\label{techniques}

In this section we review the various techniques that have been used
to study KCMs. Effective irreducibility is important for KCMs to
ensure that equilibrium properties can be predicted from the naive
Boltzmann distribution over all configurations; we sketch some
techniques for proving this in Sec.~\ref{meth:reduc}. As far as the
dynamics of KCMs is concerned, numerical simulations
(Sec.~\ref{numer:mc}) are often a convenient starting point, and
sometimes the only possible method of attack. Some properties are,
however, amenable to exact analytical solution
(Sec.~\ref{meth:exact}). Where this is not the case, a number of
approximation techniques can be used, ranging from mean field-like
decoupling schemes and adiabatic approximations to special techniques
for one-dimensional models; see Secs.~\ref{meth:mf}
to~\ref{meth:indint}. Mode-coupling approximations derived within the
projection formalism (Sec.~\ref{meth:proj}) and closely related
diagrammatic techniques (Sec.~\ref{meth:diag}) have also been
employed, and mappings to quantum systems offer scope for further
analytical work (Sec.~\ref{meth:fock}). Finally, mappings to effective
models can be helpful to understand the low-temperature dynamics of
KCMs as outlined in Sec.~\ref{meth:mappings}.

\subsection{Irreducibility proofs}
\label{meth:reduc}

The problem of the reducibility of the Markov chains that formally define
KCMs has been tackled by many authors. In general, different types of
models require different kinds of analytical or numerical techniques to
check whether reducibility effects are significant; in this section, we
sketch some of the more common approaches.

Much effort has gone into establishing the irreducibility or otherwise
of the {\bf $f,d$-SFMs}. One starts by defining what is called the
high-temperature partition (see Sec.~\ref{intro:irred}). This
partition comprises the configuration with all spins pointing up and
all other configurations that can be reached from there; the latter
are also called nucleating
configurations~\cite{FreAnd85,Reiter91}. Clearly, the configuration
with all spins down $n_i=0$ can never belong to the high-temperature
partition. This trivial reducibility is not necessarily significant,
however; as explained in Sec.~\ref{intro:irred} we only require for
``effective irreducibility'' that a typical configuration with given
energy (or equivalently up-spin concentration, if we consider the
standard SFMs without ferromagnetic interactions) belongs to the
high-temperature partition with probability one in the limit of
infinite system size.

Now consider a random configuration of the $f,d$-SFM at up-spin
configuration $c$. To find out whether this configuration belongs to
the high-temperature partition, one first flips up all mobile
down-spins. This may mobilize further down-spins, so one iterates the
procedure until a configuration with no mobile down-spins is
reached. If this final configuration has all spins up, the original
configuration belongs to the high-temperature partition.  This
cellular automaton-style rule of flipping down-spins recursively has
been studied under the name of diffusion percolation~\cite{AdlAha88},
although there one normally asks whether the final configuration contains a
spanning cluster of up-spins, rather than {\em only}
up-spins. Diffusion percolation in turn is closely related to
bootstrap percolation (BP)~\cite{ChaLeaRei79}; for a review
see~\cite{Adler91}. The relation is via a simple mapping that
exchanges the roles of up- and down-spins. From the original
configuration with a fraction $c$ of up-spins, reverse all spins to
get a configuration with up-spin concentration $1-c$. Interpreting
sites with the new up-spins (\ie\ $n_i=1$) as occupied by particles,
$m$-BP is defined by recursively removing all particles which have
fewer than $m$ occupied neighbouring sites. In spin language, this
means flipping down all up-spins that have fewer than $m$ up-spin
neighbours, \ie\ on a cubic lattice in $d$ dimensions at least
$2d-m+1$ down-spin neighbours. Reversing all spin directions again,
this is just the diffusion percolation algorithm for the $f,d$-SFM,
with $f=2d-m+1$. Thus, a configuration with up-spin concentration $c$
belongs to the high-temperature partition of the $f,d$-SFM exactly
when $m$-BP with $m=2d+1-f$ gives an empty lattice when started with
the corresponding reversed configuration that has a fraction $1-c$ of
sites occupied. As an aside, we note that if instead of the probability
of reaching an empty lattice one considers the probability for an
infinite spanning cluster of particles in the final configuration, then $m$-BP
clearly becomes a generalization of ordinary percolation, which is
included as the special case $m=0$. For $m=1$ only isolated particles
are removed compared to ordinary percolation, while for $m=2$ isolated
particles plus dangling ends of clusters of particles are removed.

Consider now a configuration of an $f,d$-SFM on a $d$-dimensional
(hyper-)cubic lattice of size $L$, with each of the $N=L^d$ spins chosen as
up with probability $c$.  Let $p(c,L)$ be the probability that such a
configuration belongs to the high-temperature partition, \ie\ that the
inverted configuration leads to an empty lattice in BP with $m=2d+1-f$. (In
Sec.~\ref{intro:irred} we had written the size-dependence of $p(c,L)$ in
terms of $N=L^d$ rather than $L$, but this should not cause confusion.)
Some trivial cases are easily understood. For the $1,d$-SFM, all spins can
be flipped up as long as there is a single up-spin in the original
configuration, so $p(c,L) = 1-(1-c)^L$ and for any $c>0$ the model is
effectively irreducible since $p(c,L\to\infty)=1$.  On the other hand, for
$f>d$ it is easy to see that $p(c,L\to\infty)=0$ for any $c<1$ and thus
these models have significant reducibility effects. The $3,2$-SFM is a
simple example: any $2\times 2$ square of down-spins can never be flipped
up whatever the state of the neighbouring spins, and for $c<1$ the
probability that such squares exist tends to one for $L\to\infty$. In the
regime $2\leq f\leq d$, it turns out that the models are effectively
irreducible. The proofs rely on the existence of what are called, in the
corresponding BP problem, large void instabilities~\cite{KogLea81}; in our
context they are large clusters of up-spins starting from which the whole
system can eventually be covered with up-spins. Taking the $2,2$-SFM as an
example, we paraphrase here an analogous argument for hard-square lattice
gases by J\"ackle \etal~\cite{JaecFroKno91,JaecFroKno91b}. Consider an
$l\times l$ square of all up-spins. A little thought shows that this can
be grown outwards---by flipping up mobile down-spins---into an
$(l+2)\times(l+2)$ square at least if there is one up-spin in each of the
four rows of length $l$ bordering the square. The probability for this is
$p_l = [1-(1-c)^l]^4$. With increasing $l$ this converges to one so quickly
that the probability
\be
p_{l\to\infty} =
\prod_{k\geq 0} p_{l+2k} =
\exp\left\{ 4\sum_{k\geq 0} \ln[1-(1-c)^{l+2k}]\right\}
\label{p_l_to_infty}
\ee
for the process to continue to infinity is nonzero. Once a
sufficiently large cluster (or ``critical droplet''~\cite{AizLeb88})
of up-spins has been established, this probability is in fact very
close to one, since for large $l$ one can approximate
\be
p_{l\to\infty} \approx 
\exp\left[ -4\sum_{k\geq 0} (1-c)^{l+2k}\right] = 
\exp\left[-4(1-c)^l/(2c-c^2)\right]
\ee
so that clusters of size above $l=\ln(2c-c^2)/\ln(1-c)$, \ie\ $l\approx
(-\ln c)/c$ for small $c$, are unstable in the sense that they will
continue to grow to infinity with high probability. Returning now
to\eq{p_l_to_infty}, the probability of reaching the all up-spin
configuration from a single up-spin ``nucleation site'', $p_{1\to\infty}$,
can be estimated by replacing the sum over $k$ by an
integral~\cite{AizLeb88}, giving
\be
p_{1\to\infty} \approx
\exp\left\{ 2 \int_0^\infty du \ln\left[1-(1-c)^u\right]\right\}
=
\exp\left\{ -\frac{2}{\ln(1-c)} \int_0^\infty dv
\ln\left[1-e^{-v}\right]\right\}
\label{p_one}
\ee
which scales as $p_{1\to\infty} \sim \exp(-{\rm const}/c)$ for small
$c$. The fact that $p_{1\to\infty}>0$ is sufficient to guarantee that
in an infinite system at least one such nucleation site will exist;
hence the original configuration belongs to the high-temperature
partitions with probability $p(c,L\to\infty)=1$.

An important proviso regarding such proofs of effective irreducibility
is whether the thermodynamic limit behaviour is reached in systems of
realistic size. To quantify finite-size effects, one can consider
$p(c,L)$ as a function of $c$. For sufficiently large $L$ this
increases steeply from zero to one in a narrow region around some
value $c_*(L)$---see Fig.~\ref{fig_irrec_sfm} above---which one could
define \eg\ by the condition
$p(c_*(L),L)=1/2$. If a system is effectively irreducible then
necessarily $c_*(L\to\infty)=0$, but the rate of this approach can be
very slow. In the example of the $2,2$-SFM above one can estimate how
large $L$ needs to be to have $p(c,L)=\order(1)$, using the condition
$cL^2p_{1\to\infty}\approx 1$ that there is of order one nucleation
site in the system. Using\eq{p_one} this gives $L\sim \exp({\rm
const}/c)$ to leading order, and inverting one has an up-spin
concentration $c_*(L) \sim 1/\ln L$ above which a finite sytem will be
essentially irreducible. For the $3,3$-SFM, one finds an even slower
convergence, $c_*(L)\sim 1/\ln(\ln L)$~\cite{EntAdlDua90}. In this
case it is clear that even for a macroscopic $L=10^{10}$ (say) the
thermodynamic limit is not yet reached and the system will show strong
reducibility effects below some nonzero $c_*(L)$.

Spin models with {\bf directed kinetic constraints} can have nonzero
thresholds $c_*\equiv c_*(L\to\infty)>0$ even for an infinite system
and are then effectively irreducible only for up-spin concentrations
$c>c_*$. This is most easily seen for models on Cayley trees, where
these thresholds can be calculated by a simple recursion. Take the
$(3,2)$-Cayley tree, where a spin is mobile if both of its neighbours
on the level below in the tree are up. We follow the arguments
of~\cite{ReiMauJaec92}; see also~\cite{ChaLeaRei79} for closely
similar reasoning regarding bootstrap percolation on Bethe
lattices. Start with a tree of $L+1$ levels, with up-spins assigned
randomly with probability $c$. Beginning
with the bottom layer, where the spins are frozen since they have no
facilitating neighbours, move upwards through the tree and flip up all
down-spins that are mobile. Call $p(c,L)$ the probability that the
spin at the top node is up at the end of this procedure. This can
happen either because the spin was originally up (probability $c$) or,
if it was originally down, because the two spins below have ended up
in the up state. Since these two spins have independent trees of depth
$L$ below them, one has the recursion
\be
p(c,L) = c+(1-c)p^2(c,L-1)
\label{Cayley_recursion}
\ee
For large tree depth $L$, $p(c,L)$ thus tends to a stable fixed point
$p(c,L\to\infty)$ of this recursion; which one is determined by the
starting value $p(c,0)=c$.  This gives $p(c,L\to\infty)=1$ for $c\geq
c_*=1/2$ and $p(c,L\to\infty)=c/(1-c)$ for $c<c_*$. The fraction of
permanently frozen spins near the top of the tree, $1-p^2(c,L\to\infty)$,
thus increases smoothly from zero as $c$ decreases below $c_*$. Above $c_*$, on
the other hand, one has $p(c,l)\approx 1$ in all layers $l$ of the tree
except for a finite number at the bottom. This means that the configuration
with all up-spins can be reached with probability close to one and (this
part of) the system is effectively irreducible.

For {\bf kinetically constrained lattice gas models}
(Sec.~\ref{model:lattice_gases}), the question of irreducibility is
normally cast somewhat differently: one asks whether there are any
particles that remain permanently blocked in their initial positions in all
configurations that are accessible via any sequence of allowed transitions,
\ie\ in all configurations within the relevant partition of configuration
space. The dynamics can then be defined as effectively irreducible if the
fraction of typical configurations that contain blocked particles vanishes
in the thermodynamic limit. The triangular lattice gas~\cite{JaecKro94} and
the hard-square lattice gas~\cite{ErtFroJaec88,JaecFroKno91,JaecFroKno91b}
have in fact been proved to be effectively irreducible in this sense, by
techniques similar to the ones outlined above.

A subtlety in determining which particles in a lattice gas are permanently
blocked is that a particle may be blocked by a sufficiently large number of
neighbouring particles which themselves are {\em not} permanently blocked. One
therefore often focuses on a subset of blocked particles, the so-called
``backbone''~\cite{KobAnd93}. This contains all particles which are
permanently frozen by {\em other frozen} particles.
Particles in the backbone remain frozen even when all mobile particles are
removed; the backbone can therefore be determined simply by iteratively
removing all mobile particles from the system. For the KA model this
procedure is closely related to bootstrap percolation; see
Sec.~\ref{res:reduc} for further details. One may of course be concerned
that the backbone ``misses'' a significant number of permanently blocked
particles, but simulations for the triangular lattice gas~\cite{JaecKro94}
suggest that this is not so: the number of particles in the backbone was
found to be a very good approximation to the number of particles that
remained blocked in long simulations of the actual dynamics.

\subsection{Numerical simulations}
\label{numer:mc}

As defined in Sec.~\ref{model:SFM}, the dynamics of most KCMs can be
described by a Markovian dynamics in continuous time as expressed by the
master equation\eq{master}. While it is possible to simulate this directly
(see below), for a ``quick and dirty'' simulation it is often convenient to
have an equivalent discrete-time formulation. We start by outlining how
the two are related. To be concrete, consider $f,d$-SFMs where the only
possible transitions between configurations are spin flips. The transition
rates can then be written in the general form $\ww(\nv\to\nv') =
\sum_i \ww_i(\nv)\delta_{\nv',F_i\nv}$ where $\ww_i(\nv)\equiv\ww(\nv\to
F_i\nv)$ and $F_i$ is the operator that flips
spin $i$, $F_i\nv = (n_1\ldots 1-n_i\ldots n_N)$. The master
equation\eq{master} then reads
\be
\deriv{t} p(\nv,t) = \sum_i [\ww_i(F_i\nv)p(F_i\nv,t) - \ww_i(\nv) p(\nv,t)]
\label{master_flips}
\ee
If any of the spin-flip rates $\ww_i(\nv)$ are greater than one, let
$\kappa$ be the inverse of the largest rate, otherwise set $\kappa=1$;
the rescaled rates then obey $0\leq\kappa\ww_i(\nv)\leq1$.  Now
consider the following discrete-time Monte Carlo dynamics where time
is advanced in steps of $\kappa/N$. At each step one of the $N$
possible transitions out of the current configuration $\nv$ is chosen
randomly; since we are dealing with spin-flips, this just means
picking a random spin to flip, $n_i$ say. The proposed transition is
accepted with probability $\kappa\ww_i(\nv)$, while with probability
$1-\kappa\ww_i(\nv)$ it is rejected and the system remains in its
current configuration $\nv$. The Markov equation for this process is
\[
p(\nv,t+\kappa/N) = \frac{1}{N}\sum_i
\left\{
\kappa\ww_i(F_i\nv)p(F_i\nv,t) + [1-\kappa\ww_i(\nv)]p(\nv,t) 
\right\}
\]
or
\be
\frac{N}{\kappa}\left[p(\nv,t+\kappa/N)-p(\nv,t)\right] = \sum_i
\left[
\ww_i(F_i\nv)p(F_i\nv,t) - \ww_i(\nv)p(\nv,t)
\right]
\label{discrete_time}
\ee
For $N\to\infty$ this becomes equivalent to\eq{master_flips}, so that the
discrete and continuous time descriptions can be used interchangeably. In
general, the discrete time dynamics is obtained by randomly selecting, at
each step, one of the possible transitions---spin flips, or moves of a
particle to a neighbouring site in the lattice gas models of
Sec.~\ref{model:lattice_gases}---and accepting the proposed move with
probability proportional to the continuous time rate for the transition.

A standard Monte Carlo simulation in discrete time is very simple to set
up, and often useful for initial exploration of the dynamics of KCMs. At
low temperatures, where relaxation timescales can become very large, such
an approach quickly runs into problems and more sophisticated approaches
are necessary~\cite{Binder95,NewBar99,LanBin00}. The key difficulty is that
in KCMs many of the transitions that are possible in principle are
forbidden by the kinetic constraints, so that a standard Monte Carlo
simulation would reject almost all proposed moves. One way around this
problem is a technique known variously as rejection-free,
continuous-time, faster-than-the-clock or
Bortz-Kalos-Lebowitz~\cite{BorKalLeb75} simulation.
In a continuous-time description, let
$\ww_i\equiv \ww_i(\nv)$ be the rates for all possible transitions out of
the current configuration $\nv$. It is then easy to show that the time
interval $\Delta t$ to the next transition is exponentially distributed
with a rate equal to the sum $\ww_{\rm tot}=\sum_i \ww_i$ of all rates, \ie\
$P(\Delta t) = \ww_{\rm tot} \exp(-\ww_{\rm tot}\Delta t)$. Values of
$\Delta t$ from this distribution can easily be sampled, so that one can go
directly to the next ``successful'' transition. It then remains to be
determined which transition actually occurs; one easily derives that the 
probability for the first transition to be the one with rate $\ww_i$ is
$\ww_i/\ww_{\rm tot}$. Sampling from this distribution can be the
rate-limiting step in the algorithm, and so it is often useful to devise
efficient methods for this. An example is provided by recent simulations of
the East model~\cite{SolEva02}: here the positions of all mobile spins in
the chain were stored in a binary tree which can be quickly searched to
determine which particular spin should be flipped in any given transition.

\subsection{Exact solutions}
\label{meth:exact}

In this section we give examples of techniques that have been used to solve
aspects of the dynamics of KCMs exactly. In the cases discussed, the
simplifying feature that makes such exact solutions possible is either a
restriction to dynamics at $T=0$, or the mean-field character of the
dynamics as in the Backgammon and oscillator models.

One of the models whose zero temperature dynamics can be solved exactly is
the $1,1$-SFM~\cite{FolRit96,SchTri97}. If the system is quenched at $t=0$
from some initial state to one with equilibrium up-spin concentration
$c\eql =1/(1+e^\beta)$, the Glauber dynamics transition rates for $t>0$
are, from\eqq{sfm_rates_formal}{Glauber}
\be
\ww(n_i \to 1-n_i) = (n_{i-1}+n_{i+1})[(1-c\eql)n_i + c\eql (1-n_i)]
\label{one_one_sfm}
\ee
Now, from the master equation\eq{master}, one easily deduces that the
average of a general observable $\phi(\nv)$ evolves in time according to
\be
\deriv{t}\lav \phi(\nv)\rav = \sum_i \lav\ww(n_i \to 1-n_i)
[\phi(F_i\nv)-\phi(\nv)]\rav
\label{phi_av}
\ee
where $F_i\nv$ is the configuration $\nv$ with spin $n_i$ flipped to
$1-n_i$.  Applying this to the $(k+1)$-spin correlation functions
$D_k=(1/N)\sum_j \lav n_j\cdots n_{j+k}\rav$ one finds, in the
zero-temperature limit where $c\eql\to 0$, the closed hierarchy
\be
\deriv{t} D_k = -2(kD_k+D_{k+1})
\label{hierarchy}
\ee
which can be solved by introducing the generating function
$G(x)=\sum_{k=0}^\infty D_k x^k/k!$. Not surprisingly, since the $T=0$
dynamics is strongly reducible---up-spins that are isolated at $t=0$ can
never flip, for example---the results for $t\to\infty$ depend strongly on
the initial conditions. For a given initial up-spin concentration $c_0$ one
finds, for example, that $c(t)\equiv D_0(t)$ converges to $c_0\exp(-c_0)$
for $t\to\infty$, rather than to the equilibrium value $c\eql=0$. It was
later shown~\cite{CriRitRocSel00} that exactly the same solution
applies to the 
asymmetric $1,1$-SFM with transition rates\eq{asymmetric_SFM}, except that
the factor 2 in\eq{hierarchy} is replaced by $1+a$. This has also been
confirmed~\cite{DesGodLuc02} via a mapping to equivalent models of (random
or cooperative) sequential adsorption~\cite{Evans93}. Looking back
at\eq{one_one_sfm}, one sees that the $T\to0$ limit corresponds to
neglecting processes occurring with rates $c\eql$; the above solution for
the dynamics will therefore also give the correct results for {\em nonzero}
temperatures on timescales shorter than $1/c\eql \approx \exp(\beta)$.

A hierarchy very similar to\eq{hierarchy} has been used to solve
exactly~\cite{PraBre01} the $T=0$ dynamics of a Glauber Ising chain in zero
field, when spin flips that leave the energy unchanged are
forbidden~\cite{LefDea01}. The only possible flips are then those causing
two neighbouring domain walls to annihilate. After a mapping to domain wall
variables via $n_i = (1-\s_i\s_{i+1})/2$, such moves correspond to two
neighbouring up-spins ($n_i=1$) flipping down simultaneously, and the
correlation functions $D_k$ as defined above obey closed equations that
differ only by numerical factors from\eq{hierarchy}. In particular, the
domain wall concentration $c=\lav n_i\rav$ converges to $c_0\exp(-2c_0)$
from an initial equilibrium state with $c=c_0$. This result can also be
obtained from a mean-field approach which becomes exact in one
dimension~\cite{DeaLef01}.

The exact solution of the $T=0$ dynamics of the asymmetric $1,1$-SFM,
described above, can actually be pushed further to calculate exactly the
probability $P(c,T)$ that the system will end up in a metastable configuration with
up-spin concentration $c$ if quenched to zero temperature from an
equilibrium state at some nonzero $T$. From this an appropriately
defined entropy of metastable configurations can be obtained since for
large systems $P(c,T)$ will be exponential, $P(c,T)\sim
\exp[N\pi(c,T)]$. In~\cite{CriRitRocSel00} the quadratic expansion of
$\pi(c,T)$ around its maximum w.r.t.\ $c$ was obtained, corresponding to a
Gaussian approximation to $P(c,T)$; more recently the full form of
$\pi(c,T)$ has also been found~\cite{DesGodLuc02}. As before, the analysis
also applies to the extreme limits of the asymmetric $1,1$-SFM, the East
model and the conventional $1,1$-SFM.

Another example where dynamical equations can be exactly solved is the
Backgammon model~\cite{FraRit95,GodBouMez95,FraRit96,GodLuc96} introduced in
Sec.~\ref{model:urn}. Most calculations have focused on the
case where the number of boxes is equal to the number of particles,
$M=N$. One defines $P_k(t)$ as the probability that a randomly selected
box contains $k$ particles. For models such as Backgammon which are in the
Ehrenfest class, this probability depends on time through a dynamical
equation of the form
\be
\frac{\partial P_k(t)}{\partial t}=f(P_k,P_{k+1},P_{k-1},P_0)
\label{hierarchy2}
\ee
where $f$ is a linear function of its arguments with coefficients
depending on $P_0$. This set of equations constitutes a closed
hierarchy of nonlinear equations, the nonlinearity appearing only
through the time-dependent coefficient $P_0(t)$. For instance, the
equation for the energy $E/M=-P_0$ is given by,
\be
\frac{\partial P_0(t)}{\partial t}=P_1(1-P_0)-e^{-\beta}P_0(1-P_0)
\label{hierarchy3}
\ee
The full hierarchy can be solved by defining a generating function
$G(x,t)=\sum_{k=0}^{\infty} x^kP_k(t)$ and solving the resulting partial
differential equation~\cite{FraRit96}. At $T=0$ one finds~\cite{GodLuc96}
\be
-P_0(t) = \frac{E(t)}{M} = -1+\frac{1}{\ln t+\ln(\ln t)}
\label{hierarchy4}
\ee
up to subdominant corrections; this solution can also be obtained
using an adiabatic approximation~\cite{FraRit95} (see
Sec.~\ref{meth:adiab}) which becomes exact for long times and at
$T=0$. At small but nonzero temperature, the energy relaxation
crosses over to exponential behaviour on a timescale whose dominant
dependence on $T$ is an Arrhenius (activated) law. Similar generating
function techniques have generally been very useful for urn models,
\eg\ in the calculation of correlation and response
functions~\cite{FraRit96,GodLuc96,FraRit97,GodLuc97,GodLuc99}. A
hierarchy similar to\eq{hierarchy2} has also been derived in a
simplified version of the Backgammon model~\cite{PraBreSan97c}.

Closed hierarchies of dynamical equations can also be derived for
oscillator models.  For the spherical Sherrington-Kirkpatrick
model~\cite{BonPadParRit96a,BonPadParRit96b} the technique is very similar
to that for the Backgammon model; for what we called the oscillator model proper in
Sec.~\ref{model:osc} the situation is even simpler since it is possible to
show that the hierarchy closes at the already at the lowest level, yielding
an exact autonomous equation for the energy $E$~\cite{BonPadRit98}. This is
similar in form to the result of an adiabatic approximation for the Backgammon
model (see Sec.~\ref{meth:adiab})---which would be exact for the oscillator
model---and reads
\be
\frac{\partial E}{\partial t}=-E^{\frac{3}{2}}\exp(-C/E)
\label{hierarchy5}
\ee
where $C$ is a constant. As a result, the energy $E(t)$ again decays to its
ground state value $E=0$ with an asymptotically logarithmic dependence on
time.

\subsection{Mean-field approximations}
\label{meth:mf}

In this section we collect some mean-field approaches to the dynamics of
KCMs; these are normally based on deriving closed dynamical equations by an
appropriate decoupling of correlations.

As an example of {\em naive} mean-field theory, which neglects all
correlations, we paraphrase here the analysis of~\cite{SchTri98} for the
relaxation of the up-spin concentration in the $f,d$-SFM. As usual, we
restrict ourselves to the non-interacting case $J=0$; a nonzero value of
$J$ has negligible effects in the interesting regime of low up-spin
concentrations. For Glauber dynamics\eq{Glauber}, the spin-flip
rates\eq{sfm_rates_formal} are
\be
\ww(n_i\to 1-n_i) = \sum_{j_1\neq \ldots \neq j_f} n_{j_1} \cdots n_{j_f}
[(1-c\eql)n_i+c\eql (1-n_i)]
\ee
Eq.\eq{phi_av} then gives for the evolution of the local up-spin
concentrations
\be
\deriv{t}\lav n_i\rav = \lav \ww(n_i\to 1-n_i) (1-2n_i) \rav = 
\sum_{j_1\neq \ldots \neq j_f} \lav n_{j_1} \cdots n_{j_f}
[-(1-c\eql)n_i+c\eql (1-n_i)]\rav
\label{rates_again}
\ee
A naive mean-field approximation decouples the average of the spin-product
on the r.h.s.\ into single-spin averages. If the system is started in
equilibrium, with $\lav n_i\rav$ uniform across the system, then this will
remain the case for all times and one obtains a simple evolution equation
for $c=\lav n_i\rav$
\be
\frac{\partial c}{\partial t} \propto c^f (c\eql-c)
\label{dsdt}
\ee
(The proportionality factor is the number of terms in the
sum\eq{rates_again}, namely $(2d)!/(2d-f)!$) Linearizing\eq{dsdt} around
equilibrium $c=c\eql$ then gives an estimate of the relaxation time, $\tau
\sim c\eql^{-f} \approx \exp(f\beta)$. It is clear, however, that this
approximation only takes cooperativity between spins into account very
crudely. Accordingly, it fails to predict the superactivated relaxation
time increase that occurs in $f,d$-SFMs for $f\geq 2$; see
Sec.~\ref{sfm:some_results}. The mean-field treatment can be extended to
analyse the relaxation of spatial fluctuations of $\lav
n_i\rav$~\cite{SchTri98} but correlation effects due to the kinetic
constraints are then still neglected.

More sophisticated mean-field approximations result if some nontrivial
correlations are kept. Consider the relaxation of a local up-spin
concentration $\lav n_i\rav$ in the East model, for example. (In
equilibrium this relaxation, for a spin that is in the up-state at
$t=0$, also determines the spin autocorrelation function; see the
discussion after\eq{eqC} below.)  From the transition
rates\eq{eq5S311} and the general result\eq{phi_av} one sees that the
time evolution of $\lav n_i\rav$ is coupled to a hierarchy of
correlations $\lav n_i n_{i-1}\rav$, $\lav n_i n_{i-2}\rav$, $\lav n_i
n_{i-1} n_{i-2}\rav$ etc~\cite{JaecEis91}. If one truncates by
neglecting all correlation functions from a given order onwards,
approximations to the autocorrelation function can be obtained by
solving the resulting system of linear equations. As explored in other
contexts, \eg\ the triangular lattice gas~\cite{JaecKro94,KroJaec94},
such approximations can also be viewed as applications of the
projection technique to a space of observables spanned by spin
products of a given order, with the memory terms neglected; see
Sec.~\ref{meth:proj}. Careful selection of the relevant set of
observables can significantly improve the results. For example, to
calculate the relaxation of a given spin $n_i$ in the East model,
Eisinger and J\"ackle considered the ``cluster probabilities'' of
having to the left of $n_i$ a domain of $k-1$ down-spins followed by
an up-spin and $m-1$ further spins in arbitrarily specified
states. Retaining these probabilities for some fixed cluster length,
\eg\ $m=6$, and all integer values $k=1,2,\ldots$, they found good
fits to simulated relaxation functions down to $c\eql=0.2$. This
approximation also revealed an interesting relation to
defect-diffusion models, with the clusters obeying an effective
diffusion equation with drift towards the spin $n_i$.

One can try to improve further on such truncation approximations by taking
neglected correlations into account through an ``effective field'' or
``effective medium''. Taking again the East model as an example, J\"ackle
and Eisinger~\cite{JaecEis91,EisJaec93} proposed the following procedure
for approximating the spin autocorrelation function: Suppose the state of
spin $n_0$ was known as a function of time $t$, and let $p$ be the vector
of probabilities for the $2^l$ configurations of the $l$ spins to the
right. Anticipating the notation of Sec.~\ref{meth:proj}, the master
equation for $p$ can be written as $\partial_t p(t) = L_1\T p(t) + n_0(t)
L_2\T p(t)$ with constant matrices $L_1\T$ and $L_2\T$; the second term
here describes transitions of spin $n_1$, which are possible only if its
left neighbour is up, \ie\ $n_0=1$. If one Laplaces transforms and
approximates the effect of $n_0(t)$ by a frequency-dependent mobility
$\Gamma(z)$, this becomes $zp(z)-p(t=0) = L_1\T p(z) + \Gamma(z) L_2\T
p(z)$. Solving this for an appropriate initial distribution $p(0)$ the
autocorrelation function of spin $n_l$ can be determined; the value of
$\Gamma(z)$ can then be deduced from the self-consistency requirement that
the same correlation function is obtained for $l=1$ and $l=2$. Somewhat
surprisingly, the resulting approximation is similar in form to a mode
coupling approximation; see\eq{Mirr_East_diag}. The same approach has also
been applied to the North-East and $(3,2)$-Cayley tree models.

\subsection{Adiabatic approximations}
\label{meth:adiab}

In this section we outline some applications of adiabatic approximations to
KCMs. These approximations are based on the assumption that a
separation of timescales occurs in the dynamics, allowing a description in
terms of separate fast and slow modes. (More generally, a whole hierarchy
of sets of modes could occur, all evolving on well-separated timescales.)
The key idea is then to assume that the slow modes evolve so gradually that
the fast modes can always equilibrate relative to the {\em instantaneous}
configuration of the slow modes. Even if a timescale separation does exist,
the model-dependent choice of slow and fast modes is not always obvious.
It requires some intuition about the physical mechanisms underlying the
occurrence of well-separated timescales; in this sense, a more complete
understanding of the validity of adiabatic approximations should ultimately
be helpful in clarifying which features of glassy dynamics are universal
and which are system-dependent.

We illustrate adiabatic approximations in this section for two models, the
East model and the Backgammon model. In the East model, the nature of slow and fast
modes is relatively easy to determine~\cite{MauJaec99}. Transitions out of
any configuration that contains at least one mobile up-spin will take place
with a ``fast'' rate of order unity, while transitions out of all other
configurations only happen with rates of $\order(c\eql)$. For small
$c\eql$, \ie\ low temperatures $T$, this gives a natural separation into
fast and slow modes. Mathematically, the latter are the occupation
probabilities $p(\nv,t)$ of all configurations with no mobile up-spins,
\ie\ with all up-spins surrounded by down-spins, while the fast modes are
the remaining $p(\nv,t)$. To eliminate the fast modes, one sets their time
derivatives in the master equation\eq{master} to zero. This is the
adiabatic approximation: fast modes equilibrate in the ``environment''
fixed by the instantaneous values of the slow modes. One obtains in this
way an effective master equation for the slow degrees of freedom. This
should in principle give a description of the dynamics which becomes exact
for low temperatures, but because of the large number of fast modes
involved has been explicitly worked out only for very small system
sizes~\cite{MauJaec99}. An interesting refinement of this method would be
to classify all slow configurations according to the number $k$ (say) of
down-spins that need to be flipped up before any of the original up-spins
can flip. Since such relaxation processes have an energy barrier of $k$ and
so require times of order $\exp(k/T) \sim c\eql^{-k}$ (see also
Secs.~\ref{sfm:some_results}, \ref{res:relax}), configurations with $k=1$
relax much more quickly than those with $k\geq 2$; within the set of slow
modes they are much faster than all others and can therefore again be
adiabatically eliminated. This process could in principle be iterated for
larger $k$ to give an effective master equation for the dynamics on a
hierarchy of increasingly long timescales.

As an aside, we mention briefly a recent analysis of KCMs on hierarchical
structures~\cite{BrePra01,PraBre01b} which is similar in spirit. As
explained in Sec.~\ref{related:hierarchical}, in these models flips of
spins in any given level $l$ are facilitated by spins in level $l-1$
below. The simplest adiabatic approximation is that the typical relaxation
timescales on the different levels, which increase as one moves up in the
hierarchy, are widely separated; for the analysis of level $l$ one can then
assume equilibrium in level $l-1$. The resulting equations can model some
non-equilibrium effects typical of glassy systems, especially with regards
to the effect of cyclic heating and cooling, but are too simple to describe
strongly cooperative behaviour.

As a second example application of the adiabatic approximation, we
consider the Backgammon model~\cite{FraRit95,GodBouMez95}; see
Sec.~\ref{model:urn}. Here a timescale separation occurs because the
probabilities $P_k$ for a randomly chosen box to contain $k$ particles
evolve very differently for $k=0$ and $k>0$. $P_0=-E/M$ is the density
of empty boxes and increases only very slowly with time. On the other
hand, the different configurations in the non-empty boxes are explored
rapidly, so that the probabilities $P_k$ ($k>0$) quickly reach an
equilibrium state compatible with the given value of $P_0$.  Consider
now the evolution equation\eq{hierarchy3} for $P_0$ (see
Sec.~\ref{meth:exact}), which for $T=0$ reads $\partial{P_0}/\partial
t=P_1(1-P_0)$. The adiabatic approximation replaces $P_1$ on the
r.h.s.\ by the value $P_1= P_1(P_0)$ that it would have in equilibrium
at the given $P_0$; in other words, $P_1(P_0)$ is the value of $P_1$
in a microcanonical ensemble with energy $E=-MP_0$.  Solving the
resulting closed equation $\partial{P_0}/\partial t=P_1(P_0)(1-P_0)$
then gives the exact~\cite{GodLuc96} long-time evolution of $P_0(t)$,
as given earlier in\eq{hierarchy4}. The adiabatic approximation thus
actually provides an exact description of the asymptotic dynamics for
the Backgammon model at $T=0$. Notice that associated with the effective
constant-energy (microcanonical) equilibrium ensemble assumed by the
adiabatic approximation is a corresponding effective temperature. This
illustrates the close connection between adiabatic dynamics and the
existence of out-of-equilibrium FDT violations (see
Secs.~\ref{intro:fdt} and~\ref{res:fdt}); a theoretical framework for
this connection is described in detail in~\cite{CriRit02b}. Finally,
let us note briefly that adiabatic methods have also been applied to
oscillator models. We already mentioned in Sec.~\ref{model:osc} that
for the oscillator model proper the adiabatic approximation is exact,
while its disordered analogue, the spherical Sherrington-Kirkpatrick
model~\cite{BonPadParRit96a,BonPadParRit96b}, requires a more
sophisticated analysis involving two slow modes.

\subsection{Methods for one-dimensional models}
\label{meth:indint}

In one dimension, additional techniques are available for analysing
kinetically constrained models. As an example, we describe here an
application to the East model~\cite{SolEva99,Evans02} of what is
variously known as the method of interparticle distribution
functions~\cite{DoeBen88}, bag model~\cite{DerGodYek91} or independent
interval approximation~\cite{KraBen97}.

Consider a quench at $t=0$ from an equilibrium state at high temperature,
with up-spin concentration $\approx 1/2$, to a low temperature $T$
corresponding to $c\eql \approx \exp(-\beta)\ll1$. The up-spin
concentration $c(t)$ will gradually decrease towards $c\eql$, with
individual up-spins becoming increasingly widely separated. It therefore
makes sense to describe the system in terms of domains.  As shown by the
vertical lines in $ \ldots1|0001|1|1|01|001|1|1|01|0\ldots$, it is useful
to define a domain as consisting of an up-spin and all the down-spins that
separate it from the nearest up-spin to the left. The length $l$ of a
domain then also gives the distance between the up-spin at its right edge and
the nearest up-spin to the left. In equilibrium, the distribution of domain
lengths and its average are
\be
P\eql(l)=c\eql(1-c\eql)^{l-1}, \quad \bar{l}\eql=1/c\eql
\label{pd_equil}
\ee
Now for small $c\eql$, the equilibrium probability of finding an
up-spin within a chain segment of {\em finite} length $l$ is
$\order(lc\eql)$ and tends to zero for $c\eql\to 0$ at fixed $l$. In
this limit the flipping down of up-spins therefore becomes {\em
irreversible} to leading order. The dynamics of the system becomes one
of coarsening by coalescence of domains: an up-spin that flips down
merges two neighbouring domains into one large domain. During such an
irreversible coarsening process, no correlations between the lengths
of neighbouring domains can build up if there are none in the initial
state. For the present model the equilibrated initial state consists
of domains independently distributed according
to\eq{pd_equil}. Therefore an independent interval approximation for
the dynamics, defined as neglecting correlations between domains,
becomes exact in the low-temperature limit. Even when not exact, the
independent interval approximation can give very accurate results,
\eg\ recently for a ``driven'' version of the East
model~\cite{Fielding02}.

The coarsening dynamics of the East model is unusual in that it
involves a hierarchy of timescales. Consider the typical rate
$\Gamma(l)$ at which domains of length $l$ disappear by coalescing with
their right neighbors.  Because domain coalescence corresponds to the
flipping down of up-spins, $\Gamma(l)$ can also be defined as
follows. Consider an open spin chain of length $l$, with a `clamped'
up-spin ($n_0=1$) added on the left.  Starting from the configuration
$10\ldots 01$, $\Gamma^{-1}(l)$ is the typical time needed to reach the
empty configuration $10\ldots 00$ where spin $n_l$ has ``relaxed''; the
relaxation process can be thought of as a path connecting the two
configurations. Call the maximum number of excited spins (up-spins except
$n_0$) encountered along a path its height $h$. One might think that
the relaxation of spin $n_l$ needs to proceed via the configuration 11\ldots
1, giving a path of height $l$. In fact, the minimal path height
$h(l)$ is much lower and given by~\cite{SolEva99}
\be
h(l)=k+1 \quad \mbox{for}\ 2^{k-1} < l \leq 2^k
\label{h_hierarchy}
\ee
where $k=0, 1, \ldots$ This result is easily understood for
$l=2^k$~\cite{EisJaec93,MauJaec99}: to relax the $2^k$-th spin
$n_{2^k}$, one can first flip up $n_{2^{k-1}}$ and use it as an
anchor for relaxing $n_{2^k}$. The corresponding path is (with
$n_{2^{k-1}}$ and $n_{2^k}$ underlined)
$1\ldots\underline{0}\ldots\underline{1}$ $\to$
$1\ldots\underline{1}\ldots\underline{1}$ $\to$
$1\ldots\underline{1}\ldots\underline{0}$ $\to$
$1\ldots\underline{0}\ldots\underline{0}$
and reaches height $h(2^k)=h(2^{k-1})+1$; the $+1$ arises because the
anchor stays up while the spin at a distance $2^{k-1}$ to its right is
relaxed. Continuing recursively, one arrives at $h(2^k) =
h(1)+k$; but $h(1)=1$ because the only path for the
relaxation of $n_1$ is $11\to 10$. A general proof~\cite{SolEva99}
of\eq{h_hierarchy} can be constructed by showing that the ``longest''
configurations that can be reached by flipping up no more than $h$ spins
have an up-spin at site $i=2^k-1$; see also~\cite{ChuDiaGra01} where
bounds on the number of configurations reachable at or below height
$h$ are derived.

The result\eq{h_hierarchy} implies that coarsening in the East model
proceeds in a hierarchical fashion.  The energy barrier for the
relaxation of spin $n_l$ is $h(l)-1$; the $-1$ comes from the
one excited spin ($n_l$) in the initial configuration. The rate for this
relaxation process is $\Gamma(l) = \order(\exp[-(h(l)-1)/T]) =
\order(c\eql^{h(l)-1})$. For $c\eql\to 0$ the dynamics thus
divides into stages distinguished by $k=h(l)-1=0, 1, \ldots$
During stage $k$, the ``active'' domains with lengths $2^{k-1}<l\leq
2^{k}$ disappear, on a timescale
$\order(\Gamma^{-1}(l))=\order(c\eql^{-k})$; different stages can be
treated separately because the relevant timescales differ by factors
of $1/c\eql$. The distribution of inactive domains ($l>2^k$) changes
only because such domains can be created when smaller domains
coalesce. Combining this with the (exact) independent interval
approximation discussed above, one finds for $l> 2^k$
\be
\frac{\partial}{\partial t} P(l,t) = \sum_{ 2^{k-1}< l'\leq 2^k }
P(l-l',t)\,\left[-\frac{\partial P(l',t)}{\partial t}\right]
\label{pd_eqn_motion}
\ee
\begin{figure}
\begin{center}
\epsfig{file=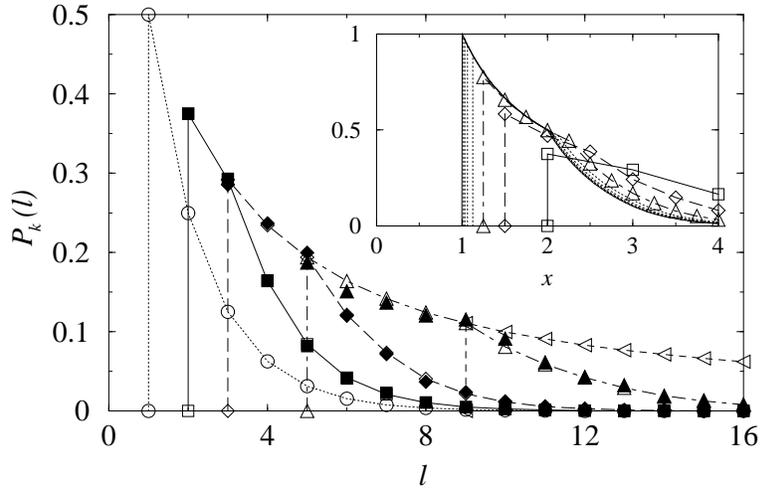, width=10cm}
\end{center}
\caption{Coarsening in the East model after a quench from the equilibrium
state with up-spin concentration $1/2$. Shown are the domain length
distributions $P_k(l)$ at the end of the various stages of the low-$T$
coarsening dynamics. Open symbols and lines: Theoretical results, for $k=0$
($\bigcirc$; initial condition), 1 ($\Box$), 2 ($\Diamond$), 3
($\triangle$). Full symbols: Simulation results for a chain of length
$N=2^{15}$ and $c\eql=10^{-4}$ ($k=1, 2$) and $c\eql=10^{-3}$
($k=3$). Inset: Scaled predictions $2^{k-1}P_k(l=2^{k-1}x)$ vs.\ $x$ for
$k=1, \ldots, 8$. Bold line: Predicted scaling
function. From~\protect\cite{SolEva99}.
\label{fig:pd}
}
\end{figure}
The term in square brackets is the rate at which active domains disappear;
$l'\leq 2^k$ because inactive domains do not disappear. This equation can
be integrated from the beginning to the end of each stage $k$, by
introducing generating functions and using the fact that all active domains
have disappeared at the end of the stage. The end result~\cite{SolEva99} is
an exact expression for the domain length distribution $P(l,t\to\infty)$ at
the end of stage $k$, which we write as $P_{k+1}(l)$, in terms of the
distribution $P_k(l)$ at the end of the previous stage. Fig.~\ref{fig:pd}
shows the results for the case where $P_0(l)$ is the equilibrium
distribution\eq{pd_equil} for up-spin concentration $1/2$. Not
unexpectedly, a scaling limit is approached for large $k$: the rescaled
distributions $\tilde{P}_k(x)=2^{k-1}P_k(l)$, with scaled domain size
$x=l/2^{k-1}$, converge to a limiting distribution $\tilde{P}(x)$ which is
independent of the initial condition and can be calculated explicitly. The
average domain length in the scaling limit is given by
$\bar{l}_k=2^{k-1}\bar{x}$, with $\bar{x}=\exp(\gamma)=1.78\ldots$ where
$\gamma$ is Euler's constant~\cite{SolEva99}. In the time-domain this leads
to anomalous coarsening with a temperature-dependent exponent, since stage
$k$ is completed on a timescale $t\sim c\eql^{-k} \sim \exp(-k/T) \sim
\exp[-\ln\bar{l}/(T\ln 2)]$ and thus $\bar{l}\sim t^{T\ln 2}$. (Such
anomalous coarsening has also been found in other models of non-equilibrium
dynamics, often without detailed balance; see~\cite{Evans02} for examples.)
By extrapolating the coarsening law to the equilibrium domain length
$\bar{l}\eql\sim\exp(1/T)$, one then also finds that for $T\to 0$ the
dominant divergence of the equilibration time for the East model is $\tau
\sim \exp[1/(T^2\ln 2)]$, \ie\ an EITS law\eq{EITS}.

\subsection{Projection and mode-coupling techniques}
\label{meth:proj}

Much of the work on the stationary dynamics of KCMs makes use of
so-called projection techniques (Sec.~\ref{sec:langevin}), which
attempt to isolate relevant slow degrees of freedom from the less
relevant fast variables. The latter end up contributing via ``memory
functions'' and in Sec.~\ref{irreducible_memory} we review the
definition of one particular memory function which is regarded as most
suitable for the analysis of systems with stochastic dynamics. The
mode-coupling approximation itself is discussed in
Sec.~\ref{meth:mca}.

\subsubsection{Projection approach}
\label{sec:langevin}

The basic ideas of the projection approach are due to
Mori~\cite{Mori65}; for a modern textbook exposition see
\eg~\cite{HanMcD86}. The key aim of the formalism is to derive exact
dynamical equations for a selected set of ``relevant'' variables, with
the contributions from the remaining ``irrelevant'' variables isolated
in a form suitable for further, approximate treatment.

Consider a system governed by Markovian dynamics in continuous time,
with a set of $S$ configurations $\nv$. All models that we discuss in
this review are of this form; for the case of an SFM, for example,
$\nv$ would be the vector formed of all the spin variables $n_i$ and
range over $S=2^N$ configurations. The basic equation governing the
dynamical evolution is thus the master equation\eq{master}
\be
\deriv{t} p(\nv,t) = \sum_{\nv'} \ww(\nv'\to\nv)p(\nv',t) - \sum_{\nv''}
\ww(\nv\to\nv'') p(\nv,t) = \sum_{\nv'} L\T(\nv,\nv')p(\nv',t)
\label{markov}
\ee
if one defines the $S\times S$ matrix $L$ (the Liouvillian operator)
with elements
\be
L(\nv',\nv) = L\T(\nv,\nv') = 
\ww(\nv'\to\nv) - \delta_{\nv',\nv} \sum_{\nv''} \ww(\nv\to\nv'')
\label{L_def}
\ee
If $p(\nv,t)$ is viewed as a time-dependent vector $p(t)$ with $S$
entries, then $\partial p(t)/\partial t = L\T p(t)$ with the formal
solution
\be
p(t) = e^{L\T t} p(0)
\label{formal_solution}
\ee
An observable of the system is just a function $a(\nv)$, which can
again be regarded as a vector. It makes sense to define a scalar
or inner product on this space of vectors which is not Euclidean but
instead reflects the equilibrium correlations between observables,
\be
(a,b) \equiv \lav ab \rav = \sum_\nv a(\nv)b(\nv)\peq(\nv)
\label{inner_prod}
\ee
where $\peq(\nv)$ is the equilibrium distribution over
configurations. Strictly speaking\eq{inner_prod} is a disconnected
correlation, and we should subtract $\lav a \rav \lav b \rav$
(compare\eq{C_eq}), but for simplicity one assumes that any nonzero
equilibrium averages have been subtracted off from all observables,
such that $\lav a\rav = 0$ etc.
%
%
In terms of the scalar product\eq{inner_prod}, time-dependent
correlation functions take the simple form
\bea
C_{ab}(t) &=& \sum_{\nv'} b(\nv')p(\nv',t|\nv,0)a(\nv)\peq(\nv)
\nonumber\\
&=& \sum_{\nv'} \left(e^{Lt}\right)(\nv,\nv')b(\nv')a(\nv)\peq(\nv) =
(a,e^{Lt} b)
\label{corr_basic}
\eea
Here $p(\nv',t|\nv,0)$ is the probability that the system is in configuration
$\nv'$ at time $t$ if it was in configuration $\nv$ at time 0;
from\eq{formal_solution} this is the $(\nv',\nv)$-element of the
matrix $\exp(L\T t)$, hence the $(\nv,\nv')$-element of the matrix
$\exp(Lt)$. From\eq{corr_basic}, if one defines for any observable $b$
its value at time $t$ as
\be
b(t) = e^{Lt} b
\label{obs_evolution}
\ee
then simply $C_{ab}(t) = (a,b(t))$. Intuitively, the element
$b(\nv,t)$ of the vector $b(t)$ can be interpreted as the average
value of $b$ at time $t$ if the system started off in configuration $\nv$ at
time $t=0$.

Now consider a set of ``relevant'' observables $a_i$. For simplicity,
assume that they all have unit variance and are uncorrelated in
equilibrium, \ie\ $(a_i,a_j)=\delta_{ij}$; the generalization to the
case of arbitrarily correlated observables will be given below. Each
$a_i(t)$ obeys the equation of motion\eq{obs_evolution}; if we Laplace
transform to $a_i(z)=\int_0^\infty dt\, a_i(t)\exp(-zt)$ this can be
written as
\be
za_i(z) - a_i = La_i(z) = L(z-L)^{-1}a_i = (z-L)^{-1} L a_i
\label{eqn_motion}
\ee
The same symbol for $a_i(t)$ and its Laplace transform $a_i(z)$ is
used here since the argument makes clear which one is meant; $a_i$
continues to denote the value of the observable at time $t=0$. The key
idea is now to project the equations of motion\eq{eqn_motion} onto the
subspace of observables spanned by the $a_i$; since the $a_i$ are
orthonormal, the appropriate projector acts as
\be
Pb = \sum_i a_i (a_i,b)
\ee
%
%
The orthogonal projector is defined as $Q=1-P$. $P$ and $Q$ obey the
usual relations for projectors, e.g.\ $P^2=P$, $Q^2=Q$,
$PQ=QP=0$. They are also self-adjoint with respect to the inner
product, \eg\ $(a,Pb)=(Pa,b)$ since both expressions give $\sum_i
(a,a_i)(a_i,b)$.

To bring the project equations into a convenient form, one now writes
$(z-L)^{-1} = (z-PL-QL)^{-1}$ in\eq{eqn_motion} and applies the matrix
equality
\be
(A-B)^{-1} = (A-B)^{-1} B A^{-1} + A^{-1}
\label{dyson}
\ee
to $A=z-QL$, $B=PL$ to get
\be
za_i(z) - a_i = (z-L)^{-1} PLa_i + (z-L)^{-1} PL (z-QL)^{-1}QLa_i +
(z-QL)^{-1}QLa_i
\ee
Carrying out the projections implied by $P$ results in
\bea
za_i(z) - a_i 
%
%
&=& \sum_j a_j(z) \Omega_{ji} +
\sum_j a_j(z) (a_j, L(z-QL)^{-1}QLa_i) \nonumber \\
& & {} + {}(z-QL)^{-1}QLa_i
\eea
where the rate matrix $\Omega$ has elements $\Omega_{jk} =
(a_j,La_k)$. Transforming back to the time-domain 
%
%
and using $Q^2=Q$ to show that $e^{QLt}Q = e^{QLQt}Q = Qe^{QLQt}Q$
results in the desired projected equation of motion,
\bea
\deriv{t} a_i(t) &=& \sum_j a_j(t) \Omega_{ji} + 
\sum_j \int_0^t dt'\, a_j(t') M_{ji}(t-t') + r_i(t)
\label{Langevin}
\eea
Here
\be
M_{jk}(t) = (a_j,LQe^{QLQt}QLa_k) \label{M_def}
\ee
is a time-dependent memory matrix (also called memory kernel) and
\be
r_i(t) = e^{QLQt}r_i, \qquad r_i = QLa_i
\label{random_force_def}
\ee
the so-called random force. Eq.\eq{Langevin} is in the form of a
generalized Langevin equation. The first term on the r.h.s.\ leads to
an exponential decay of the observables towards zero (the matrix
$\Omega_{ki}$ has only non-positive eigenvalues, because the same is
true for $L$); the second term represents a generalized friction term
with the memory kernel $M_{ki}(t)$. The name random force is used for
$r_i(t)$ because it is always orthogonal to the space of observables
being projected onto: the definition\eq{random_force_def}
implies $Pr_i(t)=0$. In particular $(a_j,r_i(t))=0$ so that the random
forces are uncorrelated with the initial values of all the observables
considered. Using this property of the random force, taking a product
of\eq{Langevin} with the different $a_k$ also gives the desired
equation for the correlation functions $C_{ij}(t)=(a_i,a_j(t))$:
\be
\deriv{t} C_{ki}(t) = \sum_j C_{kj}(t) \Omega_{ji} + 
\sum_j \int_0^t dt'\, C_{kj}(t') M_{ji}(t-t')
\ee
or, in matrix form and after Laplace transform, bearing in mind that
the initial condition is $C_{ki}(t=0)=\delta_{ki}$,
\be
C(z) = (z - \Omega - M(z))^{-1}
\ee
In the case of general observables with arbitrary equilibrium
correlations this result generalizes to
\be
C(z) = C(zC - \Omega - M(z))^{-1}C
\label{Cz}
\ee
%
%
where $C$ (in our notation, see after\eq{eqn_motion}) denotes the
correlation matrix at time $t=0$, whose elements
$C_{ij}\equiv C_{ij}(t=0)=(a_i,a_j)$ are the equilibrium correlations.

Importantly, in systems with detailed balance one can show that the
memory matrix is the correlation function of the random force.  This
follows from the fact that for such systems, the operator $L$ is
self-adjoint. (The detailed balance condition $\ww(\nv'\to\nv)\peq(\nv')
= \ww(\nv\to\nv')\peq(\nv)$ implies from\eq{L_def} that
$L(\nv',\nv)\peq(\nv') = L(\nv,\nv')\peq(\nv)$ for all $\nv$ and
$\nv'$; multiplying by $a(\nv)b(\nv')$ and summing over $\nv$ and
$\nv'$ gives the desired result $(La,b)=(a,Lb)$.)  Using also that $Q$
is self-adjoint, the definition of $M(z)$ in\eq{M_def} can thus be
written as
\be
M_{jk}(t) = (QLa_j,e^{QLQt}QLa_k) = (r_j, r_k(t))
\label{M_det_bal}
\ee
Using similar arguments one also shows that, for systems with detailed
balance, the correlation function matrix, frequency matrix and memory
function matrix are all symmetric.

The result\eq{M_det_bal} implies, in particular, that one can treat
$M(z)$ in the same way as $C(z)$, expressing it in terms of an
appropriate frequency matrix $\Omega_2(z)$ and a new, second-order
memory function $M_2(z)$. This gives for the correlation function
\be
C(z) = C\left[zC - \Omega - 
M\left(zM-\Omega_2-M_2(z)\right)^{-1}M\right]^{-1}C
\label{second_order}
\ee
where $M$ is the value of the memory matrix at $t=0$,
$M_{ij}=(QLa_i,QLa_j)$. This approach implicitly tracks the motion of
the random forces $r_i$, and so it is not surprising that the same
result for $C(z)$ would be obtained from the first-order memory
function if the space of relevant observables was enlarged to include
the $a_i$ as well as the $r_i$ (or, equivalently, the $a_i$ and
$La_i$; either way one projects onto the same space of observables).
This process can be iterated to obtain a continued fraction expression
for $C$ in terms of memory functions of increasing
order~\cite{Mori65b}.

\subsubsection{Irreducible memory function}
\label{irreducible_memory}

The projection formalism, while formally exact, hides all complexities
of the dynamics in the memory functions, and one needs to find
approximate ways of calculating these in order to make the approach
useful. In applications to microscopic models of dense supercooled
liquids (systems of classical particles obeying Newton's equations),
the relevant ``slow'' observables $a_i$ are normally chosen as Fourier
modes of the particle number density fluctuations, and the dynamics is
deterministic and time-reversible. Approximations (such as
mode-coupling, see below) are normally applied to the second-order
memory 
function. The resulting models for the correlation functions have been
much studied~\cite{BenGoetSjo84,Goetze91,GoetSjoe92} and predict \eg\
dynamical transitions---signalled by the divergence of the longest
relaxation time---as external control parameters such as the overall
particle density are varied.

For models with stochastic dynamics and detailed balance, it is less
obvious which memory function to choose as the starting point for
approximations. Above we encountered the first- and second-order
memory function; Kawasaki~\cite{Kawasaki95,Kawasaki97} suggested
another, so-called irreducible memory function, based on earlier work
on the dynamics of colloidal suspensions~\cite{CicHes87}. The idea is
to decompose the operator $QLQ$ that governs the time evolution of the
random force into two parts:
\be
QLQ = L_0 + L_1
\label{QLQ_decomp}
\ee
Here $L_0$ is defined by its action on an arbitrary vector
$b$, as
\be
L_0 b = \sum_{ij} QLa_i \Omega^{-1}_{ij} (QLa_j,b)
\label{L0_def}
\ee
while $L_1$ is defined by the relation\eq{QLQ_decomp}. Applying the
identity\eq{dyson} to $A=z-L_1$, $B=L_0$ 
%
%
to the Laplace-transform of the expression\eq{M_det_bal} for the
memory matrix then gives
\be
M_{jk}(z) = (QLa_j,(z-QLQ)^{-1}L_0 (z-L_1)^{-1}QLa_k) +
(QLa_j,(z-L_1)^{-1} QLa_k)
\ee
Calling the last term the irreducible memory function
$M\irr_{jk}(z)$ and using the definition\eq{L0_def}, this becomes
%
\be
M_{jk}(z) 
%
= \sum_{lm} M_{jl}(z) \Omega^{-1}_{lm} M\irr_{mk}(z) + M\irr_{jk}(z)
\ee
%
or in matrix form $M(z) = M\irr(z) + M(z)\Omega^{-1}M\irr(z)$. The
first-order memory function can thus be expressed in terms of the
irreducible one as $M(z) = M\irr(z)(1 - \Omega^{-1}M\irr(z))^{-1}$,
and in the correlation function matrix\eq{Cz} this
gives
%
\be
C(z)= 
C\left[zC- \Omega\left(1 - \Omega^{-1}M\irr(z)\right)^{-1}\right]^{-1}C
\label{C_Mirr}
\ee
%

A nice physical interpretation of the irreducible memory function was
given by Pitts and Andersen~\cite{PitAnd00}. They argue that a system
with stochastic dynamics (\eg\ a system of colloidal particles with
Brownian dynamics, or the much more abstract lattice gases with
kinetic constraints) must eventually be derivable from an underlying
system with deterministic, time-reversible dynamics. At long-times,
the two descriptions should give the same results for correlation
functions. This then implies that the irreducible memory function for
stochastic dynamics must be proportional to the second-order memory
function of the time-reversible description. The argument is based on
a comparison of\eq{second_order}, as applied to the time-reversible
system, with\eq{C_Mirr} when applied to the stochastic
system. Time-reversibility can be shown to imply that the matrices
$\Omega$ and $\Omega_2$ in\eq{second_order} vanish, giving
\be
C(z) = C\left[zC - M\left(zM-M_2(z)\right)^{-1}M\right]^{-1}C
\label{reversible}
\ee
For times that are long compared to the microscopic timescales of the
deterministic dynamics, the corresponding $z$ can be shown to be small
enough for the term $zM$ to be neglected~\cite{PitAnd00}. Agreement
with\eq{C_Mirr} then requires that $M_2(z) = -M\Omega^{-1}M +
M\Omega^{-1}M\irr(z)\Omega^{-1}M$. The first term is independent of
$z$ and gives a delta function-like contribution to $M_2(t)$; for
longer times, the second term shows that $M_2(t)$ of the deterministic
description and $M\irr(t)$ of the stochastic description are related
by constant factors as claimed. The upshot of this is that
approximations analogous to mode-coupling theory for dense liquids are
obtained by applying the mode-coupling approximation to the
irreducible memory function of stochastic systems.

\subsubsection{Mode-coupling approximation}
\label{meth:mca}
\label{meth:mct}

The simplest approximation for the (reducible) memory function is to
neglect it. Setting $M(z)=0$ in\eq{Cz}, the calculation is reduced to
the diagonalization of the matrix $\Omega$, and all correlation
functions become superpositions of exponentially decaying modes.
Effectively this corresponds to a mean field-like truncation of the
hierarchy of correlation functions to just those of the ``relevant
variables'' retained. This approach can describe some aspects of the
slowing down of the dynamics in kinetically constrained systems, but
is incapable of predicting
\eg\ an incomplete decay of correlation functions which would be
expected at a dynamical transition.

An improved---but still uncontrolled---approach is the mode-coupling
approximation (MCA). As an illustration, consider the East model. The
configuration $\nv$ is specified by that of all spins $n_i=0,1$, and the
matrix elements of the Liouvillian are given
by\eqq{all_rates}{eq5S311}
\be
L(\nv',\nv) = L\T(\nv,\nv') = 
\sum_i n'_{i-1}[c(1-n_i')+(1-c)n_i']\left(\delta_{\nv,F_i\nv'} -
\delta_{\nv,\nv'}\right)
\label{L_East}
\ee
Here we have abbreviated by $c\equiv c\eql$ the equilibrium concentration
of up-spins, and $F_i$ is the operator which flips spin $i$. If we are
interested in spin-correlation functions, the relevant observables are
the spin-fluctuations $\et_i=[c(1-c)]^{-1/2}(n_i-c)$, normalized such as to
obey $C_{ij} = (\et_i,\et_j)=\delta_{ij}$. Together with the unit
observable $e$ and all different products $\et_{i_1}\cdots \et_{i_m}$
($m=1\ldots N$) these observables form an orthonormal basis for the space
of all observables. In an obvious abuse of notation, products such as
$\et_j\et_k$ are here understood to be taken componentwise, \eg\
$(\et_j\et_k)(\nv) = \et_j(\nv) \et_k(\nv)$.

One can now construct the rate and memory matrices.  For an arbitrary
observable $a(\nv)$ one has, from\eq{L_East}
\be
(La)(\nv) = \sum_i n_{i-1}[c(1-n_i)+(1-c)n_i][a(F_i\nv)-a(\nv)]
\ee
and applying this to $a=\et_i$ gives
\be
L\et_i = - c\left[\et_i + [(1-c)/c]^{1/2}\et_{i-1}\et_i\right]
\ee
The rate matrix is $\Omega_{ij}=(\et_i,L\et_j) = -c\delta_{i,j}$
since $(\et_i,\et_j\et_k)=0$. The initial values of the random forces
follow as
\be
r_i = QL\et_i = L\et_i - \sum_j \et_j (\et_j,L\et_i) = -[c(1-c)]^{1/2}
\et_{i-1} \et_i
\ee
giving for the reducible memory matrix
\be
M_{ij}(t) = c(1-c)(\et_{i-1}\et_i,e^{QLQt}\et_{j-1}\et_j)
\label{M_East}
\ee
while the expression for $M\irr_{ij}(t)$ is obtained by replacing $QLQ$
with $L_1$ in the exponent on the r.h.s.\ of\eq{M_East}. The MCA can
be applied to either of 
these functions. It replaces $QLQ$ or $L_1$ by $L$, and also assumes that
the resulting fourth-order correlation function can be factorized into
pairwise contributions like $(\et_i,e^{Lt}\et_j)$. Since the spin-spin
correlation function for the East model are site-diagonal,
$C_{kl}(t)=(\et_k,e^{Lt}\et_l)=C(t)\delta_{kl}$ (see
Sec.~\ref{res:statdyn}), the only nonzero contribution to\eq{M_East}
becomes $M\mca_{ij}(t) = \delta_{ij}M\mca(t)$ with~\cite{EisJaec93}
\be
M\mca(t) = c(1-c) (\et_{i-1},e^{Lt}\et_{i-1})(\et_{i},e^{Lt}\et_{i}) 
= c(1-c) C^2(t)
\label{MCA_East}
\ee
Using this as an approximation for $M\irr_{ij}(t)$, one has
from\eq{C_Mirr}, bearing in mind that all matrices involved are
diagonal,
\be
C(z) = \left(z+\frac{c}{1 + c^{-1}M\mca(z)}\right)^{-1}
\label{C_MCA_East}
\ee
Together with\eq{MCA_East} this is a closed MCA equation for $C(t)$,
which is equivalent to a model of the glass transition studied in
detail by Leutheusser~\cite{Leutheusser84b,Leutheusser84}. One can
ask, in particular, whether on lowering $c$ a dynamical transition
occurs to a non-ergodic state where $C(t)$ no longer decays to zero
for $t\to\infty$. If $C(t\to\infty)=q$, then $C(z)\simeq q/z$ and
$M\mca(z) \simeq c(1-c)q^2/z$ for small $z$. Inserting
into\eq{C_MCA_East} and taking $z\to 0$ gives
\be
q = \left(1+\frac{c}{(1-c)q^2}\right)^{-1}
\ee
or $q/(1-q)=(1-c)q^2/c$. The largest solution in the range $0\leq
q\leq 1$ gives $C(t\to\infty)$~\cite{Goetze91}; it is easily worked
out as $q=1/2+[1/4-c/(1-c)]^{1/2}$, yielding a first-order dynamical
transition---a discontinuous jump of $q$, from 0 to 1/2---at
$c/(1-c)=1/4$. Thus, the MCA approximation applied to the irreducible
memory function of the East model predicts a spurious dynamical
transition at $c=0.2$~\cite{Kawasaki95}. As expected from the
discussion at the end of Sec.~\ref{irreducible_memory}, applying the
MCA to the {\em reducible} memory function gives even less reasonable
results: J\"ackle and co-workers found both for constrained spin
models (\eg\ the East model~\cite{EisJaec93}) and the triangular
lattice gas~\cite{KroJaec94} that unphysical divergences for the
correlation functions at long times could occur.

It should be noted that for models with stochastic dynamics, MCAs for
second- or higher-order memory functions can never predict a
non-ergodic decay of correlation functions to a nonzero value; see
\eg~\cite{LowHanRou91,PitAnd00,EinSch01}. This can be seen
from\eq{second_order}. If $M_2(z)$ is linked via a MCA to $C(z)$, then
a non-ergodic state requires that $M_2(z)$ diverge as $\sim 1/z$ for
$z\to 0$; but then $C(z)=C(zC+\order(z)-\Omega)^{-1}C$ for small $z$
which has a finite limit for $z\to 0$ rather than the assumed $1/z$
divergence (as long as $\Omega$ is nonzero). For time-reversible
dynamics, the situation is different since there $\Omega=0$; see
before\eq{reversible}.

Notice that for more complicated directed models, \eg\
the North-East model or Cayley tree models, the random force $r_i =
QL\et_i$ will contain not just second-order products of spin
fluctuations, but also higher orders such as $\et_j\et_k\et_l$. For
sufficiently simple models~\cite{Kawasaki95,PitAnd01} the coefficients
can be worked out explicitly, and the MCA then gives expressions for
the memory functions which also involve higher powers of $C(t)$. If
this procedure is too complicated, one can in addition project $r_i$
onto a subspace of observables, \eg\ the one spanned by the
second-order products $\et_k\et_k$~\cite{EinSch01}. Finally, we note
that in the context of supercooled liquids, {\em extended} MCAs have
been derived~\cite{Sjoegren80,GoetSjoe87,GoetSjoe95}. These lead to
approximations for the memory matrix of the form
$M\mca(z)[1+\Delta(z)M\mca(z)]^{-1}$, where $M\mca(z)$ is the memory
matrix in the conventional MCA, \eg\eq{MCA_East} for the East
model, and $\Delta(z)$ is a new memory matrix. The presence of a
nonzero $\Delta(z)$ ensures that in extended MCA the memory matrix
does not become singular for $z\to 0$ even if $M\mca(z)$ does, and
thus smoothes out the sharp dynamical transitions generally predicted by
conventional MCA.  The formalism of extended MCA has not yet been
adapted for models with stochastic dynamics; nevertheless,
approximations of similar form have recently been derived for
kinetically constrained models using the diagrammatic approaches
reviewed in the next section.

\subsection{Diagrammatic techniques}
\label{meth:diag}

Equilibrium correlation functions for kinetically constrained models have
also been studied using diagrammatic expansion. In fact, the first
theoretical treatment~\cite{FreAnd85} for $f,d$-SFMs was derived from a
diagrammatic expansion. We review here the formulation recently provided by
Pitts and Andersen~\cite{PitAnd01} for the East model and other
models with directed constraints; a related approach was used for the
$1,1$-SFM in~\cite{SchTri99}. The spin autocorrelation function in the East
model is site-diagonal (see Sec.~\ref{res:statdyn}), and in the notation of
Sec.~\ref{meth:proj} can be written as
\be
C(t) = (\et_i,e^{Lt}\et_i), \qquad C(z) = (\et_i,(z-L)^{-1}\et_i)
\label{C_East}
\ee
The Liouvillian is given in\eq{L_East} and can be written as $L=\sum
L_i$, with $L_i$ corresponding to spin flips at site
$i$. One can now expand the inverse in\eq{C_East}, and insert
decompositions of the identity matrix $1=\sum_\phi \phi)(\phi$, where
$\phi$ runs over the orthonormal basis vectors of the space of
all observables built up from products of the $\et_j$ (see
after\eq{L_East}). This gives for the Laplace transform of the
spin-spin autocorrelation
\be
C(z) = \sum_{k=0}^\infty \frac{1}{z^{k+1}} \sum_{i_1\ldots i_k}
\sum_{\phi_1\ldots \phi_{k-1}}
(\et_i,L_{i_1}\phi_1)(\phi_1,L_{i_2}\phi_2)\cdots
(\phi_{k-1},L_{i_k}\et_i)
\ee
Each term in this series can be represented by a diagram; the value
associated with each diagram is determined by a product of ``matrix
elements'' $(\phi,L_j\phi')$, which for the East model are easily
worked out explicitly. A closer investigation of the
structure of the diagrammatic expansion reveals that the first-order
reducible and irreducible memory functions can both be obtained as the
sum of appropriately selected subsets of diagrams~\cite{PitAnd01}. If
these subseries are summed approximately, expressions for the memory
functions result which are, nontrivially, of the same general form as
those obtained from a MCA within the projection formalism: the
irreducible memory function $M\irr(t)$ becomes a polynomial in $C(t)$.
For the East model, for example, the most straightforward
approximation yields
\be
M\irr(t) = c(1-c)C(t)
\label{Mirr_East_diag}
\ee
(This result was also obtained by J\"ackle and
Eisinger~\cite{JaecEis91,EisJaec93} using their ``effective medium
approximation''.) Compared to\eq{MCA_East}, the power of $C(t)$ on the
r.h.s.\ of\eq{Mirr_East_diag} is reduced by one, although\eq{MCA_East}
itself can also be retrieved if a different subset of the diagrams for
$M\irr(z)$ is summed. For the (3,2)-Cayley tree model and the
North-East model one obtains by the same approach {\em identical}
expressions for the irreducible memory function,
\be
M\irr(t) = 2c^3(1-c)C(t)+c^2(1-c)^2C^2(t)
\ee
again containing one power of $C(t)$ less than the results from the
MCA used by Kawasaki~\cite{Kawasaki95}.

For the East model, Pitts and Andersen~\cite{PitAnd01} pushed the
analysis even further and showed that a more sophisticated
rearrangement of the series for $M\irr(z)$ can be used to derive
approximations that are of the same form as the extended MCA for
supercooled liquids (see Sec.~\ref{meth:mca}). As expected on general
grounds from the structure of extended MCA, these improved
approximations avoid the spurious dynamical transitions
predicted for the East model by simpler approximations such
as\eq{Mirr_East_diag}. A fuller discussion of the results obtained
from the diagrammatic expansions will be given later, in
Sec.~\ref{res:statdyn}.

\subsection{Mappings to quantum systems and field theories}
\label{meth:fock}
\label{meth:rg}

It can be useful to think of the vector space of observables on the
space of configurations $\nv$ as a quantum mechanical Hilbert space. A useful
basis for this Hilbert space are the vectors $|\nv\rangle=|n_1\ldots
n_N\rangle$; $|\nv\rangle$ corresponds to the observable which is one
if each spin $i$ has the specified value $n_i$, and zero otherwise. The
vector describing the probability of being in any given configuration is then
written as $|p(t)\rangle = \sum_\nv p(\nv,t)|\nv\rangle$, and the
master equation\eq{markov} becomes
\be
\deriv{t} |p(t)\rangle = - H|p(t)\rangle
\label{quantum_markov}
\ee
The quantum Hamiltonian $H$ here corresponds to the operator denoted $-L\T$
in\eq{markov}; the minus sign is introduced so that the eigenvalues of $H$
are non-negative and its ground states just give the steady states
$|p\rangle$ of the system. Notice that the quantum mechanical Hilbert space
product is defined so that the configurations $|\nv\rangle$ are orthonormal; this
is different from\eq{inner_prod}.

We describe briefly how to construct $H$, using the East model as an
example. One essentially needs to transcribe $L\T$ from\eq{L_East}. It
is useful to adopt a particle language, with $n_i=0$ and 1
respectively corresponding to the absence and presence of a particle
at site $i$. It is then natural to define $|0\rangle$, the configuration with
$n_1=\ldots = n_N=0$, as the vacuum, and obtain other configurations by
applying suitable creation operators $\cre_i$ which act as
\be
\cre_i|\ldots n_i=0 \ldots\rangle =
|\ldots n_i=1 \ldots\rangle,
\qquad
\cre_i|\ldots n_i=1 \ldots\rangle = 0
\ee
The ``Paulion''~\cite{MatGla98} operators $\cre_i$ and their Hermitian
conjugates $\ann_i$ then commute at different sites, while at the same
site they obey anticommutation rules, $\{\ann_i,\ann_i\}$ $=$
$\{\cre_i,\cre_i\}$ $=$ $0$, $\{\ann_i,\cre_i\}=1$. The operator
$\cre_i\ann_i$ counts the number of particles at site $i$ in the usual
way, $\cre_i\ann_i|\nv\rangle = n_i |\nv\rangle$. Only one more
ingredient is needed to write down $H$: the spin-flip operator $F_i$
from\eq{L_East} becomes $\cre_i+\ann_i$ in the quantum version. Thus,
the Hamiltonian $H\equiv -L\T$ for the East model is
\bea
H &=& -\sum_i \cre_{i-1}\ann_{i-1} [c(1-\cre_i\ann_i)+(1-c)\cre_i\ann_i]
\left(\cre_i+\ann_i-1\right) \nonumber\\
&=& \sum_i \cre_{i-1}\ann_{i-1} [c(\ann_i-1)\cre_i+ (1-c)(\cre_i-1)\ann_i]
\label{H_East}
\eea
using the anticommutation relations to simplify the final
expression. Conservation of probability is reflected in the fact that
$\langle e|H=0$, where $|e\rangle = \sum_\nv |\nv\rangle = \prod_i
(1+\cre_i)|0\rangle$ is the unit or ``reference'' state; this ensures that
$\langle e|p(t)\rangle=\sum_\nv p(\nv,t)=1$ does not change in time.  As is
typical, the Hamiltonian\eq{H_East} is non-Hermitian since it is derived
purely from a dynamical problem. Since the dynamics obeys detailed balance,
however, the similarity transformation $|\nv\rangle \to
P^{1/2}\eql|\nv\rangle$ and $H\to P^{1/2}\eql H P^{-1/2}\eql$ with $P\eql =
\sum_{\nv} p\eql(\nv)|\nv\rangle\langle\nv|$ could be used to transform $H$
to an explicitly Hermitian form.

Physical observables $A$ are functions of the $n_i$, and therefore
correspond to operators which are diagonal in the basis $|\nv\rangle$,
$A(\nv,\nv')=\langle\nv|A|\nv'\rangle=A(\nv)\delta_{\nv,\nv'}$; their
expectation values are given in the quantum formulation by
\be
\lav A(t) \rav = \sum_\nv A(\nv) p(\nv,t) = \langle e|A|p(t)\rangle =
\langle e|Ae^{-Ht}|p(0)\rangle 
\ee

Above, we effectively viewed the quantum mechanical Hilbert space as a Fock
space, since it is spanned by configurations with any possible value of the
total particle number $\sum_i n_i$ between 0 and $N$. Equivalently, one
can think of the Hilbert space as the configuration space of a quantum spin
system, with $n_i=(1+\s_i)/2$ and $\s_i$ the eigenvalue of
the $z$-component $\s^z_i$ of a quantum spin operator. The particle
creation and annihilation operators then become raising and lowering
operators $\s_i^\pm = \s^x_i \pm i \s^y_i$, and the vacuum state is
the one with all spins down.

The above idea of mapping classical stochastic dynamical systems onto
quantum models was pioneered by Doi~\cite{Doi76,Doi76b} for
``bosonic'' systems, where many particles can occupy a given site, and
later generalized to the ``fermionic'' case of at most single
occupancy that is relevant to us (see \eg~\cite{GraSch80}).  An
overview of developments in the field since then and a comprehensive
bibliography can be found in~\cite{MatGla98}. As demonstrated
beautifully in recent reviews~\cite{Stinchcombe01,Stinchcombe02},
quantum mappings have proved very powerful in the analysis of many
stochastic non-equilibrium systems, particularly where the resulting
Hamiltonians are those of known (and sometimes even exactly solvable)
quantum systems~\cite{AlcDroHenRit94}. They can also form the starting
point for field-theoretic path-integral
representations~\cite{Peliti85,LeeCar95}.  Either from the latter or
directly from the real-space (lattice) quantum Hamiltonians,
renormalization group methods (see
\eg~\cite{Hu82,HooVan00,Stinchcombe01}) then also become available to
study the behaviour at large lengthscales.

For KCMs specifically, however, the benefits
of the approach largely remain to be explored. Some use has been made
of the formalism
(\eg~\cite{SchTri97,SchTri98,PigSchTri99,SchTri99,EinSch01,SchTri02})
but with few extra insights gained that would not also have been
available directly from the master equation; and in at least one case
the formal manipulations actually obscure rather than clarify the
simplifications resulting from detailed balance~\cite{EinSch01,SchTri02}.

\subsection{Mappings to effective models}
\label{meth:mappings}

The low-temperature dynamics of KCMs can often be understood by means of a
mapping to effective models. We already discussed such a mapping for
$f,d$-SFMs with $f=1$ in Sec.~\ref{sfm:some_results}, where we found that
the dynamics at low up-spin concentration $c\eql$ can be described in terms
of the diffusion of defects, in this case isolated up-spins, with an
effective diffusion constant $D\eff=c\eql/2$. Apart from diffusing,
up-spins can also ``coalesce'': when two of them are only separated by a
single down-spin, the latter can flip up and then two of the resulting
three up-spins can flip down successively. The reverse process where a
single up-spin creates a second one is of course also possible by
detailed balance. The effective low-temperature model for the $1,d$-SFMs is
thus one of diffusing up-spins which can ``react'' according to
$A+A\leftrightarrow A$, where $A$ stands for the single species of defect
``particle'' in the system. This convenient representation, in which the
kinetic constraints no longer appear explicitly, has been exploited \eg\
in~\cite{ReiJaec95,SchTri99}, and a similar description has been used for a
driven version of the $1,1$-SFM~\cite{BrePraSan00} (see
Sec.~\ref{res:driven}). Much is known about such reaction-diffusion models;
see \eg~\cite{Ben-Avraham98} for a recent list of references on the
$A+A\leftrightarrow A$ model in $d=1$. We have not specified above the
precise ratio of the reaction and diffusion rates, but its value is
expected to be unimportant in the relevant regime of small
$c\eql$~\cite{ZhoBen95}.

As explained already in Sec.~\ref{topological}, in a lattice
version~\cite{DavSheGarBuh01} of the topological froth model a similar
mapping to an effective model is also useful. At low-temperatures very
few defects ($+1,-1$-spins) exist, and it can be
argued~\cite{DavSheGarBuh01} that the dynamics is dominated by defect
pairs---dimers of adjacent $+1,-1$-spins---and isolated defects. Since
dimers can diffuse and annihilate with each other or with isolated
defects, one thus has again an effective reaction-diffusion model at
low temperatures which can be used to understand \eg\ the relaxation
of the energy, \ie\ the defect concentration, after a quench.

Other effective models for KCMs can be obtained by coarse-graining to a
continuum description; this approach has been successfully exploited to
describe the properties of lattice gases, \eg\ the KA model with and
without gravity~\cite{PelSel98,LevAreSel01}. One represents the state of
the system by a coarse-grained density field $c(z)$ which under the effect
of gravity should only depend on height $z$. Since the lattice gas is
non-interacting, the local free energy density is simply $f(c)= T[c\ln c +
(1-c)\ln(1-c)] + gcz$, with the last term accounting for the effects of
gravity. One can now postulate a standard dynamics for the conserved
density field, $(\partial/\partial t)c(z) = - \partial J(z)/\partial z$.
The current $J(z)=-\Gamma(c(z))\partial\mu(z)/\partial z$ is the product of
a local mobility $\Gamma(c)$ and the negative gradient of the chemical
potential, which is given by $\mu(z)=\delta F/\delta c(z)$ with $F=\int
dz\,f(c(z))$ the total free energy. The model is made glassy only
through the choice of the functional form of the mobility
$\Gamma(\rho)$. To model the power-law singularity of the diffusion
constant seen in simulations of the KA model~\cite{KobAnd93} (see
Sec.~\ref{ka:some_results}), this was chosen in~\cite{LevAreSel01} as
$\Gamma(\rho)=c(1-c/c\dyn)^\phi$, which tends to zero with an exponent
$\phi\approx 3.1$ as the density approaches the dynamical transition at
$c\dyn$. Being based on the behaviour of the diffusion constant in a system
at uniform density $c$, it is not obvious that this is still a good
approximation for the {\em local} mobility, especially in the interesting
high-density region where one may expect pronounced
inhomogeneities. Nevertheless, is has been shown to work remarkably well
both for the KA model under gravity~\cite{LevAreSel01} and without gravity
but with particle exchange with a reservoir allowed~\cite{PelSel98}. We
mention in passing that the dynamics of a related class of models with
density-dependent mobilities have recently been analysed
in~\cite{CorNicPicCon99,CorNicPicCon01}.

\section{Results}
\label{res}

In this section, we give a comprehensive survey of the known results on the
dynamics of KCMs, including work on related models where appropriate. We
begin in Sec.~\ref{res:reduc} with the question of (effective)
irreducibility, which ensures that naive calculations of equilibrium
behaviour apply to KCMs. The following sections are arranged to mirror the
structure of Sec.~\ref{basics}. In Sec.~\ref{res:dyntrans} we give results
for the typical relaxation timescales of KCMs and their dependence on
temperature or, for lattice gases, density; we also evaluate there the
evidence for genuine dynamical transitions in KCMs. In
Sec.~\ref{res:statdyn} we address the stationary dynamics of KCM, which
should be relevant for modelling the dynamics around the (metastable)
equilibrium of supercooled liquids. Sec.~\ref{res:noneq} is concerned with
out-of-equilibrium dynamics, including nonlinear relaxation
after quenches or crunches, hysteresis effects in heating-cooling cycles
and two-time correlation and response functions. Dynamical lengthscales in
KCMs and the evidence for dynamical heterogeneities are discussed in
Sec.~\ref{res:hetero}. In Sec.~\ref{res:landsc} we review the applicability
of energy landscape paradigms such as configurational entropies and Edwards
measures to KCMs. Finally, Sec.~\ref{res:driven} surveys some recent
results on the behaviour of KCMs under external driving, which can be used
to model \eg\ tapping experiments in granular media.

Within each subsection, we list results for the various models as far as
possible in the order in which they were introduced in
Sec.~\ref{allmodels}. First are $f,d$-SFMs and their variants with directed
constraints; where appropriate, we discuss the models with $f=1$ separately
because of their qualitatively different defect-diffusion dynamics. The
next major group of models is formed by the kinetically constrained lattice
gases, followed by the models inspired by cellular structures and the
triangle and plaquette models obtained by mappings from interacting systems
with unconstrained dynamics. Finally, results for related models such as
urn, oscillator and needle models are included where appropriate.

\subsection{Irreducibility}
\label{res:reduc}

Beginning with {\bf spin-facilitated models} with undirected constraints,
let us summarize under which conditions on $f$ the $f,d$-SFM is effectively
irreducible. Formally, this means $p(c,L\to\infty)=1$ for all $c>0$;
$p(c,L)$ is the probability that a random initial configuration with
up-spin concentration $c$ on a lattice of $N=L^d$ spins belongs to the
high-temperature partition (see Sec.~\ref{intro:irred}). In order to
understand finite-size effects, it is also useful to define the
concentration $c_*(L)$ as the one where $p(c,L)=1/2$ for given $L$;
effective irreducibility corresponds to $c_*(L\to\infty)=0$. The
irreducibility results quoted below were mostly derived within the context
of bootstrap percolation (BP). Recall from Sec.~\ref{meth:reduc} that the
$m$-BP process is defined as iteratively removing from a lattice all particles 
that have fewer than $m$ neighbours. By mapping particles to
down-spins and vacancies to up-spins we saw in Sec.~\ref{meth:reduc} that
if this process leads to an empty lattice, the corresponding configuration
in the $f,d$-SFM belongs to the high-temperature partition, provided that
$m$ is chosen as $m=2d+1-f$. As an example, the irreducibility problem for
the $3,3$-SFM corresponds to $4$-BP in $d=3$ dimensions.

As explained in Sec.~\ref{meth:reduc}, $f,d$-SFMs with $f>d$ are always
strongly reducible; for $f=1$, on the other hand, it is trivial to see that
they are effectively irreducible. Nontrivially,
Schonmann~\cite{Schonmann92} was able to prove rigorously that all models
with the intermediate values $2\leq f\leq d$ are also effectively
irreducible.  Enter~\cite{Enter87} had earlier proved the result for the
special case of the $2,2$-SFM, formalizing an earlier unpublished argument
due to Straley; Schonmann~\cite{Schonmann90,Schonmann92} gave a
generalization to BP-like models with more complicated rules. Fredrickson
and Andersen~\cite{FreAnd85} had earlier given a non-rigorous argument for
irreducibility of the $3,3$-SFM; Reiter~\cite{Reiter91} also constructed
irreducibility proofs for the $2,2$-SFM and $3,3$-SFM.

Numerical investigations of finite-size reducibility effects in SFMs
go back at least to Fredrickson and Brawer~\cite{FreBra86}, who
studied $p(c,L)$ and $c_*(L)$ in the $2,2$-SFM.  A simple linear
extrapolation of $c_*(L)$ versus the inverse linear system size
$L^{-1}=N^{-1/2}$ suggested $c_*(L\to\infty)\approx 0.04$, but
Fredrickson and Brawer~\cite{FreBra86} argued that the functional form
of this extrapolation was inappropriate since earlier
arguments~\cite{FreAnd85} had already suggested $c_*(L\to\infty)=0$.
It was later shown rigorously~\cite{AizLeb88} and confirmed by
simulation~\cite{AdlStaAha89} that for the general $2,d$-SFM, $c_*(L)$
decreases only very slowly with system size, as $c_*(L)\sim 1/(\ln
L)^{1/(d-1)}$. For other choices of $f$, the finite-size effects can
be even larger. Enter {\em et al.}~\cite{EntAdlDua90} considered the
case $f=d$; this is the ``most dangerous'' case that is still
effectively irreducible, since for $f=d+1$ and above the models are
strongly reducible. For $d=3$ the finite-size scaling of the critical
concentration was predicted to be $c_*(L) \sim 1/\ln(\ln
L)$~\cite{EntAdlDua90}; compared to $d=2$ this has an extra $\ln$ in
the denominator and this pattern continues for higher $d$, with
$c_*(L) \sim 1/\ln[\ln(\ln L)]$ for $d=4$ etc. This very slow approach
of $c_*(L)$ to zero is obviously difficult to verify numerically; for
$d=3$ initial simulations were interpreted in terms of a nonzero
$c_*(L\to\infty)$~\cite{KogLea81,AdlAha88,ManStaHee89}, but later work
showed an approach of $c_*(L)$ to zero that is consistent with the
predictions~\cite{EntAdlDua90}.

Consider next spin models with {\bf directed kinetic constraints}. For the
asymmetric $1,1$-SFM and its limit case the East model, the same argument
as for the (symmetric) $1,1$-SFM applies; all configurations except those
with all spins down belong to the high-temperature partition and
reducibility effects are unimportant. For the North-East
model~\cite{Schonmann92,ReiMauJaec92} it has been shown, via a mapping to
directed percolation, that in the thermodynamic limit a configuration will
have a finite fraction of permanently frozen spins if its up-spin
concentration $c$ is below the critical value $c_*=0.2942$. (The link to
directed percolation arises because a spin will never be flipped up if and
only if there is an infinite path starting from the chosen spin that
consists of steps towards the North or East and visits only down-spin
sites.)  For the $(a,f)$-Cayley tree models in the most strongly
constrained case $f=a-1$, one has a continuous blocking transition at
$c_*=(a-2)/(a-1)$, below which the fraction of permanently frozen spins
grows continuously from zero; this can be shown by using recursion
relations for trees of increasing depth (see Sec.~\ref{meth:reduc}). For
$2\leq f<a-1$ a blocking transition still exists, but is discontinuous.

Moving on to constrained {\bf lattice gases}, reducibility effects in the
KA model were discussed already in the original paper on the
model~\cite{KobAnd93}. As mentioned in Sec.~\ref{model:lattice_gases}, such
effects obviously depend on the parameter $m$ in the model; recall
that particles with
$m$ or more occupied neighbour sites are not allowed to move. On a cubic
lattice $m=6$ corresponds to an unconstrained system, while the case $m=3$
is strongly reducible, with any set of eight particles arranged in a cube
unable to move. For $m=4$, KA argued that the model should be effectively
irreducible in the thermodynamic limit, as follows. They focused on the
``backbone'', comprising all particles which are permanently frozen by
other frozen particles, \ie\ which remain frozen when all mobile particles
are removed (see Sec.~\ref{meth:reduc}). The backbone can thus be
determined by iteratively removing all mobile particles from the
lattice. In this process, a particle is removed if it has fewer than $m=4$
particles as neighbours, and if there is at least one free neighbour site
for which this condition would still be true after a jump to that
site. Since the first part of this criterion is just the same as BP with
$m=4$, a backbone of permanently frozen particles will remain for densities
where 4-BP does not reach the empty lattice configuration. This
implies~\cite{EntAdlDua90} that for particle densities $c\geq
1-\order(1/\ln(\ln L))$ a backbone will occur with high probability in a
system of linear size $L$; this criterion is just the obvious
transformation ($c\to 1-c$) of the one for irreducibility of the $3,3$-SFM
because the latter problem is essentially equivalent to $4$-BP. For lower
densities one expects the probability of a backbone to occur to be small,
and the system to be effectively irreducible. The theoretically expected
finite size effects are extremely strong, however: $c\approx
1-\order(1/\ln(\ln L))$ translates into a {\em double} exponential
divergence $L\sim \exp\{A\exp[B/(1-c)]\}$ of the system sizes required to
avoid a backbone at a given density. KA showed by direct simulation that up
to densities $c\leq 0.86$ for their $L=20$ system the probability for a
backbone to occur is very small ($\approx 0.007$; see
Fig.~\ref{lg_fig1} above), and that therefore
finite-size reducibility effects on their simulation results should be
negligible. (A possible caveat is that there may be particles that are
permanently frozen only by {\em mobile} neighbours, and these would not be
counted in the backbone; but simulations by J\"ackle and
Kr\"onig~\cite{JaecKro94} for the triangular lattice gas suggest that this
is a small effect.) For only slightly higher densities ($c=0.88$ and
$0.885$), they found that much larger system sizes ($L=40$ and $50$,
respectively) were required to avoid backbones; this is at least
qualitatively consistent with the theoretically expected strong increase of
$L$ with $c$.

In the KA model with particle exchange allowed at the boundary with a
reservoir at some chemical potential, or under the effect of gravity in a
simulation box of large height, reducibility effects are greatly reduced
compared to the conventional KA model. This is because particles can be
removed one by one to the reservoir, or the upper reaches of the simulation
box, and then reinserted, so that all configurations that can be
``emptied'' in this way are mutually accessible. In some cases this makes
the dynamics fully irreducible. A nice illustration is provided by a b.c.c.\
lattice where particles can move only if they have fewer than $m=5$ nearest
neighbours in their old and new positions~\cite{SelAre00}; lattice planes
can then be successively emptied starting from the top, since every
particle has at most four nearest neighbours in the lattice plane
underneath (and, due to the lattice structure, none in its own plane). For
the conventional KA setup, \ie\ a cubic lattice with $m=4$, it was argued
in~\cite{KurPelSel97} that configurations up to densities $c=1-\order(1/L)$
are mutually accessible. While there are {\em some} configurations with
such densities that can be accessed, accessibility of {\em typical}
configurations should only be possible up to lower densities
$c=1-\order(1/\ln L)$. This follows from the fact that, in a given
lattice plane at the top of the system that is to be emptied, most
particles (for $c$ close to 1) have one neighbour in the plane underneath; they can
thus be removed only if they have less than three neighbours in the
plane. The problem thus reduces to BP on a square lattice with $m=3$
which---as we know from the equivalence to the irreducibility problem in
the $2,2$-SFM---reaches the empty configuration with probability close to one only
for $1-c\sim 1/\ln L$.

For the triangular lattice gas (with two-vacancy assisted hopping), it was
shown in~\cite{JaecKro94} that no permanently blocked particles should
exist in the thermodynamic limit, at any particle concentration $c<1$. The
argument is quite similar to the irreducibility proofs outlined in
Sec.~\ref{meth:reduc}. It is based on the fact that a hexagonal ring of
vacancies can move outwards as long as there is at least one vacancy on
each of the six edges surrounding the hexagon. The probability of a local
particle configuration with a vacancy hexagon that can grow to arbitrary
size can be shown to be nonzero, and so in a thermodynamically large system
at least one such local configuration will exist with probability
one. A similar 
argument had earlier been given for the hard-square lattice
gas~\cite{ErtFroJaec88} and later refined
in~\cite{JaecFroKno91,JaecFroKno91b}.

For models inspired by {\bf cellular structures}, we are not aware of any
explicit analysis of reducibility effects. However, as explained in
Sec.~\ref{topological} these models all have dynamics of the
defect-diffusion type. By analogy with $1,d$-SFMs, reducibility effects
would therefore be expected to be irrelevant. For the {\bf triangle and
plaquette models} of Sec.~\ref{model:effective} the dynamics is clearly
irreducible since it is in one-to-one correspondence with the irreducible
(since unconstrained) spin-flip dynamics of the underlying spin system.

Finally, among the other models related to KCMs, only {\bf needle models} are not
obviously irreducible. The only case that has been addressed here is that
of needles attached at their endpoints to a square (planar) lattice, and
with their motion restricted to one side of the lattice
plane~\cite{ObuKobPerRub97}. Here it easy to see that every configuration
can be reached from any other, going via the unentangled state with all
needles orthogonal to the plane. The transformation to the unentangled
state is achieved by a series of small steps: one first ``stretches'' the
configuration in the direction perpendicular to the lattice, and then
``cuts back'' the increased needle lengths to their original value. The
overall effect is a small rotation of all needles which does not cause them
to cross, and repeated application eventually leads to the unentangled
state. For needles attached to three-dimensional lattices, the
irreducibility or otherwise of the dynamics appears to be an open problem.

\subsection{Relaxation timescales and dynamical transitions}
\label{res:dyntrans}

In this section we give results for the typical relaxation timescales of
KCMs and their dependence on temperature or, for lattice gases, density; we
also evaluate the evidence for dynamical transitions where ergodicity is
broken. As explained in Sec.~\ref{intro:irred}, our criterion for a
dynamical transition will be a divergence of an appropriate relaxation time
in the thermodynamic limit.
%

We begin with {\bf spin-facilitated models}. As explained in
Sec.~\ref{sfm:some_results}, $f,d$-SFMs with $f=1$ behave rather
differently than those with $f\geq 2$, since relaxation can occur by
diffusion of defects (isolated up-spins) through the system. This lack of
any significant cooperativity in the dynamics leads to behaviour typical of
strong glasses, with relaxation times increasing in an Arrhenius fashion as
$T$ is lowered; exemplary results for the $1,1$-SFM are shown in
Fig.~\ref{sfm_fig4} above. This expectation was confirmed in a
theoretical analysis by 
Fredrickson and Andersen~\cite{FreAnd85}, who used a diagrammatic
technique to obtain approximations to the integrated relaxation time of the
spin autocorrelation function. A later mean-field theory~\cite{SchTri98},
paraphrased in Sec.~\ref{meth:mf}, also predicted the expected Arrhenius
dependence of relaxation times. As an aside, we note that Fredrickson and
Andersen~\cite{FreAnd85} also investigated SFMs on lattices other than the
conventional cubic ones, and found that there defect-diffusion dynamics can
also occur for $f\geq 2$. This is the case for \eg\ the SFM on a triangular
lattice with $f=2$.  Here the defects are pairs of neighbouring
up-spins. Such a pair can facilitate an up-flip of a neighbouring spin; if
then one of the original up-spins flips down, the defect has effectively
rotated around one of its endpoints, and by repetition of this process can
diffuse across the lattice in a tumbling motion.

More interesting are $f,d$-SFMs with $f\geq 2$ (on the conventional
cubic lattices); we saw in
Sec.~\ref{sfm:some_results} that in these models relaxation processes
proceed in a strongly cooperative
fashion which should lead to a superactivated relaxation timescale
increase.  For $2,d$-SFMs, for example, the approximate analysis
of~\cite{FreAnd85} resulted in an integral equation for the
autocorrelation function very similar to typical mode-coupling
equations (see Sec.~\ref{meth:mca}). This predicts a divergence of the
relaxation time, and therefore a dynamical transition, at an up-spin
concentration of $c\dyn=1/\{[(3/2)^3 2d(2d-1)]^{1/2}+1\}$. Fredrickson
and Andersen argued that since their approximation was of a mean-field
type it should be reasonable at least for larger $d$. For \eg\ $d=1$
it is clearly incorrect since the $2,1$-SFM, being strongly reducible,
shows an incomplete decay of the spin autocorrelation function at {\em
any} $c$. For $d=2$, the theory fails in the opposite way: later
simulations~\cite{FreBra86,LeuDeR86} and theoretical
arguments~\cite{Reiter91} strongly suggested that there is no true
dynamical transition at any nonzero $c$. As is typical of MCA-like
theories, however, a fit of the relaxation time increase to a
power-law behaviour suggests a divergence close to the theoretically
predicted value $c\dyn$.  For larger spatial dimension, $d=3$,
simulations of the $2,3$-SFM~\cite{LeuDeR86} found agreement with the
theory of~\cite{FreAnd85} over a broader range of $c$, as expected,
although again there was no evidence of an actual dynamical
transition.  Butler and Harrowell~\cite{ButHar91,ButHar91b} also
obtained relaxation times for the $2,2$-SFM from simulations of the
persistence function (see Sec.~\ref{res:statdyn} below), finding the
expected superactivated temperature-dependence. The $2,2$-SFM and
$2,3$-SFM (with slightly modified transition rates) were revisited in
later simulations by Graham \etal~\cite{GraPicGra93,GraPicGra97}, who
also studied the $3,3$-SFM. Their data for the relaxation
times---extracted using stretched exponential fits to spin
autocorrelations---are shown in Fig.~\ref{sfm_fig5} above. Graham \etal\
fitted their results by a VTF law\eq{VTF}, with a divergence at a
nonzero temperature $T_0$. This provides a good fit over two and a
half decades in $\tau$, as does a power-law singularity at nonzero
temperature for the $2,3$-SFM data. However, extrapolations towards an
actual divergence are subject to the usual reservations; inspection of
Fig.~\ref{sfm_fig5} suggests, for example, that an EITS behaviour with
Arrhenius corrections, $\tau \sim \exp(A/T^2+B/T)$, would also fit the
data but not give any divergence at $T>0$. The absence of such a
divergence is also predicted by a recent theoretical
treatment~\cite{EinSch01} of $2,d$-SFMs, using an MCA for the second
order memory function of spin fluctuations to obtain approximate spin
autocorrelation functions. This gave a superactivated growth of the
relaxation time at low $T$, whose functional form was not however
analysed in detail. Overall, we regard the theories and simulation
data on the ``cooperative'' SFMs as compatible with the absence of a
bona fide dynamical transition at nonzero temperature. However, the
theoretical prediction of even the functional form of the temperature
dependence of relaxation timescales in these models remains an open
problem. (One plausible conjecture on the basis of the growth of
dynamical lengthscales is that the divergence of $\tau$ for small $T$
is in fact doubly-exponential, $\tau \sim \exp[A\exp(1/T)]$; see
Sec.~\ref{res:hetero}.)

Next we turn to SFMs with {\bf directed kinetic constraints}. The simplest
of these is the East model, which as discussed in Sec.~\ref{res:reduc} is
effectively irreducible at any nonzero up-spin concentration or,
equivalently, nonzero temperature. Already when the model was first
proposed~\cite{JaecEis91} it was argued that relaxation timescales should
remain finite for any $T>0$. This has recently been proved rigorously: the
longest relaxation time, obtained as the inverse of the smallest decay rate
that one would find by full diagonalization of the master equation, is
bounded between $\exp[1/(2T^2\ln 2)]$ and $\exp[1/(T^2\ln 2)]$ in the limit
of small temperatures~\cite{AldDia02}. The upper bound in this result is
also consistent with the estimate of~\cite{SolEva99}. The East model
therefore exhibits an EITS relaxation time divergence at low temperatures,
as anticipated intuitively in Sec.~\ref{sfm:some_results} on the grounds of
the cooperative nature of relaxation processes. That relaxation times in
the East model must diverge in a superactivated fashion, \ie\ more strongly
than any power of the inverse up-spin concentration
$1/c\approx\exp(\beta)$, had already been shown by J\"ackle and
coworkers~\cite{EisJaec93,MauJaec99}. They used an elegant argument based
on the fact that the relaxation necessarily becomes faster if the kinetic
constraint on the leftmost spin in a finite chain is lifted. We note briefly
that MCA approaches fail rather dramatically for the East model:
Kawasaki's~\cite{Kawasaki95} application of the MCA to the irreducible
memory function, reviewed in Sec.~\ref{meth:mca}, predicts a spurious
dynamical transition at up-spin concentration $c=0.2$.

Having seen that the $1,1$-SFM with its undirected kinetic constraint shows
strong-glass behaviour, while the East model has a much more dramatic
relaxation time increase typical of fragile glasses, it is not unexpected
that the asymmetric $1,1$-SFM which interpolates between these two extremes
shows a fragile-to-strong crossover on lowering
$T$~\cite{BuhGar01,BuhGar02}. Referring to\eq{asymmetric_SFM} in
Sec.~\ref{model:SFM}, the East model corresponds to the value $a=0$ for the
interpolating parameter, and displays cooperative relaxation on ``fragile''
timescales $\tau\sim \exp(1/T^2\ln2)$.  For any $a>0$, however, one has
the diffusion of isolated up-spins, which dominates the dynamics of the
$1,1$-SFM, as an additional relaxation process. As will be explained
shortly, the timescale for the latter is $\tau_{\rm diff} \sim (1+a^{-1})
\exp(1/T)$. This increases only in an Arrhenius (strong) fashion so that
defect-diffusion, being the faster process, dominates the relaxation at low
$T$. The crossover occurs where $\tau\approx \tau_{\rm diff}$; since the prefactor
in $\tau_{\rm diff}$ becomes large for $a\to 0$, the crossover shifts to lower
temperatures as $a$ decreases.  The derivation of the defect-diffusion
timescale $\tau_{\rm diff}$ is essentially an extension of the analogous argument
for the $1,1$-SFM given in Sec.~\ref{sfm:some_results}. Consider the rate
for diffusion of an isolated up-spin by one step to the right; the rate for
a diffusion step to the left is the same from detailed balance. The right
neighbour of the up-spin needs to flip up, which from\eq{asymmetric_SFM}
takes place at rate $c\eql\equiv c$. A successful diffusion step is only
obtained if the original up-spin then flips down before the new up-spin
does; the probability for this is $a/(a+1)$ since the rates for a down-flip
of the original and of the new spin are $a(1-c)$ and $1-c$,
respectively. This gives the overall rate of $c a/(1+a)$ for a diffusion
step, hence $\tau_{\rm diff} \sim (1+a^{-1}) \exp(1/T)$ as anticipated.

To finish off our discussion of SFMs with directed constraints, we now
discuss the North-East and Cayley tree models. These differ from all models
discussed so far in this section in that they are strongly reducible below
some nonzero up-spin concentration $c_*$; see Sec.~\ref{res:reduc}. Since
reducibility implies non-ergodicity, these models must therefore show
diverging timescales as $c$ approaches $c_*$. It is in principle possible
that a separate, and therefore nontrivial, dynamical transition could occur
at some higher $c\dyn$, but numerical studies suggest that this is not the
case and that relaxation timescales diverge only at
$c_*$~\cite{ReiMauJaec92}. In the North-East model, simulations for fairly
small lattice sizes ($L=40$) suggest a power-law divergence of the
relaxation time as $c$ approaches $c_*$, with an exponent around 5, but
possibly larger for larger lattices~\cite{ReiMauJaec92}.

A number of theories have been applied to both the North-East and Cayley
tree models and generally do predict dynamical transitions, though at
incorrect values of $c$. For the $(a,a-1)$-Cayley tree, diagrammatic
treatments~\cite{PitAnd01}, an MCA applied to the irreducible memory
function~\cite{Kawasaki95} and an effective medium
approximation~\cite{JaecSap93} have all been used. The known value of the
transition is at $c\dyn=c_*=(a-2)/(a-1)$; see Sec.~\ref{res:reduc}. The
diagrammatic method predicts a higher value, $c\dyn=(a-1)/a$ ($=2/3$ for
$a=3$, compared to the true $c_*=1/2$). The effective medium approximation
gives an even higher estimate, $c\dyn=0.690$ for $a=3$. Both are somewhat
superior to the MCA, which gives a transition at too low a value of $c$,
\eg\ $c\dyn=0.4090$ for $a=3$~\cite{PitYouAnd00}, and also incorrectly
predicts that the fraction of frozen spins jumps discontinuously to a
nonzero value below the transition. For the North-East model, all three
approaches make exactly the same predictions as for the $(3,2)$-Cayley tree
model with $a=3$. Thus, neither captures the behaviour observed in
numerical simulations~\cite{ReiMauJaec92,JaecSap93} and expected from the
relation to directed percolation, with a transition at $c\dyn=c_*\approx
0.2942$ (see Sec.~\ref{res:reduc}) and a non-analytic increase of the
fraction $q$ of frozen spins below the transition according to $q\sim
(c_*-c)^{0.25\pm0.05}$.

We next turn to relaxation timescales in kinetically constrained {\bf
lattice gases}. Kob and Andersen, in their original paper on the KA
model~\cite{KobAnd93}, determined the self-diffusion constant $D_s$ as a
function of the particle density $c$; $D_s$ was obtained from the long-time
limit of the mean-square particle displacements. For densities between
$c\approx 0.3$ and $c=0.86$, they obtained a very good fit to their data
with $D_s \sim (c\dyn-c)^\phi$, covering over three decades in $D_s$, with
$c\dyn=0.881$ and exponent $\phi=3.1$ (see Fig.~\ref{lg_fig2}
above). This suggests a dynamical
transition caused by a divergence of the diffusion timescale $1/D_s$ at
$c=c\dyn$. A singularity of Vogel-Fulcher type ($1/D_s\sim
\exp[A/(c\dyn-c)]$) could be excluded as providing a much worse fit to the
date. Relaxation times extracted from equilibrium correlation functions
also showed power law divergences at densities very close to $c\dyn$. KA
argued convincingly that their data were not affected by finite-size
effects, and that the extrapolated vanishing of $D_s$ at $c=c\dyn$ was
therefore a genuine dynamical transition. They conceded, however, that
simulations closer to or in fact above $c\dyn$ would be needed to establish
the existence of such a transition more firmly. It is intriguing that the
$c\dyn$ found by KA is quite close to the density where the (linear) system
size $L$ needed to avoid reducibility effects due to permanently frozen
particles begins to increase strongly. The theoretical expectation is that
$L$ eventually diverges as $L\sim \exp\{A\exp[B/(1-c)]\}$ (see
Sec.~\ref{res:reduc}), and this very strong increase of a lengthscale might
explain the apparent vanishing of the diffusion constant
$D_s$. In the mathematical limit $L\to \infty$, $D_s$ may remain nonzero up
to $c<1$, but its value would be so small and the system sizes required to
measure it so unrealistically large that this would be of little
physical relevance. Finally, it has been suggested that the power-law
singularity of $D_s$ might be analogous to critical slowing-down, in which
case one would expect the exponent $\phi$ to be insensitive to the precise
nature of the kinetic constraint or the lattice type. Simulations for f.c.c.\
lattices with $m=5,7,8$~\cite{ImpPel00}, and for the b.c.c.\ lattice with
$m=5$~\cite{LevAreSel01} support this hypothesis. The underlying reasons
for such apparent universality remain poorly understood, however.

For the triangular lattice gas with two-vacancy assisted
hopping~\cite{JaecKro94,KroJaec94} numerical simulations were performed of
both the self and collective diffusion constants (see
Sec.~\ref{basics2}). The self-diffusion constant $D_s$ decreases by about
four orders of magnitude as the particle concentration is increased from
$c=0$ to $c=0.77$; it can be fitted both by a power-law $D_s\sim
(c\dyn-c)^\phi$ and an exponential singularity $D_s \sim
\exp[-A/(1-c)]$. Since in the thermodynamic limit no particles are expected
to be permanently blocked (see Sec.~\ref{res:reduc}), it was argued that
the dynamical transition at $c\dyn<1$ predicted by the first fit is
spurious~\cite{JaecKro94}. However, this argument effectively assumes that
{\em irreducibility} (absence of permanently blocked particles) rules out a
dynamical {\em ergodicity} breaking transition; as explained in
Sec.~\ref{intro:irred}, this is not an obvious implication.  The
self-diffusion constant $D_s$ for the triangular lattice gas was also
obtained 
from an approximate calculation of the intermediate self-scattering
function\eq{cs}, using the projection formalism with the memory 
function set to zero.  As explained in Sec.~\ref{basics2}, the long-time
and long-wavelength limit of this quantity determines $D_s$.  The
approximation used was too simple to capture the rapid decrease of $D_s$
with increasing $c$, however, and in fact predicts a nonzero limit for
$c\to 1$~\cite{JaecKro94}. A similar approximation for the collective
diffusion constant $D$ predicts $D\sim (1-c)^2$. Extending the set of
observables included in the projection technique modifies this to $D\sim
(1-c)^3$, but even so the numerically observed decrease of $D$ with $c$ is
much more pronounced~\cite{KroJaec94}.

Next we consider models inspired by {\bf cellular structures}. As discussed
qualitatively in Sec.~\ref{topological}, these all exhibit diffusion of
appropriate defects, so that one would expect an activated temperature
dependence of relaxation times. Indeed, Davison and Sherrington considered
the relaxation time $\tau$ over which the autocorrelation function of the
local deviations $n_i-6$ from the hexagonal ground state decays to $1/e$ of
its initial value~\cite{DavShe00}, and found that it is well fitted by an
offset Arrhenius law, $\tau = A + B\exp(C/T)$. Similar behaviour is
observed in the lattice analogue of the model~\cite{DavSheGarBuh01}. The
{\bf plaquette} model also exhibits defect-diffusion and therefore
activated relaxation times; see Sec.~\ref{model:effective}. The {\bf triangle}
model, on the other hand, displays cooperative relaxation processes similar
to those in the East model. As explained in Sec.~\ref{res:relax} below, this
leads to an estimate of the relaxation timescale $\tau\sim
\exp[T^2/(2\ln2)]$~\cite{GarNew00}. This differs from the result for the
East model only through the extra factor of $1/2$ in the exponent, which
accounts for the two-dimensional nature of the model.

In models with entropic barriers such as the {\bf Backgammon model} or the
{\bf oscillator} model relaxation times remain finite at any nonzero
temperature, exhibiting only power law corrections to the dominant
Arrhenius behaviour $\tau\sim \beta^n\exp(A\beta)$; here $A$ is a
constant and $n=-2$ and $n=1/2$ in the Backgammon and oscillator models,
respectively. Relaxation times to reach the ground state at $T=0$ do
of course diverge with the system size (as $2^N$ for the Backgammon
model~\cite{Lipowski97,MurKeh97,AroBhaPra99}, or more slowly as
$N^2$~\cite{AroBhaPra99} in variants such as model C
from~\cite{GodBouMez95}), but at $T>0$ the final energy per box or
particle lies above the ground state by a finite amount and so all
timescales remain finite.

Finally, we comment on {\bf needle models}. In the model of thin needles
attached to an f.c.c.\ lattice, Renner \etal~\cite{RenLowBar95} investigated
the dependence of the rotational self-diffusion constant $D_s$ on the ratio
$l=L/a$ of needle length $L$ and lattice constant $a$. The measured values
could be well fitted by a power-law singularity $D_s\sim (l\dyn-l)^\gamma$
with $l\dyn\approx 2.7$ and $\gamma\approx 4.2$. This would indicate a
dynamical transition, though it is difficult to exclude that measurements
around $l\approx l\dyn$ would reveal a rounding of the apparent
singularity. For a similar model, with needles attached by their endpoints
to a cubic lattice, Obukhov {\em et al}~\cite{ObuKobPerRub97} argued that
there was a true dynamical transition at $l\dyn\approx 4.5$. Their evidence
for this was based on simulations of the average root-mean-square angular
displacements $\theta(t)$ as a function of time. They argued that if there
is indeed a transition, then for all $l$ near $l\dyn$, $\theta(t)$ should
show the same behaviour, up to times that diverge as $l\to l\dyn$. The
effect of an increase in length (which they implemented approximately by
freezing a small fraction of the needles) should therefore be smallest for
$l=l\dyn$, and their simulations appeared to confirm this. They also
interpreted their results as showing that $\theta(t)$ had a finite
long-time limit for $l>l\dyn$, but again it seems difficult to exclude the
alternative interpretation of a crossover to slow rotational diffusion
outside their simulation time window. For smaller lengths $l<l\dyn$,
Obukhov \etal~\cite{ObuKobPerRub97} argued phenomenologically that since a
needle interacts typically with $l^3$ others, the relaxation time scale
should increase as $\exp(l^3)$, and found some simulation evidence for
this. If there is indeed at dynamical transition then this behaviour should
cross over to a divergence at $l=l\dyn$, but this was not investigated in
detail. For the two-dimensional case of needles attached to a square
lattice, the simulation data were consistent with the relaxation time
behaviour $\tau\sim \exp(l^2)$, and no evidence of a dynamical transition
was found.

\subsection{Stationary dynamics}
\label{res:statdyn}

One of the important questions about KCMs is how good they are at
reproducing the characteristic aspects of the supercooled state.  We
therefore review in this section the results for equilibrium properties of
KCMs such as correlation, response and persistence functions.

We begin by defining the relevant quantities for {\bf spin-facilitated
models}. Many studies have analysed the spin autocorrelation function,
which using $\lav n_i\rav = c \equiv c\eql$ can be written as
\be 
C(t)=\frac{1}{N}\sum_{i} \frac{\lav n_i(t)n_i(0) \rav - c^2}{c(1-c)}
\label{eqC}
\ee
We have multiplied by a constant factor here to normalize the correlation
function to $C(0)=1$. Notice that only the $cN$ spins which are in the
up-state $n_i=1$ at time $0$ contribute nonzero averages $\lav
n_i(t)n_i(0)\rav$ in\eq{eqC}; one can therefore also write $C(t)=(\lav
n_i(t)\rav-c)/(1-c)$ where $n_i$ is any spin that is initially up.

The dynamics of the overall up-spin concentration $c(t)=(1/N)\sum_{i}
n_i(t)$ is also often of interest; in equilibrium its average is $\lav
c(t)\rav = c$ for all times. Its normalized equilibrium correlation
function is
\be 
C_c(t)=\frac{1}{N}\sum_{ij} \frac{\lav n_i(t)n_j(0) \rav - c^2}{c(1-c)}
\label{eqCM}
\ee
and is seen to be a sum of nonlocal spin correlation functions $\lav
n_i(t)n_j(0)\rav-c^2$. Finally, the persistence function $F(t)$
has also been studied; it measures the fraction of spins which, starting
from an equilibrated configuration at time $0$, have never flipped up to
time $t$.  The integral $\int_0^\infty dt\,F(t)$ gives the mean-first
passage time, \ie\ the average time after which a spin will first flip. The
persistence function and mean-first passage times can also be defined
separately for spins that were up or down in the starting configuration.

\begin{figure}
\begin{center}
\epsfig{file=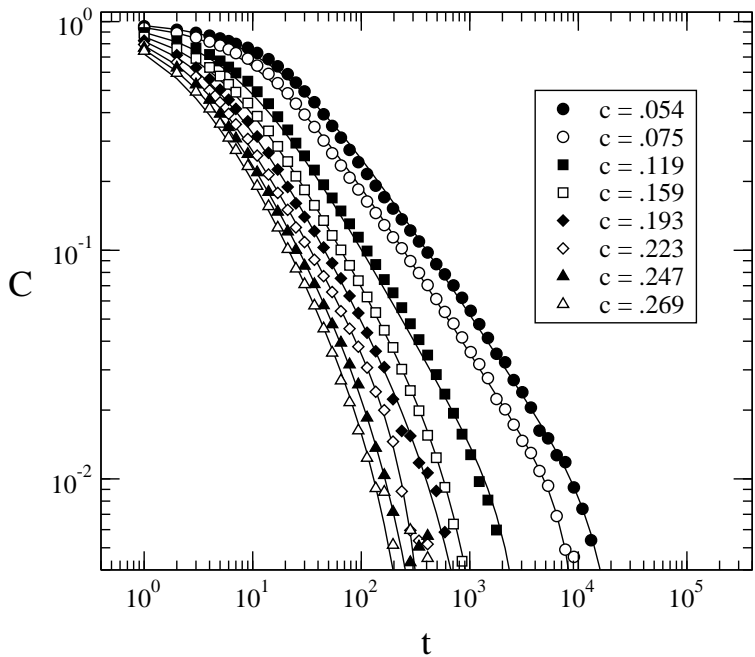, width=10cm}
\end{center}
\caption{Normalized spin autocorrelation function $C(t)$ in the
$1,1$-SFM for up-spin concentrations ranging from $c=0.269$ down to
$c=0.054$. The solid lines show fits to the form
$C(t)=(1+t/\tau_1)^{-\alpha}\exp(-at^b)$ where $\tau_1,\alpha,a$ and $b$ are
fitting parameters. As expected from the diffusive nature of the
dynamics, the exponent $\alpha$ is close to $1/2$, and $\tau_1$ is of
order $1/c$ (see text for details). From~\protect\cite{CriRitRocSel00}.
\label{stretched_FA}
}
\end{figure}

We begin by considering {\bf spin-facilitated models} with {\bf
defect-diffusion} dynamics, \ie\ $f,d$-SFMs with $f=1$. Most work has
focused on the $1,1$-SFM, though there are also a few results for
$1,d$-SFMs in $d>1$ (see below). To get some intuition, we first recap
briefly the 
discussion in Sec.~\ref{sfm:some_results} of the low-temperature, \ie\
small-$c$, dynamics of the $1,1$-SFM. We saw that up-spins occur as
isolated defects, and that these diffuse with an effective diffusion
constant $D\eff=c/2$. The typical distance between defects is $1/c$, so
that the timescale $\tau$ on which defects start noticing each other is set
by $(2D\tau)^{1/2}=1/c$, giving $\tau=c^{-3}$ as
in\eq{defect_diffusion_tau}. The spin autocorrelation function, $C(t)=(\lav
n_i(t)\rav - c)/(1-c)$ for spins with $n_i(0)=1$, simplifies for small $c$
to $C(t)=\lav n_i(t)\rav$. It is therefore just the probability that a
spin that was up at time 0 is also up at time $t$. For $t\ll\tau$, where
defects are non-interacting random walkers, this is just the return
probability of a random walk and therefore
\be
C(t)=\int_{-\pi}^\pi \frac{dq}{2\pi} e^{-2D\eff(1-\cos q)t}
\label{return_prob}
\ee
which is a function of $ct$ only, with $C(t)=1-ct$ for $ct\ll1$ and
$C(t)=(2\pi ct)^{-1/2}$ for $ct\gg 1$. This behaviour should then
cross over to a faster decay when $t$ becomes of order $\tau=c^{-3}$
and defects start interacting; in this regime $C(t)$ is already small,
of order $(c\tau)^{-1/2}=c$. Interestingly, we see here that the spin
relaxation function in the $1,1$-SFM allows three different timescales
to be defined: the {\em instantaneous} time, where $C(t)=1/e$, scales
as $c^{-1}\approx \exp(1/T)$. The longest timescale, on which defects
begin interacting is $\tau=c^{-3}=\exp(3/T)$. The {\em integrated}
timescale $\int_0^\infty dt\, C(t)$, finally, is dominated by the
$(ct)^{-1/2}$ tail of $C(t)$ up to times $t\approx \tau$, and
therefore scales as $c^{-1/2}\tau^{1/2}=c^{-2}\approx \exp(2/T)$; see
Fig.~\ref{sfm_fig4} above. All three timescales show activated
behaviour, as anticipated in Sec.~\ref{sfm:some_results}.
Fig.~\ref{stretched_FA} shows that the scaling of $C(t)$ obtained
above, \ie\ a product of $(ct)^{1/2}$ times a cutoff function on
longer timescales $t$, is qualitatively confirmed by
simulations~\cite{CriRitRocSel00}.

We now turn to detailed theoretical calculations of the spin
autocorrelation function of the $1,1$-SFM. A comprehensive analysis was
given in~\cite{ReiJaec95}. Calculations were performed within the
projection approach, with memory terms set to zero. Three different sets of
observables were considered for the projection: The first contained the
normalized fluctuations $\et_i=[c(1-c)]^{-1/2}(n_i-c)$ of local spins
around their averages, and their pairwise products; the second included in
addition some triple products. The third contained single spin fluctuations
as well as the variables $L^n\eta_i$ with $n=1\ldots 5$, which can be
motivated via a high-temperature expansion. As expected, the last
choice works 
best for large $c$ (down to $c\approx 0.3$). For lower $c$, the
approximation which includes triple spin products gives the most
accurate results 
and predicts an asymptotic decay of $C(t) \approx (2\pi
ct)^{1/2}\exp(-t\sqrt{8c^5})$. For times $c^{-1} \ll t \ll c^{-5/2}$ this
gives precisely the square-root decay as expected from the discussion
above. A treatment of the effective low-temperature model of diffusing
defects gave a similar functional form, but with the exponential cutoff
function replaced by $\exp(-\sqrt{8c^2D\eff t})$. This has the scaling
with $c^3t = t/\tau$ expected on qualitative grounds, but the
functional form is not necessarily reliable since it is derived under the
assumption that the exponent is still small and defects have just started
to interact. Information on the integrated relaxation time, which depends
on the long-time behaviour of $C(t)$, could therefore not be deduced from
this approach; within the best alternative (three-spin) approximation
of~\cite{ReiJaec95} it scaled as $c^{-7/4}$, still somewhat below the
scaling with $c^{-2}$ expected from the qualitative arguments above.  A
later analysis of equilibrium correlation functions in the
$1,1$-SFM~\cite{SchTri99} took a different approach based on a spatial
coarse-graining of the local up-spin concentrations. To produce small
concentration fluctuations, however, the coarse-graining distance must then
be of the order $1/c$ or larger, and fluctuations on such lengthscales are
no longer related to the spin autocorrelation function in an obvious way.

Interestingly, it turns out that the relaxation of the {\em overall up-spin
concentration} in the $1,1$-SFM, as determined by the correlation function
$C_c(t)$ defined in\eq{eqCM}, can be calculated exactly in the limit of
small $c$~\cite{BurDoeBen89,BenBurDoe90}. This is possible because of the
mapping of the $1,1$-SFM onto an effective $A+A\leftrightarrow A$
reaction-diffusion model; see Sec.~\ref{sfm:some_results}. An exact
calculation~\cite{BurDoeBen89,BenBurDoe90} for the latter results in
$C_c(t) = (1+2t/\tau){\rm
erfc}[(t/\tau)^{1/2}]-2[t/(\pi\tau)]^{1/2}\exp(-t/\tau)$, with a long-time
behaviour of $[\pi(t/\tau)^3]^{-1/2}\exp(-t/\tau)$. The timescale here is
$\tau = (2D\eff c^2)^{-1}= c^{-3}$ as before, so that fluctuations in the
overall up-spin concentration relax when diffusing up-spins begin to
interact (see above). In contrast to the {\em spin} autocorrelation
function, there is no decay on the shorter timescale $\sim 1/c$ for
diffusion of individual defects, because the up-spin concentration remains
unchanged while up-spins only diffuse but do not interact.

Finally for the $1,1$-SFM, we turn to the persistence function of
down-spins, which is fairly straightforward to
estimate~\cite{SchTri99}. Consider a domain of $l$ down-spins bounded by
up-spins at $t=0$. As time increases, the up-spins will have flipped spins
in a region of size $\sim \sqrt{2D\eff t}$ around each, so that only around
$l-2\sqrt{2D\eff t}$ persistent down-spins remain. Since the equilibrium
distribution of down-spin domain lengths is $P(l) \approx c \exp(-cl)$ for
low $c$, the persistence function is approximately
\be
\sum_{l\geq 2\sqrt{2D\eff t}} \left(l-2\sqrt{2D\eff t}\right) c e^{-cl} =
e^{-2c\sqrt{2D\eff t}}
\ee
and again decays on timescales scaling as $c^{-2}D\eff^{-1}\sim c^{-3} =
\tau$.

For the $1,d$-SFM in arbitrary dimension $d$, with equilibrium up-spin
concentration $c$, it was shown in~\cite{FolRit96} that $2dc$ is an exact
eigenvalue of the Liouvillian, giving the relaxation rate of an
appropriately defined staggered magnetization. The authors also found
numerically for $d=1, 2$ that exactly half this rate determines the
early stages of the decay of the spin autocorrelation functions in these
models, which are well fitted by the simple exponential
$C(t)=\exp(-dct)$. This result actually has a simple interpretation for
small $c$, where up-spins are isolated and far from each other. As for the
$1,1$-SFM, the spin autocorrelation function is then for short times just
the return probability of a random walker in $d$ dimensions with
diffusion constant $D\eff$. This is the $d$-th power of the
result\eq{return_prob} 
for $d=1$, giving for short times $C(t)=(1-ct)^d = 1-dct +
\order((ct)^2)$ consistent with the early time scaling found
in~\cite{FolRit96}.

Next we consider SFMs with {\bf cooperative dynamics}, \ie\ $f,d$-SFMs with
$f\geq 2$. As mentioned already in Sec.~\ref{res:dyntrans}, in their early
theoretical work on $2,d$-SFMs Fredrickson and
Andersen~\cite{FreAnd84,FreAnd85} predicted that the spin autocorrelation
function should show a dynamical transition at some up-spin concentration
$c\dyn$; an increasingly non-exponential shape of the correlation function
was predicted on approaching $c\dyn$ from above, while below the
correlation function should decay to a nonzero value. Simulations soon
after~\cite{FreBra86}, however, showed that this predicted transition is
spurious; instead, the spin autocorrelation functions showed stretched
exponential decays for low $c$, with stretching exponents decreasing as $c$
was lowered, and relaxation times increasing in a super-Arrhenius fashion
(see Sec.~\ref{res:dyntrans}). Similar results were later reported by
Graham \etal~\cite{GraPicGra93,GraPicGra97} for the spin autocorrelation
function in the 2,2-SFM, 2,3-SFM and 3,3-SFM (with the slight modification
that rates for allowed transitions were chosen to be independent of the
number of facilitating neighbours). The stretched exponential behaviour sets
in at low $c$ and intermediate times; the short time relaxation is
exponential. In all cases the stretching exponent $b$ stays between around
0.3 and 0.6 and decreases with $c$. Fredrickson~\cite{Fredrickson86} also
studied the autocorrelation function of the total up-spin concentration,
$C_c(t)$, and found similar behaviour, but with different stretching
exponents which were somewhat closer to 1. Harrowell~\cite{Harrowell93}
simulated the {\em persistence function} in the $2,2$-SFM and found that at
long times it was well fitted by a stretched exponential with a stretching
exponent $b$ close to 1/2, for up-spin concentrations $c$ between around
0.08 and 0.2. Additional evidence of stretching was
obtained~\cite{AleWeiKinIsr87} by analysing the power spectrum of spin
fluctuations, \ie\ the Fourier transform of the autocorrelation function
$C(t)$; see\eq{eq9S22} in Sec.~\ref{basics2}. At high temperatures the
spectrum is practically Lorentzian, with a slight broadening because even
for $T\to\infty$ the kinetic constraints still act (since $c=1/2$;
constraints only become irrelevant for $c=1$, corresponding formally to
$\beta=-\infty$). At temperatures below $T=0.5$ the power spectrum showed
$1/\omega$-noise in a large band of frequencies; from\eq{freq_FDT} this
corresponds directly to a large frequency range where the dissipative
frequency response is approximately constant, and hence to a wide spectrum
of relaxation times.

Finally, we mention simulation work~\cite{ZheSchTri99} on the spin
autocorrelation function in the $2,2$-SFM with a ferromagnetic interaction
$J$ included; see\eq{eq1S311b}. Stretched exponential behaviour is again
observed, but with parameters (relaxation time and stretching exponent)
that depend on $J$ and $T$ only through the equilibrium concentration $c$
of up-spins. This shows that the dynamical behaviour of the model in the
glassy regime is largely independent of the precise details of the energy
function and instead dominated by the effects of the kinetic constraints.

We now move on to the stationary dynamics of SFMs with {\bf directed
constraints}. In these models, it can be shown~\cite{JaecSap93} that the
directionality of the constraint implies that all nonlocal spin
correlations $\lav n_i(t)n_j(t)\rav-c^2$ vanish, so that the spin
autocorrelation function\eq{eqC} and the autocorrelation function\eq{eqCM}
of the total up-spin concentration give exactly the same information. The
proof is easiest to see in the East model: consider spin $n_i$ and a spin
to its right, $n_j$ with $j>i$. Because each spin facilitates spin-flips of
only its {\em right} neighbours, the value of $n_j(0)$ cannot affect the
state of $n_i(t)$ at times $t>0$. Hence $\lav n_i(t) n_j(0)\rav -c^2 = 0$
for $t>0$. But in detailed balance systems all correlation matrices are
symmetric (see Sec.~\ref{sec:langevin}) and so also $\lav n_j(t) n_i(0)\rav
-c^2 = 0$ for $t>0$; the two results together imply that all nonlocal
correlations vanish.

For the simplest model with directed constraints, the East model, J\"ackle
and Eisinger~\cite{JaecEis91,EisJaec93} obtained an approximation for the
spin autocorrelation function using an effective medium approximation.
Effectively the same result was derived by Pitts and
Andersen~\cite{PitAnd01} using diagrammatic methods;
see\eq{Mirr_East_diag}. The approximation predicts a spurious dynamical
transition at an up-spin concentration of $c=0.5$, with
$q=C(t\to\infty)$ increasing smoothly from zero to nonzero values. The MCA
derived by Kawasaki~\cite{Kawasaki95}, on the other hand, gives the
relation\eq{MCA_East} and, as discussed in Sec.~\ref{meth:mca}, predicts a
transition at $c=0.2$, with a discontinuous jump of $q$ from 0 to 1/2. Both
approximations can therefore only be reasonable at sufficiently large $c$,
or for short times at smaller $c$; a comparison with numerical
simulations~\cite{PitYouAnd00} shows that the effective medium
approximation is generally more accurate in these regimes.  Improved
approximations of the form of extended MCA~\cite{PitAnd01} avoid the
prediction of a spurious dynamical transition at $c>0$, and are
quantitatively more satisfactory over a larger range of times and up-spin
concentrations. However, for small $c$ they still predict a decay of $C(t)$
that is too fast and too similar to an exponential compared with numerical
simulations. It had been noticed early on~\cite{JaecEis91} that the
non-exponential behaviour is well fitted by a stretched exponential only
over a limited time range.

In Ref.~\cite{PitYouAnd00} it was also suggested that for low $T$
(\ie\ low $c$) the autocorrelation function of the East model might
exhibit scaling behaviour in the form $C(t)=\tilde C(t/\tau(T))$ with
a diverging timescale $\tau(T)$ for $T\to 0$ and a scaling function
$\tilde C$ close to a stretched exponential. It seems likely that such
scaling will indeed apply in the {\em asymptotic} long-time regime;
from arguments based on links to defect-diffusion
models~\cite{EisJaec93}, the rigorous work of~\cite{AldDia02} and
results for the out-of-equilibrium behaviour~\cite{SolEva99} the
asymptotic timescale should be $\tau(T)\sim\exp(1/T^2\ln 2)$, but the
asymptotic scaling function is expected to be a simple (not a
stretched) exponential, possibly up to power-law
factors~\cite{EisJaec93}. However, for times much shorter than
$\tau(T)$ it was shown that the correct scaling variable for the
initial decay of the autocorrelation fucntion is not $t/\tau(T)$ but
rather $\delta=[t/\tau(T)]^{T\ln 2}$ for low
$T$~\cite{SolEva99,Evans02}, giving very strongly stretched relaxation
behaviour. To be compatible with the crossover to the asymptotic
$t/\tau(T)$-scaling, the scaling function of $\delta$ would then have
to decay to zero at the finite value $\delta=1$, since $\delta>1$
gives $t/\tau(T)=\delta^{1/T}\to\infty$ for $T\to
0$~\cite{SolEva02}. Buhot and Garrahan~\cite{BuhGar01,BuhGar02} gave
an alternative derivation of the stretching exponent $T\ln 2$, by
considering the persistence function of up-spins. (For the East
model, this is essentially identical to the autocorrelation function
for $c\to 0$, since once an up-spin has flipped down the probability for it to
``reappear'' later in the same place is
$\order(c)$.) For the asymmetric $1,1$-SFM with small asymmetry
parameter $a$ they found a crossover in the persistence function from
behaviour typical of the East model ($a=0$) to that for the $1,1$-SFM,
at times around $t \sim (1+a^{-1})e^{1/T}$ where the diffusion of
up-spins enabled by the nonzero value of $a$ becomes significant (see
Sec.~\ref{res:dyntrans}).

For the more complicated SFMs with directed constraints, \ie\ the
North-East model and the $(a,a-1)$-Cayley tree models, most work has
focused on the predicting the location of the dynamical transitions,
which in these models arise from strong reducibility effects below
some up-spin concentration $c_*$ (see Sec.~\ref{res:dyntrans}). Beyond
this, almost no details on the shape of spin autocorrelation and
persistence functions are known.

Let us now consider the equilibrium dynamics of kinetically constrained
{\bf lattice gases}. Kob and Andersen~\cite{KobAnd93} simulated the
intermediate self-scattering function\eq{cs}, suitably modified to take
account of lattice symmetries (see Sec.~\ref{ka:some_results}). In
accord with the original motivation for 
defining the model, they compared their results primarily to the
predictions of MCT as applied to supercooled
liquids~\cite{BenGoetSjo84,Goetze91,GoetSjoe92,GoetSjoe95}.  No plateau at
intermediate times was found, in contrast to MCT (see
Fig.~\ref{lg_fig3} above).  This was rationalized
from the fact that in MCT the decay of correlations to the plateau is
caused by particles ``rattling'' in their cages; but in the KA model,
particles that are caged in were argued to be likely completely immobile,
so that rattling is essentially absent and any plateau would be very close
to the initial value of the correlator.  MCT predicts a power law in time
for the decay from the plateau, and in the glassy regime of high densities
the simulation results for small wavevectors (large lengthscales) were in
accord with this. But the power law exponent was not independent of density
as expected from MCT, and for larger wavevectors deviations from power law
behaviour appeared. The decay at longer times is predicted to be a
stretched exponential by MCT, and the large-wavevector data could be fitted
by this, but again with a variable stretching exponent not
expected from theory. Overall, Kob and Andersen concluded that the KA
model, in spite of having been designed to incorporate the caging effects
that MCT should be able to describe, showed surprisingly poor agreement
with MCT predictions.

\begin{figure}
\begin{center}
\epsfig{file=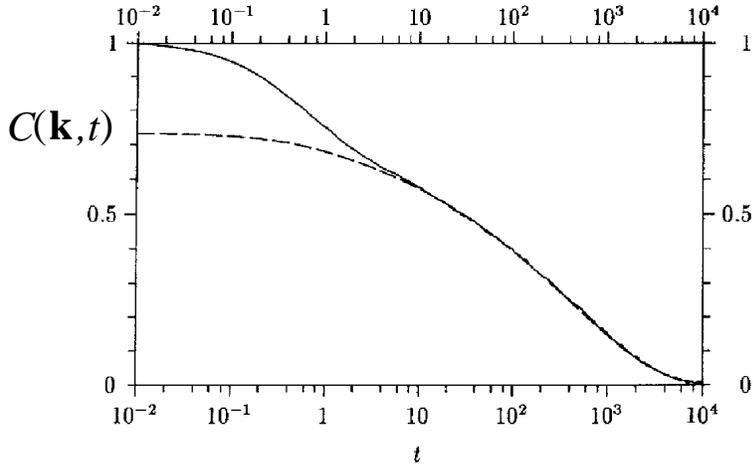, width=10cm}
\end{center}
\caption{Intermediate scattering function~(\protect\ref{c}) for the
triangular lattice gas, for particle concentration $c=0.7$ and a large
wavevector $\kv$ corresponding to lengthscales of order the lattice
spacing. Notice the two-stage relaxation; the shoulder would be
expected to grow into a plateau for even higher $c$. The dashed line
is, up to a multiplicative constant, the fraction of particles that do
not hop up between times 0 and $t$; this is seen to govern the
long-time decay of the intermediate scattering function.
\label{fig:triangular}
}
\end{figure}
For the triangular lattice gas, the autocorrelation function\eq{eqC}
has been simulated; since in this model $n_i=0$, $1$ represent a hole
and a particle, respectively, this is essentially a local density
correlation function.  Non-exponential time-dependences were found on
increasing particle concentration $c$, buth with a functional form
more complicated than a simple stretched
exponential~\cite{KroJaec94}. The correlation of Fourier-transformed
density fluctuations, \ie\ the intermediate scattering function\eq{c},
shows more structure. In particular, at high particle densities the
character of the relaxation changes as a function of the wavevector,
from a single decay at small wavevectors (large length scales) to a
two-stage decay with an intermediate plateau at large
wavevectors; see Fig.~\ref{fig:triangular}. (As explained in
Sec.~\ref{res:hetero} below, this wavevector-dependence
can be interpreted as evidence for dynamical
heterogeneity.) An MCA applied to the
reducible memory function produced satisfactory fits to the data at
low $c$ ($<0.3$), but was found to lead to unphysical divergences of
the correlation functions for larger $c$; compare the discussion in
Sec.~\ref{irreducible_memory}. Contrasting with the results for the KA
model, one notices that the triangular lattice gas exhibits two-step
relaxation processes while the KA model does not. This may be due to
the different correlation functions studied ({\em self} versus {\em
coherent} intermediate scattering function): although at least from
the MCT for supercooled
liquids~\cite{BenGoetSjo84,Goetze91,GoetSjoe92,GoetSjoe95} one would
not expect this to cause qualitative differences,
J\"ackle~\cite{Jaeckle02} hints that also for the triangular lattice
gas the self-intermediate scattering functions do not show two-step
relaxations. Another possible explanation might be that the triangular
lattice gas has genuinely different dynamics, with wider cages in
which particles can ``rattle'' while being confined by their
neighbours. But it is not obvious from the dynamical rules why this
should be the case; a closer comparison between the two models would
be desirable to clear up this puzzle.

In the topological, off-lattice version of the {\bf cellular model}
(Sec.~\ref{topological}), Davison and Sherrington~\cite{DavShe00}
considered the autocorrelation function of the local deviations $n_i-6$
from the hexagonal ground state. For the lowest $T\approx 0.25$ for which
equilibrium can be achieved, this just begins to develop a shoulder, which
one expects to broaden into a plateau for lower $T$. The region around the
shoulder could be fitted reasonably well with the prediction of the MCT for
supercooled liquids, which gives a power-law decay from a plateau
value. The plateau could be more clearly seen in the lattice version of the
model~\cite{DavSheGarBuh01}, with the timescale for the decay from the
plateau following an Arrhenius law as expected due to the activated
character of the dynamics.

In the model of thin {\bf needles} attached to an f.c.c.\ lattice, Renner {\em
et al.}~\cite{RenLowBar95} investigated equilibrium correlation functions
by extensive computer simulations. They used Newtonian dynamics, so that
the state of each needle $i$ is characterized by a its orientation,
specified by a unit vector $\uv_i$, and its angular velocity
$\omegav_i$. The autocorrelation function of the $\uv_i$ was found to
develop a shoulder for needle lengths (normalized by the lattice constant)
of $l=L/a\approx 2.5$. For larger $l$ it failed to decay completely within
the simulation time window, with the shoulder developing into a region of
very slow decay, roughly linear in $\ln t$. (Less detailed simulations for
the same model on a b.c.c.\ lattice~\cite{JCBCTFLCF02} found similar results.)
A clearer change in behaviour was seen in a carefully crafted correlation
function of the angular velocities, $\psi(t)=\lav {\cal
P}_2(\hov_i(t)\cdot\hov_i(0))\rav$, where ${\cal P}_2(x)=(3x^2-1)/2$ is the
second Legendre polynomial and $\hov_i$ is the angular velocity normalized
to unit length. The attraction of this choice is that it detects whether
the needle orientations $\uv_i$ explore the whole unit sphere (which means
that orientational caging effects are unimportant) or whether they
remain close to a 
particular orientation. In the first case also $\hov_i$ explores the whole
unit sphere, and $\psi(t)$ decays to 0 for large $t$. In the second case,
$\hov_i$---which is always orthogonal to $\uv_i$---remains
confined to the plane orthogonal to the frozen needle orientation, and
decorrelation within this plane gives $\psi(t)=1/4$ for large
$t$. Consistent with this, $\psi(t)$ was found to develop a plateau at
$\psi=1/4$ for needle lengths above $l\approx 2.7$. The change in behaviour
is smooth, and therefore is unlikely to correspond to a true dynamical
transition (see Sec.~\ref{res:dyntrans}), but nevertheless takes place over
a narrow range of $l$.

We mention finally that stretched exponential behaviour has also been found
in the energy autocorrelation function for a simplified Backgammon
model~\cite{PraBreSan97c} and for the low-$T$ Glauber dynamics of the
unconstrained ferromagnetic Ising
chain~\cite{BrePraRui94,BrePra93b,BrePra96}, in both cases in an
intermediate time window limited by exponential behaviour for early and late
times.

\subsection{Out-of-equilibrium dynamics}
\label{res:noneq}

In this section we discuss the out-of-equilibrium dynamics of KCMs, which
should be relevant for understanding the behaviour of glasses (as opposed to
supercooled liquids, which still achieve metastable equilibrium on
accessible timescales). We begin with a discussion of nonlinear relaxation
after sudden changes in \eg\ temperature
(Sec.~\ref{res:relax}). Sec.~\ref{res:heatcool} reviews results on the
behaviour of KCMs under cyclic heating and cooling. In Sec.~\ref{res:aging}
we move on to two-time correlation and response functions and effective
temperatures defined on the basis of FDT violations out of equilibrium.
Finally, Sec.~\ref{res:coarsen} briefly discusses ways of classifying
glassy dynamics in KCMs by comparing the evolution of two independent
``clones'' of a system.

\subsubsection{Nonlinear relaxation}
\label{res:relax}

\begin{figure}
\begin{center}
\epsfig{file=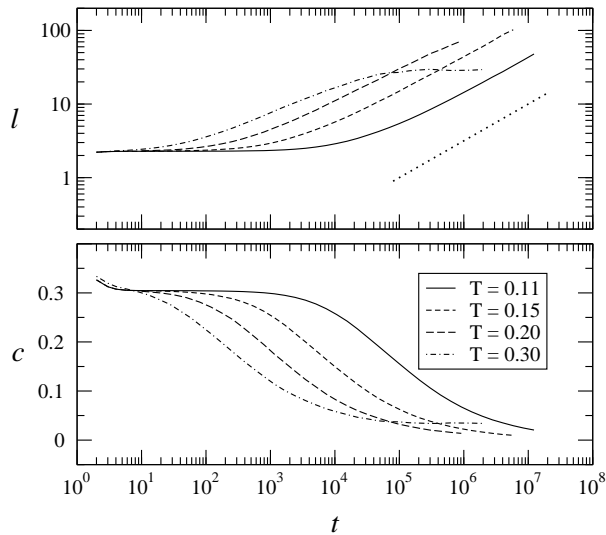, width=8cm}
\end{center}
\caption{Growth of the average domain length $\bar{l}$ (top) and
corresponding decay of
the up-spin concentration $c$ (bottom) after a quench in the
$1,1$-SFM. Results are given for several temperatures corresponding to
equilibrium up-spin concentrations (from top to bottom in the lower
plot) $c\eql\approx1.1\times 10^{-4}, 1.3\times 10^{-3}, 6.7\times
10^{-3}, 0.034$. The average domain length grows diffusively as
$\bar{l}\sim t^{1/2}$ (dotted straight line in upper
plot). From~\protect\cite{CriRitRocSel00}. \label{sfm_fig3} }
\end{figure}

We begin our discussion with {\bf spin-facilitated models}, specifically
with $1,d$-SFMs that exhibit defect-diffusion rather than cooperative
relaxation processes. In the $1,1$-SFM, the relaxation of the up-spin
concentration $c(t)$ after a quench to low $T$, and therefore low
equilibrium up-spin concentration $c\eql$, was studied
in~\cite{CriRitRocSel00}. On timescales of order unity, one has effectively
zero temperature dynamics as explained in Sec.~\ref{meth:exact}, and $c(t)$
will decay to a plateau value, \eg\ $c=(1/2)e^{-1/2}$ if the system was at
$T=\infty$, $c=1/2$ before the quench. Thereafter, relaxation takes place
via the diffusion of isolated up-spins which coalesce when they meet; as
long as $c(t)\gg c\eql$, the reverse process of one up-spin creating
another one is negligible. One thus has a process of diffusive growth of
domains of down-spins. The basic rate for this process is set by the
effective up-spin diffusion constant $D\eff = c\eql/2$, giving average
domain lengths scaling as $\bar{l}\sim(c\eql t)^{1/2}$ (see
Fig.~\ref{sfm_fig3}) and thus for the
up-spin concentration $c(t)\sim 1/\bar{l} \sim (c\eql t)^{-1/2}$. This
scaling also follows from exact results for the effective low-$c\eql$
reaction-diffusion model $A+A\to A$, see
\eg~\cite{BenBurDoe90}. Equilibrium is reached when $c=c\eql$, giving an
equilibration time scaling as $c\eql^{-3}$. Notice that this equilibration
timescale is of the same order as the longest relaxation time in the final
equilibrium state; see Sec.~\ref{res:statdyn}.

Next we consider SFMs with {\bf cooperative dynamics}. For the $2,2$-SFM,
the relaxation of the up-spin concentration after a quench from $c\eql=1$ (all
spins-up) was simulated in~\cite{FreBra86}.  Already for final up-spin
concentrations $c\eql$ around 0.3, the relaxation curves were distinctly
nonexponential, and could be fitted by stretched exponentials down to
$c\eql\approx 0.08$. Fredrickson~\cite{Fredrickson86} considered more general
changes in $c\eql$, both increasing and decreasing. For changes that were
not too large, the nonlinearities could be well described by a ``fictive
temperature'' approach (see \eg~\cite{Scherer86}), which assumes that the
instantaneous relaxation time is the equilibrium relaxation time for the
{\em current} up-spin concentration. For quenches to low $c\eql \approx 0.1$,
Graham \etal~\cite{GraPicGra93,GraPicGra97} reported that the short-time
relaxation exhibited a shoulder before crossing over into stretched
exponential behaviour. Comparing with the discussion of the $1,1$-SFM
above, this is as expected. In fact, for even lower $T$ one expects to see
a roughly $T$-independent decay onto a plateau on timescales of order
unity, reflecting the flipping down of mobile up-spins that would take
place even at $T=0$, with further decay only on the much larger
activated timescale $c\eql^{-1} \approx \exp(1/T)$ for up-flips.

Among SFMs with {\bf directed constraints}, the East model is the simplest
one. Here, the relaxation of the up-spin concentration $c(t)$ after a
quench from high to low $T$ can be understood from the analysis described
in Sec.~\ref{meth:indint}, which reveals that the dynamics takes place on a
hierarchy of timescales of order $c\eql^{-k}\approx \exp(k/T)$, with
$k=0,1,\ldots$ If $c(t)$ is plotted against the scaled time variable
$\nu=T\ln t$, then for $T\to 0$ the $k$-th stage of the dynamics shrinks to
the point $\nu=k$. In this limit the results of Sec.~\ref{meth:indint}
imply that the average domain length $\bar{l}$ will increase in a
``staircase'' fashion, with jumps at integer values of $\nu$. The up-spin
concentration $c=1/\bar{l}$ will therefore also relax in plateaux towards
$c\eql\ll 1$.  At nonzero temperature the steps between the plateaux will
be rounded and cross over into the decay predicted by the anomalous
coarsening law, $c = 1/\bar{l} \sim t^{-T\ln 2}$.
\begin{figure}
\begin{center}
\epsfig{file=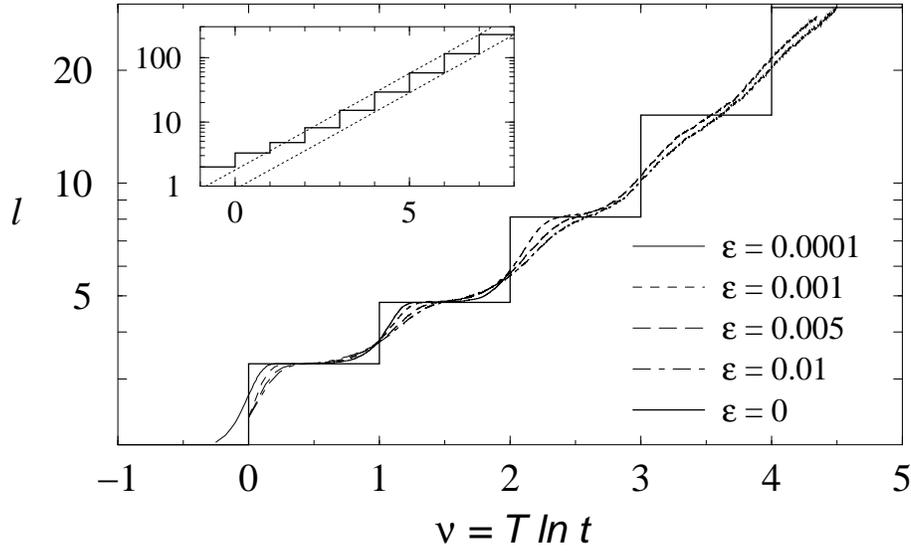, width=12cm}
\end{center}
\caption{Evolution of average domain length $\bar{l}$ in the East
model, after a quench at $t=0$ from equilibrium at $c=1/2$ to a small
temperature $T$. Simulation results for four values of
$\epsilon=\exp(-1/T)=c\eql/(1-c\eql)$ are shown, obtained from a
single run for a spin chain of length $N=2^{15}$. Note the scaled
logarithmic $x$-axis, $\nu=T\ln t$.  Bold line: Theoretical prediction
for $T\to 0$. Inset: Theory for larger $\nu$, and $\nu\to\infty$
asymptotes. From~\protect\cite{SolEva99}.
\label{fig:dbar}
}
\end{figure}
Fig.~\ref{fig:dbar} shows the results of simulations for a range of values
of $\eps=\exp(1/T)=c\eql/(1-c\eql)$. Compared to earlier
simulations~\cite{MunGabInaPie98}, the longer timescales (up to
$t=10^{10}$) reveal the plateaux in $\bar{l}$ versus $\nu$ that develop
with decreasing $T$; their values are in good agreement with the predicted
theoretical values. In~\cite{MunGabInaPie98}, the relaxation of $c(t)$ had
also been explored after {\em upward} quenches, \ie\ increases in $T$ of
$c\eql$. A strong asymmetry in the relaxation functions for upward and
downward temperature changes was found; this of course makes sense due to
the strong dependence of the relaxation time on the final temperature.

For the asymmetric $1,1$-SFM, which compared to the East model enables
facilitation also by right up-spin neighbours but at a rate reduced by the
asymmetry parameter $a$ (see\eq{asymmetric_SFM}), we already mentioned
in Sec.~\ref{res:dyntrans} that, in addition to the cooperative relaxation
processes of the East model, relaxation can proceed by up-spin diffusion on
the timescale $\tau_{\rm diff} \sim (1+a^{-1})e^{1/T}$. Buhot and
Garrahan~\cite{BuhGar01,BuhGar02} argued that the decay of $c(t)$ should
therefore cross over for $t\approx \tau_{\rm diff}$ to the diffusive domain
growth scaling $(t/\tau_{\rm diff})^{-1/2}$ typical of the $1,1$-SFM. They
confirmed this in simulations; as $a$ increases the relaxation of $c(t)$
exhibits fewer and fewer plateaux since the crossover time $\tau_{\rm
diff}$ decreases.

\begin{figure}
\begin{center}
\epsfig{file=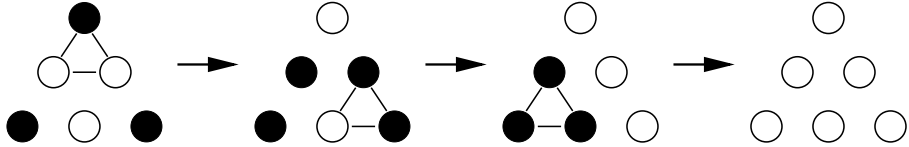, width=12cm}
\end{center}
\caption{Relaxation pathway in the triangle model. The starting
configuration has three defects (up-spins, shown as filled circles) at
the corners of an equilateral triangle of side length 2. These can be
relaxed by successively flipping the spins in three elementary
triangles as shown; the activation barrier is $\dE=1$ since one extra
up-spin is created during this process. This procedure can be iterated to
larger triangles, \eg\ up-spins in the corners of a triangle of size 4
can be relaxed by carrying out the above move sequence in three
sub-triangles of size 2, with a resulting activation barrier of
$\dE=2$. Figure from~\protect\cite{NewMoo99}.
\label{fig:triangle_relaxation}
}
\end{figure}
At this point, we briefly interrupt the usual order in which we give
results for the different KCMs and discuss the {\bf triangle model}
(see Sec.~\ref{model:effective}). The reason is that this model
exhibits, somewhat surprisingly, a time- and lengthscale hierarchy
very similar to that of the East model. Newman and
Moore~\cite{NewMoo99} showed that the relevant configurations for
relaxations at low temperatures consist of three up-spins in the
corners of equilateral triangles of side length $l=2^k$. (Recall that
the spins we are talking about here are the ``defect spins''
$n_i=(1-\sigma_i\sigma_j \sigma_k)/2\in\{0,1\}$, not the spins
$\sigma_i=\pm 1$ of the underlying interacting model, and that the
allowed transitions are simultaneous flips of the $n_i$ at the corners
of elementary {\em upward} triangles.) For $k=0$, there is no
energy barrier for flipping down the three up-spins. For $k=1$, the
spins can be flipped down by flipping three smaller triangles of unit side
length in series, as shown in
Fig.~\ref{fig:triangle_relaxation}. Since the intermediate state now
contains four up-spins, this process has an energy barrier of
one. Continuing recursively, one sees that the the minimum energy path
for flipping three up-spins in the corners of a triangle of size
$l=2^k$ is simply $k$; this is in direct analogy to the relaxation of
domains of size $l=2^k$ in the East model. The associated activation
timescales $\tau \sim \exp(k\beta) \approx c\eql^{-k}$ again become
well separated for low temperatures, and the up-spin concentration
$c=\lav n_i\rav$ after a quench shows the corresponding
plateaux~\cite{GarNew00}. The theory for the East model can be applied
to this case and, while no longer exact, provides a good approximation
to the observed plateau heights~\cite{GarNew00}. Smoothing out across
the plateaux, the typical distance between up-spins grows as
$\bar{l}\sim t^{T\ln2}$ as in the East model; since $\bar{l}\eql\sim
\sqrt{c\eql} \sim \exp(1/2T)$ in equilibrium, extrapolation of this
growth law gives an equilibration time $\tau\sim\exp(1/2T^2\ln 2)$
differing only by a factor $1/2$ from that for the East
model~\cite{GarNew00}. We had anticipated this result already in
Sec.~\ref{res:dyntrans}.

Let us return now to SFMs and consider one of the {\bf variations on SFMs}
discussed in Sec.~\ref{model:SFM_variations}: the ferromagnetic Glauber
Ising chain with the constraint that flips of spins with two up-spin
neighbours are forbidden. The coarsening behaviour of this model at $T=0$
has been studied by simulation and an independent interval approximation
combined with a scaling analysis~\cite{MajDeaGra01}. The kinetic constraint
implies that domains of up-spins cannot coalesce, because the final
down-spin between them can never be eliminated. Domain walls can therefore
be eliminated only by coalescence of {\em down-spin} domains, giving faster
growth $l_-(t) \sim t^{1/2} \ln t$ for the average length of down-spin
domains, while up-spin domains coarsen according to the conventional
$l_+(t) \sim t^{1/2}$. The up-spin concentration---which in the absence of
the kinetic constraint and at $T=0$ would remain constant in
time---therefore decays logarithmically to zero, $c(t) =
l_+(t)/[l_-(t)+l_+(t)] \sim 1/\ln t$ and this has been
likened~\cite{MajDeaGra01} to the slow compaction observed in vibrated
granular media (see Sec.~\ref{other_systems}).

Next we consider constrained {\bf lattice gases}. As pointed out in
Sec.~\ref{model:lattice_gases}, in the standard KA model the density $c$ of
particles is conserved. Nonlinear density relaxation can therefore only
be studied in variants of the model that allow for compaction under gravity
or particle exchange with a reservoir.  In the KA model with gravity (on a
b.c.c.\ lattice, with $m=5$) the relaxation of the density was studied from an
initial loose packed state of bulk density $c\approx 0.71$, obtained by
letting the particles fall from the upper half of the simulation box at
$T=0$~\cite{SelAre00}. (The equilibrium state at this temperature, by
contrast, has all particles packed at the bottom of the system to their
maximum density $c=1$; see Sec.~\ref{res:reduc} for why this maximally
dense state is accessible in spite of the kinetic constraints.) If one then
lets the system evolve at nonzero $T$, corresponding roughly to excitation
by vertical vibration of the container, $c(t)$ increases slowly. Its
time-dependence could be well-fitted by an inverse logarithmic law, $c(t) =
c_\infty - [c_\infty-c(0)]/[1+A\ln (t/\tau)]$ which has been used to
describe experimental data on granular compaction (see
Sec.~\ref{other_systems}). While the equilibrium bulk density $c\eql$ is a
decreasing function of $T$, the extrapolated long-time value $c_\infty$ of
the density first {\em increases} with $T$, goes through a maximum and then
decreases before eventually meeting the equilibrium curve. (The meeting
point occurs at $1/T\approx 0.04$. Since the lattice units were chosen such
that the height difference for a particle hop to a nearest neighbour site
in a lattice plane above or below is unity, this temperature corresponds to
a ``barometric'' excitation height of around 25; this is a substantial
fraction of the bulk height of the sample, which was $\approx 100$.) The
fact that $c_\infty<c\eql$ for lower $T$ was interpreted as evidence for a
dynamical transition, with the system failing to achieve equilibrium even
at very long times. The possibility that a slow increase of $c(t)$ towards
$c\eql$ would be observed outside the simulation time window is of course
difficult to exclude. As an aside, we note that the same model has also
been used successfully to study segregation of granular materials under
vibration: by allowing some particles to move irrespective of the kinetic
constraints, one obtains a model with two particle species. At $T>0$ the
more mobile particles then accumulate at the bottom of the simulation box,
since they can fill the holes that remain between the constrained
particles~\cite{SelAre00}.

The second variation of the KA model where density relaxation can be
investigated does not include gravity but allows particle exchange in a
boundary layer that is in contact with a particle reservoir at some
chemical potential $\mu$; see Sec.~\ref{model:lattice_gases}. A ``crunch''
then corresponds to a sudden increase in $\mu$. The relaxation of particle
density after such a crunch has been investigated by simulations and within
a coarse-grained continuum model (see Sec.~\ref{meth:mappings}). If the
final chemical potential is such that the corresponding equilibrium density
$c\eql$ is below that of the dynamical transition at $c\dyn$ (see
Sec.~\ref{res:dyntrans}), the continuum model predicts~\cite{LevAreSel01}
that the timescale for the relaxation of the density profile to the uniform
value $c\eql$ is governed by the inverse self-diffusion constant in a
homogeneous system, \ie\ $\tau \sim 1/D_s \sim (c\dyn-c\eql)^{-\phi}$ with
$\phi\approx 3.1$. As $c\eql$ approaches $c\dyn$, this timescale
diverges. For crunches to higher chemical potentials, the continuum model
predicts that the density profile relaxes towards the maximum achievable
density $c\dyn$ with a power-law time-dependence $\sim
t^{-1/\phi}$~\cite{PelSel98}, and numerical simulations are consistent with
this~\cite{KurPelSel97,PelSel98}.

The two ideas of including gravity and allowing particle exchange with a
reservoir have also been combined; the boundary layer for particle exchange
is then assumed to be at the top of the system. Density relaxations are
somewhat more complicated to predict in this situation because of the
nontrivial vertical density profile. The typical relaxation time was predicted
to diverge when the equilibrium density of the lowest, densest layer
approaches $c\dyn$, but with an exponent $\phi-2$ that is smaller than
$\phi$~\cite{LevAreSel01}. For higher reservoir chemical potentials, a
section at the bottom of the density profile relaxes to $c\dyn$ for long
times; the time-dependence for this relaxation was found from both an
asymptotic solution of the continuum model and numerical simulations as a
power law, $t^{-1/(\phi-1)}$.

We now turn to models inspired by {\bf cellular structures}; see
Sec.~\ref{topological}. In the topological froth model, glassy behaviour is
seen in the relaxation of the energy (which is proportional to the number
of defects). Starting from two different initial configurations, one
strongly disordered and one perfectly ordered, the initial configuration is
remembered at low $T$ even for the longest accessible simulation
times~\cite{AstShe99}. The time evolution of the energy relaxation from a
high-temperature, disordered state was studied in more detail for the
lattice version of the model~\cite{DavSheGarBuh01} and showed the two-step
form expected for activated dynamics, with a nearly $T$-independent decay
to a plateau on timescales of order unity, and the remaining decay taking
place only on activated (Arrhenius) timescales. From the effective
low-temperature model for this system (see Secs.~\ref{topological},
\ref{meth:mappings}) one deduces that the short-time evolution is dominated
by diffusion of defect dimers and dimer--antidimer annihilation
$A+B\to\emptyset$; the dynamics on longer, activated timescales arises
instead from the diffusion of isolated defects and defect-antidefect
annihilation, giving again $A+B\to\emptyset$. This leads to the prediction
of $t^{-1/2}$-scaling for both the short- and long-time decays, which is in
good agreement with simulations~\cite{DavSheGarBuh01}.

Finally, for models with entropic barriers such as the {\bf
Backgammon} and {\bf oscillator} models, we saw already in
Sec.~\ref{meth:exact} that the energy relaxes logarithmically slowly
after a
quench~\cite{Ritort95,FraRit95,FraRit96,GodBouMez95,GodLuc96}. This
behaviour persists for all times if the final temperature is zero; at
nonzero temperature it is observed only at intermediate times smaller
than the longest relaxation time. For the zeta urn
model~\cite{DroGodCam98,GodLuc01} with a random initial configuration,
one finds both at criticality and in the condensed regime that the
occupancies $P_k(t)$ show scaling behaviour, becoming functions of the
single scaling variable $kt^{-\frac{1}{2}}$ when both $k$ and $t$ are
large.

\subsubsection{Heating-cooling cycles}
\label{res:heatcool}

In this section we review the behaviour of KCMs under cyclic variations of
temperature (or density etc). As explained in Sec.~\ref{basics1}, such
heating-cooling cycles in real glasses show strong hysteresis effects. These
demonstrate that, as soon as a supercooled liquid falls out of equilibrium
because its relaxation times are too large to keep pace with the external
heating and cooling, it develops a strong memory of its temperature history.

To recap briefly the discussion in Sec.~\ref{basics1}, on cooling the
energy $E$ departs from the equilibrium line at some cooling-rate dependent
$T_g$; for lower temperatures, the decrease in energy with temperature $T$
is much reduced and the value of the specific heat therefore drops around
$T_g$.  When the system is heated back up, the energy increases slowly
enough with $T$ to cross {\em below} the equilibrium line, rejoining it by
a steep increase only at a higher temperature; this increase manifests
itself as a peak in the specific heat. (Notice that we use the term
specific heat here to refer to the temperature-derivative of the energy; in
equilibrium, the specific heat is also related to the amplitude of energy
fluctuations but out of equilibrium this is not the case.)

\begin{figure}
\begin{center}
\epsfig{file=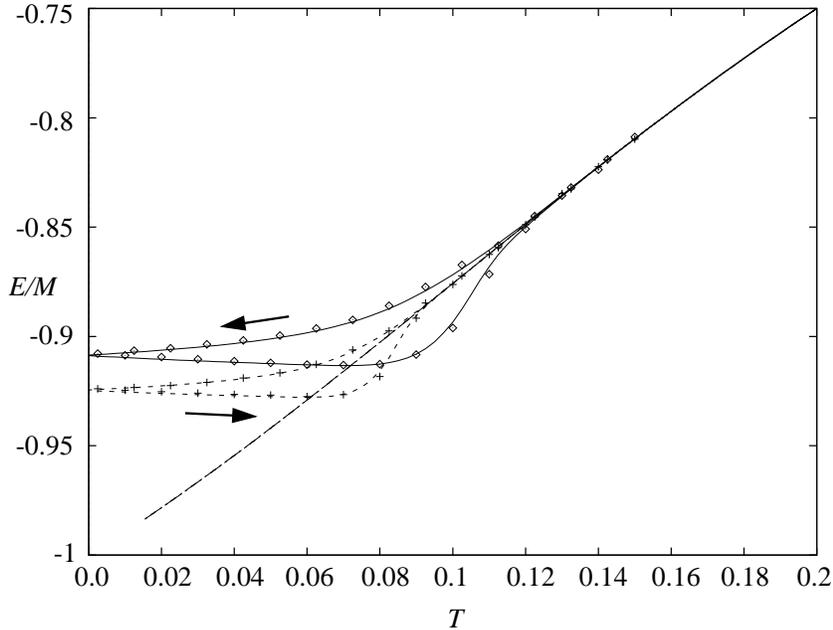, width=12cm}
\end{center}
\caption{Heating-cooling cycles in the Backgammon model at two
different cooling rates. 
%
%
For the lower cooling rate (dashed) the system follows the equilibrium
relation between $E$ and $T$ to lower
temperatures. From~\protect\cite{FraRit95}.
\label{cycle_BG}
}
\end{figure}
Many KCMs exhibit the above effects; by way of illustration we show in
Fig.~\ref{cycle_BG} typical cooling-heating cycles in the Backgammon
model~\cite{Ritort95,FraRit95}. While the temperature evolution of the
energy during the cooling process is easy to understand in terms of
the effective freezing of the slow degrees of freedom of the system,
the behaviour on reheating is less intuitively obvious. We therefore
now sketch an analysis of this phenomenon due to Brey and
Prados, who applied the concept of a ``normal curve'' for the heating
process to a number of simple models. The normal curve in general
exists for any irreducible Markov process with time-dependent
transition rates. It gives the long-time behaviour of the
time-dependent probability distribution over configurations,
independently of initial conditions, and is the analogue of the unique
stationary distribution for irreducible Markov processes with constant
transition rates~\cite{BrePra93a,BrePra93b}. For cooling processes the
dynamics becomes reducible in the limit $T=0$, even in models without
kinetic constraints, and so a normal curve does not exist; for the
heating case, however, there is no such restriction. Brey
\etal~\cite{BrePra94,BrePraRui94} studied in detail the ferromagnetic
Ising chain with Glauber dynamics. (A similar analysis can be
performed~\cite{PraBre01b} for the models with hierarchical kinetic
constraints described in Sec.~\ref{related:hierarchical}.)  They
showed that the energy during a heating process from $T=0$ can be
decomposed into two contributions, $E(T)=E_N(T)+E_p(T)$. The first
term is the normal curve for heating, constructed with equilibrium at
$T=0$ as the starting configuration. The second term describes the
correction due to the preceding cooling protocol by which $T=0$ was
reached, and vanishes in the limit of infinitely slow cooling.  Brey
\etal\ proved that the normal curve stays {\em below} the equilibrium
curve $E\eql(T)$ and coincides with it only at $T=0$ and
$T=\infty$. On the other hand, in a realistic cooling schedule
one does not reach equilibrium at $T=0$, so that $E_p(T\approx 0)$ is
positive and the total energy $E(T)$ is above the equilibrium curve at
low $T$. As $T$ is increased and the normal curve drops increasingly
below the equilibrium curve, the two effects eventually cancel and
this causes the crossing of $E(T)$ below the equilibrium curve.

The effects of cyclical heating and cooling have been studied in a number
of KCMs; we already referred to the Backgammon model above. As far as {\bf
spin-facilitated models} are concerned, Graham
\etal~\cite{GraPicGra93,GraPicGra97} studied in detail the behaviour of the
specific heat in temperature cycles for the $2,2$-SFM, $2,3$-SFM and
$3,3$-SFM. Starting from a glassy configuration obtained by quenching to
low $T$, a sharp peak in the specific heat was observed on heating, and a
much broader peak at lower temperatures on cooling; as explained at the
beginning of this section, this agrees qualitatively with experimental
observations on glasses. 

In the context of {\bf lattice gases}, the analogue of cooling runs were
studied for the grandcanonical KA model, where particle exchange with a
reservoir is allowed in a boundary layer (see
Sec.~\ref{model:lattice_gases}). The evolution of the inverse particle
density $1/c$ was simulated in slow compression runs, implemented by
decreasing the inverse chemical potential $1/\mu$ at a constant
rate~\cite{KurPelSel97}. The inverses are chosen here to emphasize the
analogy with energy and temperature in glasses; small $1/\mu$ corresponds
to low $T$, and small $1/c$ to the glassy regime of low energy. Similarly
to cooling experiments in real glasses, $1/c$ begins to deviate from the
equilibrium curve $1/c=1+e^{-\mu}$ (see\eq{grandcanonical_KA}) later and
later as the compression rate is reduced. Given that the KA model has an at
least effective dynamical transition at $c\dyn=0.881$ (see
Sec.~\ref{res:dyntrans}), where the timescale for self-diffusion appears to
diverge, one would expect that $c\dyn$ is the density that is reached in
very slow compression experiments. The results are compatible with
this~\cite{KurPelSel97}; see Fig.~\ref{lg_fig4} above. Increasing and
decreasing $1/\mu$ in 
analogy to heating-cooling cycles also leads to the expected hysteresis in
$1/c$, with lower values of $1/c$ found in the decompression phase that is
analogous to reheating.

Finally, simulations of cooling runs have been carried out in several other
KCMs. In the {\bf triangle model}, annealing runs with an exponential
cooling schedule, $T(t)=T_0\exp(-\gamma t)$ were performed, and showed the
expected deviations from the equilibrium relationship between defect
concentration $c$ and temperature $T$ when inverse cooling rates and
relaxation times became comparable~\cite{NewMoo99}. In the topological
model of {\bf cellular structures}, cooling experiments~\cite{DavShe00}
found that even for slow cooling rates the system falls out of equilibrium
at sufficiently low temperatures where relaxation timescales become very
long ($T \approx 0.2$), and similar behaviour is observed in the lattice
variant~\cite{DavSheGarBuh01}.

\subsubsection{Two-time correlation and response, and effective temperatures}
\label{res:aging}
\label{res:fdt}

As explained in Sec.~\ref{intro:fdt}, systems such as glasses which do no
equilibrate on experimentally accessible timescales show {\em aging}, which
means that their properties depend on the ``waiting time'' $\tw$ that has
elapsed since they were prepared by, for example, a quench. The
time-evolution of one-time quantities such as up-spin concentration or
particle density, discussed in Sec.~\ref{res:relax}, already testifies to
this. Often aging effects can persist, however, even when the relaxation of
one-time quantities has become so slow that their values are already
effectively constant. One then needs to consider two-time quantities such
as correlations and response functions. Since the system is out of
equilibrium, these generically violate FDT, and it has been suggested that
the FDT violation factor $X(t,\tw)$ defined by\eq{eqn:non_eq_fdt} can be
used to define an effective temperature $T\eff=T/X$.

We begin our discussion with {\bf spin-facilitated models}. As in
Sec.~\ref{res:statdyn}, let us review briefly the definitions of the
two-time quantities most frequently studied in these models. The two-time
spin autocorrelation function is, in a natural generalization of\eq{eqC},
\be 
C(t,\tw)=\frac{1}{N}\sum_{i} [\lav n_i(t)n_i(\tw) \rav - \lav n_i(t)\rav
\lav n_i(\tw)\rav ]
\label{C_twotime}
\ee
No normalizing factors have been introduced here, since the normalization
of two-time quantities is a somewhat subtle issue; see
Sec.~\ref{intro:fdt}. Comparing with\eq{corr}, it is easy to see that
$C(t,\tw)$ is (apart from a factor of $N$) the two-time correlation
function of a ``random staggered magnetization'' $\phi=(1/N)\sum_i
\epsilon_i n_i$, with the signs $\epsilon_i=\pm 1$ chosen randomly for each
$i$~\cite{Barrat98}. Imposing the constraint $\sum_i \epsilon_i=0$
simplifies matters by making $\lav \phi(t)\rav=0$ for all $t$. The associated
response function is obtained by adding a term $-Nh\phi$ to the energy
function $E$; if the field is increased from zero to a small constant
value $h$ at time $\tw$, then the normalized change in $\phi$,
$\lav\phi(t)\rav/h$, gives the two-time step response function $\chi(t,\tw)$
for $t>\tw$. As emphasized in
Sec.~\ref{intro:fdt}, in an out-of-equilibrium situation two-time
correlation and response are nontrivial functions of their two time
arguments, whereas in equilibrium they depend only on $t-\tw$.

As in the case of stationary dynamics, one may also be interested in the
two-time correlations of the overall up-spin concentration; by analogy
with\eq{eqCM}, but again without normalization, this is
\be 
C_c(t,\tw)=\frac{1}{N}\sum_{ij} 
[\lav n_i(t)n_j(\tw) \rav - \lav n_i(t)\rav \lav n_j(\tw)\rav]
\label{Cc_twotime}
\ee
The corresponding perturbation in the energy function, which defines the
response $\chi_c(t,\tw)$, is $-Nhc=-h\sum_i n_i$. For the standard SFMs
where $E=\sum_i n_i$, this effectively changes temperature from $T$ to
$T/(1-h) = T+hT+\ldots$, so that $\chi_c(t,\tw)$ can also be thought of as
measuring the response of the up-spin concentration to small temperature
changes. 

After these preliminary definitions we turn to results for SFMs with
undirected constraints. All the work on two-time quantities that we are
aware of has focused on $1,d$-SFMs with their defect-diffusion dynamics.
Simulations in $d=1,2$ considered a quench from $T=\infty$ ($c\eql =1/2$)
to small $T$ and $c\eql$ and measured the spin correlation function
$C(t,\tw)$, normalized by the equal-time value at the earlier time,
$C(\tw,\tw)$~\cite{FolRit96}. A strong dependence on $\tw$ was observed for
waiting times of order unity, while in the regime $1\ll \tw \ll c\eql^{-1}$
the effect of $\tw$ was negligible. This makes sense in light of the
discussion in Sec.~\ref{res:relax}. For times of order unity the system
evolves through the flipping-down of mobile up-spins; as emphasized in
Sec.~\ref{meth:exact}, in this regime one expects the exactly solvable
$T=0$ dynamics to correctly predict the dynamics, and this was indeed found
in~\cite{FolRit96}. Further evolution of the system requires diffusion of
isolated up-spins, and so only takes place once $\tw$ becomes of order
$1/D\eff \sim c\eql^{-1}$. Even for $\tw$ of this order and larger,
however, simulations showed aging effects on $C(t,\tw)$ to be rather
weak~\cite{FolRit96,CriRitRocSel00}. For the $1,1$-SFM, one might expect
that in this time regime, where the model exhibits growing domains of
down-spins, the two-time correlations should collapse when plotted as a
function of the ratio of typical domain lengths at the early and late
times, $\bar{l}(\tw)/\bar{l}(t)\approx c(t)/c(\tw)$. The simulations
of~\cite{CriRitRocSel00,CriRitRocSel02} did not show this, but the values
of $\tw$ accessed may not have been large enough to see the expected
scaling.

In the $1,1$-SFM, the response function $\chi(t,\tw)$ conjugate to the spin
autocorrelation is found to be non-monotonic as a function of the later
time $t$~\cite{CriRitRocSel00,CriRitRocSel02,BuhGar02b}. This may appear
surprising, but a nice intuitive justification for this behaviour was given
in~\cite{BuhGar02b}, for the regime of times long compared to the initial
fast relaxation processes (see above). For low up-spin concentrations
$c(t)$, up-spins are far apart, as they would be in equilibrium if $c\eql$
is small. Since only up-spins and their neighbours are mobile and can
therefore contribute to the response, $\chi(t,\tw)$ should be
proportional to $[c(t)/c\eql]\chi\eql(t-\tw)$, where $\chi\eql(t-\tw)$ is
the equilibrium response at the final temperature after the quench. This
form fits simulation data very well~\cite{BuhGar02b}; the non-monotonicity
arises since $\chi(t,\tw)$ is a product of two factors, one ($c(t)$)
decreasing with $t$ and one ($\chi\eql(t-\tw)$) increasing. The behaviour
of the autocorrelation function can be rationalized with a similar approach.
Intriguingly, these scaling relations suggest that $C(t,\tw)$ and
$\chi(t,\tw)$ are related by the trivial {\em equilibrium} FDT, even though
\eg\ the up-spin concentration $c(t)$ is still far above its equilibrium
value $c\eql$. Simulation results indeed showed that a plot of
$T\chi(t,\tw)$ versus $C(t,t)-C(t,\tw)$ (see Sec.~\ref{intro:fdt}) gives a
straight line of slope one through the origin~\cite{BuhGar02b}. Earlier
attempts~\cite{CriRitRocSel00} at FDT plots using the disconnected
correlation function $(1/4)\lav(2n_i(\tw)-1)(2n_i(t)-1)\rav$, chosen in
such a way as to be automatically normalized to unity at $t=\tw$, had
produced rather counterintuitive non-monotonic relations between response
and correlation.

An interesting twist to the apparently trivial FDT relations in the
out-of-equilibrium dynamics of the $1,1$-SFM is provided by recent
work on defect 
(domain-wall) dynamics in the ferromagnetic Ising chain with unconstrained
Glauber dynamics~\cite{MaySol02}. The dynamics of these defects is rather
similar to those in the $1,1$-SFM, except that rather than coalesce they
annihilate when they meet. Indeed, appropriate scaling plots of domain-wall
autocorrelation and response functions look rather similar to those
in~\cite{BuhGar02b}, and the FDT plot becomes a straight line at long
times. This suggests again that the FDT violation factor is $X(t,\tw)=1$ and
equilibrium FDT holds. However, when plotted \eg\ against $t/\tw$, one
finds that $X(t,\tw)$ is a nontrivial function and generically $<1$. This
apparent paradox is resolved by noticing that significant FDT violations
only occur on timescales $t-\tw$ of order $\tw$, where the autocorrelation
function has already decayed to such a small fraction of its equal-time
value that FDT violations are not visible either in the FDT plot or the
scaling collapse. A detailed investigation of $X(t,\tw)$ for the $1,1$-SFM
for similar effects in the regime $t-\tw \sim \tw$ should therefore be
worthwhile.

We finish off our discussion of the $1,1$-SFM by briefly mentioning results
for the response $\chi_c(t,\tw)$ of the up-spin concentration to small
temperature changes~\cite{FolRit96}. As is typical of activated dynamics,
this response function is actually negative, and much larger in absolute
value than the equilibrium response. The apparently counterintuitive
negative sign can be understood by considering \eg\ a small decrease in
$T$: this slows down the relaxation of $c(t)$ to lower values, giving {\em
larger} values of $c(t)$ rather than smaller ones as in equilibrium.

Next we turn to SFMs with {\bf directed constraints}, in particular the
East model. The two-time spin autocorrelation function $C(t,\tw)$ and
corresponding response $\chi(t,\tw)$ were simulated
in~\cite{CriRitRocSel00}. Because of the hierarchy of well-separated
timescales that dominates the out-of-equilibrium dynamics (see
Sec.~\ref{meth:indint}), $C(t,\tw)$ exhibits plateaux, and a naive plot
against $t/\tw$ does not give a reasonable collapse of the curves. On the
other hand, plotting $C(t,\tw)/C(\tw,\tw)$ against the domain length ratios
$\bar{l}(\tw)/\bar{l}(t)=c(t)/c(\tw)$ gave good scaling collapse, as
expected from the coarsening character of the out-of-equilibrium dynamics
at low $T$. In fact, since in the limit $T\to 0$ domains simply coalesce
irreversibly, one would predict $C(t,\tw) = c(t)[1-c(\tw)]$ and thus
$C(t,\tw)/C(\tw,\tw)=c(t)/c(\tw)$, and this is compatible with the data
of~\cite{CriRitRocSel00}. 

The response function $\chi(t,\tw)$ was found to be monotonic in $t$ for
the East model, in contrast to the results for the $1,1$-SFM; this may be
because the much slower decrease in time of $c(t)$ is not sufficient to
produce noticeably non-monotonic behaviour. FDT plots were also considered
in~\cite{CriRitRocSel00}, but have to be regarded with caution since they
were constructed using a disconnected correlator as for the $1,1$-SFM. They
consisted to a rough approximation of two straight line segments, but with
no obvious limit plot being approached for long times; the FDT violation
factor $X$ in the non-equilibrium sector became rather small ($\approx
0.1$) for low temperatures. This is not inconsistent with the coarsening
character of the model, which in $d>1$ would be expected to give $X=0$ (see
Sec.~\ref{intro:fdt} and~\cite{CorLipZan01}). Since the East model is
$d=1$-dimensional, however, this comparison is not conclusive.

At this point we consider the {\bf triangle model}, because of the
similarities of its out-of-equilibrium dynamics to that of the East model.
Interestingly, the response function $\chi_c(t,\tw)$ of the up-spin
concentration of the ``defect spins'' $n_i=(1-\s_i \s_j\s_k)/2$ was found
to be non-monotonic in $t$~\cite{GarNew00}, and led to corresponding
non-monotonic FDT plots. The origin of this behaviour can be understood from
the plateaux in the evolution of $c(t)$, which occur between the relaxation
timescales $\tau_k\sim \exp(k/T)$. The perturbation conjugate to $c$ is
essentially an increase in temperature, which reduces the $\tau_k$ but
leaves the heights of the plateaus in $c(t)$ unaffected. The response is
therefore largest around the transitions between the plateaux, and small in
between. Garrahan and Newman~\cite{GarNew00} argued that this argument
should also apply to the {\em local} response function $\chi(t,\tw)$, and
conjectured on this basis that $\chi(t,\tw)$ in the {\em East model} should
also exhibit non-monotonic behaviour, at lower temperatures than those
simulated in~\cite{CriRitRocSel00}. Notice that in the triangle model,
because of its derivation via a mapping from a system of interacting spins
$\s_i=\pm 1$ (see Sec.~\ref{model:effective}), there are further
correlation and response functions that one can
consider~\cite{NewMoo99,GarNew00}. One intriguing observation is that the
two-time autocorrelation function of the $\sigma_i$ (rather than the $n_i$)
seems to scale neither with $t/\tw$, nor with the ratio of typical
distances between defects $\bar{l}(\tw)/\bar{l}(t)$; the reasons for this
are not presently understood~\cite{GarNew00}.

\begin{figure}
\begin{center}
\epsfig{file=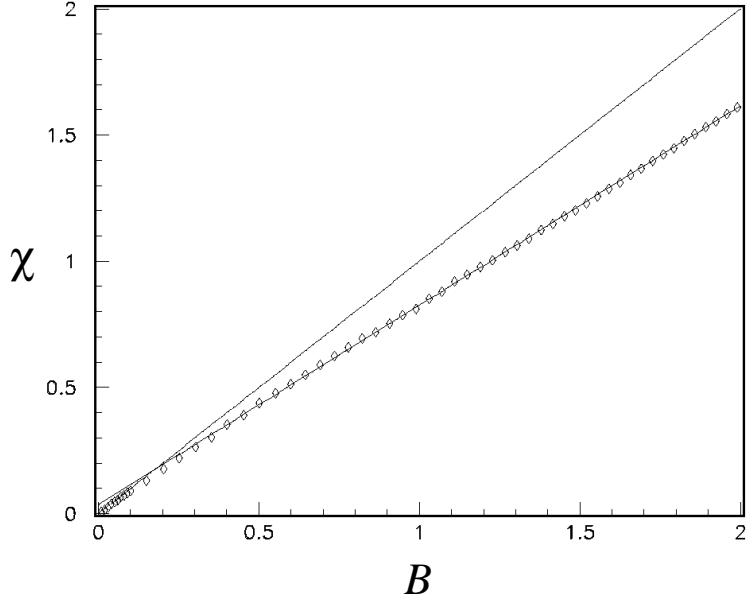, width=10cm}
\end{center}
\caption{Out-of-equilibrium FDT plot for the grandcanonical KA model:
simulation results for waiting time $\tw=10^5$ after a ``crunch'' to
chemical potential $\mu=2.2$. Symbols show a parametric plot, obtained
by varying the final time $t$, of the two-time response (of particle
displacements to random forces) against the corresponding two-time
correlation (mean-square particle displacement $B(t,\tw)$). The plot initially
follows equilibrium FDT, indicated by the line through the origin, but
then crosses over onto a second straight line with slope smaller by an
FDT violation factor of $X\approx0.79$. From~\protect\cite{Sellitto98}.
\label{fig:KA_FDT}
}
\end{figure}
We next review results on two-time quantities in constrained {\bf
lattice gases}. As explained in Sec.~\ref{res:relax}, one needs to
consider the recent variations on the KA model that include gravity or
a particle reservoir to study these out-of-equilibrium quantities. As an
analogue of two-time correlation functions, the average squared
particle displacement $B(t,\tw) = \lav [\rv_a(t)-\rv_a(\tw)]^2 \rav$
has been studied; the corresponding response function that would be
related via FDT in equilibrium is obtained by applying a random force
to each particle and measuring the displacement in the direction of
the force. In the grandcanonical KA-model, aging effects on $B(t,\tw)$
were studied after a crunch, \ie\ an increase of the chemical
potential $\mu$ to a point where the equilibrium density $c\eql$ would
be above the critical value
$c\dyn$~\cite{KurPelSel97,PelSel98,Sellitto98}. (Since particle
exchange only acts in the boundary plane, this increase of chemical
potential is performed slowly, rather than near-instantaneously as in
a conventional temperature quench, to avoid inhomogeneities across the
sample.) For large $\tw$, it was found that $B(t,\tw)$ becomes to a
good approximation a function of the scaled time difference
$(t-\tw)/\tw$ and increases roughly logarithmically, indicating very
slow anomalous diffusion. A qualitative explanation for this is
provided~\cite{PelSel98} by supposing that $B(t,\tw)\sim \int_{\tw}^t
dt'\,D_s(t')$, with $D_s(t')$ a time-dependent self-diffusion constant which
depends on density according to $D_s(t) \sim (c\dyn-c(t))^\phi$. Since,
as described in Sec.~\ref{res:relax}, the density approaches the
critical value as $c\dyn-c(t) \sim t^{-1/\phi}$ this gives $D_s(t)\sim
t^{-1}$ and thus directly the observed logarithmic increase
$B(t,\tw)\sim \ln(t/\tw)$ of the particle displacements. Remarkably,
an FDT plot of the conjugate response versus $B(t,\tw)$ was of a
simple ``mean field'' form (see Sec.~\ref{intro:fdt}), consisting of
two approximately straight line segments~\cite{Sellitto98}; this is
shown in
Fig.~\ref{fig:KA_FDT}. The FDT violation factor was $X\approx 0.79$ in
the non-equilibrium sector (though its dependence on $\tw$ does not
appear to have been investigated), and it was later shown that this
value can be understood from an appropriately defined Edwards measure;
see Sec.~\ref{res:landsc}. Similar FDT results were also found for
compaction under gravity at constant number of
particles~\cite{Sellitto01}.

As mentioned in Sec.~\ref{res:relax}, the KA model connected to a particle
reservoir has also been considered when subject to gravity, with the
contact layer with the reservoir at the top of the system. For sufficiently
high reservoir chemical potential $\mu$, the system develops a dense zone
at the bottom where the particle density slowly approaches the critical
value $c\dyn$. Using a continuum model, it was
argued~\cite{LevAreSel01}---in accord with
simulation~\cite{Sellitto01}---that in this dense zone the mean-square
particle displacement scales as $B(t,\tw) \sim
\tw^{-1/(\phi-1)}-t^{-1/(\phi-1)}$, where $\phi\approx 3.1$ is the exponent
for the divergence of the inverse self-diffusion constant at the critical
density. Intriguingly, the exponents here are {\em negative}, implying that
for $t\to\infty$ the displacement saturates to a constant value (which
itself tends to zero for $\tw\to\infty$). Notice also that, in contrast to
the case without gravity discussed above, the aging here is not
``simple'', \ie\ $B(t,\tw)$ is not a function of the scaled time
difference $(t-\tw)/\tw$ alone.

Moving on to KCMs inspired by {\bf cellular structures}, Aste and
Sherrington~\cite{AstShe99} studied the two-time {\em persistence} function
in the topological froth model, defined as the fraction of cells that have
not been involved in any moves between $\tw$ and $t$. While for high
temperatures this is $\tw$-independent and decays exponentially with
$t-\tw$, for low temperatures ($T<1$) simulations show aging effects. These
can be qualitatively understood~\cite{AstShe99} from the fact that most
moves are due to the diffusion of pairs of pentagonal and heptagonal cells,
whose concentration decreases with $\tw$. The two-time correlation function
for local deviations from the hexagonal ground state configuration, $\lav
(n_i(t)-6)(n_i(\tw)-6)\rav$, along with the conjugate response function,
was simulated in~\cite{DavShe00}. This correlation function decays to a
plateau within times $t-\tw$ of order unity, while the remaining decay
takes place on timescales growing with $\tw$. The response function is
non-monotonic in $t$ at fixed $\tw$. This can be understood by arguments
similar to those for the $1,1$-SFM above; the decay in the response at long
times again arises from the decrease in the number of defects that drive
the dynamics, which in this case are pairs of 5- and 7-sided cells. The
behaviour of the lattice version of the model is qualitatively
similar~\cite{DavSheGarBuh01}. Interestingly, however, if response and
correlation are normalized properly (see Sec.~\ref{intro:fdt}) by the equal
time correlator at the {\em later} time $t$, the resulting FDT plot becomes
the trivial equilibrium one~\cite{SheDavBuhGar02}. The physical reasons for
this remain to be understood.

Of the models that arise via a mapping from underlying interacting
spin systems with unconstrained dynamics we have already dealt with
the triangle model above. Two-time quantities in the {\bf plaquette
model} have recently been considered in~\cite{BuhGar02b}, focusing on
the correlation and response functions for the defect spins $n_i= (1-
\s_i \s_j \s_k \s_l)/2$ (see Sec.~\ref{model:effective}). Recall that
the elementary transitions between configurations are simultaneous
flips of four of the $n_i$ in the corners of an elementary lattice
square. This implies in particular that pairs of n.n.\ defects $n_i=1$
can diffuse unidirectionally---pairs in the $x$-direction can diffuse
along the $y$-direction and vice versa---and pairs of defects in
diagonally opposite corners of lattice squares can oscillate. Both of
these processes are fast, taking place on timescales of order one
since they involve no change of the energy $E=\sum_i n_i$, and
determine the behaviour of $C(t,\tw)$ and $\chi(t,\tw)$ for $t-\tw$ of
order unity.  On longer timescales, diffusion of isolated defects
takes over. This proceeds by an isolated defect creating a freely
diffusing defect pair, which must then be absorbed by another isolated
defect. The activation energy for creating the pair is $\dE=2$, and
the overall timescale for this process scales as $\exp(2/T)/c(t)$; the
factor $1/c(t)$ gives the typical probability that the defect-pair
will indeed be absorbed by a {\em different} isolated defect, rather
than the original one.  On the basis of these considerations, good
scaling collapse of response and correlation functions in the two
different time-regimes could be obtained~\cite{BuhGar02b}. Remarkably,
an FDT plot of $T\chi(t,\tw)$ versus $C(t,t)-C(t,\tw)$ gave data
collapse onto a master plot for a range of different $t$ and $\tw$,
and consisted of two straight line segments. Buhot and
Garrahan~\cite{BuhGar02b} gave a plausible argument for the location
of the breakpoint between these two segments, but the value of the FDT
violation factor $X$ in the non-equilibrium sector remains to be
understood.

Let us finally turn to models related to KCMs, beginning with {\bf urn
models}. For the Backgammon model different type of correlation functions
have been considered.  In the original paper on the Backgammon
model~\cite{Ritort95}, the energy-energy correlation function was simulated
at $T=0$, finding simple aging scaling with $(t-\tw)/t_w$. These results
were later confirmed by numerical integration of the exact master equation
solution in~\cite{FraRit96} and by asymptotic expansion
techniques~\cite{GodLuc96,GodLuc97} which showed the existence of
subdominant logarithmic corrections to the simple scaling. The
energy-energy autocorrelation function does not show the existence of a
fast relaxation process analogous to the $\beta$-relaxation in supercooled
liquids and glasses. Such a separate fast process does appear, however, in
the autocorrelation function $C(t,\tw)$ of the local number of particles
per box. This correlation function and its associated response was
considered in~\cite{FraRit97}, using a numerical integration of a truncated
hierarchy of dynamical equations; a detailed analytical solution was
subsequently given by Godr\`eche and Luck~\cite{GodLuc99}. The main
findings are that, for large $\tw$, $C(t,\tw)$ develops a pronounced
plateau, as does the corresponding response; the long-time decay from this
plateau again shows simple aging scaling. The FDT violation factor
$X(t,\tw)$ when plotted against $C(t,\tw)$ for fixed large $\tw$ is well
approximated by a piecewise constant function, equal to one for values of
$C(t,\tw)$ above the ($\tw$-dependent) plateau and to a smaller value for
smaller $C$. This second, nontrivial value of $X$ tends to one
logarithmically as $1-{\rm const}/\ln^2\tw$, however, so that there is no
nontrivial limit plot; also, $X$ does not correspond to a ratio between the
actual temperature $T$ and the temperature defining the effective
equilibrium state found in the adiabatic analysis of the dynamics (see
Sec.~\ref{meth:adiab}).

For the zeta-urn model~\cite{GodLuc01} the FDT violation factor $X(t,\tw)$
is found to become asymptotically a nontrivial function of the ratio
$t/\tw$. Of interest is particularly the limit $X_\infty$ obtained for
$t/\tw\to\infty$, which is related to universal amplitude ratios in
critical dynamics. Along the critical line in the phase diagram of the urn
model (see Sec.~\ref{model:urn}) one finds that $4/5<X_{\infty}<1$. This
contrasts with analogous results for ferromagnetic Ising models at
criticality (see \eg~\cite{GodLuc02}) where $0<X_{\infty}<1/2$ and is
more similar to the Backgammon model where
$X(t,\tw)\to 1$ in the limit of large $t$.

For the {\bf oscillator model} (see Sec.~\ref{model:osc}), it is natural to
consider the two-time autocorrelation function of the oscillator positions
$x_i$, and the corresponding response. One
finds~\cite{BonPadRit98,Nieuwenhuizen98b} that these display simple scaling
with logarithmic corrections: defining $g(s)=s (\ln s)^{3/2}$, one has
$C(t,\tw)={g(\tw)}/{g(t)}$ for the correlation and
$R(t,\tw)=-\partial\chi(t,\tw)/\partial\tw = {g(\tw)}/[t{g(t)}]$ for the
impulse response.  The effective temperature derived via the FDT violation
can easily be computed and gives the equipartition result $T_{\rm
eff}=2E(t)$ in the long-time limit; the simplicity of this result has no
counterpart in any of the other models.

\subsubsection{Coarsening versus glassiness}
\label{res:coarsen}

We conclude this section with a brief discussion of an interesting
quantity for characterizing the qualitative nature of
out-of-equilibrium dynamics. This is the overlap between two
``clones'' of a system evolving under different realizations of the
stochastic noise in the dynamics. Specifically, one imagines that the
system has aged until $\tw$ and is in configuration
$\nv^{(1)}(\tw)$. One then makes a copy
$\nv^{(2)}(\tw)=\nv^{(1)}(\tw)$ and lets the two clones evolve
independently for $t>\tw$. The quantity of interest, introduced in the
context of the spherical SK model~\cite{CugDea95} and analysed in
detail in~\cite{BarBurMez96}, is then the ``clone overlap''
\be
Q_{\tw}(t)=\frac{1}{N}\sum_i
[\langle n_i^{(1)}(t)n_i^{(2)}(t)\rangle - 
 \langle n_i^{(1)}(t)\rangle \langle n_i^{(2)}(t)\rangle]
\ee
The averages are both over the configuration at the starting time
$\tw$ and over the subsequent stochastic evolution; only the former
couples the two clones. The decay of $Q_{\tw}(t)$ tells one how fast
the clones separate in configuration space as they evolve, and is to
be compared with the two-time correlation function $C(t,\tw)$ defined
in\eq{C_twotime}, which measures how much each clone has decorrelated
from its configuration at time $\tw$. In equilibrium, because of
detailed balance, the forward evolution by time $\Delta t = t-\tw$ of
one clone is equivalent to backward evolution by the same time, so
that
\be
Q_{\tw}(\tw+\Delta t) = C(\tw+2\Delta t, \tw)
\label{Q_C}
\ee
Both quantities are of course functions of $\Delta t$ only because of
TTI; for exponentially decaying correlation functions\eq{Q_C} gives
$Q_{\tw}(\tw+\Delta t) \propto C^2(\tw+\Delta t,\tw)$.

Barrat \etal~\cite{BarBurMez96} proposed the name ``type I'' for
systems for which $Q_{\tw}(t)$ remains large while $C(t,\tw)$ decays
to zero. Intuitively, this corresponds to the system ``falling down a
gutter'' in configuration space, where the two clones remain similar
even though they have moved far from their starting point at $\tw$. A
number of coarsening systems display this behaviour, with in fact
$S_\infty=\lim_{\tw\to\infty} \lim_{t\to\infty} Q_{\tw}(t) > 0$ while
the analogous limit for $C(t,\tw)$ vanishes. The intuitive reason for
this is that on large lengthscales and at low $T$ most coarsening
models behave essentially deterministically~\cite{Bray94}---with the
ferromagnetic Ising chain with Glauber dynamics an obvious
exception---so that the two clones stay closely correlated in their
evolution while moving far from their configuration at time $\tw$. In
``type II'' systems, on the other hand, $Q$ and $C$ decay on the same
timescale, and this can be interpreted as true ``glassy'' evolution
resulting from a rugged energy landscape in which the two clones
quickly begin to following different routes.

Only a few studies exist of the clone overlap in KCMs. One reason for
this is that the limiting quantity $S_\infty$ is not useful for KCMs:
barring dynamical ergodicity breaking, the system will eventually
equilibrate to its trivial Boltzmann distribution and the clones then
decorrelate, giving $Q_{\tw}(t\to\infty)=0$ and hence
$S_\infty=0$. (In an infinitely large coarsening system, on the other
hand, $S_\infty$ can be nonzero since equilibrium is never reached.)
One therefore has to look at finite times and consider whether
$Q_{\tw}(t)$ decays more slowly (type I) or on the same timescale
(type II) as $C(t,\tw)$.

One-dimensional spin-facilitated models were studied
in~\cite{CriRitRocSel00}. Numerical results for the East model showed
plateaux in the $t$-dependence of $Q_{t_w}(t)$ where $Q$ was larger
than expected from the equilibrium relation\eq{Q_C}, indicating a
resemblance to coarsening (type I) systems. For the $1,1$-SFM, on the
other hand, the equilibrium relation\eq{Q_C} was found to be valid to
a good approximation, showing that $Q$ and $C$ decay on the same
timescale and the dynamics is therefore of type II. Similar type II
behaviour has also been observed for the lattice model of the
topological froth~\cite{DavSheGarBuh01} and a disordered version in
$d=3$ of the plaquette model~\cite{AlvFraRit96}. Results for the
evolution of the grandcanonical KA model~\cite{Sellitto02} after a
crunch to large reservoir chemical potential likewise suggest type II
behaviour: even though the particle density $c$ approaches the value
$c\dyn$ where the (effective) dynamical transition takes place,
$Q_{\tw}(t)$ always tends to zero as $t\to\infty$. A direct comparison
with $C(t,\tw)$ would however be needed to make this argument more
conclusive.

The implications of these results for KCMs in general remain unclear:
the fact that the limiting quantity $S_\infty$ cannot be used makes a
clear-cut distinction into type I and II dynamics on the basis of the
clone overlap difficult. One possibility would be to look at the
so-called anomaly in the two-time response function $\chi(t,\tw)$
conjugate to $C(t,\tw)$. This anomaly can be defined as $A(\Delta
t)=\chi(\Delta t,0)-\lim_{\tw\to\infty}\chi(\tw+\Delta t,\tw)$ and
measures the difference in step response between an aging system and
an equilibrium system. Barrat \etal~\cite{BarBurMez96} suggested that
type I (coarsening) and II (glassy) dynamics should correspond
respectively to a zero and nonzero long-time value $A(\Delta
t\to\infty)$ of the anomaly.  Studying the behaviour of $A(\Delta t)$
for KCMs could therefore help to clarify whether a classification into
coarsening versus glassy behaviour is meaningful for these models.

\subsection{Dynamical lengthscales, cooperativity and heterogeneities}
\label{res:hetero}

As explained in Sec.~\ref{intro:cooperativity}, an important question
in glassy dynamics is whether the increase in relaxation timescales is
linked to a growth in a dynamical lengthscale. Such a lengthscale
could arise from cooperativity in the dynamics; if the dynamics is
spatially heterogeneous, then the size of the heterogeneities (\ie\
the size of regions within which the dynamics is approximately
homogeneous) also defines a length. In this section we report on
the various attempts in the literature at defining dynamical
lengthscales for KCMs. Notice that the absence of a growing {\em static}
(equilibrium) lengthscale is trivial in KCMs, since equilibrium
correlations are ruled out by the non-interacting energy
functions used.

We begin the discussion with {\bf spin-facilitated models}.  A first
category of lengthscale definitions is based on irreducibility
considerations: as we saw in Secs.~\ref{meth:reduc}
and~\ref{res:reduc}, at low temperatures KCMs are effectively
irreducible only for systems above a given size. We call such
lengthscales ``irreducibility lengths''; they have also been referred
to as percolation lengths because of the link to bootstrap
percolation, \eg\ in~\cite{NakTak86b,NakTak86}, or cooperativity
lengths~\cite{ErtFroJaec88,Frobose89,JaecFroKno91,%
ReiMauJaec92,SapJaec93}. Consider for example an $f,d$-SFM. As
discussed in Sec.~\ref{meth:reduc}, one can define the probability
$p(c,L)$ that a randomly chosen equilibrium configuration in a sytem
of linear size $L$ belongs to the high-temperature partition; this
means that the all-up spin configuration can be reached by a series of
transitions respecting the kinetic constraints. A characteristic,
$c$-dependent irreducibility length $L_*(c)$ can then be defined as
that $L$ for which $p(c,L)$ has a given value, say
$p(c,L)=1/2$~\cite{ErtFroJaec88,Frobose89,JaecFroKno91}.  This is just
the inverse function of the critical concentration $c_*(L)$ defined in
Sec.~\ref{meth:reduc}. $L_*(c)$ could also be obtained as the inverse
function of a somewhat differently defined critical concentration,
$c_*(L)=\int_0^1 dc\, c[dp(c,L)/dc]$~\cite{NakTak86b}. However, since
the derivative $dp/dc$ is non-negligible only in the narrow $c$-range
where $p(c,L)$ increases steeply from $0$ to $1$ (compare
Fig.~\ref{fig_irrec_sfm} above), the two definitions
are essentially identical. Closely related is the definition of an
irreducibility lengthscale proposed in~\cite{ReiMauJaec92}: instead of
measuring the probability that a randomly chosen configuration belongs
to the high-temperature partition, let $f(c,L)$ be the average
fraction of down-spins that remain after all mobile spins have been
flipped up iteratively. Setting $f(c,0)=1$, one can define
$q(c,L)=f(c,L-1)-f(c,L)$, the probability that a down-spin remains
immobile for system size $L-1$ but not for size $L$. If (as is the
case for \eg\ the 2,2-SFM) the system is effectively irreducible, then
$f(c,L\to\infty)=0$ for any $c>0$ and thus $\sum_{L=1}^\infty
q(c,L)=1$. An average lengthscale can then be defined as $\sum_L L
q(c,L)$~\cite{ReiMauJaec92}.  Whatever method is used, one typically
finds irreducibility lengths that diverge very quickly as temperature
is lowered. For the $2,2$-SFM, for example, the critical up-spin
concentration for effective irreducibility decreases only
logarithmically with system size, $c_*(L)\sim 1/\ln L$; see
Sec.~\ref{res:reduc}. This gives $L_*(c) \sim \exp(A/c)$ or, with
$c\approx \exp(-1/T)$, a {\em doubly} exponential divergence $L_*(T)
\sim\exp[A \exp(1/T)]$ of the irreducibility length as the temperature
is decreased. Irreducibility lengths can also be defined for models
with directed kinetic constraints, \eg\ the North-East model and the
$(3,2)$-Cayley tree, and then diverge at the up-spin concentration
$c_*$ below which even infinite systems are strongly
reducible~\cite{ReiMauJaec92}.

\begin{figure}
\begin{center}
\epsfig{file=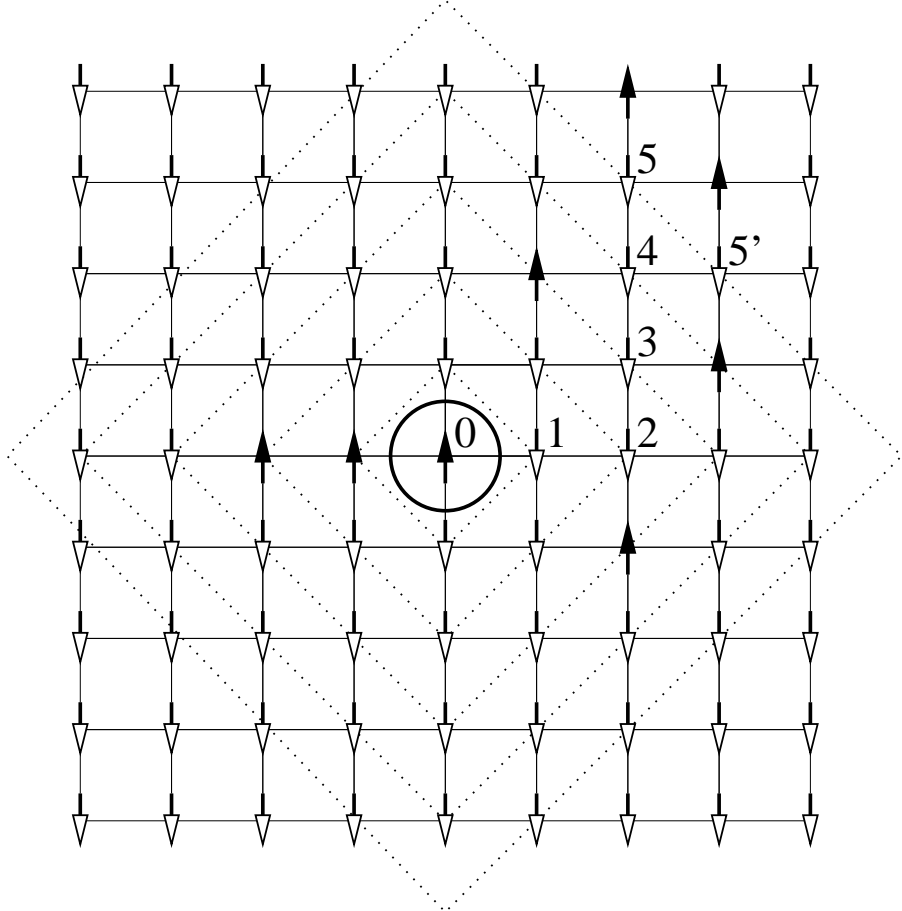, width=12cm}
\end{center}
\caption{Definition of local irreducibility (or cooperativity) length
in the $2,2$-SFM, for the circled spin in the centre. Dotted lines
show n.n.\ shells around this spin. If the spins in the first n.n.\ shell
are held fixed, the centre spin cannot be flipped. The same is true if
the spins in the second shell are held fixed while other spins are
allowed to move. Continuing outwards, only once spins in the fifth
n.n.\ shell are allowed to flip can the centre spin be flipped down,
by the spin-flip sequence $5$ (or $5'$) $\to 4 \to 3 \to 2 \to 1 \to 0$. The
cooperativity length is therefore $l=5$.
\label{coop}
}
\end{figure}

More local approaches to defining an irreducibility length have also
been proposed, \eg\ by Sappelt and J{\"a}ckle~\cite{SapJaec93}. They
defined the length $l(i,\nv)$, for a given spin $i$ and configuration
$\nv$, as the size (measured in number of n.n.\ shells) of the
smallest region around spin $i$ within which other spins have to be
flipped to make the spin mobile. Fig.~\ref{coop} shows an
example for the $2,2$-SFM. The spin in the centre of the configuration
shown has $l(i,\nv)=5$ because in order to make it mobile one needs to
flip some spins in the 5th n.n.\ shell, but none that are further
away.  In addition to an average lengthscale, which is comparable
to the global irreducibility lengths defined earlier, this method
yields a whole distribution of lengthscales. Sappelt and
J{\"a}ckle~\cite{SapJaec93} found that it had two distinct maxima, one
for small $l$ (in fact at $l=1$) and a second broad one around the
average value of $l$. The spatial distribution of $l(i,\nv)$ should
also be able to give insights into the origin of dynamical
heterogeneities, but has not to our knowledge been analysed.

The above definitions of irreducibility lengths all share the feature
that they take the dynamics of KCMs into account only through the
presence or absence of kinetic constraints: they measure how big a
system (or a region where motion is allowed) needs to be for all or
most spins to become mobile {\em eventually}, but do not consider what
the timescales required would be. It is therefore not immediately
obvious whether and how these lengthscales are related to typical
relaxation times in KCMs. Possible connections have nevertheless been
investigated; \eg\ for the $f,2$-SFM with $f=1.5,2,3$ (the last case
being strongly reducible) typical relaxation times were found to
increase roughly exponentially with the average of the local
irreducibility length $l(i,\nv)$ defined above~\cite{WilTetDes99}. In
the $2,2$-SFM, Nakanishi and Takano~\cite{NakTak86b} also found a
stronger-than-power-law timescale increase with the irreducibility
length $L_*(c)$, albeit using an unconventional definition of
relaxation time as the {\em longest}---as opposed to typical, \eg\
integrated---relaxation time for up-spin concentration fluctuations.

Definitions of lengthscales closely related to irreducibility lengths
but now accounting for the timescales involved in relaxation have also
been proposed. Schulz and Schulz~\cite{SchSch98} analysed
cooperativity in the $2,2$-SFM by randomly selecting a lattice site
and then running the dynamics, allowing spin flips only in
increasingly large regions around the chosen spin. The smallest region
within which a relaxation of the spin occurred within some (long)
fixed {\em time interval} was then defined as the spin's cooperatively
rearranging region. The size of this region is clearly a dynamical
analogue of the quantity $l(i,\nv)$ discussed above, and in fact
at least as large as the latter. As expected, the size distribution
of the cooperative regions broadened towards larger values as $T$ was
lowered~\cite{SchSch98}. It could be fitted with two exponentials,
corresponding to small and large regions in broad qualitative
agreement with the results of~\cite{SapJaec93} described above; with
decreasing $T$ the fraction of large regions as well as their average
size increased, the latter in a superactivated fashion.

Among other possible tools for defining dynamical lengthscales,
nonlocal dynamical correlations are obvious candidates. For the
$2,2$-SFM, Fredrickson and Brawer~\cite{FreBra86} numerically
simulated equilibrium correlations between different spins, $\lav
n_i(0) n_j(t)\rav - c^2$. These decay to zero for $t\to\infty$, but
are also zero at $t=0$ since different spins are uncorrelated in
equilibrium. Fredrickson and Brawer~\cite{FreBra86} found that the
onset of nonzero dynamical correlations was fast, while their decay
was much slower and took place on the same timescale as the decay of
the spin autocorrelation function. Interestingly, they also observed
that dynamical correlations were significant only within a
relatively short spatial range, \eg\ of order five lattice spacings
even for the relatively small up-spin concentration of $c\approx 0.08$.

Next we review studies of dynamical heterogeneities, which have again
mainly focused on the $2,2$-SFM. In an early
study~\cite{AleWeiKinIsr87} the fluctuations of the up-spin
concentration were analysed. In a large system, these should be
Gaussian, but for the small systems ($L=16\ldots 32$) simulated
in~\cite{AleWeiKinIsr87} non-Gaussian fluctuations were detected in a
number of higher-order time correlation functions, suggesting
nontrivial spatial correlations due to cooperative dynamics.

\begin{figure}
\begin{center}
\epsfig{file=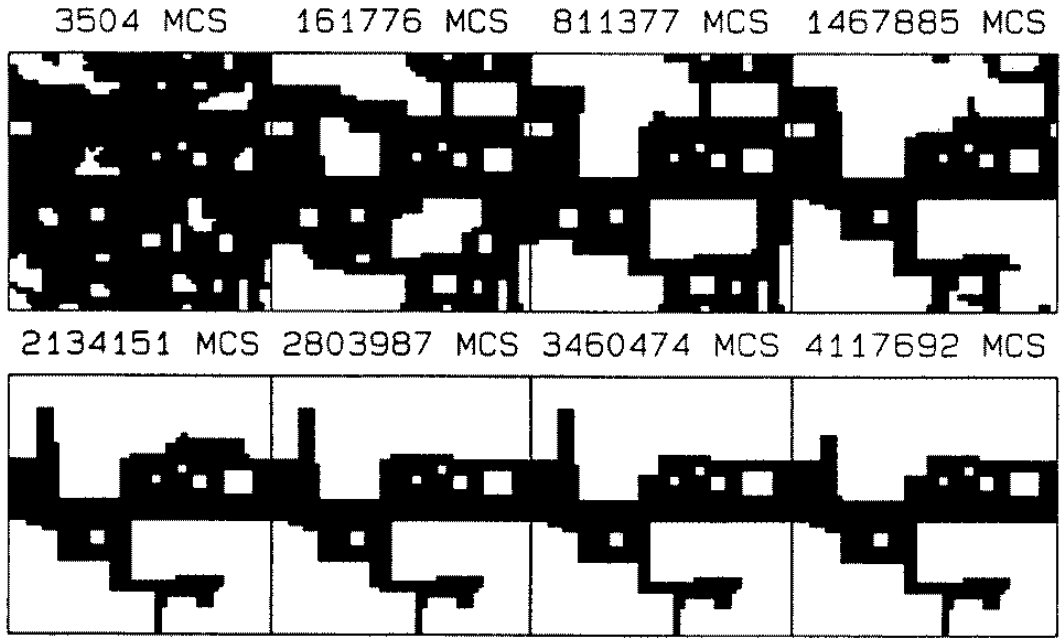, width=15cm}
\end{center}
\caption{Dynamical heterogeneities in the $2,2$-SFM. Starting from an
equilibrium configuration at up-spin configuration $c\eql\approx
0.083$ the plots show, for a series of increasing times (measured in
Monte Carlo steps, MCS), as black those spins which have never
flipped. The lower four plots in particular demonstrate the existence
of very long-lived regions of frozen
spins. From~\protect\cite{ButHar91}.
\label{fig:ButHar}
}
\end{figure}
A more direct analysis of heterogeneities in the $2,2$-SFM was given
by Butler and Harrowell~\cite{ButHar91}. They started the system from
a random equilibrium configuration and then recorded for each spin how
often it flipped within a time interval $t$. Obviously (if the system
is large enough for reducibility effects to be negligible) then all
spins should eventually flip infinitely often as
$t\to\infty$. However, Butler and Harrowell found long-lived regions
of spins that did not flip even for very long times $t$, implying
pronounced dynamical heterogeneity; see Fig.~\ref{fig:ButHar}. The
remaining sites of the 
lattice, \ie\ those containing mobile spins, they classified as either
active or inactive depending on whether these spins were able to make
the surrounding spins mobile on the timescales $t$ considered. The
long-time relaxation is dominated by this ``propagation of mobility''
from the active sites, while inactive sites occur as islands of mobile
spins confined by immobile down-spins and do not contribute
significantly to the relaxation except at short times. The typical
distance $\xi$ between active sites then provides an intuitively
appealing lengthscale characterizing the heterogeneity of the
dynamics. It is, however, difficult to make the definition of an
active site precise; in their simulations, Butler and
Harrowell~\cite{ButHar91} chose as active sites those spins which were
mobile at time $t=0$ and whose eight surrounding spins had all flipped
after some suitably chosen time interval. (This time interval must be
neither too short---otherwise no sites would be classified as
active---nor so long that inactive sites are counted because mobility
from active sites has already been propagated toward them.) Butler
\etal\ found a 
convincing power-law relationship between the relaxation
timescale---measured by the mean-first passage time, \ie\ the integral
of the persistence function---and the distance between active sites,
$\tau \sim \xi^\delta$ with exponent $\delta \approx 7.6$ over six
decades in $\tau$. This fit to the observed dependence of $\tau$ on
$T$ (or $c$) is better than one based on the Adam-Gibbs
relation~\cite{FreBra86}; see Sec.~\ref{res:landsc}. Butler and
Harrowell also estimated $\xi$, or rather the concentration $1/\xi^2$
of active sites, theoretically and found good agreement with the
simulated values~\cite{ButHar91}. Intriguingly, however, their
calculation turns out to be very similar to that of the concentration
of nucleating sites occurring in irreducibility proofs for the
$2,2$-SFM (Sec.~\ref{meth:reduc}). This suggests that $\xi$ should be
related to the irreducibility length $L_*(c)$, and in fact Butler and
Harrowell speculate that these two lengths might diverge in a similar
fashion as $c\to 0$. If this is so, then using $L_*(c) \sim \exp(A/c)$
and the power-law relating $\tau$ with $\xi \sim L_*$ one would
predict $\tau \sim \exp(A'/c)$ for small $c$, corresponding to an
extremely strong, doubly-exponential increase $\tau \sim
\exp[A'\exp(1/T)]$ of the relaxation time with temperature. However,
the simulation results were obtained in the regime where the
lengthscales are still small, with \eg\ $\xi\approx 7$ for the lowest
$c\approx 0.08$ in qualitative agreement with the correlation function
results of~\cite{FreBra86} described above.

In a companion paper, Butler and Harrowell also considered a more
direct operational definition of a dynamical lengthscale for the
$2,2$-SFM~\cite{ButHar91b}. This is obtained by adding free surfaces
to the $2,2$-SFM and defining the cooperativity length as the
lengthscale over which deviations from bulk relaxation behaviour are
observed. The free surfaces are implemented by adding two rows of
facilitating up-spins on opposite boundaries of the square lattice,
while maintaining periodic boundary conditions in the other
direction. The persistence time of spins near the surface is
small---and shows a simple Arrhenius dependence on temperature---but
grows to the bulk value in the layers further from the surface. Pinned
surfaces consisting of down-spins, on the other hand, give persistence
times that {\em decrease} into the bulk. The distance from the surface at
which bulk behaviour is reached defines a dynamical lengthscale and,
encouragingly, turns out to be similar for free and pinned surfaces.
It increases by a factor of three while the persistence time increases
by four orders of magnitude; again, a power-law relationship $\tau\sim
\xi^{\delta}$ was observed with $7.0<\delta<7.6$~\cite{FolHar93}.
Extrapolating naively, Butler and Harrowell~\cite{ButHar91b} then
also estimated that the typical relaxation time increases of $\sim
10^{12}$ observed on supercooling glass-forming liquids would
correspond to a growth of the cooperativity length by a relatively
modest factor of around $3^3\approx30$.

In a later study, Foley and Harrowell~\cite{FolHar93} further analysed
dynamical heterogeneities in the $2,2$-SFM by measuring the spatial
correlations of the first passage times averaged over different
regions of the lattice. (For a visualization of the local, unaveraged
first passage times see also~\cite{Harrowell93}; a more recent study
of kinetic structures in SFMs is~\cite{KahSchDon98}.)  Starting from
an equilibrium configuration, they measured for each spin $i$ the time
$\tau_i$ at which it first flips. They then defined, for any given
region of linear size $l$, the average of the $\tau_i$ in that region
as $\tau(l)$, and considered the moments
\be
m_q(l)=\frac{\lav [\tau(l)-\tau]^q \rav}{\lav [\tau(1)-\tau]^q \rav}
\ee
Here the averages are over all regions of size $l\times l$ in the
numerator, and over all regions of size $1\times1$, \ie\ all lattice
sites, in the denominator; $\tau=\lav\tau(1)\rav=\lav\tau(l)\rav$ is
the average first passage (or first flip) time for the whole
lattice. The moments $m_q(l)$ thus 
measure the fluctuations in the average first passage time of regions
of size $l$, scaled so that $m_q(l)\approx 1$ corresponds to times
$\tau_i$ which are fully correlated within regions of size $l$. The
decrease of $m_q(l)$ with $l$ can thus be used to define a correlation
length $\xi_q$, for which Foley and Harrowell found two main
results~\cite{FolHar93}. First, it is not possible to identify a
single such length scale since the value of $\xi_q$ depends significantly
on the order $q$ of the moment considered; this could suggest a
multifractal structure of the spatial correlations in the
dynamics. Secondly, they again observed a power-law relation between
timescales and dynamical lengthscales, $\tau \sim \xi_2^\delta$,
though with an exponent $\delta\approx 12$ that is rather larger than
for the lengthscales derived from the concentration of actives sites.

More recently, the ratio $Q$~\cite{Heuer97} of the lifetime of
heterogeneous regions to their local relaxation time has also been
measured, in a modified version of the $2,2$-SFM where multi-state
spins are used to model orientational degrees of freedom. $Q$ can be determined
from an appropriately generalized persistence function, and was found
to be of order unity~\cite{HeuTraSpi97}. This could in fact have been
expected on the basis of the results of Butler and
Harrowell~\cite{ButHar91,ButHar91b} for the $2,2$-SFM: the timescales
for propagation of mobility, which limits the lifetime of
heterogeneities, are of the same order as typical relaxation
timescales. At present it therefore seems that SFMs cannot model the
values of $Q\gg 1$ observed in some recent experiments (see
Sec.~\ref{intro:cooperativity}).

Finally, we mention a very recent approach to the study of
heterogeneities in SFMs, proposed by Garrahan and
Chandler~\cite{GarCha02}. They map the non-equilibrium trajectories of
a system on a $d$-dimensional lattice onto the statics of a
$d+1$-dimensional spin system. This space-time view provides an
interesting geometrical framework for understanding dynamical
heterogeneities.  For example, since in the 1,1-SFM and the East model
spatial domains of down-spins can only be ``invaded'' from their left
(or, for the 1,1-SFM, right) boundary, their two-dimensional space-time
representations always give closed regions, separated by interfaces
formed by up-spins. Since only neighbours of up-spins are mobile, this
shows geometrically that mobile sites will ``follow each other
around'', in interesting correspondence with simulation results for
Lennard-Jones systems~\cite{DonDouKobPliPooGlo98,DonGloPooKobPli99}.
%
%

Garrahan and Chandler~\cite{GarCha02} also studied dynamical
heterogeneities in the $1,1$-SFM and East model quantitatively, by
considering the time-averaged magnetizations $m_i(t)=(1/t)\int_0^t
dt'\,[2n_i(t')-1]$. Slow spins that do not flip have the maximal value
$(=1)$ of the ``heterogeneity'' $m_i^2(t)$, while fast spins give
lower values. For $t\to0$ and $t\to\infty$ there are no spatial
correlations between the $m_i^2(t)$, but at intermediate $t$ of the
order of typical relaxation times, nontrivial spatial structure can
appear. This can then be used to define a lengthscale for dynamical
heterogeneities, which increases slowly with decreasing temperature
$T$~\cite{GarCha02}. The $k$-dependence of the structure factor
(Fourier transform) $S(k)$ associated with the correlations $\langle
m_i^2(t)m_j^2(t)\rangle - \langle m_i^2(t)\rangle \langle
m_j^2(t)\rangle$ also shows nontrivial features; \eg\ in the East
model, the hierarchical nature of relaxation processes leads to
space-time regions of up-spins with a fractal structure, giving a
power-law decrease $S(k)\sim k^{-\ln 3/\ln 2}$ for intermediate $k$.

Next we review studies of dynamical lengthscales and heterogeneities
in {\bf constrained lattice gases}. An irreducibility length can be
defined if, instead of the fraction $f(c,L)$ of permanently immobile
spins in SFMs one considers the fraction of particles in the backbone
(see Sec.~\ref{meth:reduc}; recall that the backbone contains all
particles that are permanently frozen in place due to the presence of
other such particles). For the triangular lattice
gas~\cite{JaecKro94}, simulation results showed a growth of this
length for particle concentration $c\to 1$ that could be fitted by an
exponential divergence $\sim \exp[-{\rm const}/(1-c)]$. Following
earlier work on the hard-square lattice
gas~\cite{ErtFroJaec88,JaecFroKno91}, J\"ackle and Kr\"onig then
compared this timescale-independent definition of a lengthscale with
dynamical quantities, by measuring the diffusive displacements of
particles in finite-size lattices~\cite{JaecKro94}. Strong deviations
from the limiting behaviour for large systems were found, \eg\ up to
$L=15$ at particle concentration $c=0.7$; this length is of similar
order of magnitude as the irreducibility length $L_*\approx 8$ for
this $c$. The finite-size effects on diffusion are visible already for
small particle displacements, and thus genuinely due to cooperative
dynamics rather than the trivial upper limit on displacements imposed
by the finite lattice. Similar size effects appear in correlation
functions measured on lattice strips of finite width~\cite{DonJac96},
both for translational motion and for orientational degrees of freedom
in the appropriately extended model (see
Sec.~\ref{model:lattice_gases}). Intriguingly, it was observed
in~\cite{ErtFroJaec88} (for the hard-square lattice gas) that the
irreducibility length $L_*$ is substantially larger than the distance
over which particles need to diffuse before the mean-square
displacement becomes linear in $t$. This shows that the irreducibility
length is a rather subtle measure of the cooperative nature of the
dynamics, and cannot simply be thought of as the size of a cage within
which particles are trapped until they can diffuse freely.

J\"ackle and Kr\"onig~\cite{KroJaec94} further studied dynamical
heterogeneities in the triangular lattice gas by considering nonlocal
dynamical correlations, as measured via the dynamic structure
factor\eq{c}. As explained in Sec.~\ref{basics2}, the latter should
decay as $\exp(-D\kv^2 t)$ for long times and small wavevectors $\kv$,
reflecting the diffusive nature of the dynamics; $D$ is the collective
diffusion constant. For larger $\kv$, the observed long-time decay
rates will deviate from $D\kv^2$. The onset of these deviations at
wavevectors of length $k_c$ (say) then defines a length scale $1/k_c$
below which the dynamics is heterogeneous; this was found
in~\cite{KroJaec94} to increase with $c$, but the precise form of this
dependence was not analysed.

Very recently, heterogeneities in the KA-model have also been
studied~\cite{FraMulPar02}. Motivated by results for mean-field spin
glasses~\cite{DonFraParGlo99,FraDonParGlo99,FraPar00}, the
fourth-order correlation function
\be
C_4(t) = \frac{1}{N c^2(1-c)^2}\sum_{ij} \left(
\lav n_i(t)n_i(0)n_j(t)n_j(0)\rav -
\lav n_i(t)n_i(0)\rav \lav n_j(t)n_j(0)\rav \right)
\label{Cfour}
\ee
was simulated. This can also be written as the scaled variance
$C_4(t)=N(\lav q^2(t)\rav - \lav q(t)\rav^2)$ of the ``overlap''
between configurations a time $t$ apart,
\be
q(t) = \frac{1}{Nc(1-c)} \sum_i \left(n_i(t) n_i(0) - c^2\right)
\label{qt}
\ee
By definition, $q(0)=1$ and thus $C_4(0)=0$. As $t$ increases, $q(t)$
will decay, approaching $q(t\to\infty)=0$ for times long enough for
the system to have lost all memory of its initial
configuration. However, for particle concentrations $c$ close to the
(at least effective) dynamical transition $c\dyn=0.88$ in the KA
model, one would suspect that the system remains trapped near its
initial configuration for a long time. This will give a nonzero value
of $q(t)$ which will also fluctuate strongly between dynamical
histories started off at different initial configurations $n_i(0)$,
leading to a large value of $C_4(t)$. Consistent with this, it was
found in~\cite{FraMulPar02} that the simulated $C_4(t)$ exhibited a
maximum at finite $t$, before decaying again as the system finally
loses memory of its initial configuration%
\footnote{%
In the simulations of~\cite{FraMulPar02} it appears that, at least
for the smaller values of $c$ investigated, $C_4(t)$ decays to values
below unity for $t\to\infty$. On the other hand, from the
definitions~(\protect\ref{Cfour},\ref{qt}) one calculates, using that
$n_i(0)$ and $n_i(t\to\infty)$ are uncorrelated equilibrium
configurations, that $C_4(t\to\infty)=1$. The origin of this
discrepancy is unclear to us. However, more recent simulations
confirm the theoretical expectation $C_4(t\to\infty)=1$ (J J Arenzon,
private communication), also in inhomogeneous systems as long as one
focuses on approximately homogeneous subregions~\cite{AreLevSel03}.}%
. The maximum becomes higher and shifts to larger $t$ as $c$ is
increased towards $c\dyn$, reflecting the fact that the system is
trapped more strongly, and for longer times, at higher $c$. Similar
results have been found in frustrated lattice gases~\cite{FieDeCon02}
and Lennard-Jones glasses in $d=2$
dimensions~\cite{MelRamGouKleMou95}. In the spherical $p$-spin glass,
a rough analogue of $C_4(t)$ can be shown to have a maximum which
actually diverges as the dynamical transition of this mean-field model
is approached; this is related to the divergence of an appropriately
defined static spin-glass susceptibility below the
transition~\cite{DonFraParGlo99,FraDonParGlo99,FraPar00} and suggests
a corresponding diverging lengthscale.  In the KA model, the
definition\eq{Cfour} of $C_4(t)$ as a sum over all pairs of lattice
sites likewise suggests that the observed increase in the maximum
value of $C_4(t)$ reflects a growing lengthscale over which the
dynamics is heterogeneous because configurations remain dynamically
correlated. How this length is related to others defined \eg\ for the
triangular lattice gas (see above) is not obvious, and a closer
investigation of this issue would seem worthwhile. A
wavevector-dependent generalization of $C_4(t)$, obtained by including
a factor $\exp[i\kv\cdot(\rv_i-\rv_j)]$ in the definition\eq{Cfour},
could be helpful in defining the lengthscale for dynamical
heterogeneities more precisely. Such a quantity would be closely
related to the structure factor of dynamical heterogeneities
considered by Garrahan and Chandler~\cite{GarCha02}. This can be seen
from the fact that \eg\ the fourth-order contribution to their
correlation $\langle m_i^2(t)m_j^2(t)\rangle - \langle m_i^2(t)\rangle
\langle m_j^2(t)\rangle$ is proportional to $\int_0^t dt_1\, dt_2\, dt_3\,
dt_4\, (\lav n_i(t_1)n_i(t_2)n_j(t_3)n_j(t_4)\rav - \lav
n_i(t_1)n_i(t_2)\rav \lav n_j(t_3)n_j(t_4)\rav)$, which one would
expect to behave qualitatively similarly to the terms under the sum
in\eq{Cfour}.


\subsection{Energy landscape paradigms}
\label{res:landsc}

In this section we review studies investigating the application of
energy landscape paradigms such as configurational entropies and
Edwards measures (see Sec.~\ref{intro:landscape}) to KCMs.

\begin{figure}
\begin{center}
\epsfig{file=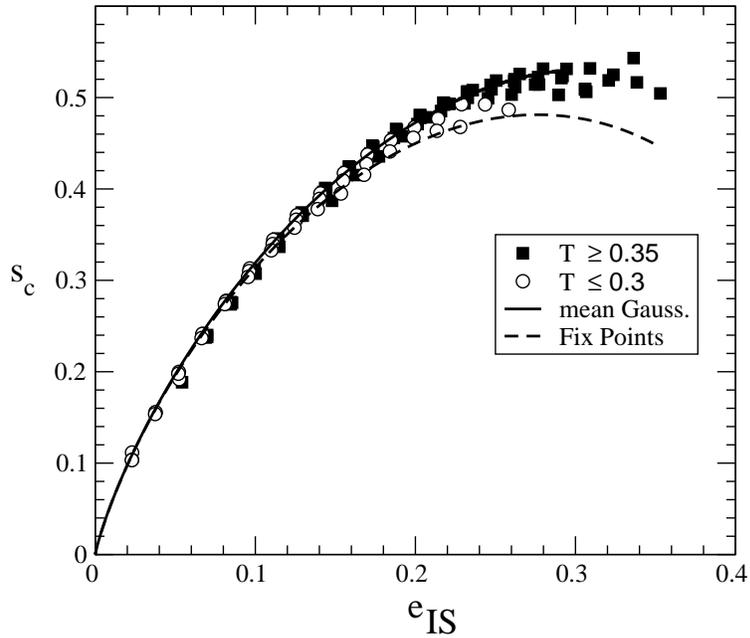, width=10cm}
\end{center}
\caption{Stillinger-Weber configurational entropy in the $1,1$-SFM for
a chain of 64 spins. The lower curve gives the result from a direct
count of configurations which are frozen at $T=0$, and the symbols are
estimates derived from the actual dynamics at different temperatures
(assuming that the free energy of the inherent structures is
independent of their energy $e_{\rm IS}$). The upper curve is obtained
by an approximation based on integrating the temperature dependence of
the average value of $e_{\rm IS}$. From~\protect\cite{CriRitRocSel00}.
\label{sc_FA}
}
\end{figure}

The usefulness of the {\bf Stillinger-Weber configurational entropy}
was studied in~\cite{CriRitRocSel00} for the East model and the
$1,1$-SFM. It was argued that the SW entropy is not relevant for
understanding glassy effects in these models. The key observation is
that all reasonable definitions of a SW-like configurational entropy
are independent of the asymmetry parameter $a$ which interpolates
between the two models (see\eq{asymmetric_SFM}), while the actual
dynamics varies dramatically between the limits of $a=0$ (East model) and $a=1$
($1,1$-SFM). The natural definition of an inherent structure
(IS) is as a configuration that is frozen at $T=0$ (see
Sec.~\ref{intro:landscape}); in such a configuration, all up-spins are
isolated. The number of such configurations with a given up-spin
concentration or equivalently energy $e_{\rm IS}$ is easily
counted~\cite{CriRitRocSel00} and its 
logarithm gives the configurational entropy $Ns_c(e_{\rm
IS})$, shown in Fig.~\ref{sc_FA}. (The
configurational entropy for the triangle model can be obtained by
similar reasoning~\cite{GarNew00}.) One might hope that a
configurational entropy calculated over inherent structures of a given
{\em free energy} $f$, rather than {\em energy} $e_{\rm IS}$, may have
better properties; but in the low-temperature regime $f\approx e_{\rm
IS}$ since entropic contributions from the size of the basins around each
IS are negligible, and so no significant differences are expected. As
an alternative, it might be interesting to study the
timescale-dependent definition of a configurational entropy over
metastable states proposed by Biroli and Kurchan~\cite{BirKur01} (see
Sec.~\ref{intro:landscape}). For this one would anticipate clear
differences between the $1,1$-SFM and the East model, arising from the
fact that in the latter there is a whole hierarchy of well-separated
timescales on which metastable states could be defined.

The applicability of the Adam-Gibbs~\cite{AdaGib65} relation $\tau
\sim \exp({\rm const}/Ts)$ between typical relaxation times $\tau$
and the thermodynamic entropy $s$ has also been investigated for KCMs.
Underlying this relation is the assumption that there is a connection
between a lengthscale characterizing cooperative dynamics and the
(inverse of the) thermodynamic entropy $s$. Sappelt and J\"ackle
argued that such a connection cannot exist in general, and certainly not in
KCMs~\cite{SapJaec93}. This is most obvious in models such as the
North-East model or the (3,2)-Cayley tree: at the up-spin
concentration $c_*$ where these models become strongly reducible, the
irreducibility length $L_*$ diverges (see Sec.~\ref{res:hetero}) while
the entropy per spin, $s=-c\ln c -(1-c)\ln (1-c)$, is a smooth
function of $c$. Even in other KCMs, the divergence of an
appropriately defined cooperativity length for $c\to 0$ is much more
pronounced than the weak logarithmic divergence of the
entropy~\cite{SapJaec93}. Even though the foundations of the
Adam-Gibbs relation are therefore uncertain in KCMs, Fredrickson and
Brawer found that for the $2,2$-SFM the relation holds to
a reasonable accuracy over a range of around six decades in the
relaxation time~\cite{FreBra86}, while for the $2,3$-SFM it appears to
be violated~\cite{Fredrickson88}. As explained in
Sec.~\ref{res:hetero}, however, Butler and Harrowell~\cite{ButHar91}
later showed that there were small but systematic deviations from
Adam-Gibbs even for the $2,2$-SFM, and that the temperature-dependence
of $\tau$ is better rationalized by a power-law link to an
appropriately defined lengthscale.

In the last few years, the applicability of {\bf Edwards measures} to
the description of glassy dynamics in KCMs has received growing
attention. (This includes work on KCMs driven into non-equilibrium
stationary states by external forcing, which is discussed in
Sec.~\ref{res:driven} below.) For SFMs, one needs to decide which
configurations to regard as ``blocked''. A natural definition is to
use again configurations where no spins can move at $T=0$, \ie\ which
contain no mobile up-spins; these are identical to the inherent
structures in SFMs discussed above. The Edwards measure is then a
uniform measure over the subset of these configurations with the
desired values for specified observables such as the energy. For the
$1,1$-SFM and the East model, as well as the interpolating
asymmetric model, a recent analysis of the $T=0$
dynamics~\cite{DesGodLuc02} shows that the results are not well
described by averages over an Edwards measure constrained to have the
correct up-spin concentration. Spin-spin correlations, for example,
fall off super-exponentially in the final configurations actually
reached by the $T=0$ dynamics whereas a flat average over blocked
configurations gives an exponential decay.

More successful is the application of Edwards measures to constrained
lattice gases; here the definition of the blocked configurations to be
included in the Edwards measure is straightforward. In the
grandcanonical version of the KA model, where particle exchange with a
reservoir is allowed, it was shown
in~\cite{BarKurLorSel00,BarKurLorSel01} that the FDT violations
observed after an increase of the chemical potential into the
non-equilibrium region (see Sec.~\ref{res:fdt}) were well predicted by
an approopriate Edwards measure. Specifically, the effective
temperature deduced from the slope of the out-of-equilibrium part of
the FDT plot (Fig.~\ref{fig:KA_FDT}) agrees numerically with that
found from the entropy of 
blocked configurations, evaluated at the density $c(\tw)$
reached during the aging process. The structure factor, \ie\ the
spatial correlations of density fluctuations which develop during
aging---but are absent in equilibrium---was likewise well predicted by
the Edwards measure approach.

The geometrical organization of the blocked configurations which the
Edwards measure focuses on has recently also been
investigated~\cite{FraMulPar02}. The authors generated blocked
configurations by an annealing process in the number of mobile
particles. They then moved a randomly chosen particle to a different
location, ran the dynamics of the KA model from this starting
configuration and monitored whether the system became blocked again
after a few transitions or whether it remained substantially
unblocked, with many particles being mobilized.  They found that the
``unblocking probability'' is high at low densities, but tends to zero
at the density of the (effective) dynamical transition, $c\dyn \approx
0.88$. Above this density, blocked configurations are therefore stable
with probability one. Measurements of the overlaps between these
configurations showed that at all densities there are blocked
configurations arbitrarily close to other blocked configurations. This
contrasts with results for mean-field spin glasses (\eg\ $p$-spin
spherical models), where in the regime corresponding to high density
there is a minimum distance between the analogues of blocked
configurations~\cite{FraMulPar02}. This observation suggests that the
geometrical organization underlying glassy dynamics in KCMs is rather
different from that in mean-field models, and clearly deserves
further investigation.

\subsection{Driven stationary states}
\label{res:driven}

In this final results section we review recent work on KCMs in
out-of-equilibrium stationary states generated by external
driving. Much of this research is motivated by attempts to understand
the behaviour of granular materials under steady tapping or vibration.

We begin again with {\bf spin-facilitated models} and their variants.
The effects of driving by repeated excitation or ``tapping'' have been
studied in, 
for example, the $1,1$-SFM~\cite{BrePraSan99}. A tap corresponds to
evolution at nonzero temperature $T$ for some short time interval
$t_{\rm tap}$, after which the system is allowed to relax fully under
zero temperature dynamics.  A possible motivation for this dynamics
comes from granular media; one regards the chain of spins as a cross
section through a granular medium, with up- and down-spins
corresponding to low- and high-density regions or holes and particles,
respectively. The tap typically generates up-spins (holes) while the
subsequent zero temperature relaxation can only flip spins down, thus
filling holes with particles. The kinetic constraint of the $1,1$-SFM
imposes the restriction that isolated holes cannot be filled. Using
the fact that the zero temperature dynamical equations of the
$1,1$-SFM can be exactly closed~\cite{FolRit96,SchTri97}, the dynamics
of the joint tapping and relaxation process can be solved to lowest
order in $t_{\rm tap}$ and $\exp(-1/T)$, and a logarithmic decay of
the particle density $1-c$ is found over a wide time
interval~\cite{BrePraSan99}. This behaviour is reminiscent of that
found in parking-lot models and other adsorbtion-desorbtion
models~\cite{Evans93}. The response of the particle density to sudden
changes in tapping intensity also displays interesting memory
effects~\cite{BrePra01b,BrePra02} but these are not specific to
tapping dynamics and occur even in \eg\ the Glauber Ising
chain~\cite{BrePra02b}. Hysteresis effects from cyclic variations of
the tapping intensity have likewise been investigated~\cite{PraBreSan00}.
Further work on the tapped $1,1$-SFM showed that in the limit of short
or weak taps, an effective master equation can be used to describe the
evolution of the system from tap to tap, without needing to consider
the intermediate excited states generated by the
tapping~\cite{BrePraSan00}. The steady state of the resulting model
can be described by an Edwards measure at the appropriate up-spin
concentration, where all metastable or frozen configurations---those
in which only isolated up-spins occur---contribute with equal
probability~\cite{BrePraSan00}.

The effects of different kinds of tapping dynamics in the $1,1$-SFM
were further investigated in~\cite{BerFraSel02}, where the analysis
was also extended to the East model. A ``tap'' was either a single
Monte Carlo sweep of the system at some $T>0$ (this corresponds to a
fixed tap duration of $t_{\rm tap}=1$), or a random flip applied to
each spin independently with some probability $p<1/2$; in between taps
the system relaxes again at zero temperature. Thermal and random
tapping as defined in this way gave quite different results, \eg\ in
terms of the magnetization reached in the stationary state. These
differences persisted even for small tapping intensity ($T$ or $p\to
0$). A flat Edwards measure over frozen configurations with the
correct up-spin concentration was found to describe other aspects of
the stationary state, such as the distribution of domain lengths, only
for thermal tapping at moderate intensity---consistent with the
results of~\cite{BrePraSan00}---while systematic deviations occurred
for random tapping.

Recently, a driven version of the East model has also been proposed
that is motivated by {\em
rheological} considerations~\cite{Fielding02}:
``soft'' glasses such as dense emulsions and colloidal suspensions can
be driven into non-equilibrium steady states by shear
flow~\cite{LiuNag01}. A modified version of the
East model with three-state spins was used, but the third state turned
out to be irrelevant for the qualitative behaviour.  The
``rheological'' driving was implemented by adding {\em unconstrained}
up-flips from $n_i=0$ to $n_i=1$ to the model, at a rate $\dot\gamma$
which can loosely be thought of as the shear rate. This modification
breaks detailed balance. One may expect, however, that the stationary
state reached after a long time should be similar to that obtained
after aging (without shear) for a time $\tw\sim 1/\dot\gamma$. The
steady state under shear can be worked out using the same domain
picture as for the aging case; see Sec.~\ref{meth:indint}. An
independent interval approximation again needs to be made, though in
the driven case it is not clear whether this becomes exact for $T\to
0$ as it does in the aging case. The theory nevertheless provides a
good description of simulation results. There is also the expected
close (though not perfect) match between the domain length
distributions for the aging and sheared cases when $\tw$ and
$\dot\gamma$ are related by $\tw=1/\dot\gamma$~\cite{Fielding02}.

In the Glauber Ising chain with ``falling'' dynamics (only
energy-decreasing moves are allowed) driven by random taps,
Lef{\`e}vre and Dean~\cite{LefDea01} calculated a number of
observables (energy fluctuations, correlation functions and domain
size distributions) exactly within the Edwards measure and observed
very good agreement with numerical simulations. As an aside, we note
that for the same ``tapping
and falling'' dynamics in ferromagnets on random graphs of fixed
connectivity $r>2$ ($r=2$ gives the Ising chain),
Dean and Lef\`evre~\cite{DeaLef01} found in simulations a first order
phase transition in the stationary behaviour. For $p$ below some
threshold $p_c$, the energy $E$ in the stationary state equalled the
ground state energy, jumping by a finite amount to some $E^*=E(p_c+0)$
as $p$ crosses the threshold. This was interpreted in terms of a
change in behaviour of the Edwards entropy $S(E)$: in an approximate
calculation $S(E)$ is concave above $E^*$ and therefore gives, in a
thermodynamic formalism, locally stable states of energies $E>E^*$,
while for $E<E_*$ the entropy $S(E)$ is convex so that states with
these energies are unstable.

Finally, driven steady states have also been studied in the context of
the {\bf KA model} connected to two reservoirs at unequal chemical
potential $\mu$~\cite{Sellitto02b}. The difference in chemical
potentials sets up a particle current between the two reservoirs, and
a nontrivial density profile which can be well reproduced using a
continuum model (see Sec.~\ref{meth:mappings}). An interesting feature
is that the system may show ``negative resistance'': if the particle
densities of the two reservoirs are both increased by the same factor,
the current may decrease. This occurs because the decrease in mobility
with increasing density can overwhelm the increase in the density
difference which drives the current.

\section{Conclusions and outlook}
\label{conclusion}

In this review we have discussed the glassy dynamics of kinetically
constrained models (KCMs). Their characteristic feature is that they
have trivial, normally non-interacting, equilibrium behaviour.  The
existence of slow glassy dynamics can thus be studied without any
``interference'' from an underlying equilibrium phase transition.  A
further advantage of KCMs is that they introduce explicitly, via
constraints on the allowed transitions between configurations, the
cooperative character of the dynamics whose origin in more realistic
glass models we do not yet fully understand. In our discussion of KCMs
we have included spin-facilitated Ising models (SFMs) and their
variants; constrained lattice gases; models inspired by cellular
structures; the triangle and plaquette models obtained via mappings
from interacting systems without constraints; and finally related
models such as urn, oscillator, tiling and needle models. We now
summarize the results and assess how good KCMs are at modelling glassy
dynamics in physical systems such as structural glasses. Avenues for
future research are also discussed.

Broadly speaking, KCMs fall into two classes. The first one contains
the $1,d$-SFMs, the cellular models and the plaquette model, all of
which can be analysed in terms of appropriately defined defects that
diffuse and react which each other. Typical relaxation timescales show
an activated temperature-dependence, so that these KCMs model
``strong'' glasses. The reaction-diffusion picture provides a fairly
full understanding of the dynamics both in and out of equilibrium,
including \eg\ the shape of correlation and response functions.  Some
open questions remain, however, especially with regard to
fluctuation-dissipation theorem (FDT) violations and the description
of out-of-equilibrium dynamics in terms of an effective temperature.
Urn and oscillator models also fall into the category of ``strong
glass'' KCMs, but due to their lack of spatial structure require
different conceptual tools to understand the dynamics, in particular
the notion of entropic barriers which slow down the dynamics at low
temperatures.

The second class of KCMs contains all remaining models, in particular
the constrained lattice gases and SFMs with directed constraints or
with facilitation by $f>1$ spins. These show genuinely cooperative
dynamics which cannot be broken down into the motion of localized
defects. Their relaxation times diverge in a superactivated fashion as
temperature decreases, so that they model ``fragile'' glasses. The
cooperative nature of the dynamics means that these models are much
less well understood than the defect-diffusion KCMs. It also implies
that {\bf reducibility} effects become a serious concern. While arguments
developed \eg\ for bootstrap percolation show that in the
thermodynamic limit most models become effectively irreducible in the
sense that almost all configurations are dynamically accessible, the
system sizes required can be extremely large in the glassy regime (low
up-spin concentrations for SFMs, or high particle concentrations for
lattice gases). 

Related though distinct is the question of {\bf dynamical transitions} where
ergodicity is broken because of diverging relaxation timescales. The
only cooperative KCM for which strong evidence for such a transition
exists is the KA lattice gas, in which relaxation times appear to
diverge when the particle concentration approaches
$c\dyn\approx0.88$. Even here finite size effects are difficult to
exclude, however, since around this concentration reducibility effects
also become strong for the system sizes that are accessible in
numerical simulations. In the absence of analytical arguments the
existence of a dynamical transition is therefore likely to remain
conjectural. However, from a more pragmatic point of view the more
important question is {\em why} relaxation times diverge as quickly as
they do around $c=c\dyn$, not whether they are truly infinite or
finite but extremely large at higher densities. This remains an
essentially open issue, as does the origin of the conjectured
universality in the timescale divergence near $c\dyn$.
In SFMs, the ``fragile'' timescale divergence with decreasing
temperature $T$ also remains poorly understood, except for the
simplest cases such as the East model where an EITS law $\tau \sim
\exp(A/T^2)$ has been found.

Closely related are the issues of
{\bf heterogeneous dynamics} and {\bf dynamical lengthscales}. KCMs
are ideal for 
the study of these effects, having only trivial static correlations
so that all effects are directly due to the dynamics. Direct evidence
for dynamical heterogeneities has been found in the $2,2$-SFM and more
recently also in the East model. There has also been some success in
identifying dynamical lengthscales and relating these to the observed
relaxation times. But more needs to be done, both at the analytical
and the numerical level, to identify an unambigously defined
cooperativity length and understand how its growth affects the
dynamics. This is particularly important now that much more data on
heterogeneities in experimental systems are becoming available.

Many of the cooperative KCMs are found in simulations to have
stretched exponential {\bf relaxation functions at equilibrium}. A
quantitative theoretical understanding of these effects remains to be
achieved, again with the possible exception of the East model
where there are at least plausible conjectures for the stretching
behaviour at low temperatures. Mode-coupling approximations, the most
successful of which are based on approximations to the irreducible
memory functions, generally perform rather badly, predicting \eg\
spurious dynamical transitions. Recent diagrammatic expansions offer
some improvements, giving results formally analogous to those of the
extended mode-coupling theory for supercooled liquids, but still
predict relaxations which are too close to exponential deep in the
glassy regime. A better understanding of the physical nature of these
approximation techniques will be essential for progress in this
direction. Adiabatic approximations, by contrast, are based directly on a
physically intuitive separation into fast and slow degrees of freedom
and have been used with some success in the analysis of cooperative 
KCMs. A clear example of this is the East model, where the timescale
separation (involving in fact a whole hierarchy of widely separated
times) becomes exact in the low-temperature limit and gives a fairly
full understanding of the out-of-equilibrium behaviour after a deep
quench.  Further exploration of such techniques both for KCMs and more
general glass models should therefore be fruitful.

For cooperative KCMs in general our understanding of the
{\bf out-of-equilibrium dynamics} is still only at the beginning. While some
general qualitative features such as apparent freezing in cooling runs
and hysteresis in heating-cooling cycles are well understood, more
complex effects such as the behaviour of two-time correlation and
response functions still present puzzles. There is no simple picture
as yet of the observed FDT violations, for example: in the KA lattice
gas and the plaquette model, simple mean field-like FDT plots
consisting to a good approximation of two straight lines are observed,
while other KCMs show more complex behaviour including non-monotonic
FDT relations. The robustness of these results to the choice of
observable, the existence of a well-defined effective temperature
$T\eff$ and its connection to an appropriately defined configurational
entropy also remain to be clarified. The KA model is the most
encouraging in this sense: the $T\eff$ from an appropriate FDT plot
agrees with that derived from an effective equilibrium description in
terms of a flat ``Edwards measure'' over blocked configurations. In
other models, however, no such simple correspondence is found and much
work remains to be done to understand these results in a wider context.

Finally, KCMs without detailed balance are also beginning to be
explored. We touched on these when discussing stationary states
reached by {\bf external driving} which can model the tapping of granular
materials or shear flow of ``soft'' glassy materials. Recent work has
focused on the suitability of Edwards measures for describing the
resulting stationary states, but a coherent picture is yet to emerge.
More generally, the similarities in the phenomenology of granular
materials (which are effectively at temperature $T=0$) and ``thermal''
glasses suggest that detailed balance is not a key ingredient in
glassy dynamics. One may for example expect that slowly driven and
aging systems behave in similar way, and for a driven version of the
East model this has indeed been confirmed. Future work on new KCMs
without detailed balance will no doubt deepen our understanding of
driven glassy systems. 

Overall, we believe that the simplicity of KCMs and their ability to
combine slow dynamics with trivial equilibrium behaviour make them
prime candidates for further progress in the issues at the heart of
current research in glassy dynamics.

{\bf Acknowledgments:} We are grateful to all our collaborators, much
of whose work has been described in this review, including Luis
Bonilla, Andrea Crisanti, Martin Evans, Suzanne Fielding, Silvio Franz, Adan
Garriga, Luca Leuzzi, Peter Mayer and Paco Padilla. We are also
indebted to many other colleagues, too numerous to mention, for
enlightening discussions on many of the topics covered in this
review. Particular thanks are due to all participants in the SPHINX
workshop on {\em Glassy behaviour of kinetically constrained models}
held in Barcelona in March 2001, for an exciting and productive
meeting. PS acknowledges financial support through Nuffield grant
NAL/00361/G. FR has been supported by the Spanish Ministerio de
Ciencia y Tecnolog\'{\i}a Grant BFM2001-3525.

\section{List of abbreviations}

\begin{tabbing}
KWW:zz\=\kill
b.c.c.	\>body-centred cubic\\
BP	\>bootstrap percolation\\
EITS	\>exponential inverse-temperature square\\
f.c.c.	\>face-centred cubic\\
FDT	\>fluctuation-dissipation theorem\\
IS	\>inherent structure\\
KA	\>Kob-Andersen\\
KCM	\>kinetically constrained model\\
KWW	\>Kohlrausch-Williams-Watts\\
MCA	\>mode-coupling approximation\\
MCT	\>mode-coupling theory\\
n.n.	\>nearest neighbour\\
SFM	\>spin-facilitated model\\
	\>($f,d$-SFM: spin-facilitated model on $d$-dimensional lattice
with $f$ facilitating spins)\\ 
SW	\>Stillinger-Weber\\
TTI	\>time-translation invariance\\
VTF	\>Vogel-Tamman-Fulcher
\end{tabbing}

\bibliographystyle{unsrt}
\bibliography{review}

\end{document}